# Artificial Intelligence based Sensor Data Analytics Framework for Remote Electricity Network Condition Monitoring

Sirojan Tharmakulasingam

A dissertation submitted in fulfillment
of the requirements for the degree of

**Doctor of Philosophy**

School of Electrical Engineering and Telecommunications

The University of New South Wales

August 2020

# ABSTRACT

Rural electrification demands the use of inexpensive technologies such as single wire earth return (SWER) networks. There is a steadily growing energy demand from remote consumers, and the capacity of existing lines may become inadequate soon. Besides, the existing SWER networks are very inefficient and experience poor voltage regulation. Furthermore, high-impedance arcing faults (HIF) from SWER lines can cause catastrophic bushfires such as the 2009 Black Saturday event. Replacing SWER lines by cables as recommended by the Royal Commission comes at an astronomical cost and service providers are not able to comply with.

As a solution, reliable remote electricity networks can be established through breaking the existing systems down into microgrids, and existing SWER lines can be utilised to interconnect those microgrids. The development of such reliable networks with better energy demand management will rely on having an integrated network-wide condition monitoring system.

As the first contribution of this thesis, a distributed online monitoring platform is developed that incorporates power quality monitoring, real-time HIF identification and transient classification in SWER network. Informative features are extracted from the current & voltage signals, and Artificial Intelligence (AI) based classification techniques are developed to classify faults and transients. The proposed approach demonstrates higher HIF detection accuracy (98.67%) and reduced detection latency (115.2 ms).

Secondly, a remote consumer load identification methodology is developed to detect the load type from its turn-on transients. An edge computing-based architecture is proposed to facilitate the high-frequency analysis for load identification with the minimised data transmission. Computationally efficient load identification methodologies are developed to enable their real-time deployment on resource-constrained devices. The proposed approach is evaluated in real-time, and it achieves an average accuracy of 98% in identifying different loads.

Finally, a deep neural network-based energy disaggregation framework is developed to separate the load specific energy usage from an aggregated signal. A generative approach is applied to model energy usage patterns. The proposed framework is evaluated using a real-world data set. It improves the signal aggregate error by 44% and mean aggregate error by 19% in comparison with the state-of-the-art techniques.

# ACKNOWLEDGMENTS

First and foremost, I would like to express my most profound appreciation to my primary supervisor, Prof. Toan Phung. Prof. Toan, accept my endless gratitude for your timely guidance, continuous motivation, valuable suggestions and constructive reviews of all the manuscripts that I have produced during my Ph.D. candidature. I am equally grateful to my joint supervisor, Prof. Eliathamby Ambikairajah. Prof. Ambi, your unfailing support, guidance and encouragement undeniably shaped my research to a high standard. I owe a big debt of gratitude for your timely support and continuous attention to my research progress. It has been an absolute privilege to work with you both, and this thesis would not have been possible without your supervision.

I acknowledge the tuition fee scholarship from UNSW, research stipend and top-up scholarship from Tyree Foundation, Australia for my Ph.D. studies. The support I received from the School of Electrical Engineering and Telecommunication, UNSW cannot be forgotten. Further, I would deeply appreciate Mr Zhenyu Liu for his technical support and detailed instructions during the experiments. I also thank my research colleague Shibo Lu for fruitful technical discussions, suggestions and motivations during my research and experiments.

A warm word to my close friends who are always there for me: Arunkumar, Anu Raghavi, Arunan, Anusuya, Navaroshan, Tharshini, Kaavya, Gajan and Kawsihen – thank you for making my Ph.D. days very memorable and fun-filled.

Words cannot express my gratitude to my father Tharmakulasingam and my mother Kamalarani for your continuous encouragement and moral support. Appa & Amma, your countless sacrifices bring me up to this doctoral degree. This Ph.D. is a tribute to you both.

I also place on record, my sense of gratitude to all, who have lent their hands in this venture.

# CHAPTER 1

# 1. INTRODUCTION

## 1.1. Research Overview and Motivation

The electricity grid extension to remote areas demands the use of inexpensive electrification technologies in order to ensure economic viability. Conventional three-phase lines and single-phase lines are not economical for rural electrification. Hence the single wire earth return (SWER) networks are used to distribute the single-phase power to remote consumers from the main grid. At present, there are more than 200,000 km of SWER lines throughout Australia [1]. However, there is a steadily growing energy demand from existing consumers, and the capacity of existing lines may become inadequate soon.

Besides, there are several problems associated with the SWER line system. It is very inefficient because of the use of galvanised steel conductors, its voltage regulation is very poor because of the line impedances, and its fault monitoring is very minimal. Another major problem with long-distance remote electricity transmission lines like SWER is the bushfire risks result from arcing faults. Five out of fifteen most destructive fires of 2009 Black Saturday bushfires in Victoria, Australia, were caused by High-impedance arcing faults (HIF) [2]. After the Black Saturday Bushfires in 2009, the Victorian Bushfires Royal Commission recommended replacement of all SWER lines.

The cost involved in the replacement of existing SWER lines over 200,000km, with more efficient and high capacity conductors is enormous. Thus, distribution network service providers are not likely to be able to comply with the complete replacement. Alternatively, reliable remote electricity networks can be established through breaking the existing systems down into microgrids with their internal renewable sources, and existing SWER lines can be utilised to interconnect those microgrids. However, the development of such reliable remote electricity networks with better energy demand management will rely on having a sophisticated condition monitoring system. It needs to obtain real-time details of grid condition, consumer energy usage and detect faults online. This thesis explores such real-time monitoring techniques that can enhance the remote electricity networks and its energy demand management.

Reliable operation of remote electricity distribution networks heavily depends on transmission line monitoring devices that can detect faults online and transmit alarm signals to smart relays/control stations. HIF is a common issue in medium voltage networks such as SWER lines, and it is challenging to be detected by conventional protective relays because it does not draw a fault current large enough to trip the protection relays [3]. Furthermore, HIF identification requires rapid detection results to avoid catastrophic bushfires. For instance, the time to ignite is around 200 milliseconds for HIF current (the type of fault is 'wire on the ground', and the object is soil) from 1 to 10 amps [4]. Stringent time constraints associated with such time-critical fault identifications urged to minimise the data transmissions to a remote location for data processing. Thus, the line monitoring devices should have local intelligence to detect these time-critical faults and send an alarm signal to the adjacent smart relays to isolate the faulty region as early as possible.

On the other hand, local renewable microgrids seemed promising to address the increasing energy demand from existing remote consumers. In order to ensure the economic viability of microgrid operation, the consumer energy demand needs to be appropriately managed [5]. Therefore, the consumer energy usage patterns need to be obtained and analysed in order to generate near real-time feedbacks for optimal demand-side management [6]. As part of electricity grid modernisation, smart meters are being installed in consumer locations that can improve visibility in electrical energy consumption patterns and measurements. Real-time / near real-time analysis of smart meter data enables the load management and load shift from peak demand hours to off-peak periods, thus alleviating the load burden on the remote microgrids during high demand times.

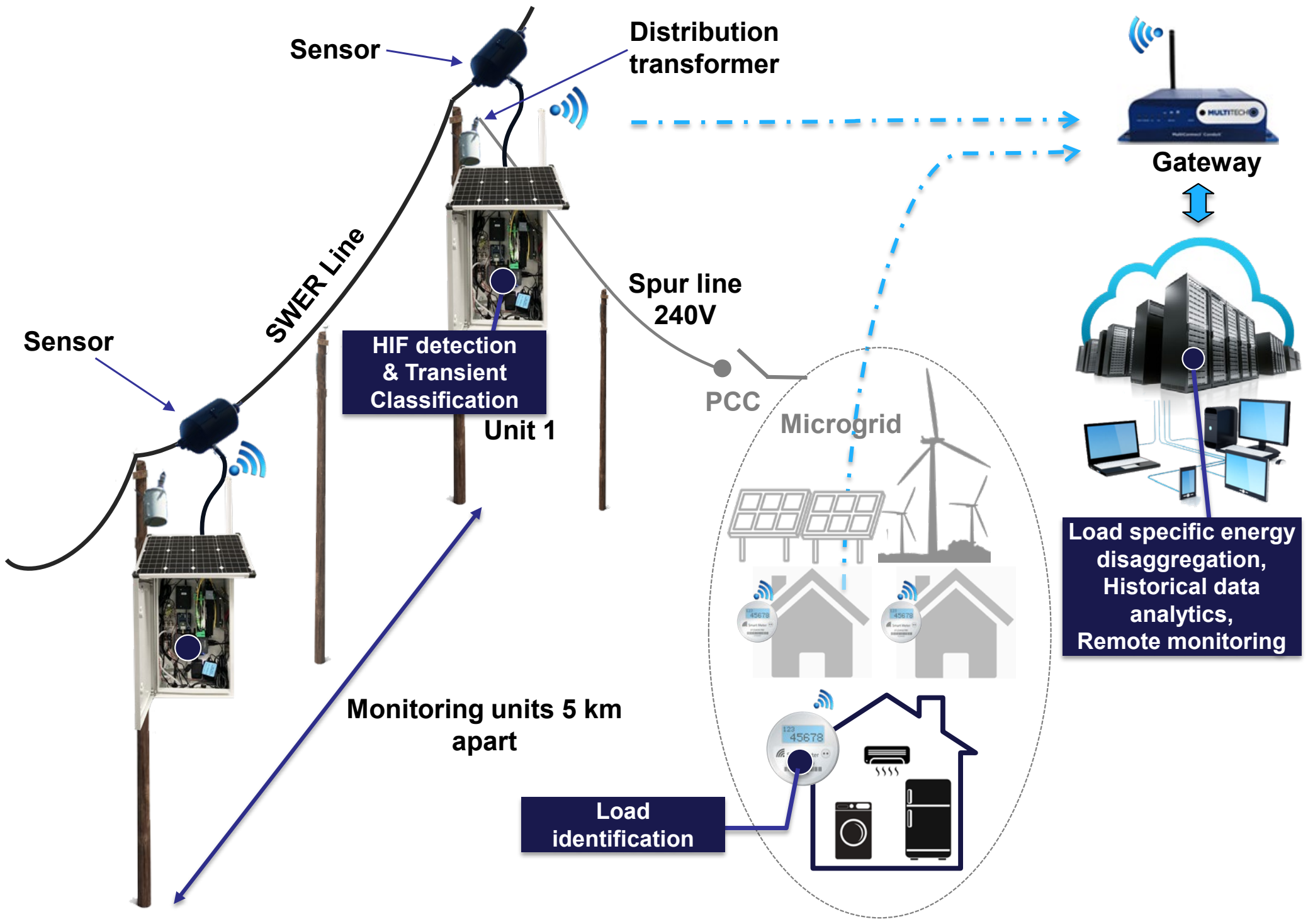

Fig. 1.1. Smart monitoring system for remote electricity networks

In this context, the research described in this thesis provides original contributions to the condition monitoring system for remote electricity networks. The novel contributions of this work are the development of Artificial Intelligence (AI) based condition monitoring techniques and optimisation for its real-time deployment as summarised below.

## 1.2. THESIS CONTRIBUTIONS

1. **A distributed online monitoring platform is developed that incorporates power quality monitoring, real-time High Impedance Fault (HIF) identification and transient classification in Single Wire Earth Return (SWER) network.** In order to carry out these tasks, a hierarchical data analysis framework is proposed which can continuously analyse the current & voltage signals and calculates electrical parameters such as RMS values and harmonics information. These can be directly used for threshold-based power quality disturbance detection. Simultaneously, a transient detection algorithm is proposed and implemented to isolate the transients. In order to differentiate the faulty conditions against normal switching transients, discriminative features are extracted using advanced signal processing techniques. Artificial Intelligence (AI) based classification techniques are proposed to classify faults and transients from the extracted features. The proposed methodologies can be listed as follows:

a. A feature map is calculated based on the entropy of wavelet coefficients up to 3 levels of Wavelet packet decomposition to capture the variations of HIF from the healthy conditions.

b. A short-time FFT based feature map is formulated that can be extracted with reduced computational complexity compared to the wavelet packet entropy-based feature, which struggles due to the resource constraints in the real-time deployment.

c. A light-weight Convolutional Neural Network (CNN) structure is developed, that facilitates the automated higher -level feature extraction, HIF discrimination and transient classification.

d. Optimisation techniques such as hardware-based parallelism, pipelining and timed loops are adapted to reduce the latency, increase the throughput and reduce the jitter respectively for the real-time deployment of the proposed algorithms.

e. A hierarchical data analytics architecture is intended to facilitate different latency requirement applications in which edge computing paradigm is suggested for time-sensitive applications (low latency) in the electricity grids.

f. The proposed approach is validated with the laboratory experiments in real-time and an extensive validation is carried out with the high-power arcing data collected from an industrial high-power testing station.

2. **A distributed, consumer load identification methodology is developed to detect the load type from its turn-on transients.** High-frequency components of the current signal are required to extract the load specific signatures. An edge computing based deployment architecture is proposed to facilitate the high-frequency component analysis with the minimised data transmission. The primary aim of this contribution is to develop computationally efficient methodologies for the sub-tasks of load identification such as event detection, feature extraction and load classification to enable their real-time deployment on resource-constrained embedded devices such as smart meters. The proposed techniques can be summarised as follows:

a. An investigation on the impact of sampling frequency and digitisation resolution of the signal in the load classification accuracy and amount of data generation is carried out to develop an economical load identification system.

b. An empirical estimate of RMS value based event detection method is developed to detect the high-power load switching transients with reduced computational complexity.

c. A wavelet decomposition based event detection approach is leveraged to isolate the turn-on transients of low-power devices. Furthermore, it can separate the transient states from steady-states in an input signal.

d. A nine-dimensional feature vector is introduced for load type identification, which encompasses the average energies of octave scale sub-band decomposition of FFT spectrum and jumps in active and reactive power between adjacent steady states.

e. A lightweight, fully connected neural network is proposed for load type classification that can be implemented in real-time on resource-constrained embedded hardware.

3. **A deep neural network based energy disaggregation framework is developed to disaggregate the load specific energy usage from an aggregated signal which is captured in the main panel level**. The aggregated signal is forwarded to the disaggregation framework along with the load type labels – identified in the load identification process, to estimate the energy consumption of each load. The main aids of this framework can be summarised as follows:

a. A convolutional variational autoencoder (CVAE) that contains a stochastic encoder and generative decoder is proposed to model the energy usage pattern and estimate the load specific energy usage.

b. A Kullback-Leibler divergence based penalty is introduced with the loss function in order to distribute all encodings evenly around the centre of the latent space. It improves the load specific energy estimation by eliminating the discontinuities from the latent space.

c. Each load type of energy consumption pattern is independently modelled using separate networks. The incoming signal is directed to the corresponding networks based on the load type label for near-real-time energy disaggregation. This approach improves system scalability.

d. The proposed framework is evaluated with the energy consumption data of five different devices from a real-world data set (UK-DALE) along with the standard error measures. The evaluation results are compared with the state-of-the-art techniques.

## 1.3. Organisation of the Thesis

The remainder of the thesis is organised as follows.

**Chapter 2** provides an overview of remote electricity networks, significant challenges in remote electricity distribution, recent upgrades and its requirement. This chapter also reviews the state-of-the-art techniques to address the major problems in remote electricity networks such as online power quality monitoring, real-time HIF detection and consumer load management.

**Chapter 3** details the digital signal processing and AI-based techniques, which are leveraged in this thesis to enhance the remote electricity grid condition monitoring applications. Furthermore, application-specific timing constraints, input signal resolutions and hardware requirements are investigated, and the limitations of existing monitoring techniques are highlighted.

**Chapter 4** proposes a distributed online monitoring platform for power quality monitoring, real-time HIF identification and transient classification in SWER network. Feature extraction through applying signal processing techniques and deep learning based fault identification methodologies are described. Furthermore, a hierarchical data analytics approach is outlined that can facilitate different fault identification latency requirements. Optimisation techniques for real-time execution of the proposed methodologies are explained. Experimental system validation and the fault identification results are reported in this chapter.

**Chapter 5** introduces a distributed, consumer load identification methodology to detect the load type from its turn-on transients. Computationally efficient transient analysis for load identification sub-tasks such as transient detection, feature extraction and load type classification are described. Effect of sampling frequency and digitisation resolution on load identification are discussed. An edge computing based deployment architecture is outlined to enable privacy-preserving load analysis using high-frequency components inside smart meters. Finally, experiment setup, real-time system testing, and the load classification results are summarised in this chapter.

**Chapter 6** proposes a deep neural network based energy disaggregation framework to separate load specific energy usage from the main panel level aggregated power signal. The network architecture for load-specific energy usage modelling, penalty term and loss function to train the network are described in detail. The proposed framework is evaluated with a real-world data set (UK-DALE), and the results are reported against standard error measures.

**Chapter 7** concludes the thesis with a summary of the research contributions and presents potential research directions to follow up from this thesis.

## 1.4. LIST OF PUBLICATIONS

During the course of this thesis project, a number of publications have been made based on the work presented here and are listed below for reference.

Patent

1. **T. Sirojan**, S. Lu, B. T. Phung and E.Ambikairajah, “Apparatus and process for real-time detection of high-impedance faults in power lines”. No. PCT/AU2019/051219.

Journal Articles

2. **T. Sirojan**, S. Lu, B. T. Phung, D. Zhang and E. Ambikairajah, "Sustainable Deep Learning at Grid Edge for Real-time High Impedance Fault Detection," in *IEEE Transactions on Sustainable Computing*, doi: 10.1109/TSUSC.2018.2879960.
3. **T. Sirojan**, S. Lu, B. T. Phung, D. Zhang and E. Ambikairajah, “Real-time Smart Monitoring System for Overhead Transmission Lines in Rural Electrification Schemes,” *IET Science, Measurement & Technology*. (under review).
4. S. Lu, **T. Sirojan**, B. T. Phung, D. Zhang and E. Ambikairajah, " DA-DCGAN: An Effective Methodology for DC Series Arc Fault Diagnosis in Photovoltaic Systems," *IEEE Access*, vol. 7, pp. 45831-45840, 2019.
5. A. Arunan, **T. Sirojan**, J. Ravishankar and E. Ambikairajah, "Real-Time Adaptive Differential Feature-Based Protection Scheme for Isolated Microgrids Using Edge Computing," in *IEEE Systems Journal*, doi: 10.1109/JSYST.2020.2986577.
6. S. Lu, R. Ma, **T. Sirojan**, B.T. Phung and D.Zhang, “Lightweight Transfer Nets and Adversarial Data Augmentation for Photovoltaic Series Arc Fault Detection with Limited Fault Data," in *IEEE Transactions on Industrial Electronics*. (under final revision)

Conference Papers

7. **T. Sirojan**, S. Lu, T. Phung and E. Ambikairajah, "Embedded Edge Computing for Smart Meter Data Analytics," *2nd IEEE International Conference on Smart Energy Systems and Technologies*. 2019.
8. **T. Sirojan**, T.Phung and E. Ambikairajah, "Enabling Deep Learning on Embedded Systems for IoT Sensor Data Analytics : Opportunities and Challenges," *6th IEEE International Conference on Information and Automation for Sustainability (ICIAfS)*. 2018.
9. **T. Sirojan**, T. Phung and E. Ambikairajah, "Deep Neural Network based Energy Disaggregation," *6th IEEE International Conference on Smart Energy Grid Engineering*. 2018.
10. **T. Sirojan**, S. Lu, B. T. Phung, D. Zhang and E. Ambikairajah, " High Impedance Fault Detection by Convolutional Deep Neural Network," *IEEE International Conference on High Voltage Engineering and Application (ICHVE)*. 2018.
11. **T. Sirojan**, T. Phung and E. Ambikairajah, "Intelligent edge analytics for load identification in smart meters," *IEEE Innovative Smart Grid Technologies - Asia (ISGT-Asia)*, 2017, pp. 1-5.
12. R. Song, S. Lu, **T. Sirojan**, B. T. Phung and E. Ambikairajah, "Power quality monitoring of single-wire-earth-return distribution feeders," *IEEE International Conference on High Voltage Engineering and Power Systems (ICHVEPS)*, 2017, pp. 404-409.

# CHAPTER 2

# 2. CHALLENGES IN REMOTE ELECTRICITY NETWORKS AND MITIGATION TECHNIQUES: A REVIEW

## 2.1. INTRODUCTION TO REMOTE ELECTRICITY NETWORKS

Remote electrification is the process of bringing electrical power to rural areas. Typically, electrification begins in urban areas and gradually extends to rural areas. The extensions of national electricity grids to the rural areas are often technically difficult, costly and inefficient due to the remoteness and sparse population densities. Furthermore, electricity distribution using two-wire single-phase lines and three-wire three-phase lines are suitable for densely populated urban areas. They are not economical for the sparsely distributed rural loads. Therefore, SWER power lines are introduced to ensure the economic viability of rural electrification. It has been reported that the SWER lines can save approximately 30% on the capital costs of conventional three-phase lines and 50% on single-phase systems [1].

### 2.1.1. SWER NETWORK

Internationally, SWER has proven to be a low cost and reliable electricity transmission technology for servicing low consumer densities in remote communities. SWER supplies single-phase power to rural loads from the primary grid with a single transmission line. The earth is used as the return path of the single-phase current to eliminate the need for a neutral wire.

SWER technology was invented in New Zealand in 1925 and soon gained prominence as the preferred rural electrification scheme in New Zealand and Australia. SWER has also been used in Africa, Brazil, Canada and the United States of America. Typically, SWER systems are the remote end of radial extensions of the main grids and thus consist of long feeder lengths. SWER feeders energised from three-phase substations through isolating transformers, and It commonly operates at 12.7 kV or 19.1 kV [2]. The isolating transformer is placed to electrically isolate the three-phase network from the SWER feeders, which provides earth fault protection [3]. Low-cost electrification to the sparely populated remote loads can be achieved with the spur lines tapped off from the long radial SWER lines through the distribution transformers as shown in Fig. 2.1. But the downside is that SWER is generally suitable for small rural loads such as lightning, household appliances and irrigation pumps in farmlands.

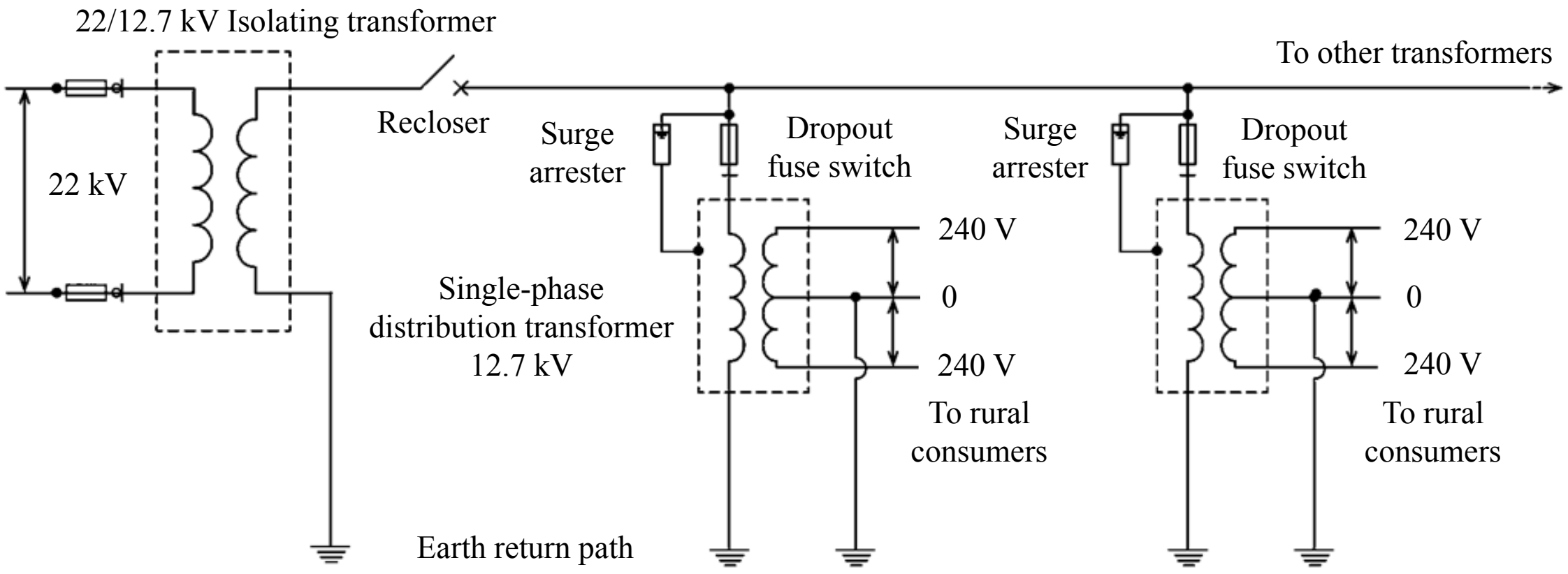

Fig. 2.1. SWER transmission structure

## 2.2. MAIN CHALLENGES IN REMOTE ELECTRICITY NETWORKS

Since the SWER networks seemed like the lowest cost way of delivering power to the rural communities, the majority of the remote electricity networks in Australia and New Zealand used SWER feeders. There are more than 200,000 km of SWER feeders have been installed throughout Australia and in operation for more than 50 years. There are a few problems associated with the existing SWER networks, as detailed below.

### 2.2.1. Power quality issues

Typically, SWER networks consist of very long feeder lengths with high impedance. Thus, losses in SWER systems are high, and its voltage regulation is inferior because of voltage drops along its long transmission feeders. Consumer loads are generally light in SWER networks, and load density is around 0.5 kVA/km with an average demand of 3.5 kVA per consumer [4]. Lower load density along long feeders also resulting in the current having leading power factor. Furthermore, dynamic loading variations cause oscillations in the supplied voltage. On the other hand, damages due to lightning, wildlife and trees are common in remote electricity networks. These are the potential causes of power quality problems such as dips, swells, rapid changes and harmonic distortions in supply voltage and current, which violates the power quality standards.

### 2.2.2. Bushfire risks due to high-impedance arcing faults

HIF generally occurs when an energised overhead SWER conductor contacts with a poor conductive surface such as overgrown tree branches or falls onto the ground, touching the sand or soil. The major problem with the remote electricity networks is the bush fire risk from arcing faults resulting from HIF in SWER feeders and any remote system. There are two types of SWER line HIF that can initiate bushfires: a) fallen energised overhead conductor creates high voltage electric arcs near vegetation (usually dry grass) [5] b) live overhead conductor touches tree branches (vegetation conducting high voltage current) [6]. In rural electricity networks, a SWER line with HIF can remain energised on earth or vegetation for an extended period, during which it produces high-temperature arcing. These types of HIF in SWER feeders have created catastrophic fires in 2009 Victorian Black Saturday Bushfires, Australia. The worst fire on Black Saturday initiated when a SWER feeder fell to the ground and ignited dry grass. Another two significant fires were triggered by the loosen SWER feeders that contacted nearby vegetation. These bushfires caused the death of 173 people and massive losses of property [7].

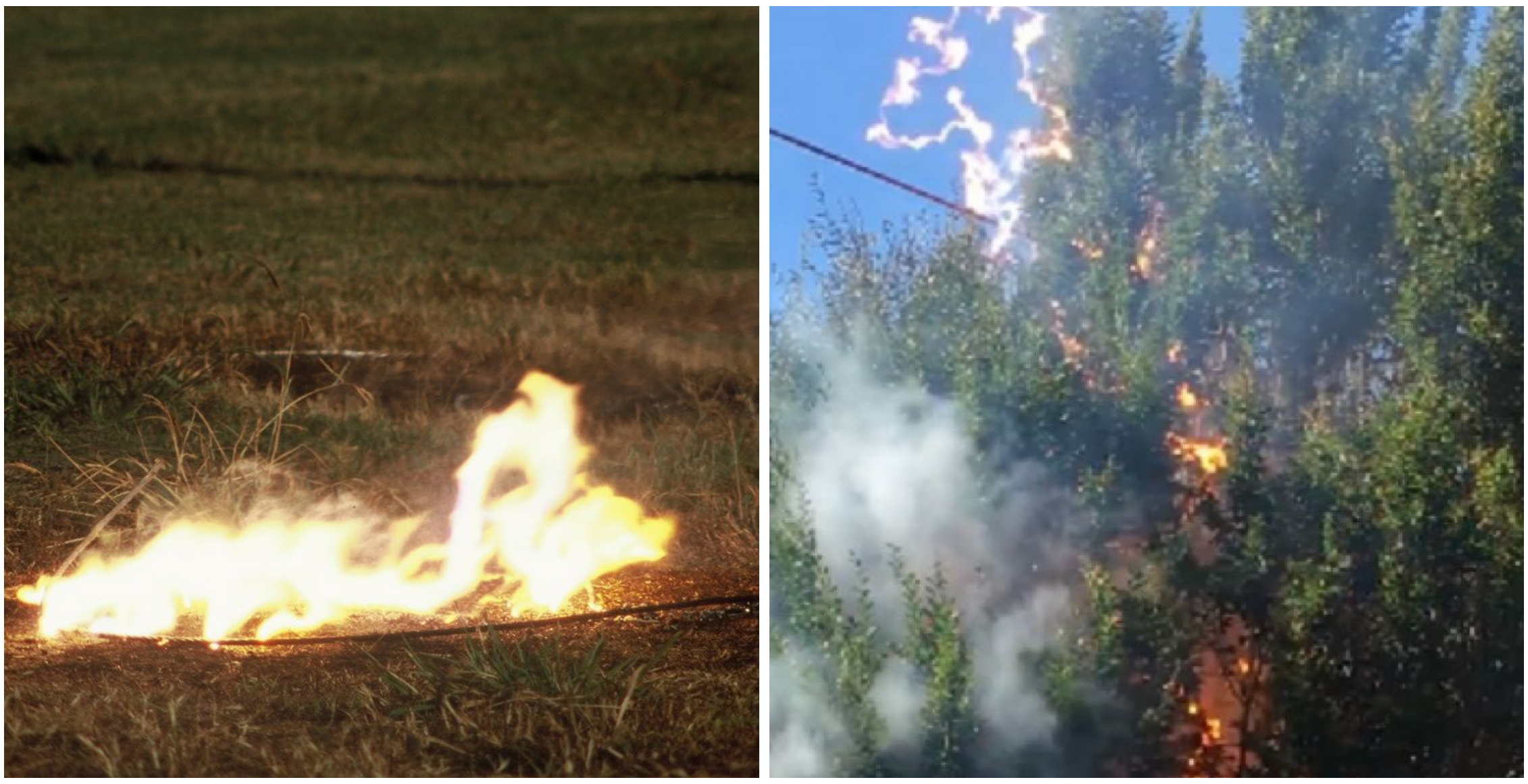

(a) (b)

Fig. 2.2. Types of HIF that can trigger catastrophic bushfires (a) broken overhead conductor on dry grass [5] (b) live overhead conductor touches tree branches.[6]

After the Black Saturday bushfires, the Government of Victoria initiates multiple research programs to investigate on how remote powerlines start fires and to explore mitigation methodologies to eliminate powerline fire risk. Consequently, some safety actions have been proposed for SWER feeders such as selective undergrounding, use of more efficient insulated conductors and the installation of protection relays to identify and isolate the faulty SWER feeders when the fire risk is high.

Protection relays can facilitate rapid isolation of faulty region only when there is a significant increase in the SWER line current. On the other hand, protection relays fail to detect low current faults since the fault current cannot be discriminated from healthy variations in consumer load current. When an energised SWER conductor falls on the ground or touches with vegetation, poor conducting surface contact causes the resulting fault (HIF) to draw relatively small current that is not enough to trip the protection relays. Disastrously, rapid detection and isolation of HIF on SWER feeders cannot be achieved with the conventional protection relays, and it is a remaining research gap in the SWER line bushfire safety tool kit.

### 2.2.3. The rapid increase in energy demand and its management

Since the SWER networks are single-phase and single-conductor systems, current carrying capacity is limited. The existing capacity of these SWER feeders may become inadequate soon since more demanding loads are rolled-out in remote areas. In order to cope with increasing energy demand, SWER feeders need to be replaced with more efficient and higher capacity conductor systems. However, the replacement of higher capacity conductors to the entire, long-spanning SWER networks will not be economically viable. Thus, remote electricity distributors are hesitant to comply with the SWER network-wide replacement. Alternatively, it is possible to establish extended, isolated microgrid systems that combine a group of interconnected rural consumers in a microgrid system. Local renewable energy power sources such as solar, wind, fuel cell, gas turbine, and Small Modular Reactors (SMR) can be leveraged for the microgrid establishment. The existing SWER feeders can interconnect the remote microgrids that can enable energy sharing from microgrids with excessive renewable generations [8]. Anyhow, the increasing energy demand management in rural areas with local renewable energy microgrids requires a smart energy demand management framework that can optimise the microgrid operation.

## 2.3. Mitigation steps – towards reliable remote electricity networks

As part of smart grid upgrades, utilities are installing smart sensors to enhance the electricity grid condition monitoring. Recently, Energy Australia has spent around $170 million to install 12,000 sensors on its electricity distribution network [9]. The data

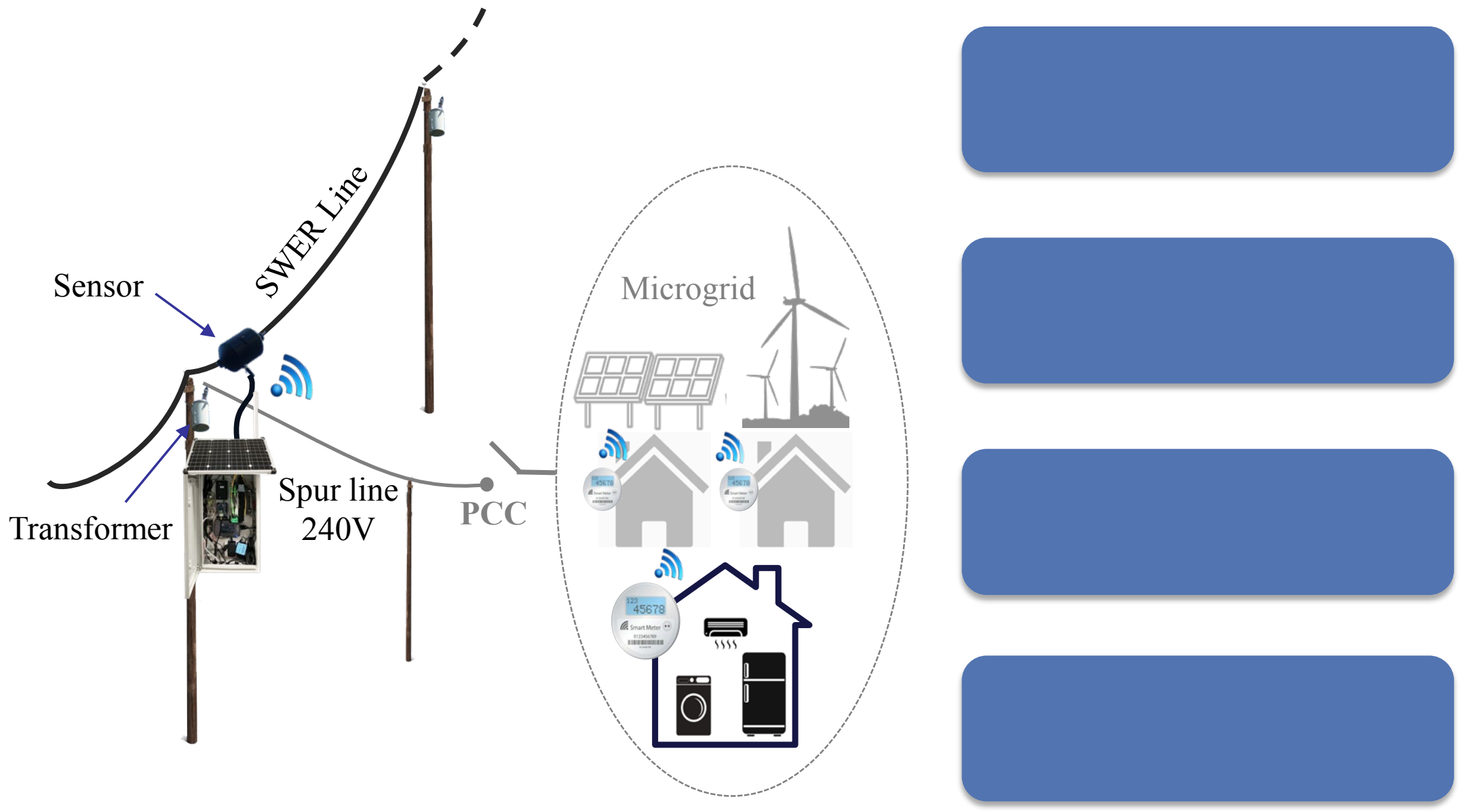

Fig. 2.3. Condition monitoring steps for relaibale remote electricity grids

collected from those sensors can be utilised for condition monitoring applications which can mitigate the challenges in remote electricity networks, as highlighted in Fig. 2.3.

### 2.3.1. ONLINE POWER QUALITY MONITORING

Since SWER networks experience a wide variety of variations in the electricity supplied to the remote consumers, power quality monitoring functionalities are required, that can provide appropriate protection to SWER network equipment and remote consumer loads. Majority of the power quality disturbances in the SWER systems are caused by the consumer loads that are connected to the same network. Thus, it is necessary to monitor the voltage and current inputs and the disturbances generated by the consumer loads.

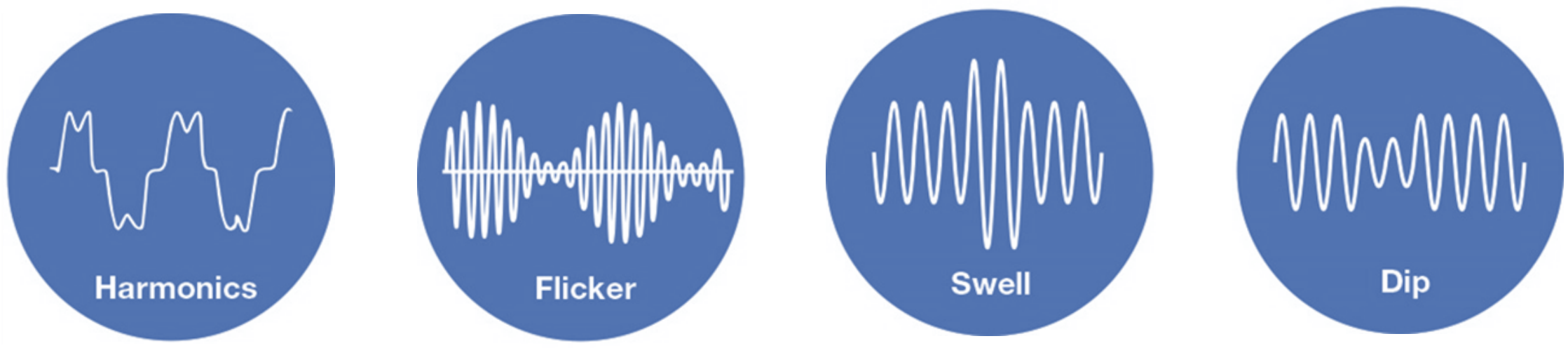

Fig. 2.4. Major power quality disturbances [10]

However, subtle power quality disturbances generally pass through conventional protection devices and contribute to equipment degradation over time [10]. These disturbances can cover dips, swells, rapid variations and harmonic distortions, as shown in Fig. 2.4. SWER network infrastructure will benefit from clean power and extended life when the remote electricity providers integrate the advancement of technologies such as high-resolution time-frequency signal analysis that can facilitate the precise online detection of power quality events.

#### 2.3.1.1. EXISTING POWER QUALITY MONITORING TECHNIQUES

Increased usage of electronic equipment introduces non-stationary states which are no longer easily analysed with the time-domain signals. Researchers have used various signal processing schemes such as wavelet packet analysis for the separation of the higher-order harmonics from the fundamental component [11][12][13]. Continuous wavelet analysis based algorithm is formulated in [14] to identify the frequency variations and other harmonics types such as sub-harmonics and inter-harmonics. The Gabor-Wigner transform, another time-frequency analysis technique has been introduced to detect voltage sag, swell, fluctuations and harmonics with higher clarity[15]. A combination of Fourier, wavelet and short-time correlation transform is derived to enhance the performance of power quality disturbance detection with the rule-based expert systems [16][17]. An empirical-mode decomposition technique is applied with the Hilbert-Huang transform to identify voltage spikes and notches from distorted waveforms using probabilistic neural networks [18]. Achlerkar et al. [19] proposed a variational mode decomposition to extract central frequencies, relative energy ratios, instantaneous amplitudes and zero crossings that can be used as features for a decision tree based power quality disturbances detection. Pujiantara et al. [20] proposed a fast Stockwell transform for real-time power quality analysis with reduced computational complexity. A low-complexity feature set that includes mean, variance, energy and min-max values of the filtered voltage signal is identified for the considerable dimensionality reduction of raw signals during long-term power quality monitoring [21]. An adaptive filter based low-cost signal processing technique is introduced for the real-time estimation of harmonic content in voltage, and current waveforms [22].

Even though the proposed time-frequency analysis techniques produce substantial advancements on power quality monitoring, their accuracies and computational complexities heavily rely on the input signal resolution. However, the required input signal resolution and the constraints in acquiring high-resolution signals from the electricity grids in real-time is not focused on those research works. Chen et al. [23] proposed a high-resolution technique for flicker detection with a down-sampling method. They have sampled the raw signal at 7680 Hz, and their experimental results have demonstrated that the frequency resolution of a signal is prominent in the accuracy. On the other hand, it is essential to keep track of real-time and long-term power quality attributes to ensure the quality control and preventive maintenance of electrical equipment. Most of the previous research works have failed to do a feasibility analysis on long-term and real-time implementation of the proposed power quality monitoring methodologies and its resource requirements in a practical application. Grigorescu et al. [24] developed a power quality monitoring system that can support both real-time and long-term monitoring. They have sampled the raw signals at 12800 Hz for better accuracy. However, the high-resolution data processing and its computational resource requirement are not investigated in the context of real-time implementation. Zhang et al. [25] proposed an online power quality monitoring system over the internet and pointed out that raw data transmission through the internet is not economical. However, they have not recommended any solutions to handle the high data rate to support long-term power quality monitoring. Bi et al. [26] introduced an online power quality monitoring system based on ethernet, and their experimental results show that a shared 10Mbps ethernet connection could not fulfil the monitoring system demand. Thus, the proposed solution is not scalable and not suitable for long-term power quality tracking.

Power quality indices need to be analysed on distributed locations of the electricity networks in order to mitigate network-wide power quality problems. The distributed power quality monitoring is challenging in the remote electricity networks since most of them are remote end of radial extensions from the main grids, and it spans on broad geographical areas. In addition, there are significant limitations in the availability of communication technologies in rural areas. Di Bisceglie et al. [27] proposed a fully decentralised cooperative sensor network architecture for voltage quality monitoring. The proposed architecture is implemented by forming a mesh network using a Zigbee/IEEE 802.15.4 communication protocol. However, the physical communication range of Zigbee protocol and mesh network architecture is neither suitable nor economical for remote electricity networks. Gómez et al. [28] have developed a cloud computing-based web services framework that can facilitate the analysis of incoming power quality data from the power quality meters that are distributed along the electricity network. Even though it supports online power quality monitoring, the data transmission cost, data resolution requirement and communication aspects of the proposed framework are not investigated, which are essential for an efficient power quality monitoring system and its practical implementation. Hence, the resource constraints in the rural electrification networks need to be addressed while developing an online power quality monitoring system.

### 2.3.2. REAL-TIME HIF IDENTIFICATION

HIF contains several non-linear characteristics that need to be studied for precise fault identification. Hence, several HIF models have been developed, which can mimic typical HIF features. A basic HIF model is presented in Fig. 2.5. The model contains two variable DC sources $V_p$ and $V_n$ that are connected to two diodes $D_p$ and $D_n$ along with two variable resistance $R_p$ and $R_n$. This arrangement of random magnitude variation in unequal DC sources models the asymmetric nature of HIF. Typically, the HIF fault current remains below 10% of the full load current of a feeder [29]. Since HIF does not draw a sufficient fault current to trip the conventional protection relays, HIF detection on SWER feeders has been one of the challenging problems for the rural electricity providers. The small increment in fault current during HIF in SWER feeders always coincides with the current variations during the consumer load changes and other switching transients in terms of magnitude change in time-domain signals. Thus, HIF identification requires a granular level time-frequency analysis of the high-resolution signals.

On the other hand, the time delay in HIF detection process is critical since bushfires can be triggered within a short amount of time (i.e., 200 ms for HIF current from 1 to 10 A [30]) due to the sustained arcing at the contact surface. However, the granular level signal analysis and communication of HIF detection results are resource-intensive and time-consuming compared to the simple logic checks in the overcurrent protection schemes. Thus, there exists a trade-off among the computational complexity, HIF detection accuracy and execution time (latency) of an algorithm in a real-world application.

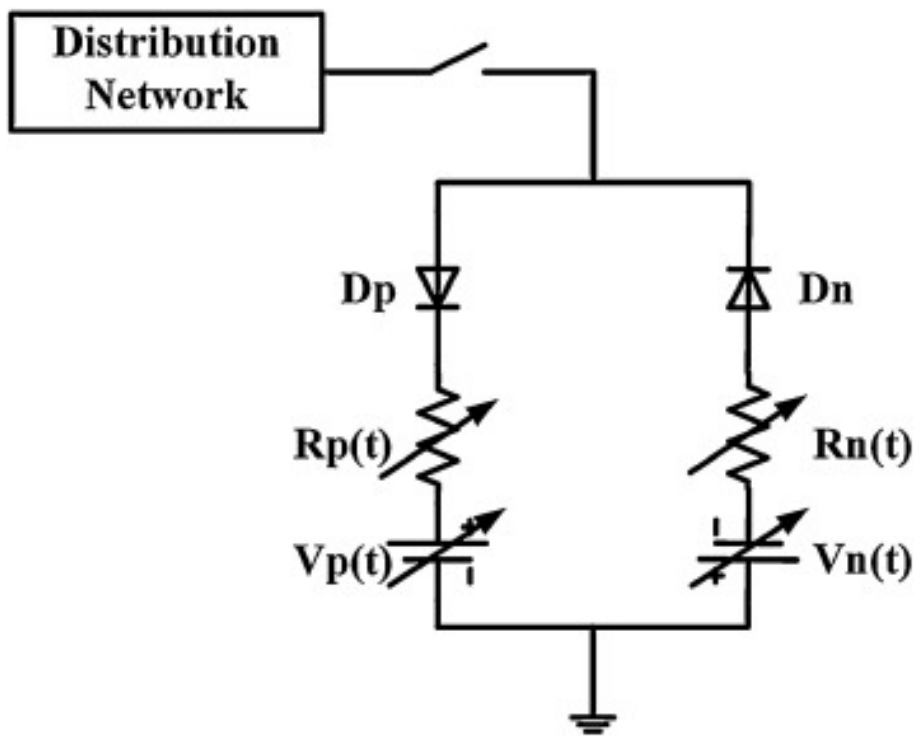

Fig. 2.5. Basic HIF model [29]

#### 2.3.2.1. HIF Background & Existing HIF detection techniques

Typically HIF appears at the primary side of the electricity distribution networks (15kV-25kV) [31]. It has been estimated that around 5-10% of the electricity distribution faults are HIF [32]. Previous researches demonstrate that the over-current protection relays did not detect 25-32% of the downed conductor faults that can be considered as HIF [33][34]. Hence, researchers have investigated the HIF signal characteristics and the challenges

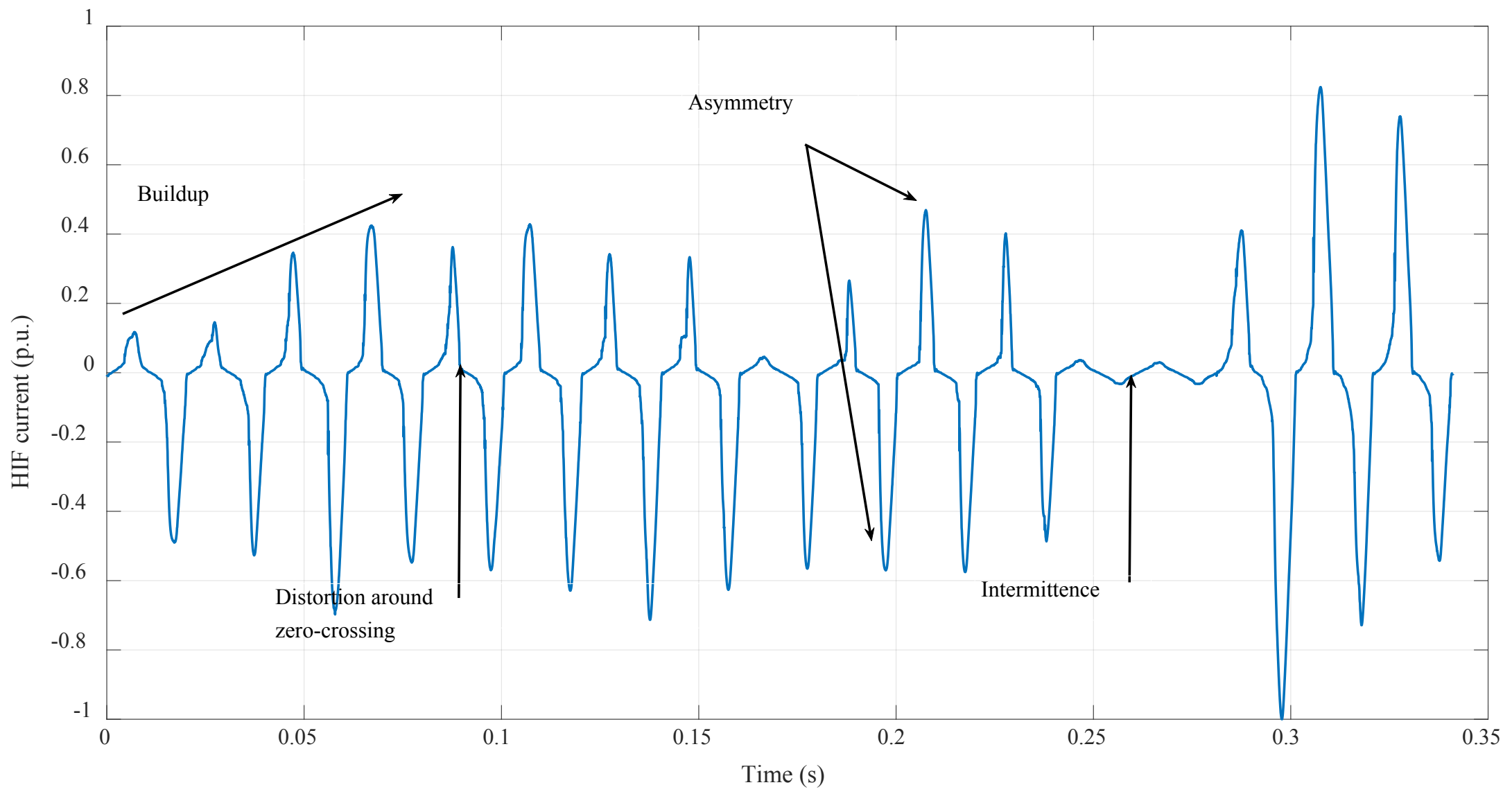

Fig. 2.6. Typical HIF current waveform

associated with HIF detection and isolation.

Fig. 2.6. shows a typical HIF current waveform generated from the laboratory experiment. It also highlights the general HIF characteristics: (1) intermittence, (2) asymmetry, (3) fault current build-up and (4) distortion around zero-crossing. The HIF current mostly generates a few conduction cycles followed by some non-conducting periods, and this property is defined as the intermittence nature of HIF current [35]. The rapid voltage

variation in the faulty conductor introduces the asymmetries in the current waveform that can be observed in the variable peak values and shapes of positive and negative half-cycles, as shown in Fig. 2.6 [36]. HIF current also reveals the gradual escalation of its magnitude, and this property is known as the current build-up [37]. Furthermore, it also contains significant distortions around zero-crossings. Besides these common characteristics, high-frequency components of the HIF current waveform have shown randomness and non-stationary variations during the fault [38]. Researchers have transformed the HIF signal into different analysis domains such as time-domain, frequency-domain and time-frequency domain to identify unique sets of properties that can differentiate HIF. The different target domains are capable of mathematically representing different HIF characteristics. These domain-specific HIF features are summarised and reviewed below.

**Time-domain analysis:** It can capture the temporal variations of HIF signals. Gautam et al. [29] proposed a mathematical morphology(MM) based time-domain signal processing technique which tracks the shape of the HIF voltage signal. The reported results have demonstrated that MM based methods are capable of identifying insignificant shape variations in time-domain signals, as shown in Fig. 2.7. Siadatan et al. [39] introduced a time-domain analysis method based on chaotic and duffing functions that can distinguish

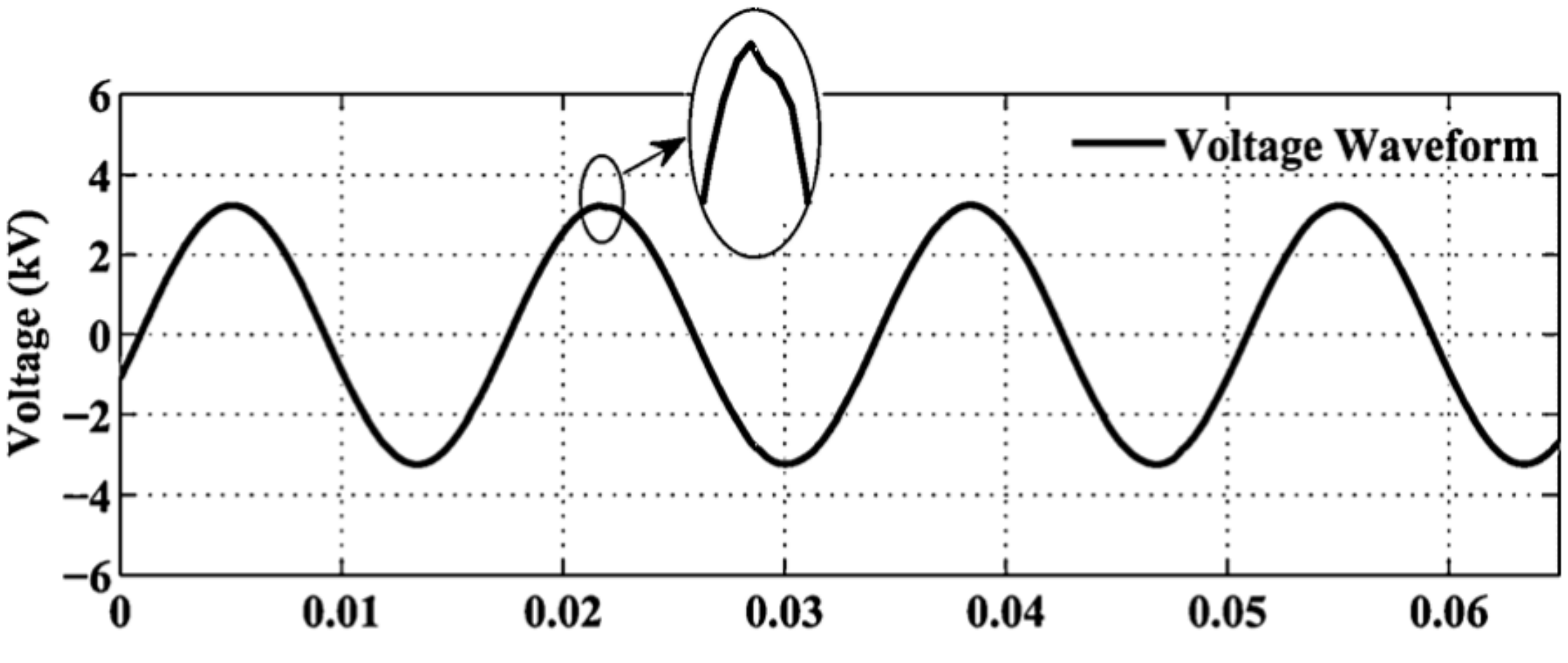

Fig. 2.7. Voltage waveform during HIF [29]

HIF in a noisy signal via observing non-linear state variations. Hou et al. [40] demonstrated that the zero-sequence magnitude of current and voltage waveforms could be used to represent the irregularities of HIF. Faridnia et al. [41] derived twelve unique indices by applying correlation functions (autocorrelation and partial-autocorrelation) to the current signal, voltage signal and their derivations. The reported results have demonstrated the ability of correlation functions in discriminating HIF from no-fault conditions. Mamishev et al. [42] proposed a temporal characterisation method for HIF current signal using root-mean-square (RMS) values. The chaotic properties (randomness) of HIF are analysed and modelled through fractal geometry concepts that allow the distinction between HIF and switching transients. Even though the time-domain analysis has some success stories in HIF detection, it struggles when the HIF current is very minimal. In such cases, the frequency-domain information is required to achieve precise HIF discrimination.

**Frequency-domain analysis:** Typically, HIF signals contains low-frequency as well as high-frequency contents since they are often associated with an electric arc. Hence frequency component analysis can be categorised into two classes: (1) low-frequency spectral analysis and (2) high-frequency spectral analysis.

The low-frequency spectral analysis focuses on the patterns in lower-order harmonics and sub-harmonics. Lee et al. [43] proposed a HIF detection algorithm based on the fundamental and third harmonic of the fault signal. It is a two-terminal numerical algorithm that requires the measurements from both sides of the transmission lines. However, communication aspects and associated delays are not investigated. Hence, the practical application of this approach to a real-time HIF identification is not feasible. Soheili et al. [44] proposed a Fourier based approach that uses a combination of third and even order harmonics to extract unique HIF signatures to improve the fault identification accuracy while reducing the computational burden. Shahrtash et al. [45] utilised the total energy of odd, even and inter-harmonics up to 400 Hz as the features for a decision tree based classifier to distinguish HIF from normal operations. The low-frequency energy comparison between HIF and other system disturbances such as short-circuit fault, capacitor switching and load switching is reported in Fig. 2.8 [46]. It shows a clear deviation of normal short-circuit fault condition from others. However, only small variations in the energy can be observed between HIF and other disturbances, which make the HIF identification more challenging. Aucoin et al. [47] proposed a HIF detection methodology using burst noise intensity at lower frequencies near fundamental and lower order harmonics. They claimed that the fault signal at very low frequencies is not likely to get attenuated from external sources. Snider et al. [48] reported that the first and third harmonics of residual voltage $V_r$ and current $I_r$ along with the second harmonics of residual admittance $\Upsilon_r$ and residual power $P_r$ form an optimal feature set for HIF identification. Furthermore, they have commented that the lower-order harmonics of residual quantities can capture the non-symmetrical properties while filter out the symmetrical components. However, the low-frequency noise and asymmetric variations during switching transients might exhibit similar characteristics as HIF, which makes these fault detection processes more challenging and error-prone. Thus, researchers attempted to evaluate the impact of high-frequency spectral analysis on HIF identification.

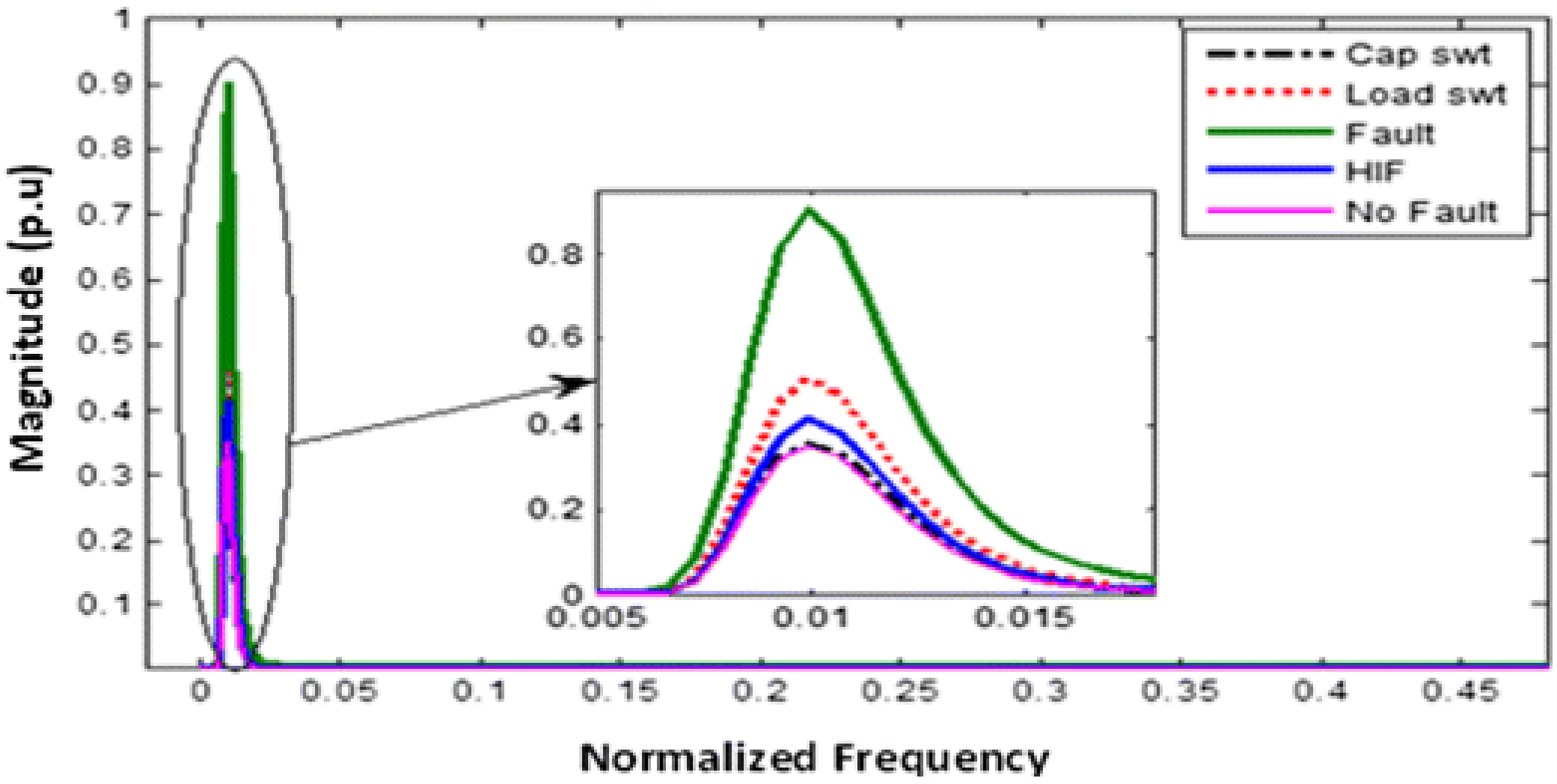

Fig. 2.8. Low-frequency energy comparison of different distubances including HIF [46]

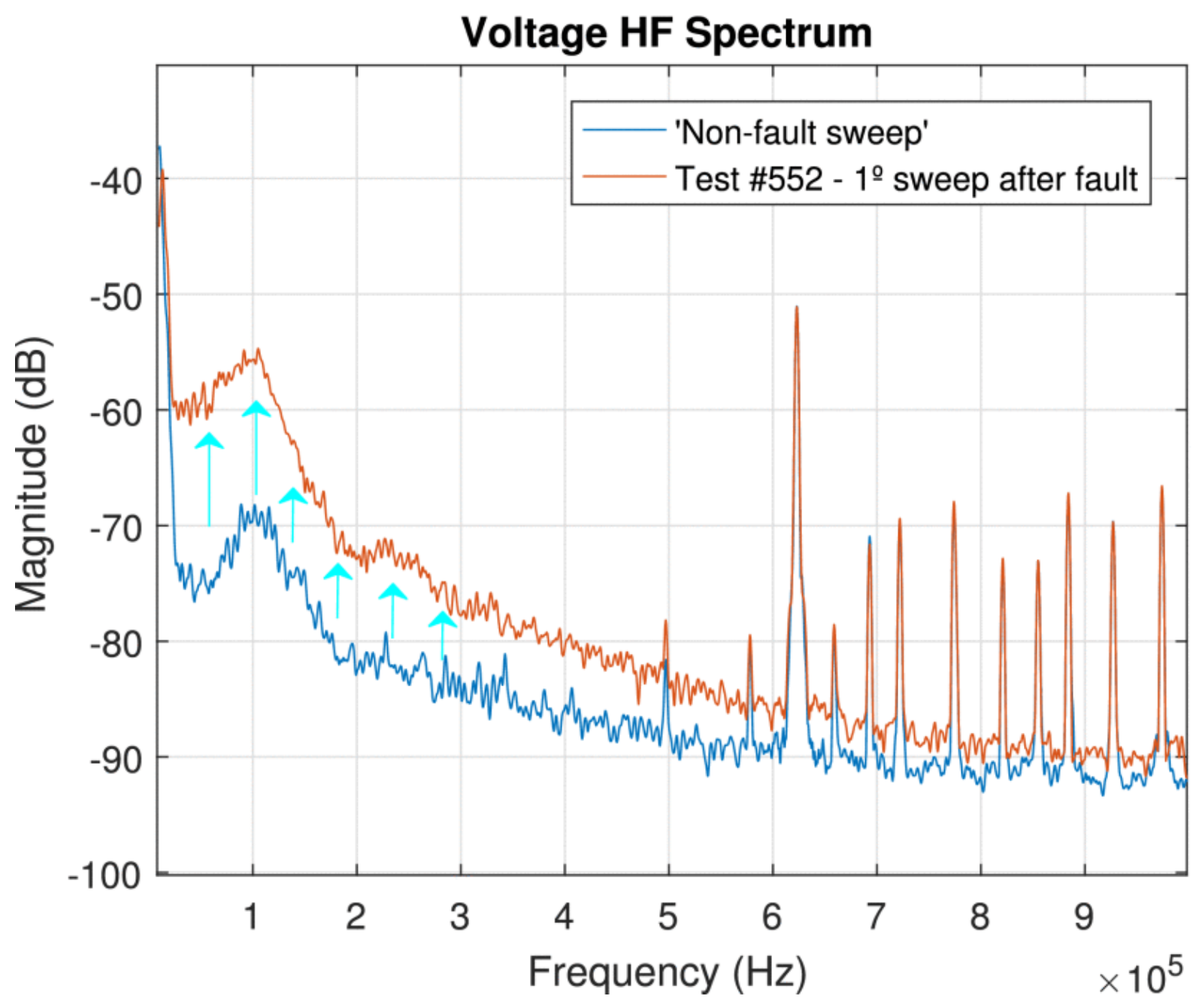

Fig. 2.9. High frequency spectrum comparison of voltage waveform during Non-fault (blue) HIF(red) [51]

The high-frequency spectral analysis offers more extensive exposure for frequency variations, especially in higher frequency bands. Wali et al. [49] proposed a high-frequency power spectrum based feature for HIF representation. Cui et al. [50] introduced a stochastic HIF monitoring approach using time-synchronised μ-PMUs that can capture the signals with high sampling frequencies. Kalman filters are employed in their approach to decompose higher-order harmonic coefficients from the high-resolution signals. Gomes et al. [51] extracted the high-frequency contents from the voltage signals during the faults, as shown in Fig.2.9, to enhance the HIF classification performance. They have claimed that the proposed approach demonstrates high accuracy regardless of the magnitude of fault current and the presence of real noise. Cui et al. [52] computed a compelling feature set through a feature ranking algorithm using Fourier transform and Kalman filter estimation. Bahador et al. [53] derived a relationship between the high-frequency information in the magnetic field and the HIF location. The input waveform is sampled at a very high sampling frequency(50 kHz) to facilitate the early location of HIF. However, the feasibility of processing such high-frequency data in real-time and its computational power requirements are not analysed, which are the primary limitations in practical implementations. Furthermore, intermittence is a prominent and unique characteristic of HIF, and it introduces rapid changes in frequency contents. Hence, it is essential to temporally localise the frequency components in order to capture the variations, which can be unique signatures of HIF. Therefore, several studies have focused on time-frequency domain analysis, as summarised below.

**Time-Frequency domain analysis:** It captures both the frequency information along with its time of occurrence, as visualised in Fig. 2.10. Hence, this time-frequency domain analysis is more suitable to analyse HIF waveforms that exhibit time-varying spectrum. Lima et al. [54] proposed a short-time Fourier analysis based time-frequency representation of second, third and fifth harmonics of HIF current to identify HIF occurrence as well as to distinguish HIF from similar disturbance types such as capacitor bank switching and feeder energisation. Mengda et al. [55] proposed a time-frequency clustering of HIF signal based on Stockwell transformation, which can precisely identify HIF event. Zhang et al. [56] introduced a feature sequence based on the wavelet time-frequency entropy that is calculated on each half cycle of HIF current. Wavelet time entropy is leveraged with a sliding window to identify the HIF start-up, as shown in Fig. 2.11. Based on the reported results, the proposed time-frequency entropy outperforms traditional harmonics based methods. However, the computational complexity of the proposed approach is not analysed. A multiresolution pyramidal Hermite transform is adopted to obtain multiple resolutions of frequency components that can improve the HIF identification with the reduced computational burden [57]. The Neuro wavelet algorithm is introduced in [58] for HIF detection in extra-high voltage transmission lines. The input current signals are sampled at 20 kHz to analyse the frequency components up to 10 kHz for precise discrimination of HIF from other fault types. Qi et al. [59] outlined that the sampling frequency, the mother wavelet and the level of decomposition play an influential role in wavelet analysis, which affects HIF identification accuracy and computational complexity. Daubechies wavelet family is one of the widely used, proper orthogonal mother wavelets for HIF discrimination due to its powerful performance and easy implementation [60][61]. It has been reported that about 40% of the HIF detection techniques in the literature are wavelet-based [62][63][64][65]. Its primary limitations are (1) narrow high-frequency support, (2) loss of feature resolution and (3) subjectivity to the choice of decomposition levels and mother wavelet [32].

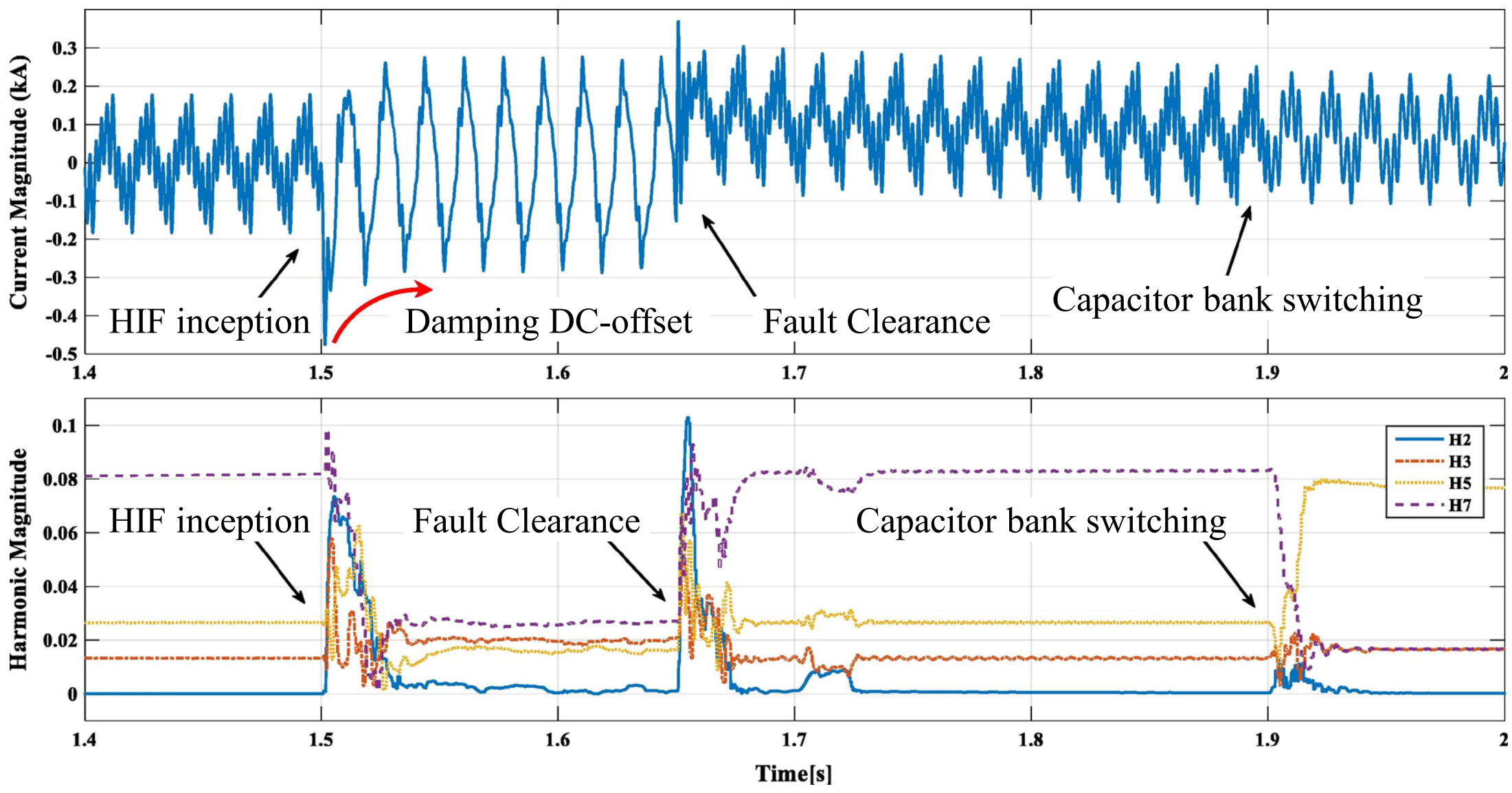

Fig. 2.10. Harmonic variations during faults and transients [44]

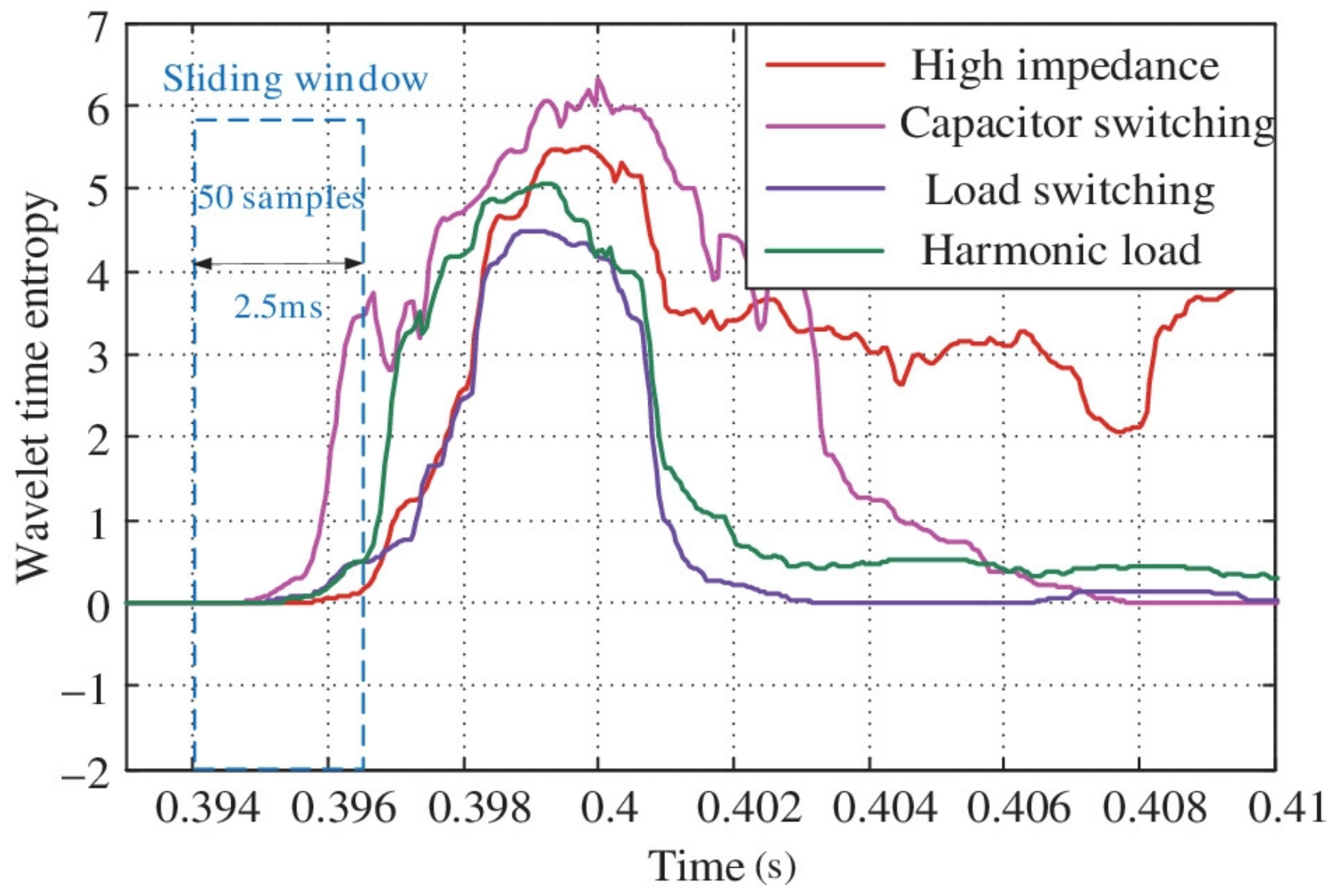

Fig. 2.11. Wavelet time entropy feature for power system events start-up identificaion [56]

The extracted information from different domains such as time-domain, frequency-domain, time-frequency domain and hybrid domain combinations are used to separate the HIF state from a healthy state. Several techniques have been developed to find the partition using simple thresholds, as shown in Fig. 2.12 [54] [66]. Even though these threshold-based techniques are fairly easy to implement and execute in real-time, they are not capable of detecting complex variations associated with HIF and often ends up with either false-positives or true-negatives. Hence, more complex algorithms are required to distinguish HIF from other disturbances. Fuzzy subtractive clustering model

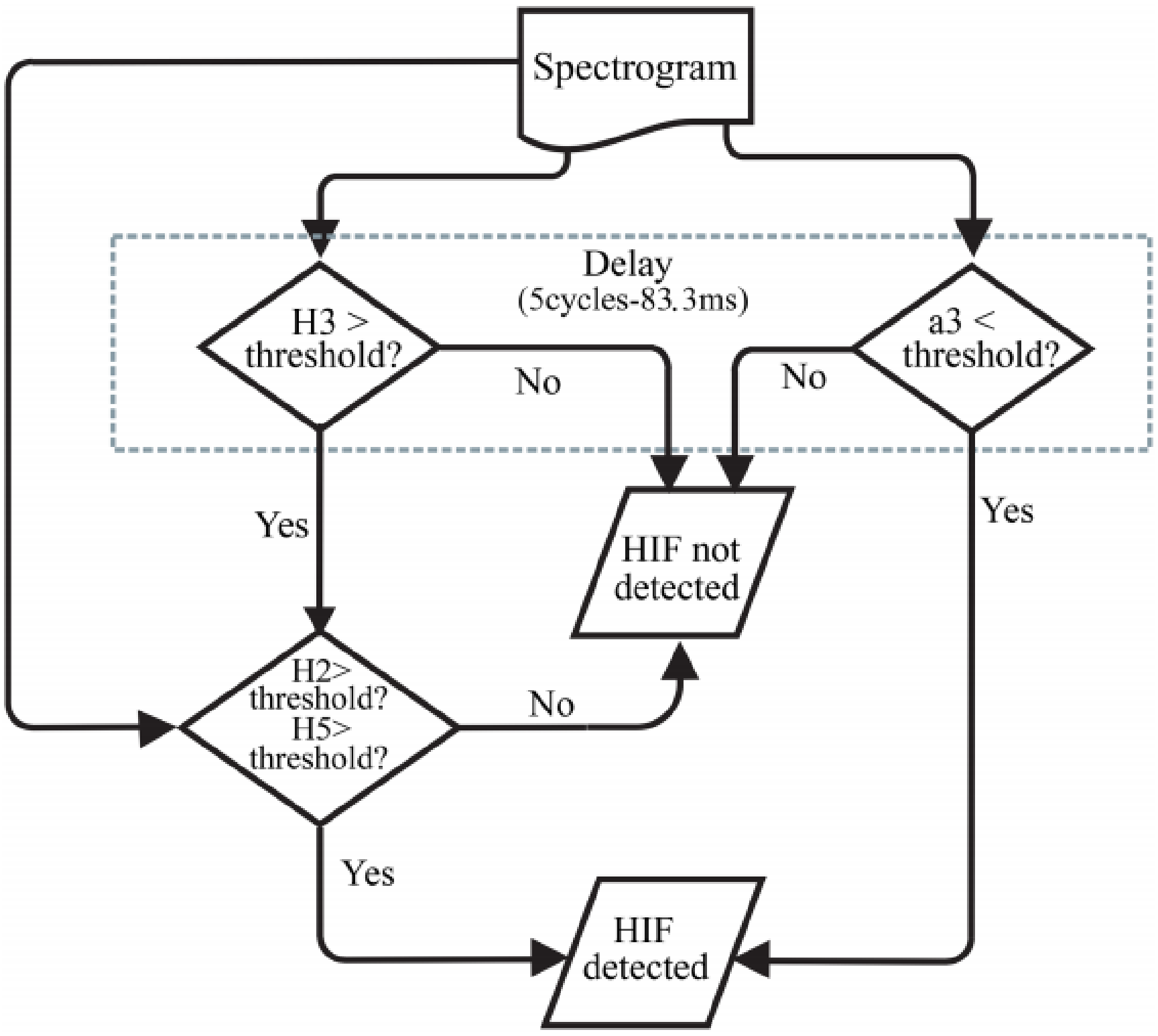

Fig. 2.12. Simple threshold based HIF detection [54]

[67][68][69][70], support vector machines [71][72][73][74] and neural networks [75][76][77][78] demonstrate better performance on HIF detection compared to other techniques such as decision trees[79][80][81], moving sum approach[82], genetic algorithm[83][84], Bayesian framework[50][85], mathematical morphology[86] and rule-based algorithms[87].

However, most of the published HIF identification techniques did not address the limitations of real-world implementation, such as real-time high-frequency data processing feasibilities, evaluation of computational requirements, assessment of HIF identification latency on microprocessors (digital relays) and communication bandwidth requirements. The HIF detection scheme proposed in this thesis aims to fill the research gaps mentioned above, with the recent advancements in data science and information technology.

### 2.3.3. REMOTE CONSUMER LOAD IDENTIFICATION

Besides the fault detection in rural electricity networks, energy demand management is also a challenging task, especially with a limited amount of generation. Since there is a steadily growing energy demand from rural consumers, renewable energy microgrids are getting established in remote areas, and its energy management systems strive to use energy resources efficiently and save energy. The key attributes of the economic rural microgrid operation are optimal consumer load management and balancing the local demand with the integration of local power generation. The backbone SWER network supplies the energy to the rural microgrids when there is a deficient, and it also transfers the energy when excessive local renewable generation[8]. Thus, efficient usage of energy resources in rural microgrids require load management through consumer load identification. The data from the smart meters in the consumer sites can be leveraged to identify the loads. Extraction of informative features and load identification from the main panel level current and voltage signals can be achieved by applying signal processing and artificial intelligence techniques.

#### 2.3.3.1. Existing Load Identification Techniques

Electricity load identification is also known as load monitoring can be broadly categorised into two types: (a) Intrusive Load Monitoring (ILM) and (b) Non-intrusive Load Monitoring (NILM), as shown in Fig. 2.13.

**ILM:** It represents a distributed sensing approach where the sensors need to be attached to each appliance that requires monitoring. Generally, ILM is implemented with smart plugs. Each consumer loads are connected through separate smart plugs, and then middleware platforms are employed to coordinate and integrate the data streams from each socket. The middleware platforms also facilitate the real-time management of the connected household devices. Furthermore, it is possible to identify the load type and connected location. Radio-frequency identification (RFID) tags and communication protocols such as X10, ZigBee and Modbus are widely used to establish ILM network [88][89][90]. Typically, each appliance is denoted as a service in the middleware platforms to facilitate the automatic load identification and self-integration [91][92].

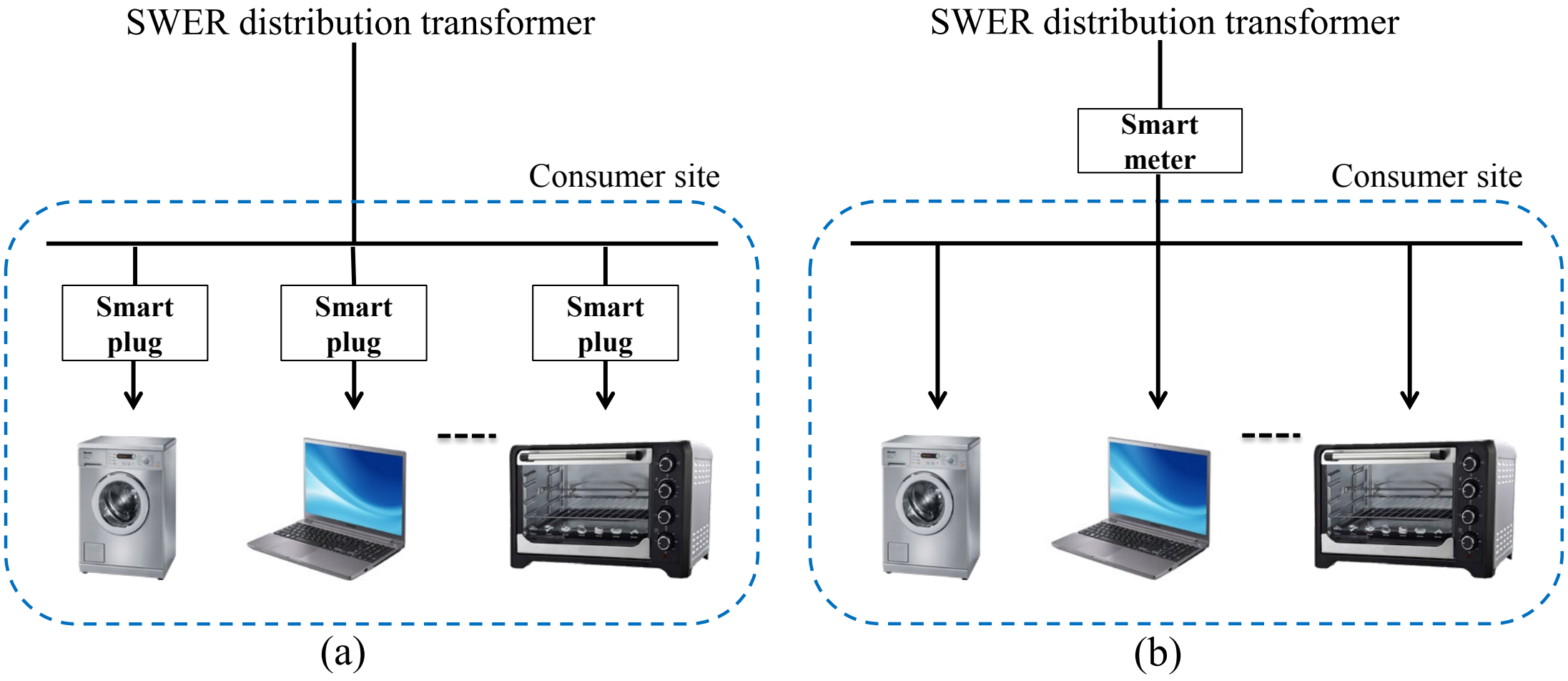

Fig. 2.13. Load monitoring types (a) ILM (b) NILM

Appliance identification and status monitoring are relatively straightforward in ILM since each load are registered as a separate service in the analysis platform. However, ILM cannot be applied to the devices that are directly connected with the live wire (i.e. devices not connected through smart plugs). Thus, sensors or smart plugs need to be integrated with each device, which makes the ILM as not an economic one. Furthermore, ILM consumes more communication bandwidth due to the data transmissions from each appliance. Hence, researchers have attempted to explore economic solutions for load identification, which does not require more sensing and communication infrastructures.

**NILM:** In contrast to the ILM, NILM can be achieved with a single point of sensing, typically with the smart meters that are installed in each consumer locations. NILM can be defined as a process of analysing the changes in current and voltage signals entering into a consumer site to extract the specific load signatures. The extracted signatures from the aggregated load data in the meter panel level can be leveraged to determine the appliances that are being used in a consumer site. It was initially proposed in 1992 by Hart [93]. However, NILM encountered several barriers, such as the failure of discriminating appliances with similar power consumptions, difficulties in identifying loads with time-variable patterns and requirements of high-resolution data. The recent developments in Advanced Metering Infrastructure (AMI) enables bi-directional communication between utilities and customers and facilitates NILM to leverage high-resolution smart meter data. Furthermore, the increased availability and reduced cost of faster analog to digital (A/D) converters and high-speed microprocessors enhance the capabilities of smart meters [94]. These advancements transform the NILM towards a modern era. Hence, several researchers have focused on the NILM based techniques for load identification.

The raw current and voltage waveforms from the smart meters can be used to compute the power metrics for energy metering as well as to detect load specific variations for appliance identification. The analysis process to detect load signatures can be divided into two types: (1) steady-state analysis and (2) transient state analysis.

**Steady-state analysis:** Steady-state operations of the appliances are studied to derive unique steady-state characteristics. The derived properties should be stable during steady-state operation of the consumer load. Real power and reactive power of the devices are commonly used steady-state properties for NILM, as shown in Fig. 2.14 [95][96]. Even though it is exposing the discriminations of the devices that have different power consumption ratios, these properties fail to differentiate the appliances which have similar power ratings, as indicated by an overlapping cluster in Fig. 2.14. RMS values of the current and voltage signals along with the phase difference, are leveraged to overcome the drawbacks of power-based properties, which enhance the load identification accuracy [97][98]. Lam et al. [99] introduced a V-I trajectory-based technique to separate the appliances into distinct clusters. Gupta et al. [100] extracted unique properties from the steady-state voltage noise generated by the load operation. The major drawback of this approach is the noise interference from the wiring infrastructure and the monitored environment. Wu et al. [101] compared the steady-state harmonic components of five different household appliances and demonstrated the variations, as visualised in Fig. 2.15. Liu et al. [102] summarised a set of steady-state current decomposition techniques that can improve the performance of appliance identification. Those techniques are derived from the source separation methods based on steady-state features. Admittance based steady-state features are proposed in [103], that can improve the precision of NILM. Bouhouras et al. [104] proposed the x-axis projections of lower-order harmonic vectors as the appliance specific signature, and their experimental results demonstrate that the accuracy of this approach is improved when the harmonics phase angle is included into the computations.

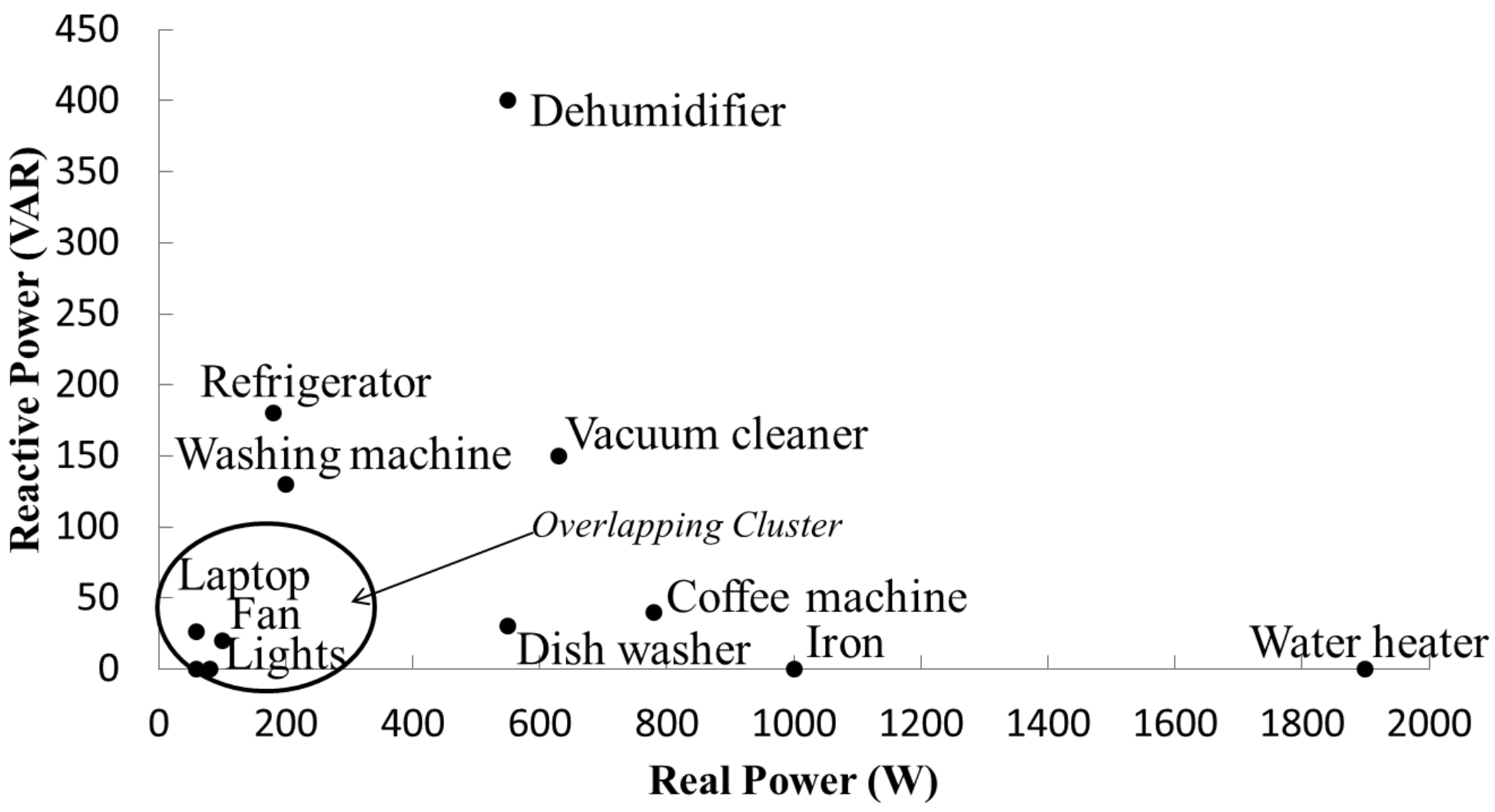

Fig. 2.14. Load distribution with real power and reactive power [95]

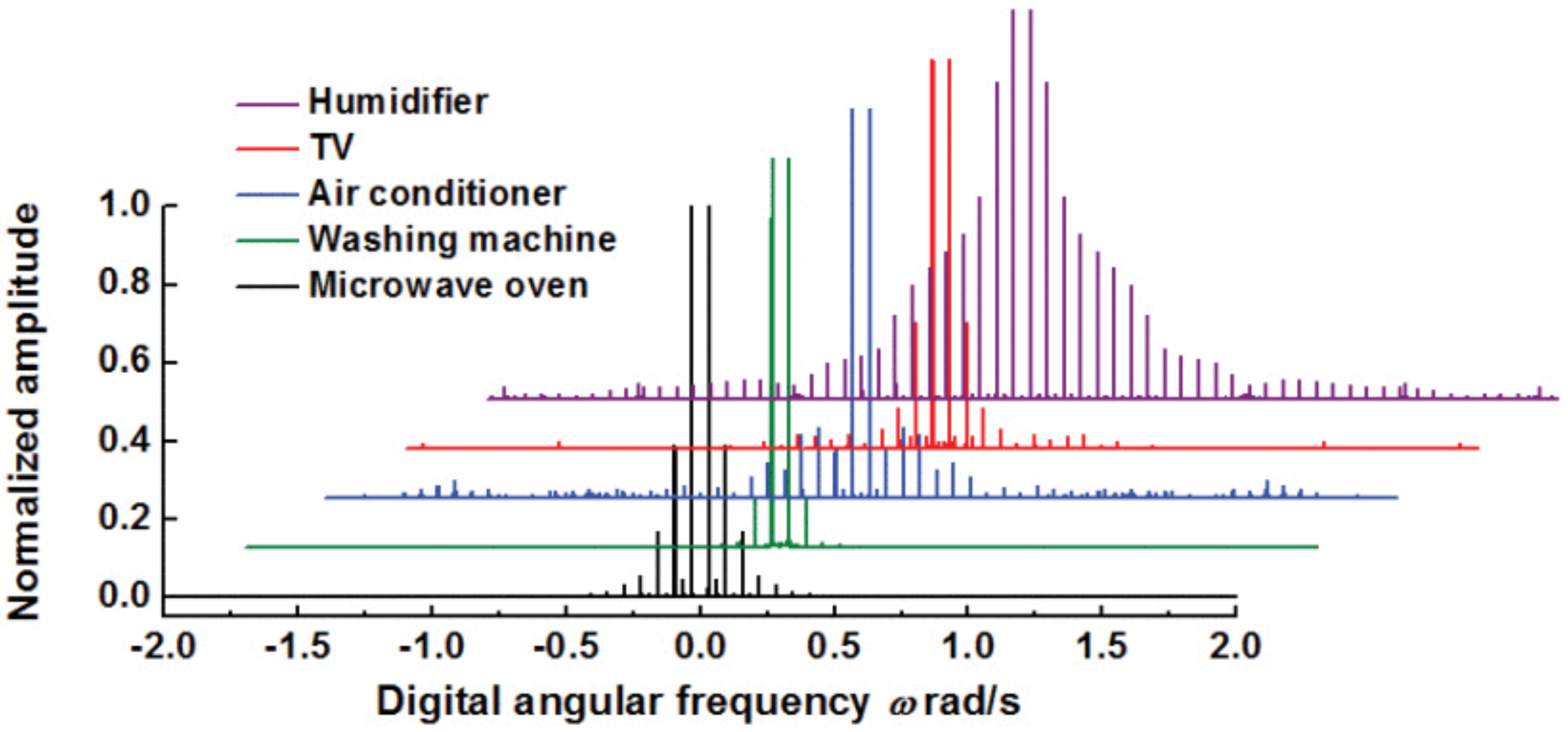

Fig. 2.15. Comparison of steady-state harmonic components [101]

Typically, steady-state properties are extracted from the low-sampling signals since the time-variations of those properties are relatively low. The most prominent issue with steady-state signatures is that when there are loads with similar features or the combinations of different loads matches with a new load, the load identification process produces inaccurate results. Moreover, these steady-state properties are not appropriate for multi-state appliance identifications. Hence, transient-state analysis is focused.

**Transient-state analysis:** It represents the waveform analysis during appliance switching. It has been reported that the majority of appliances exhibit unique transient characteristics and those transient properties demonstrate minor overlappings compared to steady-state signatures [95]. But, it requires high-sampling input waveforms to capture high-speed transients. Meehan et al. [105] compared the transient-state signals with the steady-state signals for the typical household appliances. For instance, the temporal current waveforms of a microwave during transient and steady states are visualised in Fig. 2.16. The visual representation proves that the transient-state contains more variations (high information gain) compared to steady-state.

Meziane et al. [106] proposed a set of novel turn-on transient features for electrical load identification and clustering. The feature set is based on the amplitude modulation, which describes the current magnitude variation from the load turn-on until it reaches a steady-state. Even though high-sampled (100 kHz) waveforms are used to extract unique signatures, its practical implementation possibilities are not investigated. Davies et al. [107] proposed a load specific transient classification technique based on deep neural networks. They have extracted transient features from the raw current waveform sampled up to 1.2 kHz and demonstrates that the load identification performance significantly improves with signal sampling rates. V-I trajectory-based transient signatures are extracted in [108] and [109] from 30 kHz and 100 kHz raw current and voltage signals for appliance identification, respectively. Ancelmo et al. [110] extracted the damping characteristics of initial transient current, usually named as inrush current to categorise the electrical loads, as shown in Fig. 2.17.

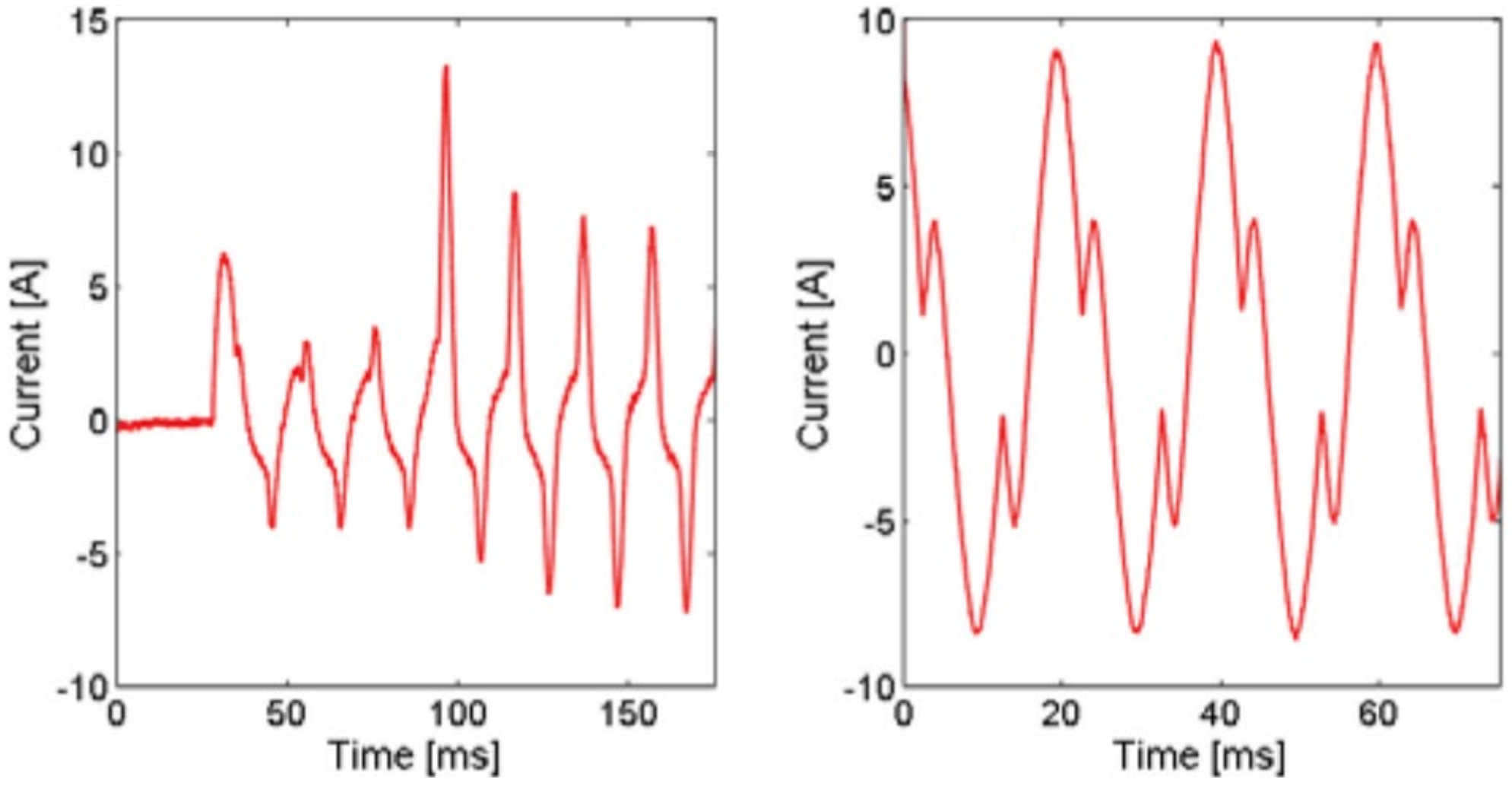

(a) (b)

Fig. 2.16. Current waveform of microwave. (a) Transient state (b) Steady state [95]

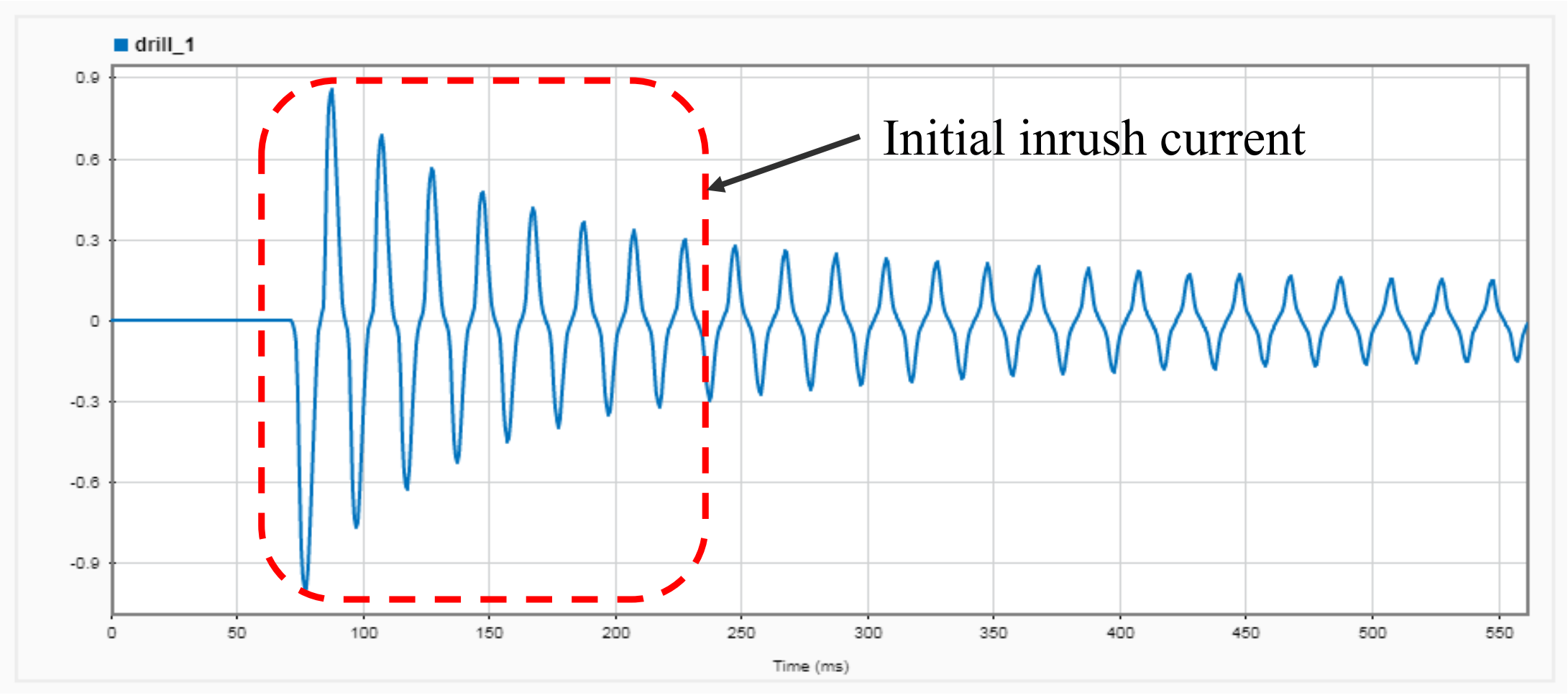

Fig. 2.17. Initial inrush current of an electric drill

Green et al. [111] proposed a framework based on a combination of geometric and statistical methods, which analyses the transient shapes and higher-order harmonics for NILM. Liu et al. [112] leveraged the transient power waveforms as the load-specific characteristics and measured its similarity with the existing patterns using dynamic time wrapping algorithm to classify the load transients. Active power transient curves are used in [113] for household appliance separation. Spectral envelopes of transient signals are employed to identify the non-linear loads such as variable speed drives [114]. Duarte et al. [115] extracted switching voltage transients based feature vector for load monitoring using wavelet transformation. Novel orthogonal high-order wavelet filters are developed in [116] for transient signature extraction from raw signals. Fisher linear discriminative analysis and correlation analysis are proposed in [117], to select an optimal transient-state feature set from the high-frequency signatures extracted from the raw signals that are sampled at 20 MHz.

As described above, the load specific signatures can be extracted from steady-state analysis or transient-state analysis or a combination of both. After the signature exploration, researchers have attempted to solve the load identification problem using two different approaches: (1) optimisation approach and (2) pattern recognition approach.

**Optimisation approach:** It aims to find a combination of electrical appliances which can produce an aggregated signature similar to the extracted features from the signal at the meter panel level. It can be achieved through formulating an objective function to minimise the difference between the selected aggregated signature and the extracted feature values. The objective function can be mathematically represented as below[118]:

$$min \vec{x}\, \varepsilon_j = g_j\left(\sum_{i=1}^{R}\left(x_i f_{i,j}\right), \varphi_j\right) \tag{2.1}$$

$$= \sum_{k=1}^{N}\left(\hat{y}_{(k|j)} - y_{(k|j)}\right)^2 \tag{2.2}$$

where $x_i$ – $i^{\text{th}}$ appliance, $f_{i,j}$ – feature $j$ of the $i^{\text{th}}$ appliance, $\varphi_j$- feature $j$ of the unknown aggregated load, R- total number of appliances, N- total number of points in feature $j$, $\hat{y}_{(k|j)}$ – feature $j$ extracted from the known load signatures and $y_{(k|j)}$ – feature $j$ extracted from the unknown aggregated load. In the case of only one device is turned on, the direct difference between the unknown load and each known appliances are calculated. The device with a minimal difference can be considered as the newly turned on load. But, when an unknown signature corresponds to more than one load, the optimisation becomes more complicated. Researchers have proposed Integer Programming (IP) [119][120] to solve this optimisation problem. Even though this approach performs better for the devices with known signatures, it struggles to identify unknown patterns. Moreover, this approach is not scalable since the computational complexity is exponentially increased with the number of devices.

**Pattern recognition approach:** It aims to identify the appliances by matching the detected features with the learned signatures one-by-one. It deviates from the optimisation approach since multiple appliance presences are not matched simultaneously in the pattern recognition process. Each appliance status (on / off) is determined without considering the state of other devices. This process can be mathematically explained as follows:

$$\vec{y_j} = \sum_{i=1}^{R} \left(x_i f_{i,j}\right) \; and \; \vec{y_j} \in \Re^N \tag{2.3}$$

$$x_i \in \{0,1\} \; and \; \vec{x} \in z^R \tag{2.4}$$

where $\vec{y_j}$ – unknown appliance feature $j$ extracted from the aggregated signal and $x_i$ - $i^{\text{th}}$ appliance status: 0 – Off, 1 – On. Researchers have applied several pattern recognition techniques for NILM such as neural networks [121][122][123][124], k-nearest neighbour algorithm [125][126][127] and Bayes classifier [128][129][130]. The pattern recognition approach is comparatively more robust and scalable, especially in identifying appliances with variable signatures.

The load identification accuracy heavily depends on the information gain of extracted signatures. As described above, steady-state signatures often overlap with other appliance's patterns. Hence a majority of the research works use very high-frequency signals (typically from 30 kHz – 20 MHz), for precise extraction of discriminative transient signatures. However, in practice, it is very challenging to process such high-resolution data in real-time, especially on embedded hardware such as smart meters. Therefore, a recommended sampling frequency and digitisation resolution need to be derived experimentally. The recommended parameters should provide sufficient load identification accuracy while ensuring the feasibility of real-time execution on low-power embedded hardware. The load identification methodology proposed in this thesis aims to fulfil the requirements mentioned above, with the recent technological advancements.

### 2.3.4. Near real-time load specific energy disaggregation

Consumer load specific energy disaggregation significantly contributes to reducing peak energy demand which is critical in rural microgrid operation through providing near real-time actionable feedback. It enables the possible consumer loads to shift from peak demand hours to off-peak period. Such a reduction in peak energy demand not only allows cheaper local energy generation but also reduce the energy requirement from backbone SWER networks to meet critical peak demand. Furthermore, the report released by the American Council for an energy-efficient economy states that appliance specific near real-time feedback paves the way to potential energy savings up to 12%, as shown in Fig. 2.18. Therefore, feedbacks generated through energy disaggregation facilitates the survival of existing SWER networks while providing an economical solution for increasing energy demand from remote consumers. In order to disaggregate load specific energy from an accumulated energy profile from the meter panel level, the energy usage pattern of individual appliances needs to be modelled. The overview of energy disaggregation process is outlined in Fig. 2.19.

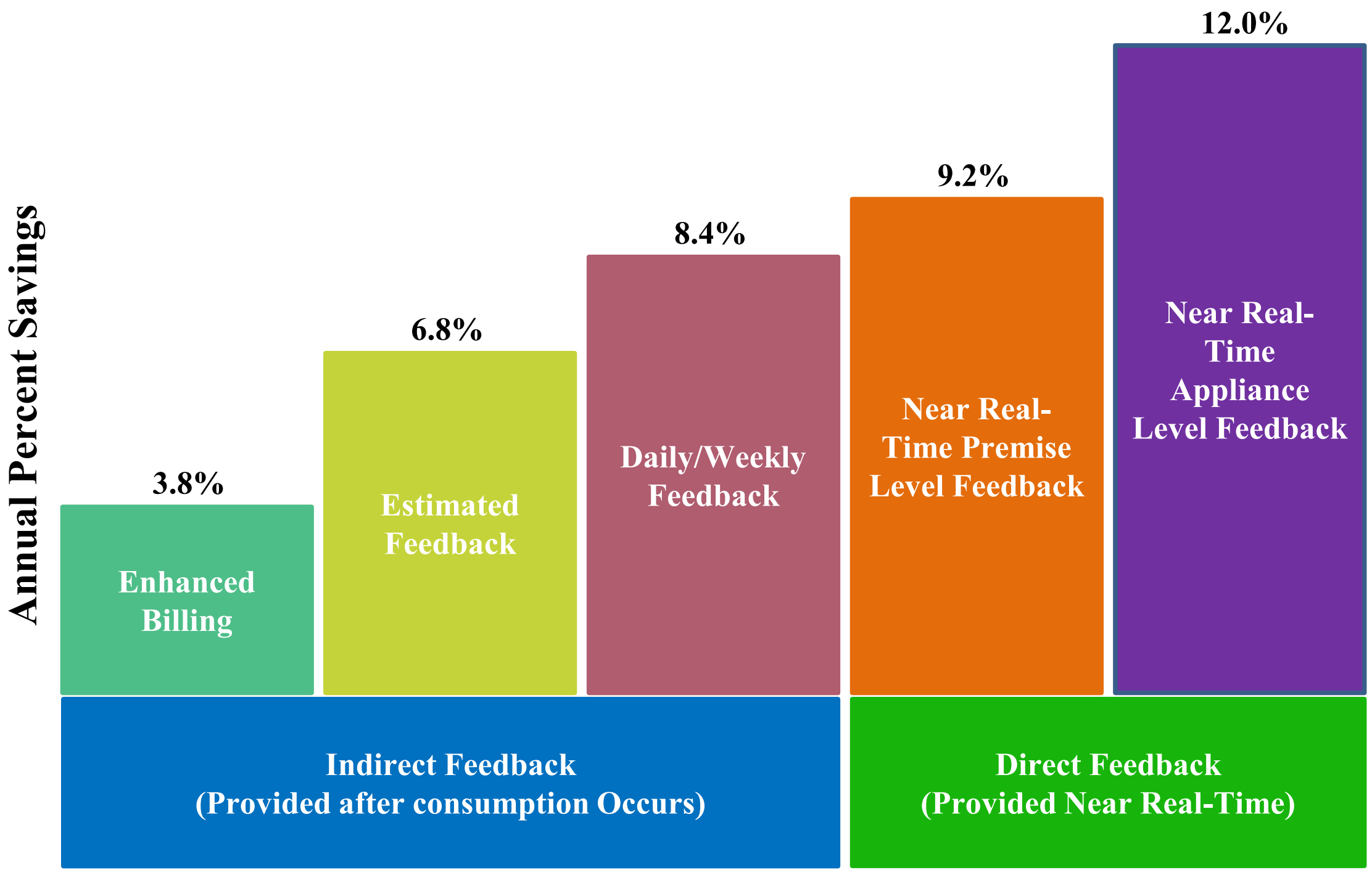

Fig. 2.18. Energy savings by feedback type

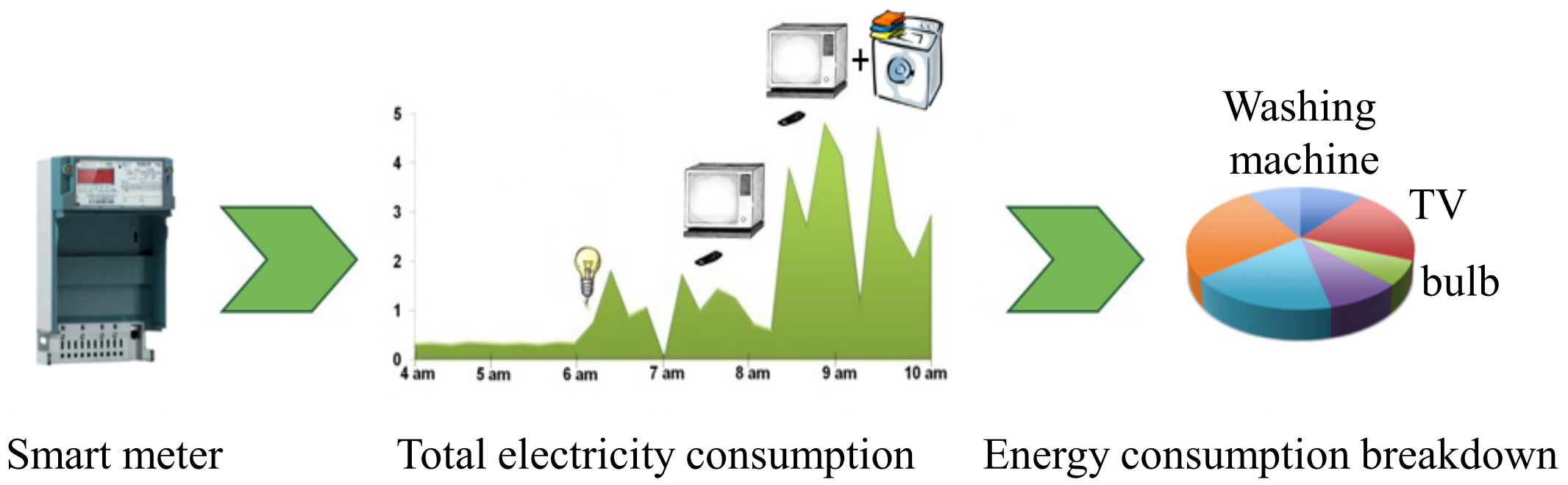

Fig. 2.19. Overview of load specific energy disaggregation

#### 2.3.4.1. Existing Energy Disaggregation Techniques

The primary goal of energy disaggregation problem is to divide the total electricity consumption into appliance specific energy usage. It can be mathematically represented as follows:

$$y[n] = \varepsilon[n] + \sum_{i=1}^{L} x_i[n], \;\; n \in \{t_0, t_1, \ldots\ldots T\} \qquad (2.5)$$

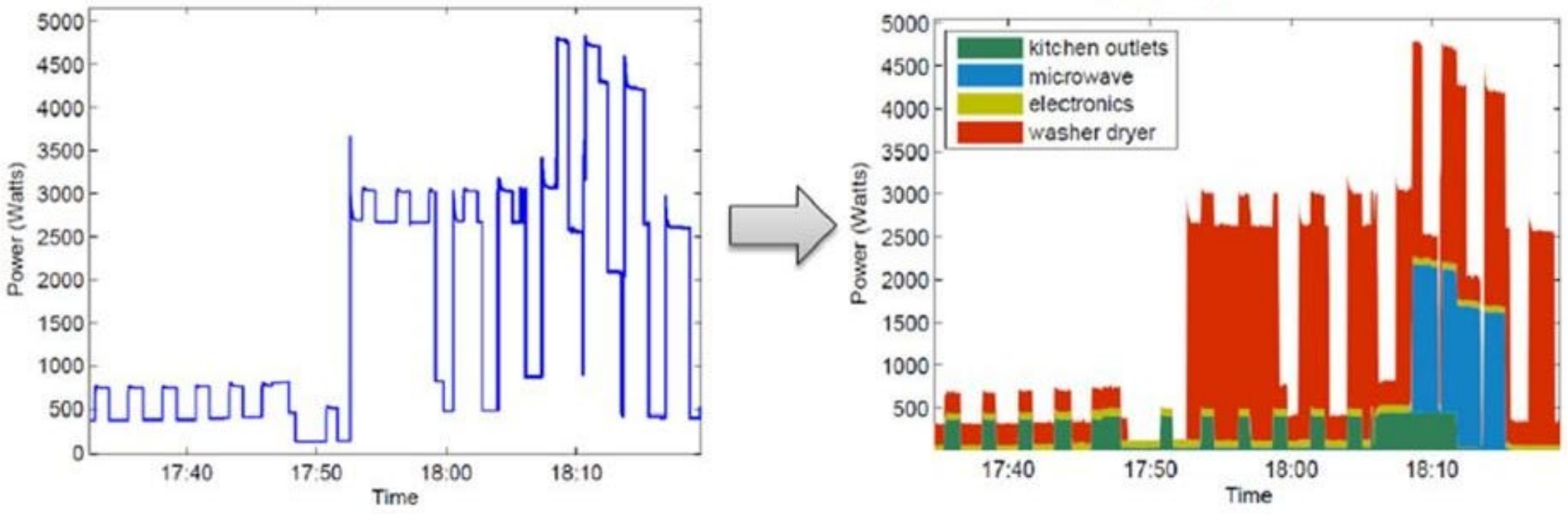

(a) (b)

Fig. 2.20. Load specific energy disaggregation process (a) total power (b) appliance specific power [131]

where $y[n]$ – total power at time $n$, $x_i[n]$ – power consumption of an $i^{\text{th}}$ appliance at time $n$ and $\varepsilon[n]$ – noise at time $n$. The energy disaggregation process is graphically explained in Fig. 2.20 [131]. Several state-of-the-art algorithms have been proposed to accomplish load specific energy separation.

Hidden Markov Model (HMM) is an approach chosen by several researchers for energy disaggregation[132][133]. HMM is used to model the energy consumption time series and represent the state of the appliances that are not directly observed. There are two states in HMM: (1) Observable state and (2) Hidden state. In the context of energy disaggregation, the observable state models the aggregate energy consumption, and the hidden states model the states of individual devices. Each hidden state is characterised using a probability distribution that corresponds to all the possible outputs [134]. A simple HMM can be represented as follows:

$$\lambda = \{S, O, P_o, A, B\} \tag{2.6}$$

where $S = \{s_1, s_2 \ldots., s_N\}$ – a finite set of an appliance's hidden state, $O = \{O_1, O_2, \ldots., O_T\}$ – a finite set of total power consumption observation, $P_o$ – initial probabilities, $A$ – probability matrix $(N \times N)$ for state transition from one state to the next state, $B$ – probability matrix $(N \times T)$ to detect a particular observation at the future state and the $A$, $B$ can be defined as:

$$A[i,j] = p(S_t = j \mid S_{t-1} = i) \tag{2.7}$$

$$B[j,t] = p(O_t = t \mid S_t = j) \tag{2.8}$$

where $\sum_i A[i,j] = 1$ and $\sum_j B[j,t] = 1$. The first step in the HMM application for energy disaggregation is to learn the model parameter $\lambda$ from the observed aggregated energy consumption $O$. After that, HMM can derive the optimal sequences of the hidden state $S$ from the learned parameter $\lambda$ and the given observations $O$. This process is known as learning and inference. A bunch of algorithms such as Viterbi decoding [135][136] and Baum-Welch expectation-maximisation algorithm [137][138][139] have been proposed to accomplish this task. However, these algorithms are not scalable and highly complex in space and time.

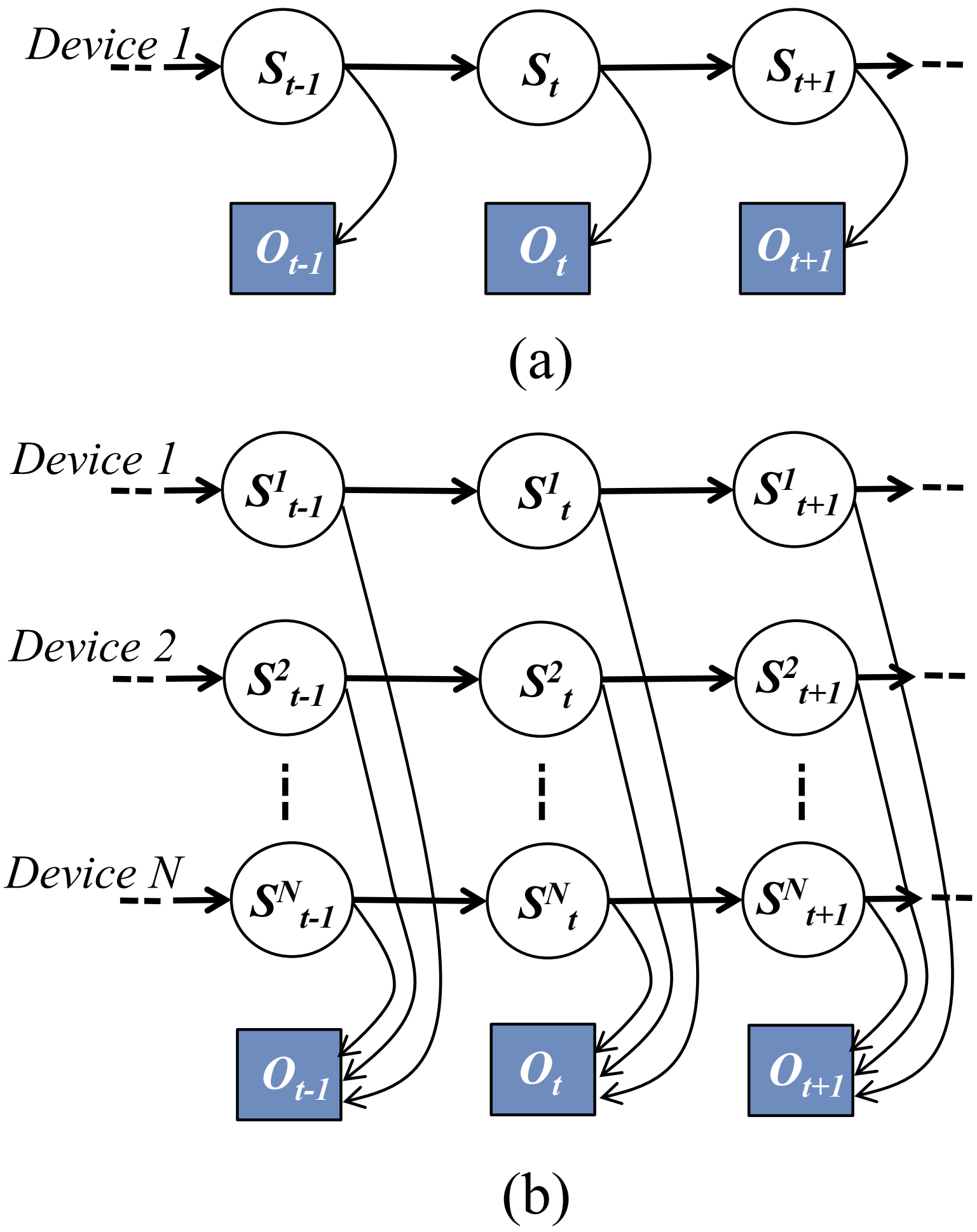

Fig. 2.21. Variations of HMM (a) Simple HMM (b) FHMM

Several variations of HMM such as Factorial HMM (FHMM) and super-state HMM (SSHMM) are used for energy separation. FHMM is a direct extension of HMM by adding multiple independent hidden state chains $S = \{S^1,\ S^2,\ \ldots.S^L\}$, where $S^i$ is the set of states of device $i$ and it can be represented as $S^i = \{s^i_1,\ s^i_2,\ \ldots.s^i_N\}$. Fig. 2.21 compares the internal structure of HMM and FHMM. The observations in FHMM are related to numerous hidden variables. Hence, FHMM is widely used to model the energy consumption of individual appliances independently. However, FHMM is often suffering from the local optima, and its computational complexity is high [95]. Kolter et al. [140] introduced an additive factorial approximation technique, which not only reduces the computational load of FHMM but also mitigates the local optima. Nevertheless, its performance on energy disaggregation of electronics loads is not satisfactory.

SSHMM is another variation of HMM, and it has been applied to energy breakdown. It varies from HMM by introducing super-states, which are the combinations of possible states on each appliance [141]. Makonin et al. [142] proposed a new Viterbi algorithm variant to compute sparse matrices with a large number of super-states efficiently. The sparse Viterbi algorithm preserves dependencies between loads and facilitates the energy breakdown for multi-state loads. Nashrullah et al. [143] applied a median data filter to the SSHMM to minimise the number of generated super-states, that reduces the computational complexity. The HMM-based variations demonstrate better performance only for the appliances with well-defined and controlled states. However, all the varieties of HMM struggle on energy breakdown for uncontrolled, multi-state and variable devices.

Apart from HMM-based models, Graph Signal Processing (GSP) is an emerging technique, which is widely applied to the emergy breakdown applications. Generally, GSP deals with the data defined by a graph. In energy disaggregation problem, the aggregated power consumption data is a signal $Y = [y_1, y_2, \dots \dots, y_n]$, where $y_i$ represents the power consumption at time instance *i.* In GSP algorithms, a weighted undirected graph is formulated to model the relationships and patterns that are hidden in the signal. The graph can be mathematically represented as follows [144]:

$$G = \{V, A\} \tag{2.9}$$

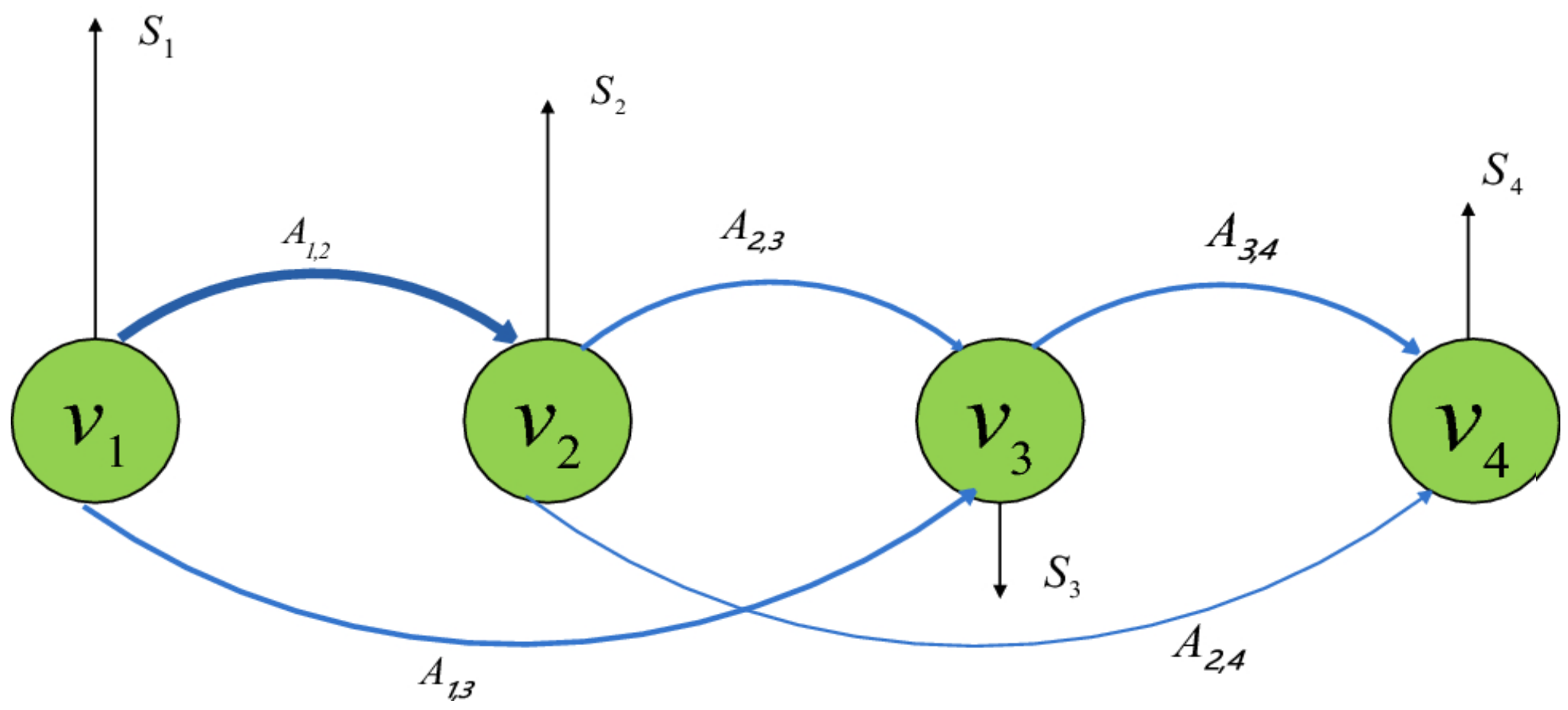

Fig. 2.22. Simple GSP model [145]

where $V = \{v_1, v_2, \ldots\ldots, v_m\}$ set of vertexes, $A$ is the weighted adjacency matrix, which represents the edges that connect the graph vertexes and $S$ depicts the mapping from $V$ to the output labels, as shown in Fig.2.22. Each element $y_i$ in the aggregated power signal is mapped to a vertex $v_i \in V$. Each weight $A_{ij}$ of the connecting edge between vertex $v_i$ and $v_j$ denotes the correlation strength between $y_i$ and $y_j$. Gaussian kernel weighting functions are widely used in the literature, and it can be explained as follows:

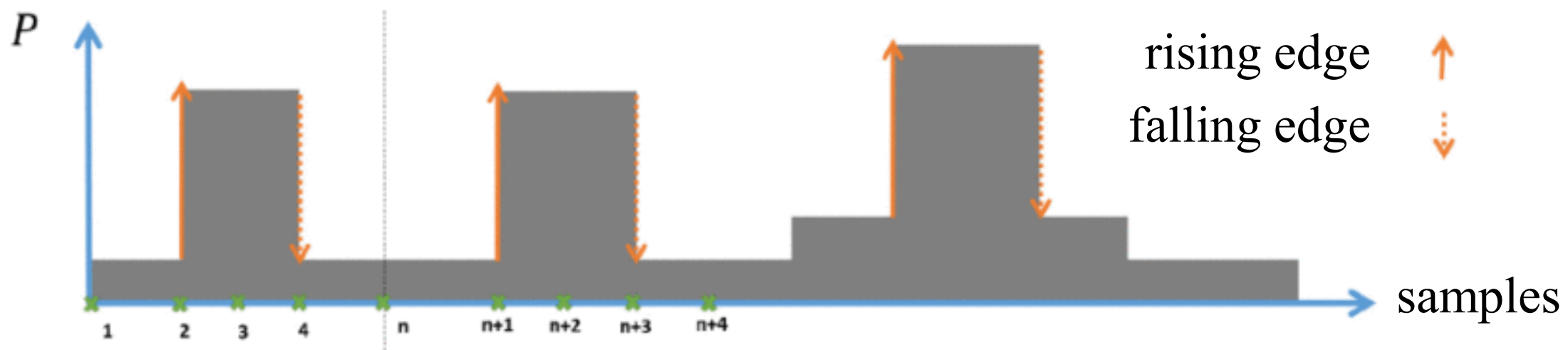

(a)

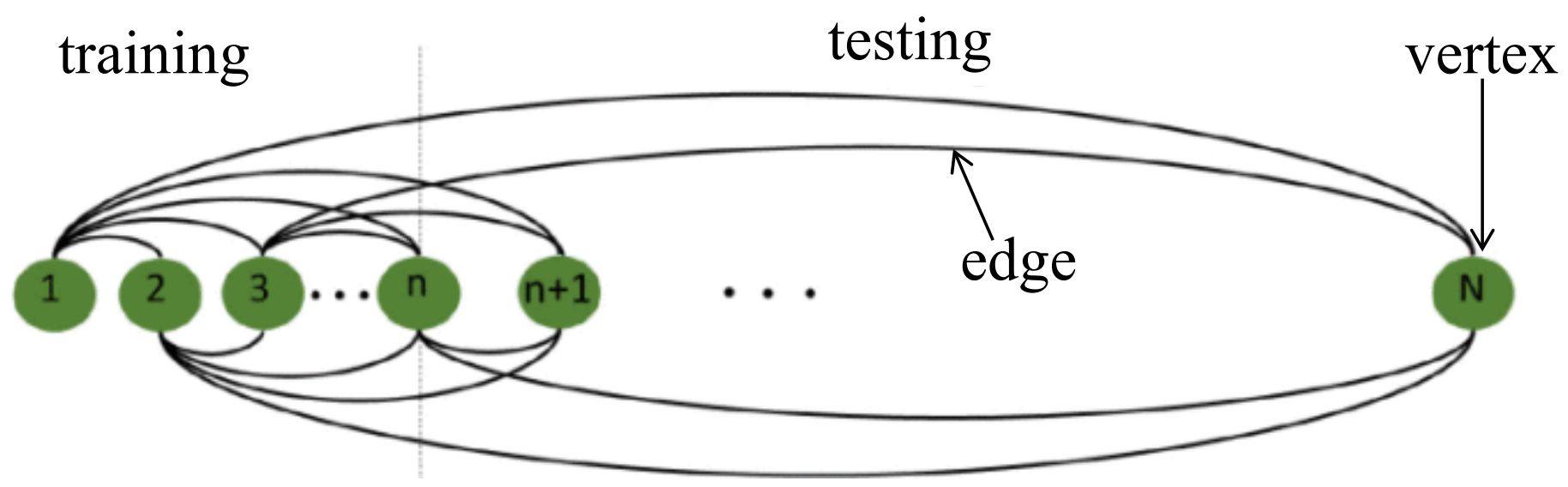

(b)

Fig. 2.23. GSP based energy disaggregation (a) aggregated power signal (b) corresponding GSP model [145]

$$A_{i,j} = exp\left[-\frac{\left(y_i - y_j\right)^2}{\varepsilon^2}\right] \tag{2.10}$$

where $\varepsilon$ is a scaling coefficient. A GSP based modelling process is outlined in Fig. 2.23 [145]. According to the power consumption signal shown in Fig. 2.23 (a), the weight of the edge between second and fourth vertex takes a large value since those two vertexes represent the events of the same device. Conversely, the weight of the link between second and third vertex is insignificant since there are no direct correlations. Hence, GSP is a powerful technique to represent the load specific energy consumption. However, most of the conventional GSP based algorithms suffer from large training overhead and computational complexity.

He et al. [145] leveraged the piecewise smoothness of the power signal to overcome the shortcomings in the conventional GSP approaches. It aims to find a smooth graph signal via variation minimisation. Besides, a simulated annealing technique is employed for further refinement of the proposed GSP-based energy breakdown technique. Zhai et al. [144] proposed a new graph learning algorithm to choose a suitable graph for energy consumption representation of an appliance. Kumar et al. [146] introduced regularisation methods to enhance the smoothness of the graph signal, which reduces the computational complexity of GSP methodologies. Zhao et al. [147] proposed an unsupervised GSP-based technique to disaggregate the loads from the low sample-rate signals. The proposed approach outperforms HMM-based methods and robust to noisy data. A semi-supervised GSP-based filtering and feature matching approaches are introduced to improve the event-based energy disaggregation tasks [148].

GSP-based approaches perform better only when the average power consumption of each device is distinct enough from the energy traces of other devices. Most of the proposed GSP-based techniques require manual load labelling after energy breakdown.

Furthermore, GSP based approaches often fail to disaggregate the loads with variable power consumptions.

In recent years, several deep learning techniques have been introduced into the field of energy disaggregation. Deep learning algorithms can automatically learn the relevant features from the aggregated power signal in order to model the energy consumption pattern of an appliance. Berg et al. [149] proposed a deep learning based neural energy decoder to disaggregate the power. It is achieved via identifying additive sub-components of the aggregated signal with an unsupervised learning approach. The performance of the proposed approach is validated with the signal sampled at high-frequency (12kHz). However, this approach is only applicable to the appliances with two-states.

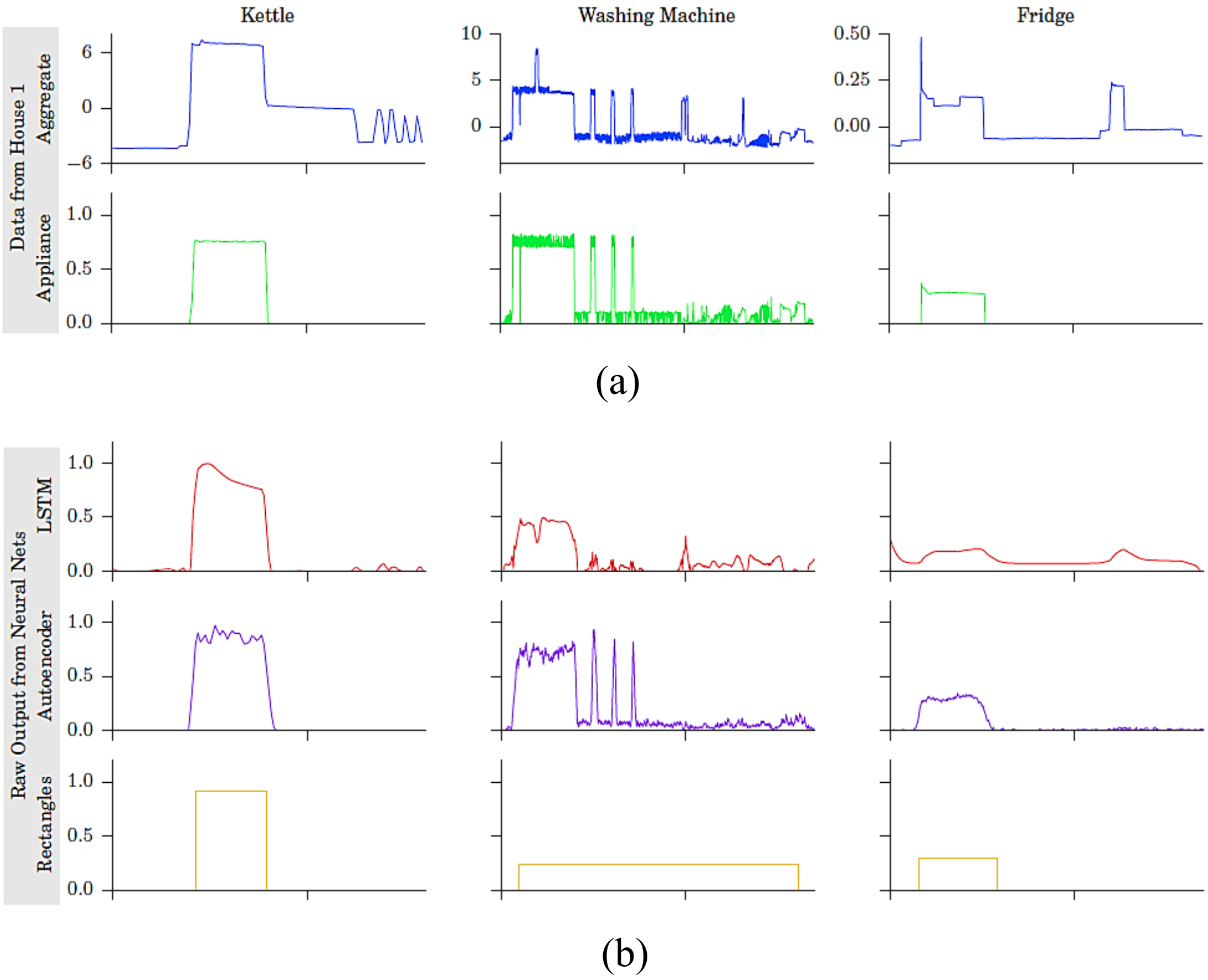

Fig. 2.24. Energy disaggregation performance comparison (a) aggregatted signal and appliance specific ground truth (b) predicted appliance specific energy consumption. [151]

An ensemble of neural networks has been introduced to address load specific energy breakdown from high-resolution current and voltage signals [150]. Kelly et al. [151] adapted three neural network architectures such as long short-term (LSTM) memory cell, denoising autoencoders (DAE) and rectangles, to perform energy disaggregation on power signals captured at low sampling frequencies. The predicted results are summarised in Fig. 2.24. Khodayar et al. [152] proposed a deep temporal dictionary learning approach along with an optimisation program to capture the non-linear temporal variations in the energy signals. Singh et al. [153] introduced a deep sparse coding technique for multi-level dictionary learning based approaches since they seem more promising to address the energy disaggregation problem.

A combination of FHMM with deep neural network (DNN) is used in [154] to separate the source devices from the aggregated signals. A Gaussian distribution is employed to model each device, whereas DNN models the aggregated power signal. Chen et al. [155] proposed a convolutional sequence to sequence model for power consumption disaggregation. The features are extracted using gated liner unit convolutional layers for device-specific energy model formation. A pre-processing stage is introduced in [156] to enhance the household power disaggregation. Kaselimi et al. [157] adapted a Bayesian-optimised bidirectional LSTM to identify the individual contribution of loads in the aggregate demand. Furthermore, a non-causal model is recommended to characterise the energy consumption of the multi-state appliances. An extensive comparison of state-of-the-art deep learning based energy disaggregation techniques can be found in [158].

However, the methodologies discussed above have just scratched the surface of the deep learning concepts. There are a vast number of techniques available in the deep learning community, which have huge potential to enlight the filed of energy disaggregation. Hence, it is certainly worthwhile to investigate in this direction. Furthermore, a huge amount of long-term energy consumption data is required to design, develop, test and benchmark more precise energy disaggregation algorithms. The following section summarises the existing real-world energy consumption datasets.

#### 2.3.4.2. Energy Datasets

Appliance based energy consumption modelling and disaggregation task need the aggregate energy demand data from a consumer site along with the ground truth power consumption of individual loads. Robust algorithm development requires real-world data collected from noisy environments.  State-of-the-art energy disaggregation algorithms are compared on different public datasets, and the results are reported in Fig. 2.25.

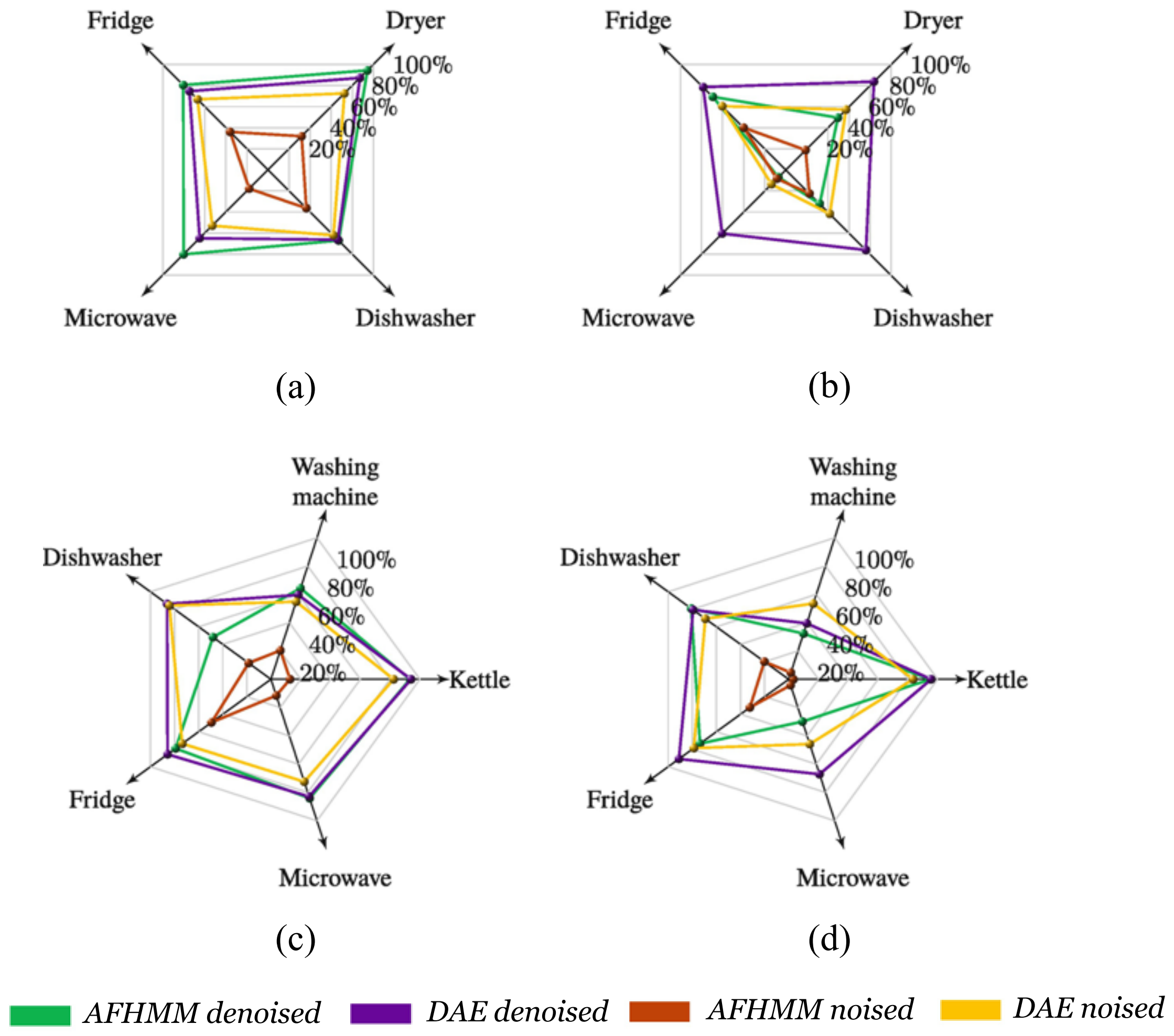

Fig. 2.25. Performance of disaggregation algorithms on different datasets (a) REDD seen (b) REDD unseen (c) UK-DALE seen (d) UK-DALE unseen. [158]

**Reference Energy Disaggregation Dataset (REDD):** It is the first and freely available energy dataset released by Massachusetts Institute of Technology [159]. REDD contains both high-frequency (15kHz) and low-frequency (0.5-1 Hz) data, recorded for a few weeks from 6 six different homes in the USA.

**United Kingdom- Domestic Appliance-Level Electricity Dataset (UK-DALE):** It is the first open-access UK energy dataset released by Imperial College [160]. It contains the aggregated active power demand (sampled at 44.1 kHz) and power consumed by individual loads (sampled at 1/6 Hz) from 5 different homes. These power consumption data is recorded for long-term (655 days). Hence, UK-DALE is more suitable and recommended for comprehensive algorithm development. The energy disaggregation framework proposed in this thesis is validated with this dataset.

Besides these, several other public datasets are available such as BLUED [161], DRED [162], AMPDS [163], REFIT [164] and ECO [165] for power breakdown applications. Based on the state-of-the-art energy breakdown results reported in Fig. 2.25, it is evident that there is a huge room for improved energy disaggregation algorithms. An improved energy disaggregation approach proposed in this thesis, which outperforms the state-of-the-art methods reported in the literature.

## 2.4. GENERALISED IMPLEMENTATION PROCEDURE OF MITIGATION STEPS

The implementation procedure of the mitigation steps such as power quality monitoring, HIF detection, load identification, and energy disaggregation, to establish a remote reliable electricity network can be broadly divided into three stages: (1) Data acquisition, (2) Feature extraction, and (3) Classification / Modelling / Decision making as shown in Fig. 2.26.

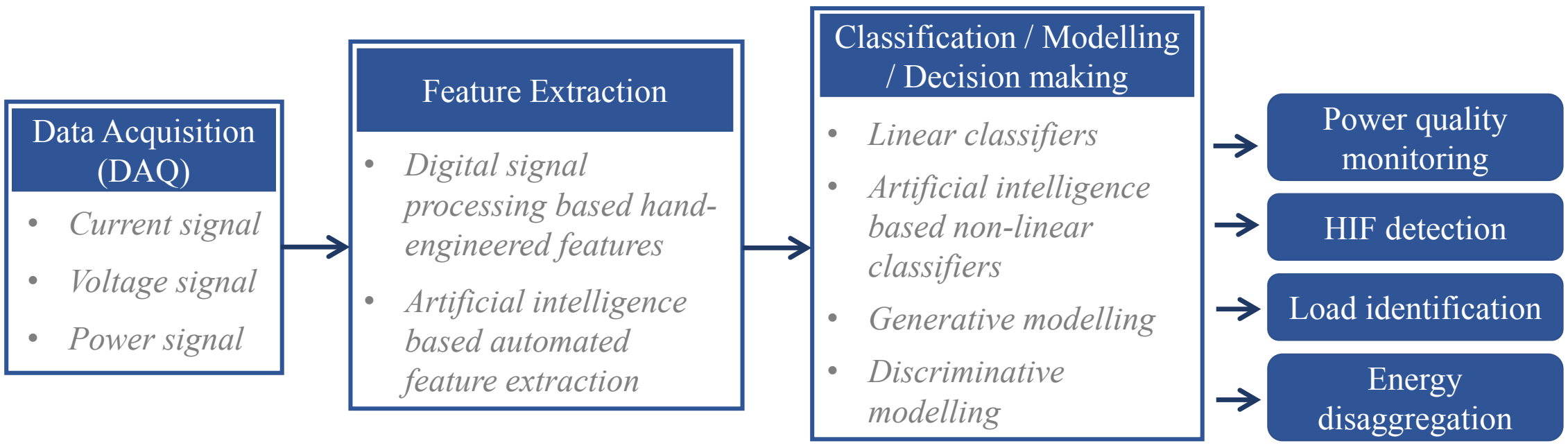

Fig. 2.26. Generalised implementation steps of electricity grid condition monitoring

### 2.4.1. Data acquisition

Data acquisition (DAQ) is the process of measuring analog signals such as current, voltage and power with a digital system. DAQ hardware digitalises incoming analog signal into digital representation so that any digital systems can interpret them. Since analog signals continuously vary over time, an Analog to Digital converter (ADC) is used to take periodic samples of the signal at a predefined rate named as the sampling rate. Another key specification in the selection of a digitiser is the resolution which is the ability to identify discrete values within the operating input range of the device. Signal sampling rate and resolution play a vital role in capturing all the information from a continuous-time signal and determines the information richness of the captured signal.

### 2.4.2. Feature extraction

Feature extraction is the process of extracting useful information from the digitised information-rich signals. Since most of the digitised signals are infinite, researchers typically separate them into small subsets for feature extraction and this process is named as windowing/framing. Informative features are extracted from the framed signal by applying feature engineering techniques that can select or combine the information in discrete time instances. The feature extraction process effectively transforms the higher dimensional data into a manageable dimensional feature vector, while preserving the data to describe the original signal completely. This process is also known as dimensionality reduction since it reduces the amount of redundant data for a given monitoring application.

Typically, the process of constructing explanatory features can be categorised into two approaches: (1) Hand-engineered feature extraction, (2) Automated feature extraction. In the context of the electricity grid condition monitoring, hand-engineered feature extraction often refers to the application of digital signal processing algorithms along with the domain knowledge to build features. On the other hand, automated feature extraction transforms the manual workflow through Artificial Intelligence (AI) based techniques that can automatically extract essential elements from the raw signals. State-of-the-art signal processing and AI techniques that can be used for feature extraction are introduced in the next chapter.

### 2.4.3. Classification, Modelling and Decision making

The extracted informative features are used to develop predictive models and classifiers. The desired goals can be achieved through deriving a correlation function between the input feature vector and output. Linear models or non-linear models can model these correlations. A linear combination of the input features is leveraged to calculate the output score in the linear models. On the other hand, non-linear models are required to improve accuracy when a decision-making task cannot be approximated well with the linear hyperplanes. Learning process to establish these models can be broadly categorised into two types: (1) Supervised learning, (2) Unsupervised learning. Supervised learning aims to build a model from the known input data associated with a target label to generate reasonable predictions for the unknown data. Diversely, unsupervised learning used to derive inferences from the data without any target labels. It is commonly used for cluster analysis to learn useful properties such as hidden patterns or groupings of a dataset.

On the other hand, there are two extensive classes of approaches for determining the parameters of linear or non-linear models: (1) Generative approach and (2) Discriminative approach. The generative approach aims to learn each possible output independently and determine the more likelihood output for the input features. Contrarily, the discriminative approach seeks to determine the differences that can separate each output rather than learning the properties of each output. Thus, generative approaches are often used when there are requirements to derive a model from a data distribution. In contrast, discriminative methods are more suitable to find the boundaries that separate the input data into different output classes. State-of-the-art AI-based non-linear modelling approaches that can be used for classification and decision making are described in the next chapter.

## 2.5. CHAPTER SUMMARY

An introduction of the remote electricity grids and the role of SWER networks in rural electrification are summarised in this chapter. The significant challenges such as power quality issues, bushfire risks due to high impedance arcing faults, the rapid increase in energy demand and its management in remote electricity networks are discussed. The alleviation steps such as power quality monitoring, real-time HIF identification, remote consumer load identification and near real-time load specific energy disaggregation are outlined that can facilitate the establishment of reliable rural electricity networks. A comprehensive review of the existing state-of-the-art monitoring solutions is reported under each mitigation step. Finally, the generalised alleviation steps implementation procedure which contains DAQ, feature extraction, and classification/ modelling/decision making are summarised. The following chapter provides a detailed description of the digital signal processing techniques and AI-based approaches that are used in this thesis. Furthermore, it highlights the limitations on the real-time deployment of monitoring techniques to facilitate the development of algorithms with real-time execution feasibility.

# CHAPTER 3

# 3. DIGITAL SIGNAL PROCESSING AND AI-BASED TECHNIQUES FOR ELECTRICITY GRID CONDITION MONITORING, AND ITS REAL-TIME IMPLEMENTATION LIMITATIONS

## 3.1. Introduction

The modern sensing and digitising technologies enable the information-rich signal acquisition from the distribution feeders as well as the consumer premises. State-of-the-art remote electricity network condition monitoring applications heavily rely on digital signal processing and AI-based techniques to transform the information-rich signals into a required output that can enhance the network operations [1]. This transformation process contains several phases such as feature extraction, modelling, classification and decision making. In this chapter, the digital signal processing and AI-based techniques used in the proposed remote electricity grid monitoring methodologies are briefly introduced. Then, the real-time implementation requirements for each monitoring applications such as power quality monitoring, HIF detection, consumer load identification and appliance specific energy usage monitoring are summarised. Finally, the hardware deployment architectures and its resource limitations are studied to identify the bottlenecks in the existing condition monitoring techniques.

## 3.2. Digital Signal Processing Techniques

The power system signals like current, voltage and power are continuous-time signals that are analog in nature. These analog signals need to be digitised by the ADC through quantisation in order to manipulate them for information extraction mathematically. During the digitisation, the continuous signals are transformed into a sequence of digits that represents discrete samples of a continuous variable. Digital signal processing represents the manipulation of those discrete samples.

In digital signal processing, engineers and researchers typically analyse the digitised signals in different domains such as time-domain, frequency-domain and time-frequency (hybrid) domain based on the target application requirements. The signal analysis domain needs to be chosen appropriately since each domain representation can express different essential characteristics of a signal. The signal domain transformation functions and processing methods used in this thesis are outlined below.

### 3.2.1. Fourier Analysis

The Fourier analysis is a mathematical transformation technique that transforms a time-domain signal $x(t)$ into a function of frequency domain $X(\omega)$. Fourier analysis for continuous-time signals can be categorised into two divisions: (1) Periodic signals and (2) Aperiodic signals. Conversion process that is used for periodic time signals is named as Fourier series, whereas the method used for aperiodic signals are called as the Fourier transform[2].

The Fourier series synthesis equation:

$$x(t) = a_0 + \sum_{n=1}^{\infty} a_n \cos(\omega t n) - \sum_{n=1}^{\infty} b_n \sin(\omega t n) \tag{3.1}$$

where $a_0$, $a_n$ and $b_n$ coefficients hold the amplitudes of the direct current, cosine and sine waves, respectively. Thus, the Fourier series analysis equation:

$$a_0 = \frac{1}{T} \int_{-T/2}^{T/2} x(t)dt \tag{3.2}$$

$$a_n = \frac{2}{T} \int_{-T/2}^{T/2} x(t) \cos\left(\frac{2\pi tn}{T}\right) dt \qquad (3.3)$$

$$b_n = \frac{-2}{T} \int_{-T/2}^{T/2} x(t) \sin\left(\frac{2\pi tn}{T}\right) dt \qquad (3.4)$$

where $x(t)$ and $T$ are the time domain signal and period of the signal, respectively.

The Fourier transform synthesis equation:

$$x(t) = \frac{1}{\pi} \int_{0}^{\infty} Re\, X(\omega) \cos(\omega t) - Im\, X(\omega) \sin(\omega t)\, d\omega \qquad (3.5)$$

where $Re\, X(\omega)$ and $Im\, X(\omega)$ are the real and imaginary parts of the frequency spectrum, respectively. Thus, the Fourier transform analysis equation:

$$Re\, X(\omega) = \int_{-\infty}^{+\infty} x(t) \cos(\omega t)\, dt \qquad (3.6)$$

$$Im\, X(\omega) = -\int_{-\infty}^{+\infty} x(t) \sin(\omega t)\, dt \qquad (3.7)$$

Even though the Fourier analysis states that a time-domain signal $x(t)$ is formed by summing an infinite number of scaled sine and cosine waveforms, processing unlimited number of time-frequency instances are not feasible with the digital systems. Thus, a finite-duration discrete sequence $x[n]$ needs to be sampled from the continuous-time domain signal $x(t)$ where $n$ is the number of samples. The sampling frequency of a digital signal is the number of samples per second.

Nyquist-Shannon sampling theorem establishes a sufficient condition for a sampling frequency that allows a discrete sequence to capture all the properties from a continuous-time signal of finite bandwidth. The theorem states that the acceptable sample rate for a perfect reconstruction of a time-domain signal x(t) which contains no frequencies higher than $B$ Hz, should be greater than $2B$. Thus, it is essential to band limit the signal before digitising in order to avoid the imperfections known as aliasing during signal reconstruction. Finite-bandwidth signals can be obtained by considering the limited bandwidth that contains most of the energy of the signal [3]. In order to make sure the frequency content of an input signal is limited, an analog low pass filter (also known as anti-aliasing filter) is generally placed before the sampling process.

Discrete Fourier Transform (DFT) can be used to convert the finite-duration, finite-bandwidth discrete-time signals into the frequency domain. DFT accepts the discrete-time input $x[n] = (x[0], x[1], \ldots\ldots, x[N-1])$ and returns the same number of frequency outputs $X[m] = (X[0], X[1], \ldots\ldots, x[N-1])$. The transfer function can be defined as follows:

$$X[m] = \sum_{n=0}^{N-1} x[n] e^{-2\pi j m n / N} \tag{3.8}$$

where $N$ is the number of input samples, and $m$ is the frequency band number $m = 0, 1, \ldots\ldots, N-1$. The frequency band size determined by the frequency resolution $\Delta f = \frac{f_s}{N}\ Hz$, where $f_s$ is the sampling frequency. The computed $X[m]$ values are complex and contain the information regarding the amplitude (energy) and phase of a certain frequency range sine waves. Direct evaluation of $X[m]$ requires $O(N^2)$ operations since there are $N$ number of outputs in $X[m]$, and each output involves the addition of $N$ terms. In order to reduce this complexity, the Fast Fourier Transform (FFT) algorithm is introduced that can compute the same results in $O(N \log N)$ operations through eliminating trivial processes. The Cooley-Tukey algorithm which operates in divide and conquers approach is the well-known and widely used FFT technique in power system analysis. This FFT implementation not only reduces the computational complexity but also eliminates excessive memory usage, which is critical in embedded devices such as smart relays and electrical pole-mounted monitoring devices.

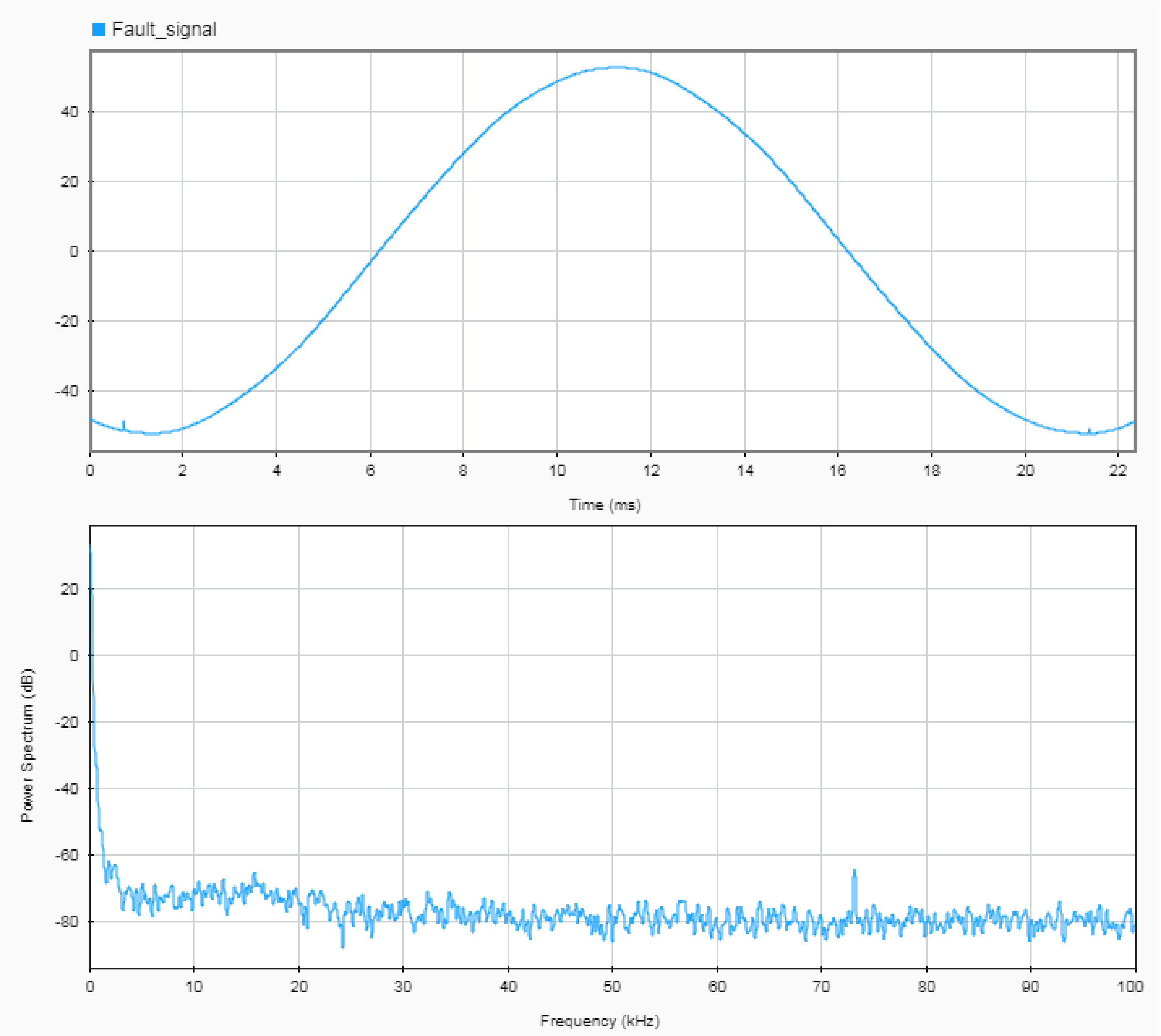

Fig. 3.1. Fault signal and its frequency spectrum computed by FFT

Fig. 3.1. shows a fault current signal sampled at 200 kHz and its frequency spectrum that visualises the signal energy distribution over the frequency range (0 – 100 kHz) based on Nyquist sampling theorem. These energy distribution patterns can be used as informative features for condition monitoring applications. However, the frequency spectrum does not preserve any time information. Thus, it is difficult to identify the precise time of fault occurrence, which is essential in condition monitoring applications.

The short-time Fourier transform (STFT) is developed to achieve a balance between time and frequency through sliding a short moving window along with the time series and calculating the FFT of the short frame [4]. The resulting expansion enables the visual representation of the frequency spectrum of a signal as it changes with time known as a spectrogram. Fig. 3.2. visualises the spectrogram (time-frequency representation) of the same fault signal as in Fig. 3.1. where the time information was missed. In the spectrogram visualised in Fig. 3.2., the x-axis represents time, the y-axis represents frequency, and the third dimension indicated by colour intensity represents the amplitude of a specific frequency at a particular time. However, it is not feasible to achieve precise time resolution simultaneously with accurate frequency resolution with STFT since it has a fixed resolution. A broad time signal window provides better frequency resolution but results in poor time resolution. On the other hand, a lean time signal window offers better time resolution but fails to provide good frequency resolution, as shown in Fig. 3.3. Thus, the STFT window length needs to be decided based on the target application resolution requirements.

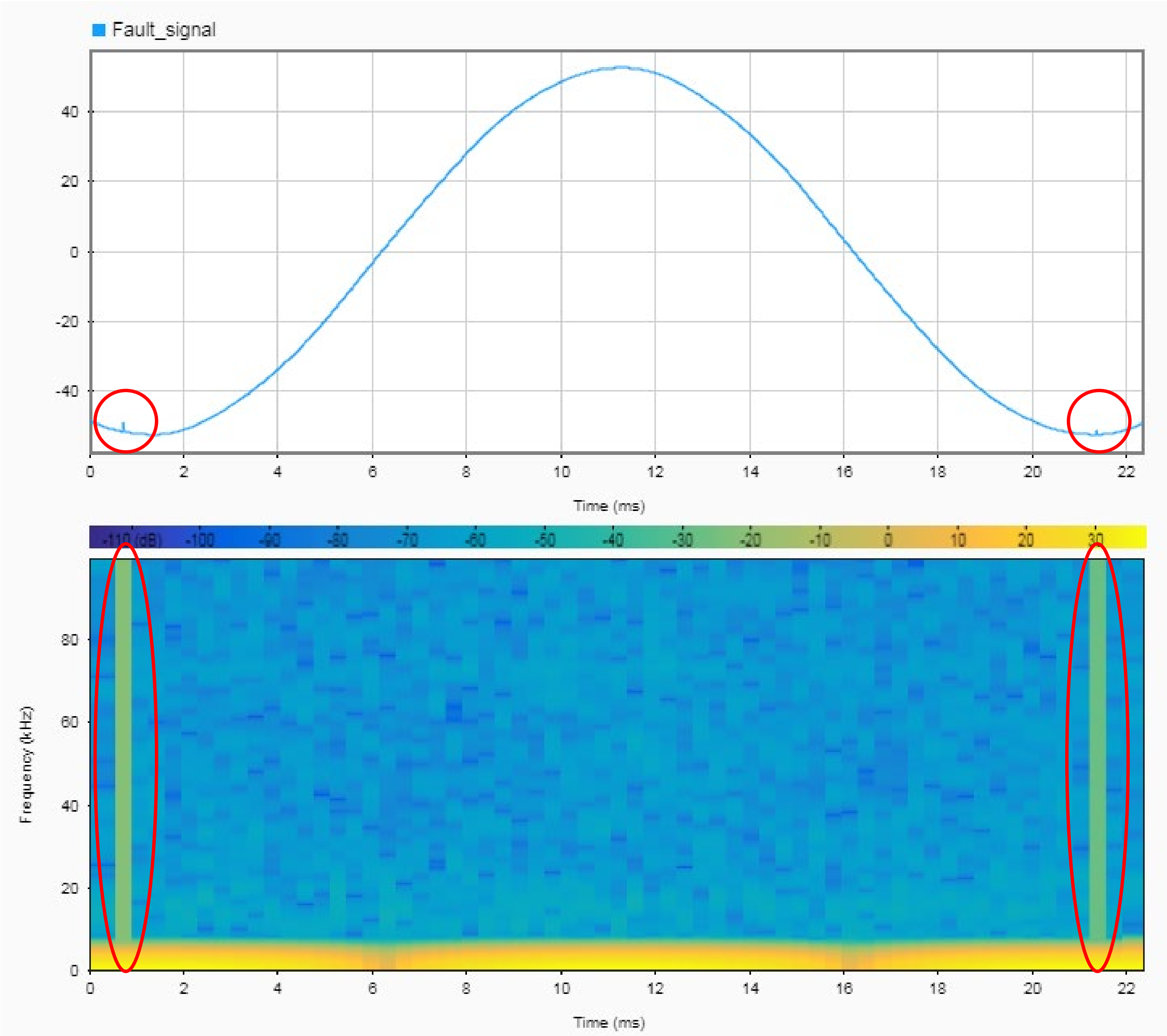

Fig. 3.2. Fault signal and its time - frequency representation computed by STFT

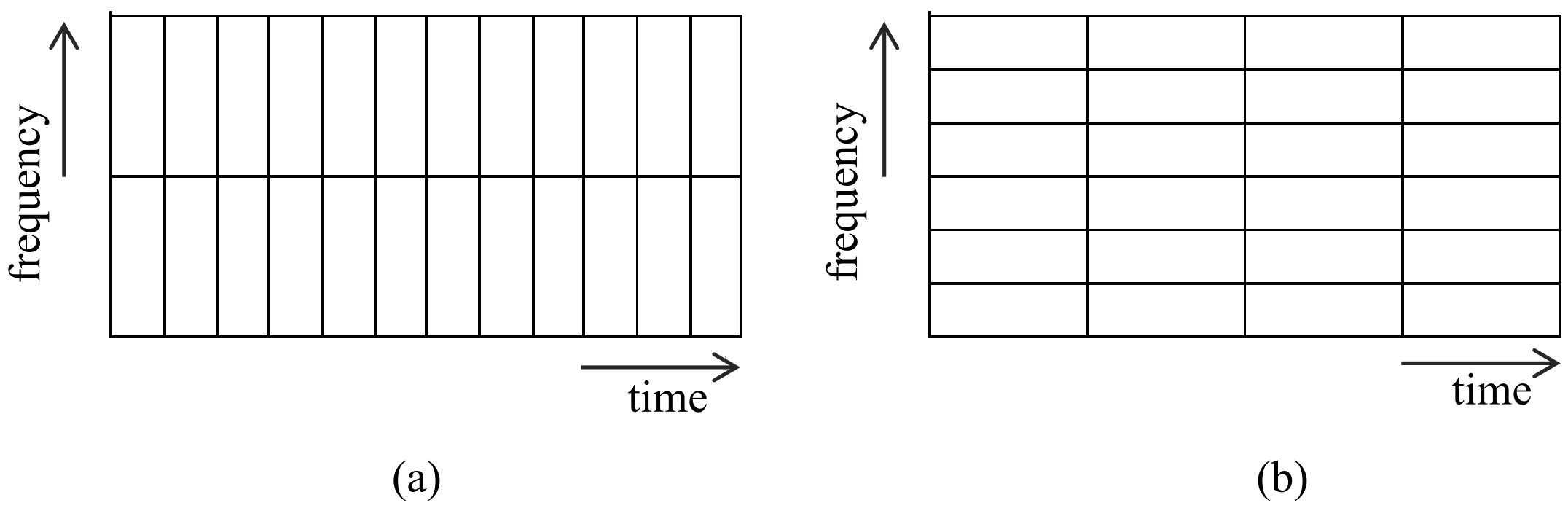

Fig. 3.3. STFT resolution comparison (a) precise time resolution (b) precise frequency resolution

### 3.2.2. Wavelet Analysis

Wavelet analysis is introduced to overcome the fixed time-frequency resolution, which is considered as a limitation with the Fourier analysis. It enables multiresolution analysis such that, it is possible to analyse a signal at different frequencies with different resolutions, whereas in STFT, all frequencies can be analysed with a constant resolution, as shown in Fig. 3.4.

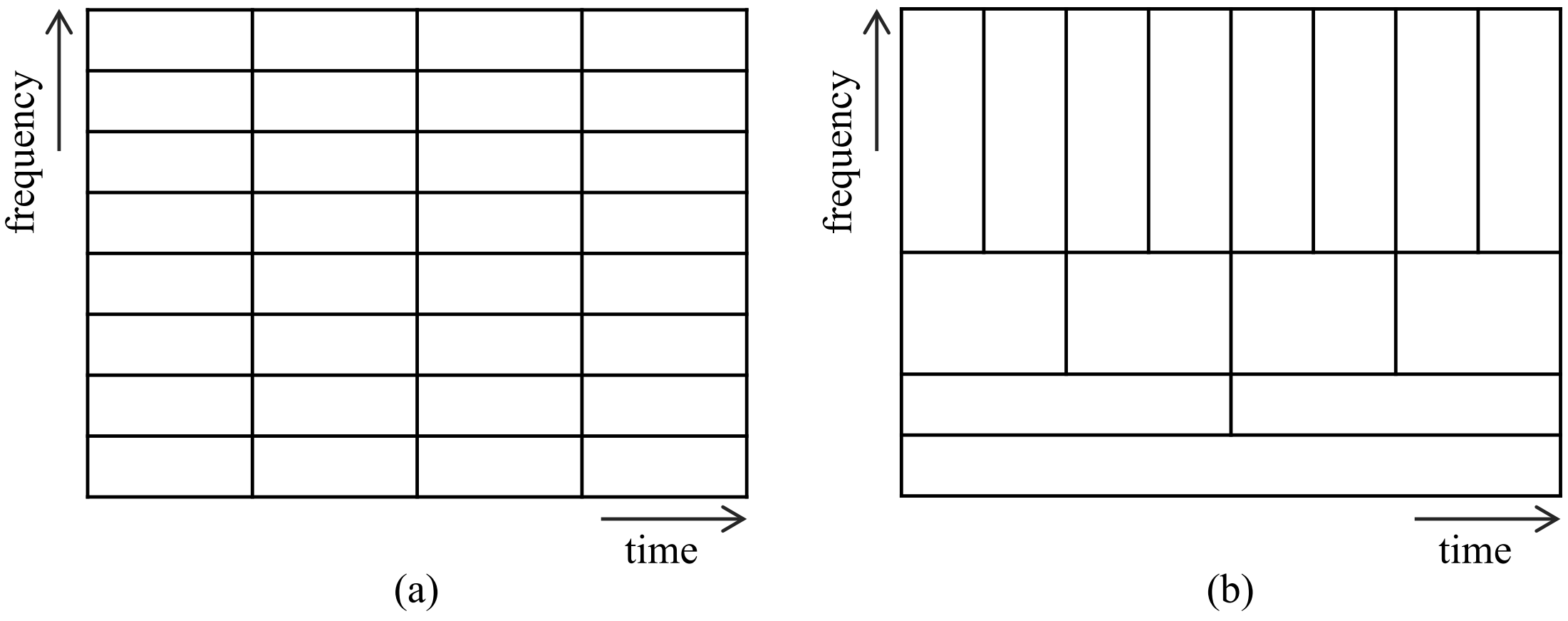

Fig. 3.4. STFT vs Multiresolution analysis comparison (a) STFT (b) Multiresolution time-frequency plane

In condition monitoring applications, generally, the low frequencies such as fundamental component and lower order harmonics exist for the entire duration of a signal. Thus, poor time resolution and precise frequency resolution are allocated for lower frequencies, as outlined in Fig. 3.4. (b). On the other hand, precise time resolution and poor frequency resolution are assigned to higher frequencies since high-frequency transients persist only for a short amount of time.

The Wavelet analysis computes the correlation between the analysed signal and a wavelet function $\psi(t)$ which is a wave-like oscillation. It can be scaled and shifted to extract intrinsic signal properties in a different time and frequency resolutions. The wavelet function $\psi(t)$ is also known as mother wavelet, which has an amplitude starts with zero, then oscillates and terminates at zero, as shown in Fig. 3.5.

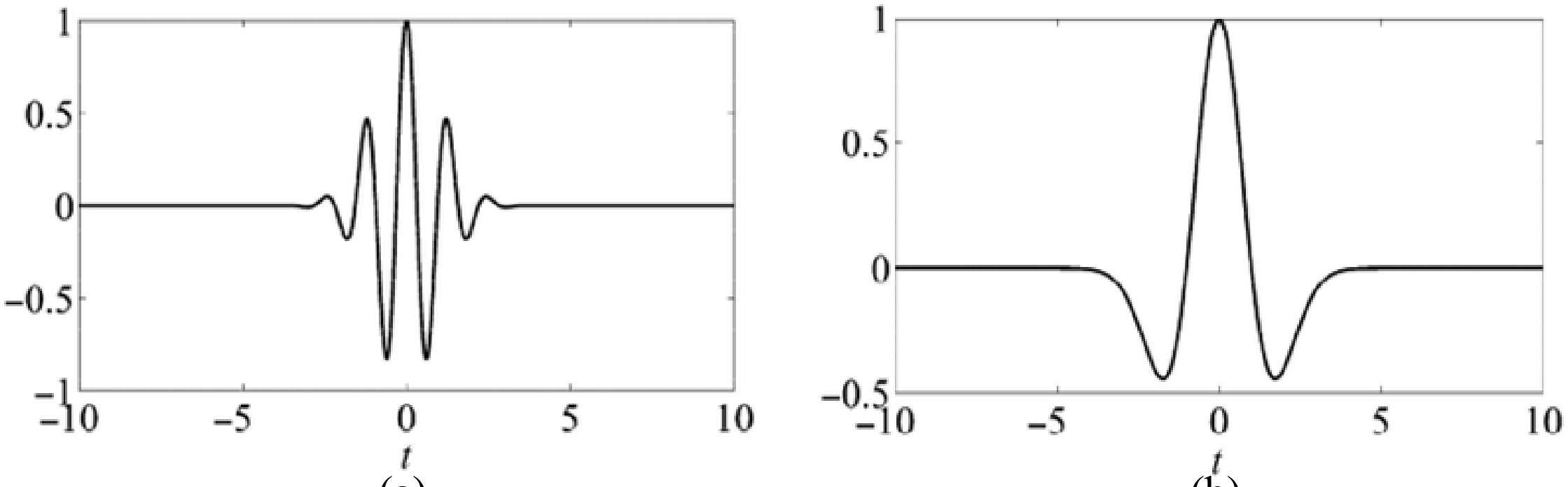

(a) (b)

Fig. 3.5. Mother wavelets (a) Morlet (b) Mexican hat

The mother wavelet function $\psi(t)$ should satisfy the following mathematical criteria:

$$E = \int_{-\infty}^{\infty} |\psi(t)|^2 dt < \infty \tag{3.9}$$

where $E$ represents energy, and a wavelet should have finite energy.

$$C_{\psi} = \int_{0}^{\infty} \frac{|\psi(\omega)|^2}{\omega} \, d\omega \; < \; \infty \tag{3.10}$$

where $\psi(\omega)$ is the Fourier transform of the wavelet function $\psi(t)$, $\psi(0) = 0$.

Such mother wavelet function $\psi(t)$ is shifted with a translation parameter $\tau$ and dilated with a scale factor $s$ to extract wavelet coefficients $X_{CWT}(\tau, s)$ of the continuous-time signal $x(t)$ that can be used as the features for condition monitoring applications. The continuous wavelet transformation (CWT) to compute the wavelet coefficients $X_{CWT}(\tau, s)$ can be defined as follows:

$$X_{CWT}(\tau, s) = \frac{1}{\sqrt{|s|}} \int_{-\infty}^{\infty} x(t)\psi^{*}\left(\frac{t-\tau}{s}\right) dt \tag{3.11}$$

where $\psi^{*}$ represents the complex conjugate of the mother wavelet in case of a complex wavelet. In each scale $s$, the signal energy is normalised through the division of wavelet coefficients by $\sqrt{s}$ to maintain the same energy range throughout the scales [5]. Mother wavelet function $\psi(t)$ always associates with a centre frequency $f_c$ in each scale where the scale $s$ is inversely proportional to that frequency. Thus, lower frequencies correspond to a larger scale, whereas higher frequencies correspond to a smaller scale. A change in scale $s$ not only changes the wavelet centre frequency but also alters the frame length, as shown in Fig. 3.6. On the other hand, shifting parameter $\tau$ represents the location of the mother wavelet in the time domain. Thus, the wavelet can be shifted along with the signal by varying $\tau$. The resulting wavelet coefficients are two-dimensional representation in which the time axis can be computed through running $\tau$ for a constant scale $s$, and the scale (frequency) axis can be calculated by changing scale $s$ for a constant $\tau$. Therefore, each wavelet coefficient is associated to a scale and a time instance.

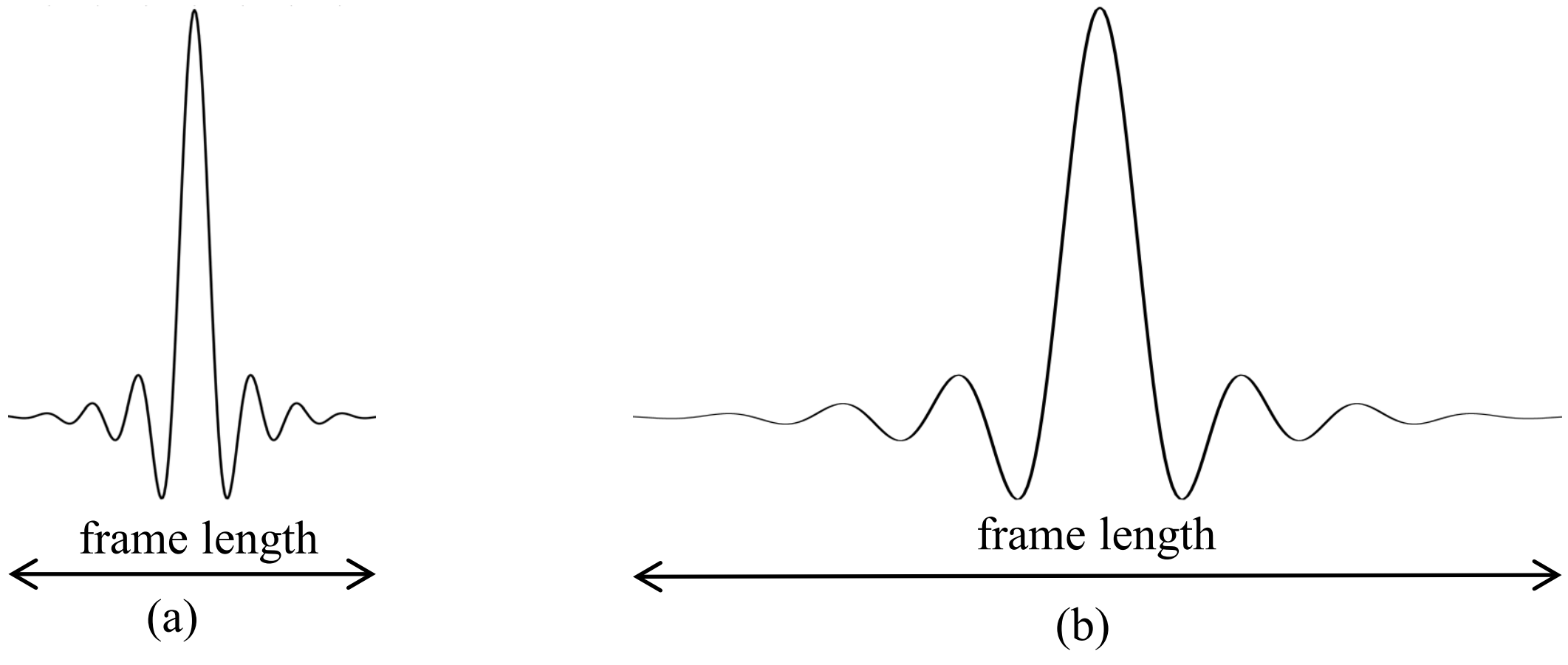

Fig. 3.6. Meyer mother wavelet in different scales (a) small scale (b) large scale

However, computation of wavelet coefficients at every possible scale and time instance is a huge computation-intensive task since it continuously dilates and contracts the mother wavelet along with the signal. Furthermore, it generates a massive amount of redundant data which makes the data analysis process tedious in real-time condition monitoring applications. The computational complexity can be reduced by choosing a subset of discrete scales and time instances [6]. This discretisation procedure can be expressed in mathematical terms such that the discrete scale can be denoted as $s = s_0^j$, and the discrete translation as $\tau = ks_0^j\tau_0$ where $s_0 > 1, \tau_0 > 0, j, k \in \mathbb{Z}$. Typically, dyadic scales and positions based on powers of two ($s_0 = 2^1, \tau_0 = 2^0$) are used to make the signal analysis as efficient as well as accurate [7]. Hence, the mother wavelet function is given as follows:

$$\psi_{j,k}(t) = s_0^{-j/2}\psi\left(s_0^{-j}t - k\tau_0\right) \tag{3.12}$$

The discretised version of CWT (DCWT) can be derived using the discrete mother wavelet function $\psi_{j,k}(t)$. The derivation can be formulated as follows:

$$X_{DCWT}(j,k) = \int x(t)\psi_{j,k}^*(t)\, dt \tag{3.13}$$

where $X_{DCWT}(j,k)$ is the wavelet coefficients at a discrete scale $j$ and location $k$. The necessary and sufficient condition for this wavelet coefficients $X_{DCWT}(j,k)$ can be outlined as follows:

$$A\|x\|^2 \le \sum_j \sum_k |X_{DCWT}(j,k)|^2 \le B\|x\|^2 \tag{3.14}$$

where $\|x\|^2$ represents the total energy of the signal $x(t)$, $A > 0$, and $B < \infty$.

Even though the DCWT enables the efficient computation of the CWT through discrete sampling, it still produces highly redundant data as far as the signal reconstruction is concerned. On the other hand, discrete wavelet transform (DWT) produces sufficient details for analysis and synthesis of the original time-domain signal with a significant reduction in computation time [7]. DWT implementation process is considerably easier than CWT since it is based on the concept of multiresolution filter banks and wavelet filters. Digital filtering techniques are used in DWT to obtain the time-frequency representation of a discrete-time signal. The signal is analysed at different frequency bands using digital filter banks.

A filter bank contains several filters that can decompose a time-domain signal into different frequency bands, as shown in Fig. 3.7. A discrete-time signal $x[n]$ is the input signal to the wavelet filter bank sampled at $f_s$. In the first level of signal decomposition, $x[n]$is filtered through $H[n]$ and $L[n]$ which are high-pass and low-pass filters, respectively. The filtered output from both filters contains the same amount of sample points as the input signal, even though each of them only represents half of the frequency components in the original signal. Thus, the output signals are downsampled by a factor of 2, as shown in Fig. 3.7. The high-pass and low-pass filter outputs are also known as details ($d$) and approximations ($a$) of the signal $x[n]$, respectively.

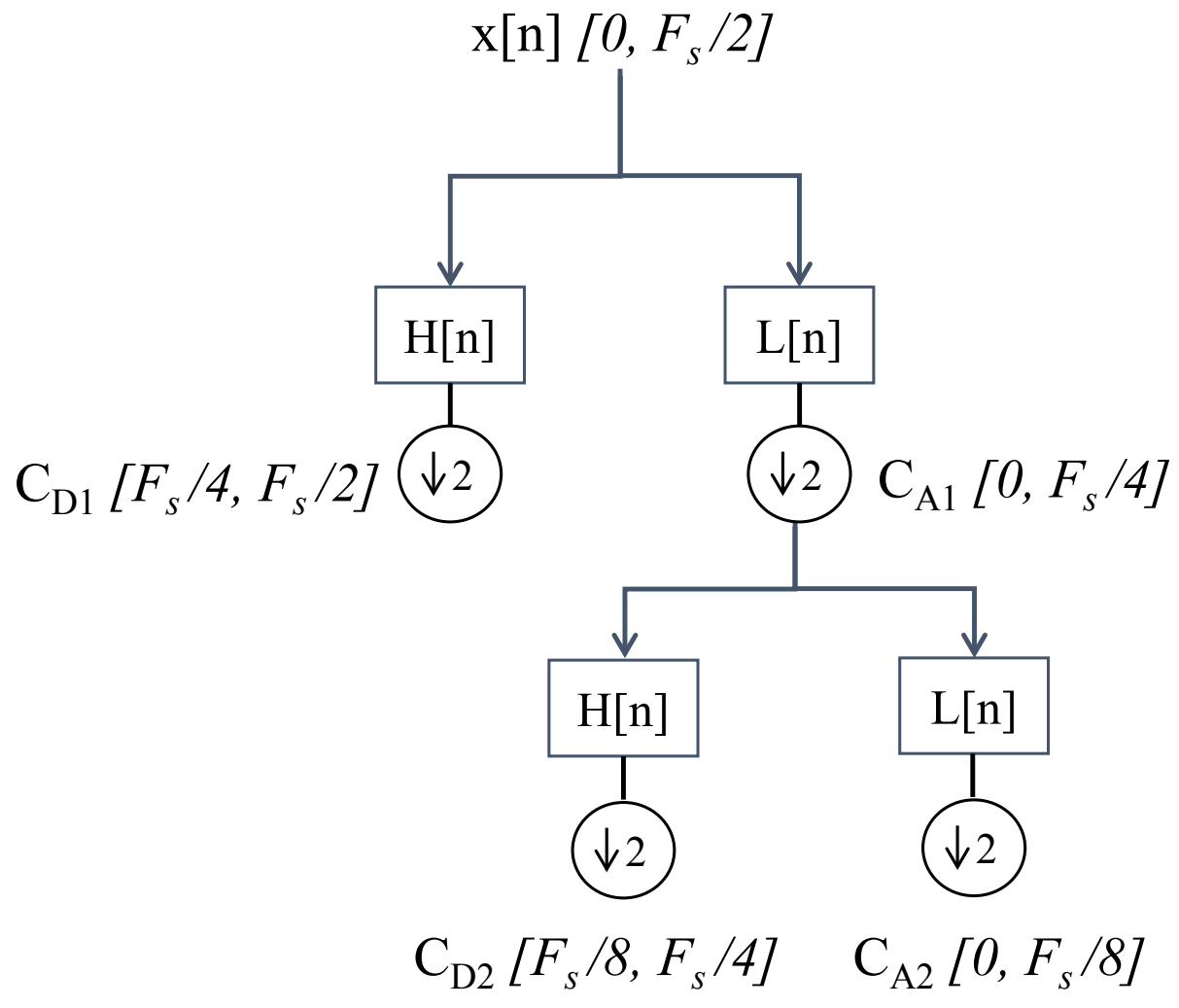

Fig. 3.7. Wavelet decomposition using filter banks

After each level of decomposition, the high-pass filter output provides a detailed frequency visualisation of the upper half frequency content of the previous level low-pass filter output. Such implementation facilitates the band-pass representation of a signal. The wavelet function used in CWT that associated with a centre frequency acts as a band-pass filter during the convolution of mother wavelet with the time domain signal. Similarly, in DWT, cascading operations of low-pass filtering, downsampling, and high-pass filtering also act as a band-pass filter.

For instance, a fault signal sampled at 200 kHz and its 3-level decomposition using filter banks are visualised in Fig. 3.8. The high-pass filter outputs $d_1, d_2$ $and$ $d_3$ produce the band-pass output of $[50 - 100\ kHz], [25 - 50\ kHz]$ $and$ $[12.5 - 25\ kHz]$, respectively. The low-pass filter outputs $a_1, a_2$ *and* $a_3$ contain information about the mean behaviour of the signal which can be used for original signal reconstruction. The original signal can be reconstructed as follows:

$$x_s[n] = a_k[n] + \sum_{j=1}^{k} d_j[n] \tag{3.15}$$

$$x_s[n] = a_3[n] + d_1[n] + d_2[n] + d_3[n] \tag{3.16}$$

where $j$ represents the decomposition level.

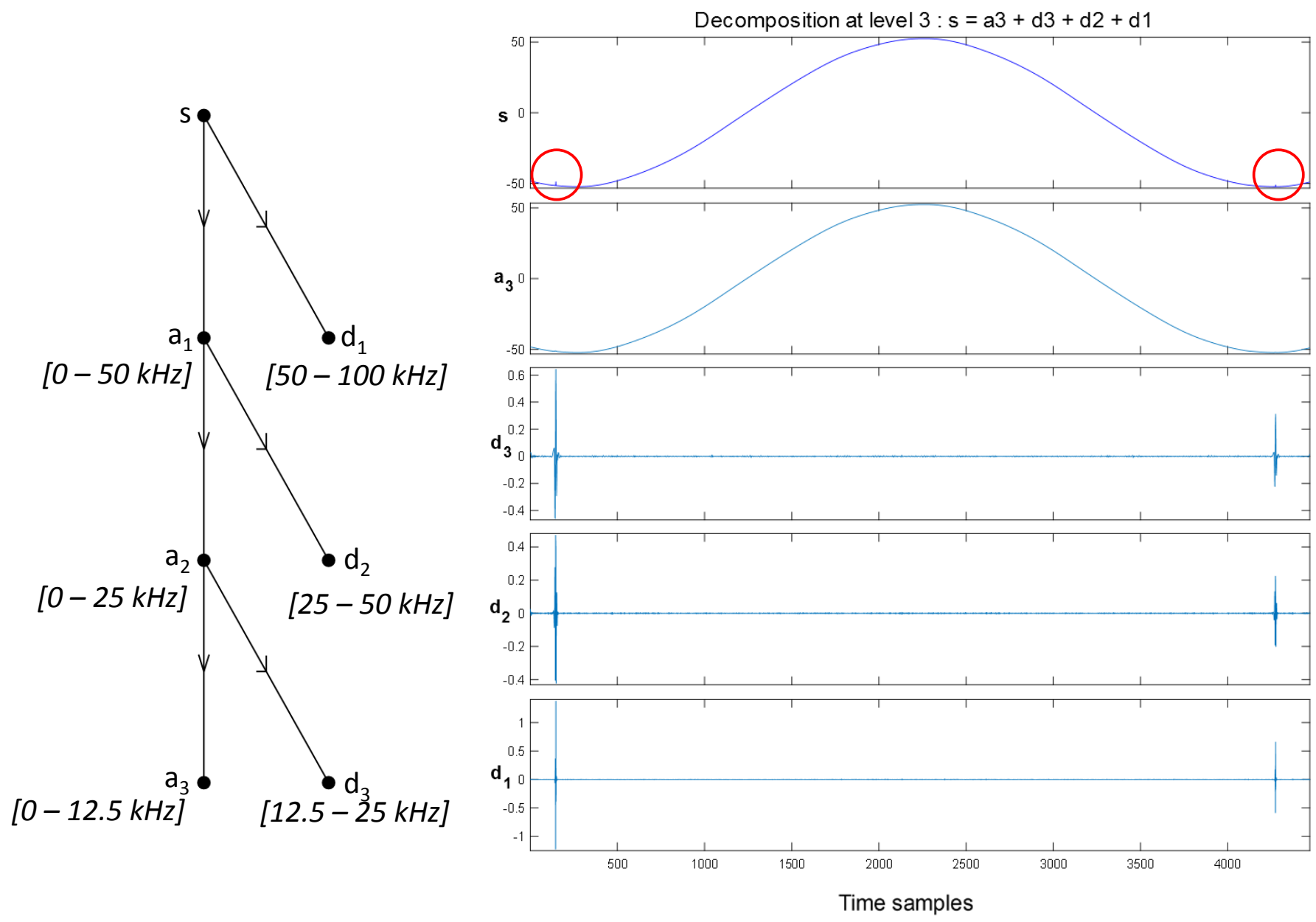

Fig. 3.8. 3-Level decomposition of a fault signal using DWT

In DWT, only the approximation coefficients $a_j$ are further decomposed at each level whereas the detail coefficients $d_j$ that contain high-frequency content are remaining untouched at each level. High-frequency component analysis is essential for several condition monitoring applications such as fault analysis and transient classifications which target hidden high-frequency information. For such applications, the frequency resolution of the decomposition filters used in DWT might not be sufficient to extract the hidden signatures in high-frequency bands. The required frequency resolution for high-frequency contents can be achieved with the wavelet packet transform (WPT), which is a linear combination of wavelets. WPT decomposes the approximate as well as detail coefficients to reach the required level of frequency resolution, as shown in Fig. 3.9.

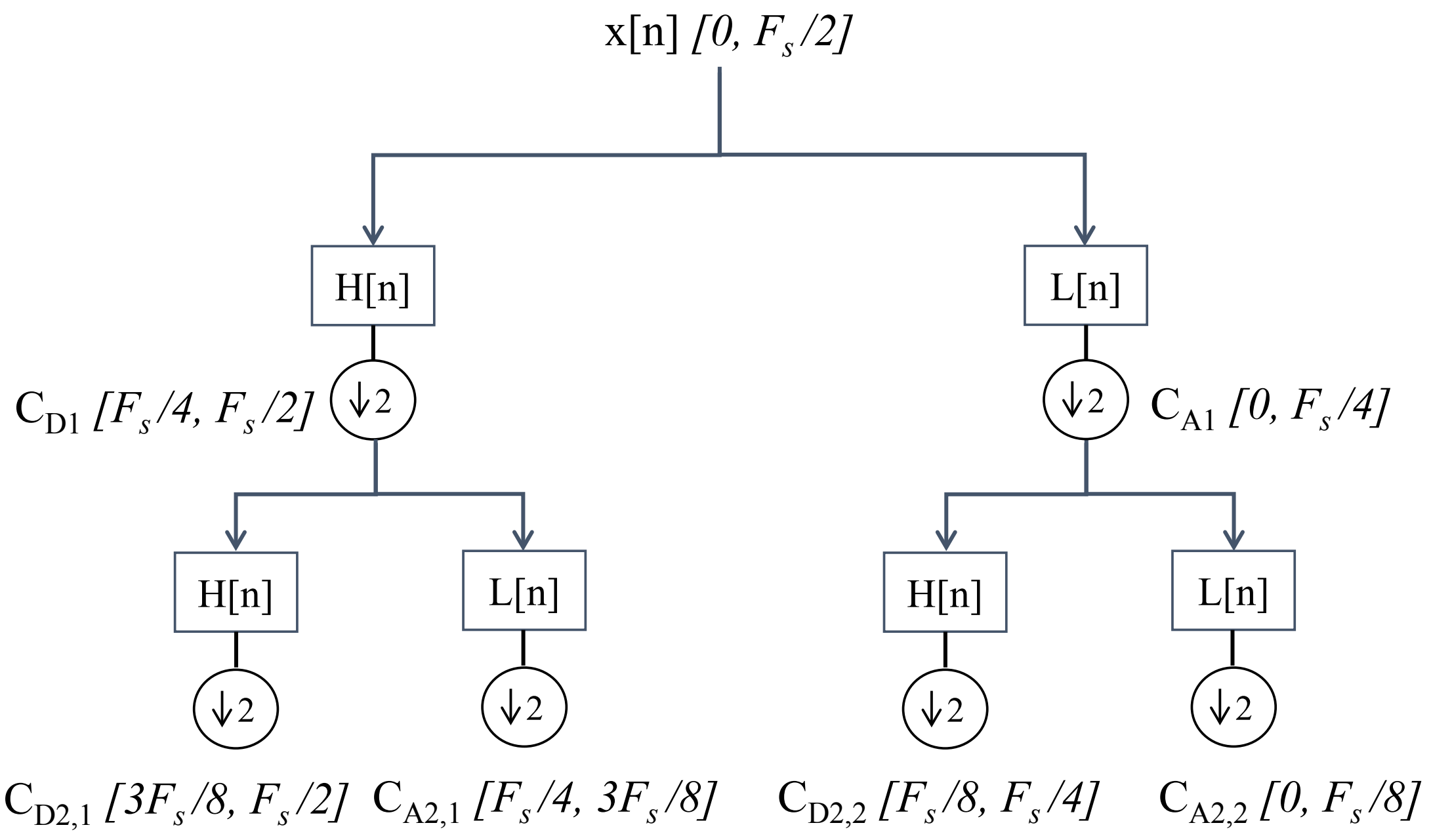

Fig. 3.9. Wavelet packet decomposition tree

A wavelet packet can be mathematically represented as below:

$$\psi_{j,k}^{i}[n] = 2^{-j/2}\psi^{i}\left[2^{-j}n - k\right] \tag{3.17}$$

where $i, j,$ and $k$ are the modulation parameter, dilation parameter, and translation parameter, respectively. Modulation parameter $i = 1, 2 \ldots\ldots, j^{l}$ where $l$ is the decomposition level in the wavelet packet tree. The wavelet $\psi^{i}[n]$ can be formulated as a recursive function in terms of high-pass $H[k]$ and low-pass $L[k]$ discrete filters as follows:

$$\psi^{2i}[n] = \frac{1}{\sqrt{2}}\sum_{k=-\infty}^{\infty} H[k]\psi^{i}\left[\frac{n}{2} - k\right] \tag{3.18}$$

$$\psi^{2i+1}[n] = \frac{1}{\sqrt{2}}\sum_{k=-\infty}^{\infty} L[k]\psi^{i}\left[\frac{n}{2} - k\right] \tag{3.19}$$

Time-domain signals are always the input for wavelet analysis. The resulting wavelet coefficients can be manipulated and leveraged for several applications in the context of electricity grid condition monitoring. These include signal denoising, signal compression, feature extraction, and transient property analysis. Signal denoising, compression, and feature extraction are achieved by representing the original signal with the reduced number of wavelet coefficients based on thresholding [8]. Transient property analysis can be performed by the high-frequency sub-band decompositions of a transient signal.

As a final note of wavelet analysis, the success of signal decomposition and analysis heavily relies on the choice of mother wavelet function in all different forms of wavelet analysis such as CWT, DCWT, DWT, and WPT. Furthermore, the required resolutions can be achieved with the deeper levels of decomposition, but the computational complexity and processing time will increase. Thus, the mother wavelet type and the number of decomposition levels are the two primary hyperparameters that need to be tuned appropriately based on the application requirements in the context of wavelet analysis for a real-world problem.

## 3.3. Artificial Intelligence Based Techniques

AI, also known as machine intelligence, is the simulation of human intelligence processes by machines, especially computing systems. In practice, domain experts and specialists make decisions based on the domain knowledge gained from their past experiences, especially under time pressure [9]. They select an experience that worked before for similar situations and derives solutions for the new problems [10]. Based on this concept, human experts can make correct decisions by following similar facts and inferences.

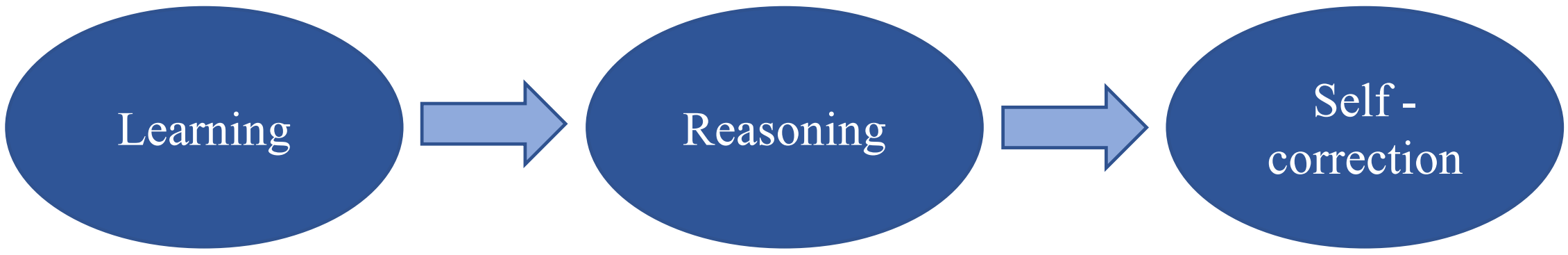

Fig. 3.10. Cognitive steps of AI

AI systems are adopting this human decision-making process into a machine implementable form through three cognitive steps: (1) learning, (2) reasoning, and (3) self-correction, as outlined in Fig. 3.10. The learning process is a fundamental building block that accumulates knowledge and experience of AI systems through observations. AI algorithms capture information that is also known as features from the observations and structure them into a reusable form of knowledge model. This learning process can be broadly categorised into two types, as shown in Fig. 3.11. (a) supervised, and (b) unsupervised. In the supervised learning process, the observations are labelled so that input-output correlations can be formulated for each label (ground truth) to make predictions on new input. The labelled observations can be denoted as $\{(x_1, y_1), \ldots\ldots, (x_N, y_M)\}$ where $x_i$ is the feature vector of the $i$-th observation and $y_i$ is its ground truth. The learning process seeks a correlation function $g: X \rightarrow Y$, where $X$ is the observable variable, $Y$ is the target variable, and $g$ can be formulated as follows:

$$g(x) = arg\max f(x, y) \quad (3.20)$$

where $f$ is a scoring function $f: X \times Y \rightarrow \mathbb{R}$ and $g$ are defined to return the target value $(y)$ which gives the highest score.

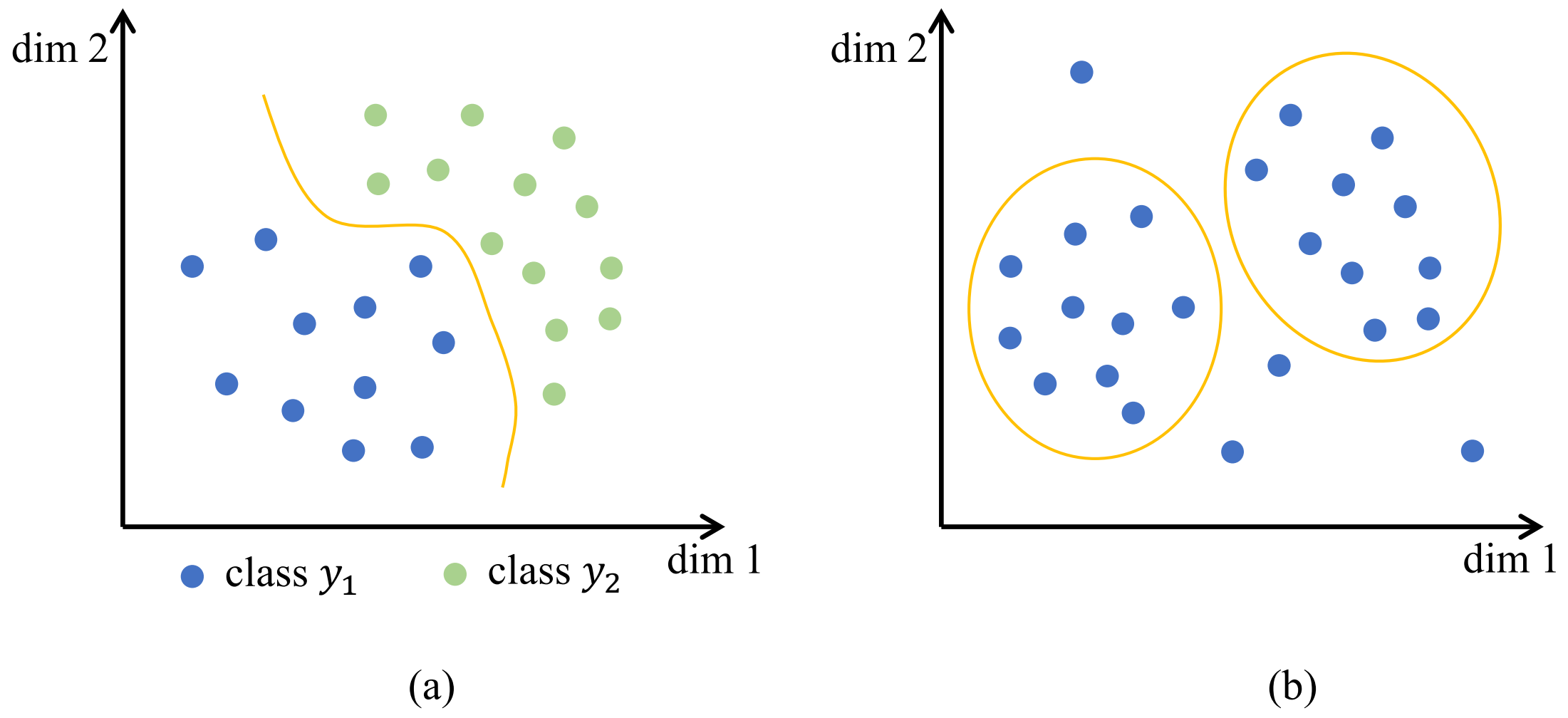

Fig. 3.11. Types of learning process (a) supervised (b) unsupervised

On the other hand, the unsupervised learning process only receives a sequence of inputs $\{x_1, x_2, x_3 \ldots\ldots x_N\}$, but obtains neither target labels, nor rewards from its environment [11]. The unsupervised learning process can be leveraged to learn the similarities and differences in the data and prune unstructured noise. The most common applications of unsupervised learning are clustering and dimensionality reduction.

The knowledge modelling approach can be divided into two categories: (1) generative and (2) discriminative. The generative modelling approach aims to model all the dependencies and intrinsic properties of the data by learning the joint probability distribution $P(X, Y)$ where $X$ is the observation variable and $Y$ is the target variable. Generative modelling can be symbolically explained as follows:

$$P(X, Y) = P(X \cap Y) = P(Y \cap X) = P(Y)P(X|Y) \tag{3.21}$$

where $P(Y)$ is the distribution of the target variable and $P(X|Y)$ is the distribution of observations for a given target label. Generative models estimate parameters of $P(X|Y)$ and $P(Y)$ directly from the training observations. $P(Y)$ can be derived as follows:

$$P(Y) = \sum_{x} P(Y, X = x) \tag{3.22}$$

Followed by the training, generative models can predict the conditional probability $P(Y|X)$ based on Bayes theorem as specified below:

$$P(Y|X) = \frac{P(X|Y)\, P(Y)}{P(X)} \tag{3.23}$$

$$P(X) = \sum_{y} P(X, Y = y) \tag{3.24}$$

Since the generative models learn the joint probability distribution $P(X, Y)$, it is more informative than discriminative models. It can be used to generate new samples similar to existing observations. Hence, generative models are widely used for data augmentation.

In contrast to generative learning, the discriminative models directly learn the conditional probability distribution $P(Y|X = x)$ from training data. They directly discriminate the target value $Y$ for any given observation $X$ by identifying decision boundaries rather than learning the entire data distribution, as shown in Fig. 3.12. This learning type not only makes the learning task easier but also leads to better learning result. Discriminative models often perform better on classification tasks when they have given a reasonable amount of training data. Furthermore, supervised learning is required to obtain discriminative models. On the other hand, generative models are learnt using supervised or unsupervised learning. The suitable modelling type can be determined based on the application requirement and available dataset characteristics.

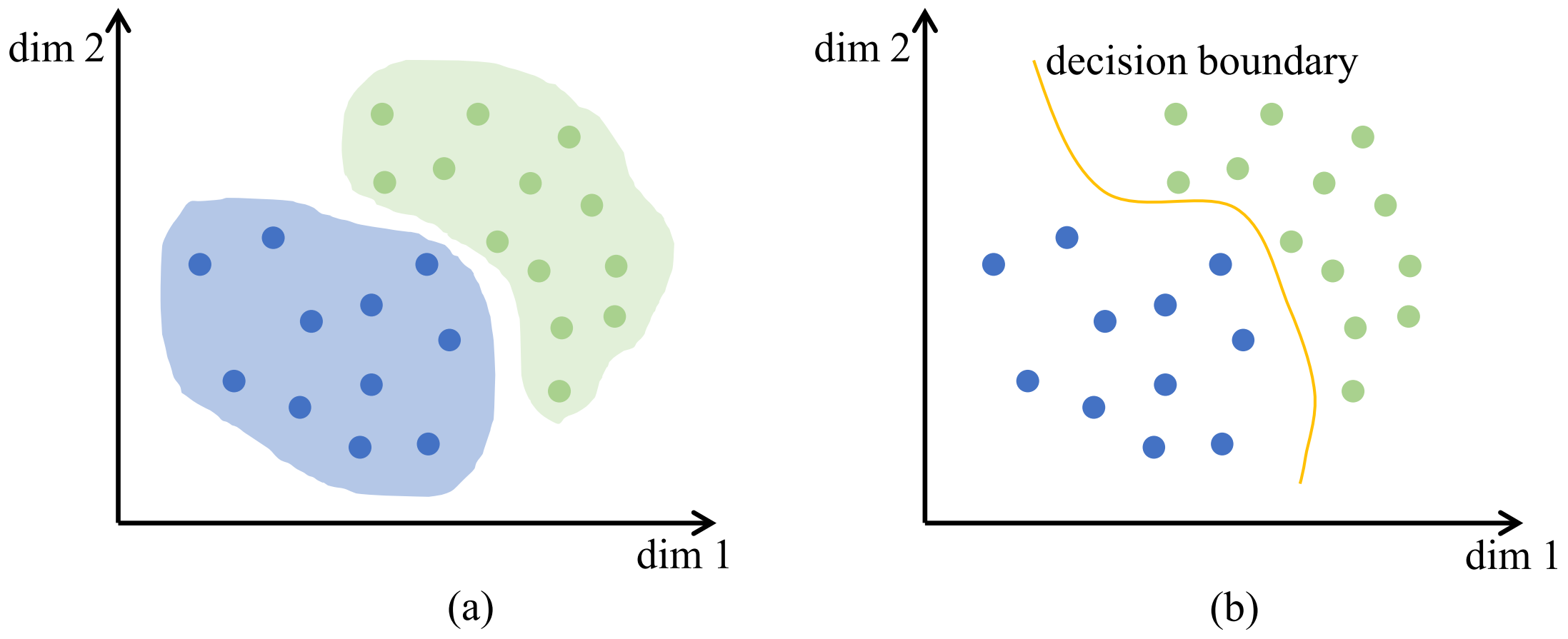

Fig. 3.12. Types of learning approach (a) generative (b) discriminative

Once the knowledge modelling has been completed, the second step in the cognitive process is reasoning. It is the procedure of making predictions and deriving logical conclusions from the modelled knowledge. The method of reasoning can be categorised as either (a) deductive, or (b) inductive. In deductive reasoning, the truth of the context guarantees the validity of the conclusion. Diversly, in the inductive reasoning, the fact of the context only supports the conclusion rather than providing a guarantee, as shown in Fig. 3.13. In the condition monitoring applications, inductive reasoning is commonly practised, in which, observations are collected, and tentative models are created to describe and predict future behaviour. This process can be continued until the presence of anomalous data forces the developed model to be revised [12]. The self-correction step in the cognition process of an AI system is introduced to fulfil this requirement. It represents the fine-tuning process of an existing knowledge model.

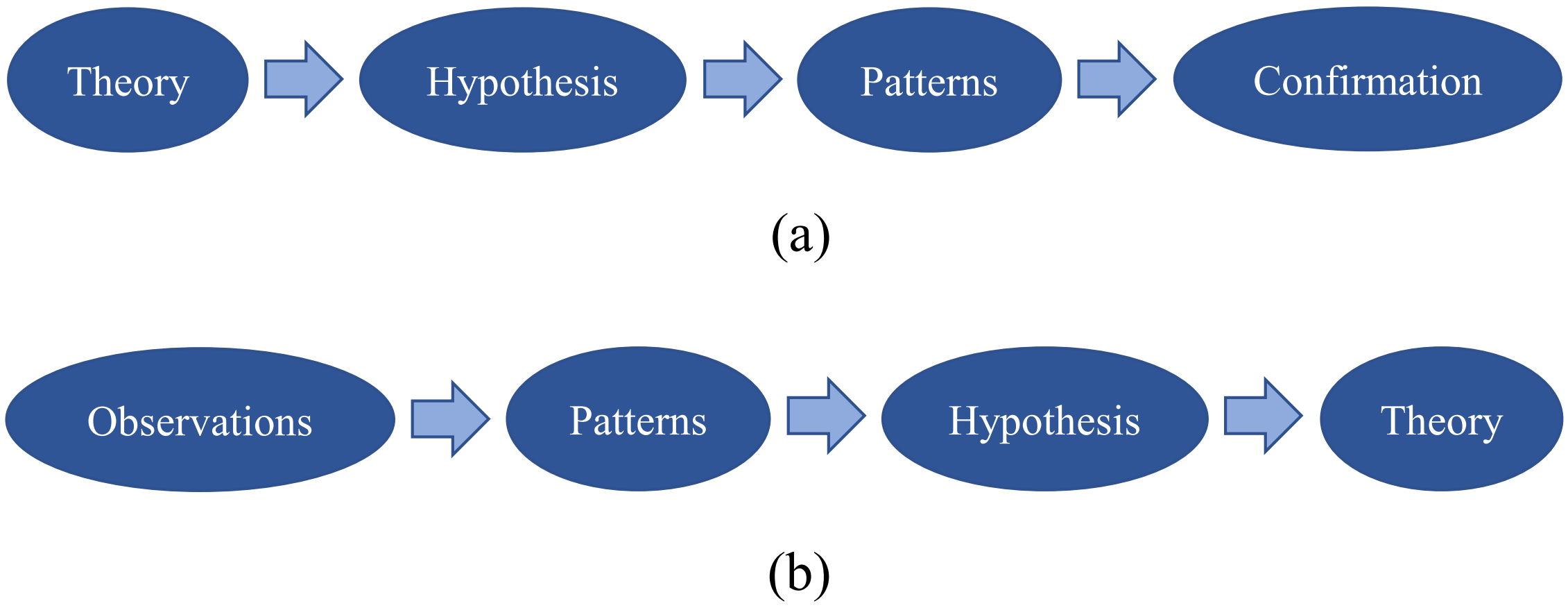

Fig. 3.13. Types of reasoning (a) deductive (b) inductive

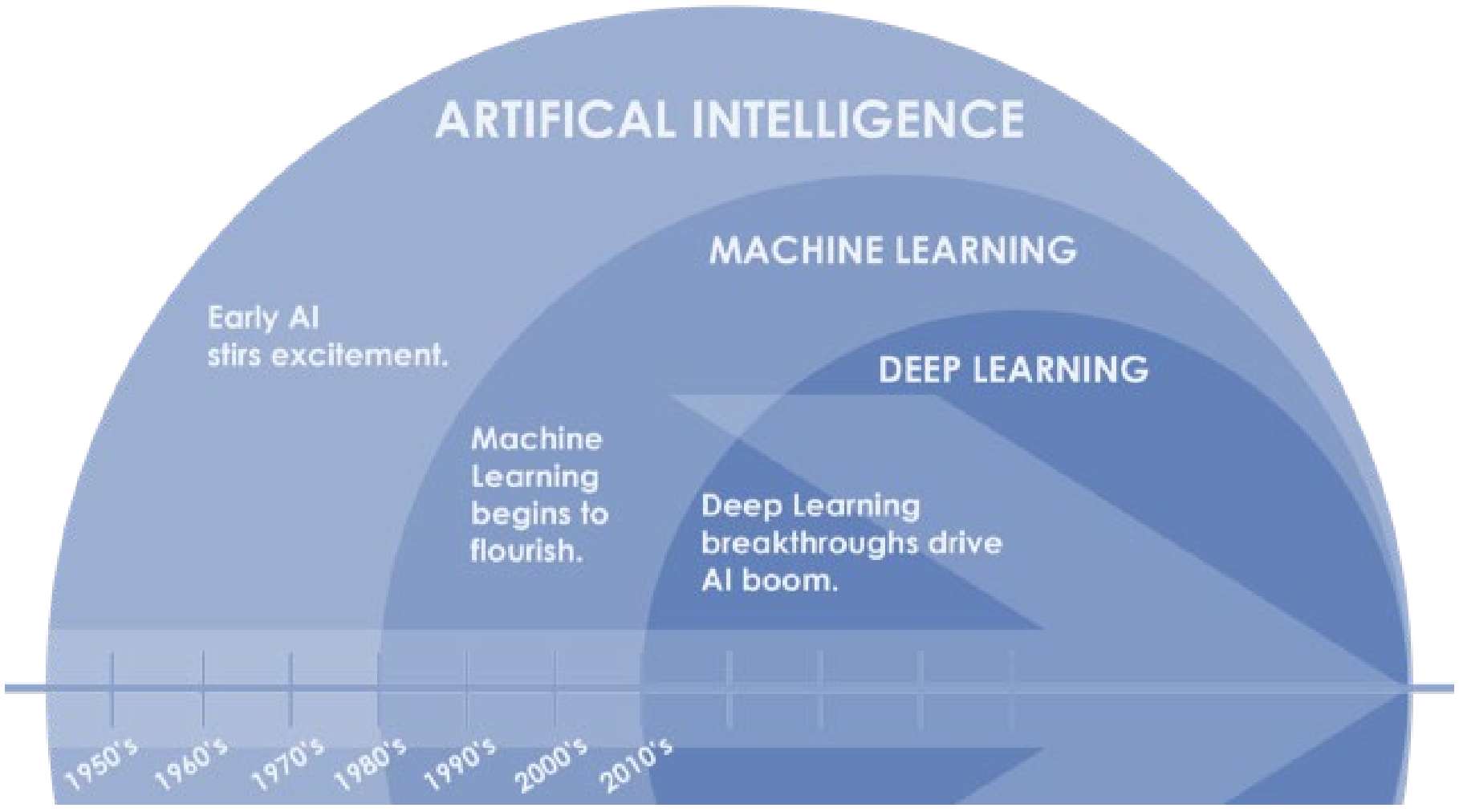

Fig. 3.14. Roadmap of AI [13]

The term AI was first introduced in 1956, and it is the broader concept that includes all the intelligent systems from Good Old-Fashioned AI (GOAI) to futuristic technologies. In the mid-1980s, machine learning grew out of the quest for AI, that applies computational statistics to enable learnings and predictions in machines. By 2010, AI's outlook spectacularly transformed by the introduction of deep learning, which is driving today's AI explosion [13]. The complete roadmap of AI is summarised in Fig. 3.14. Deep learning is based on artificial neural networks that imitate the structure of a human brain. The AI-based techniques have been used in this thesis for feature extraction, modelling, classification and decision making in condition monitoring applications. Those techniques are briefly introduced below.

### 3.3.1. Artificial Neural Networks (ANN)

ANN is an information processing structure that is inspired by the biological nervous system in the human brain. ANN structure is composed of a large number of processing elements, also known as neurons. These processing elements are highly interconnected such that they can work together in unison to model complex patterns and to solve specific problems. Neurons in the human brain receive sensory inputs from the environment through dendrites, process the inputs and produce the output via axons, as shown in Fig. 3.15. (a). Researchers have transformed this biological concept into an artificial neuron, also known as the perceptron which receives input, concatenates the input with their internal states, thresholds with an activation function and produce the output, as shown in Fig. 3.15. (b).

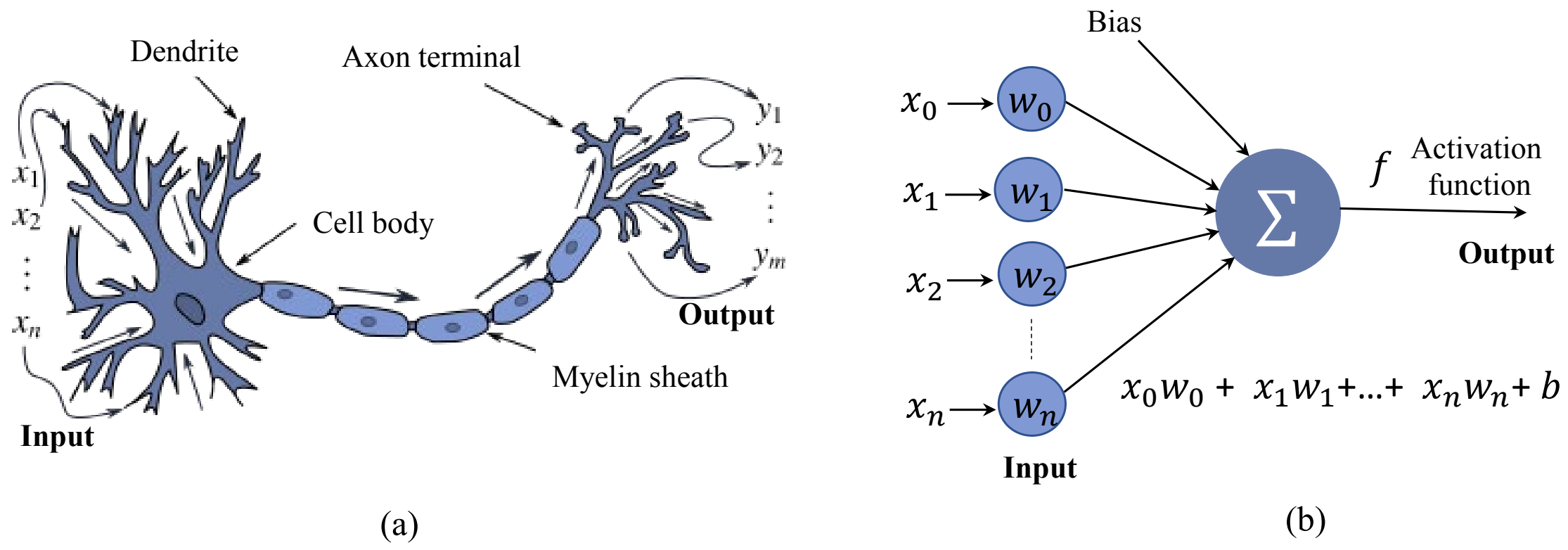

Fig. 3.15. (a) A biological neuron (b) A perceptron inspired by biological neuron

The inputs to a perceptron can be denoted as a set of independent variables $\{x_0, x_1, x_2 \ldots\ldots x_n\}$, which is a feature representation (feature vector) of a single observation. Each of these vector elements is multiplied by a corresponding connection weight that can be represented as $\{w_0, w_1, w_2 \ldots\ldots w_n\}$ . The significance of each input values in output will be decided based on the weightage scheme, which is determined during network training. The weighting process can be mathematically represented as follows:

$$\sum x_i.w_i = x_1.w_1 + x_2.w_2 + x_3.w_3 \dots \dots \dots . + x_n.w_n \quad (3.25)$$

In addition to the weights, bias $b$ is another parameter associated with perceptron. Bias acts like an intercept constant added with the linear equation. It helps the network to derive a model that can fit well for the given observations. Thus, bias is added with the sum of products as follows:

$$Output_{intermediate} = \sum x_i.w_i + b \quad (3.26)$$

As the final step to produce the output of a perceptron, an activation function is applied to $Output_{intermediate}$. Fig. 3.16 reports some of the widely used activation functions in the condition monitoring literature. A neural network is formed when a collection of perceptrons are interlinked through an activation function that introduces non-linearity into the perceptron output and network model. Activation function determines not only the individual perceptron output but also its accuracy and computational efficiency of training a model. Furthermore, it has a significant impact on the neural network's ability to converge and the convergence speed. The network convergence is a state where the network has gained enough knowledge to appropriately respond to training patterns with minimised error margin. These non-linear activation functions enable the neural networks to learn and model complex correlation functions that map an input to an output.

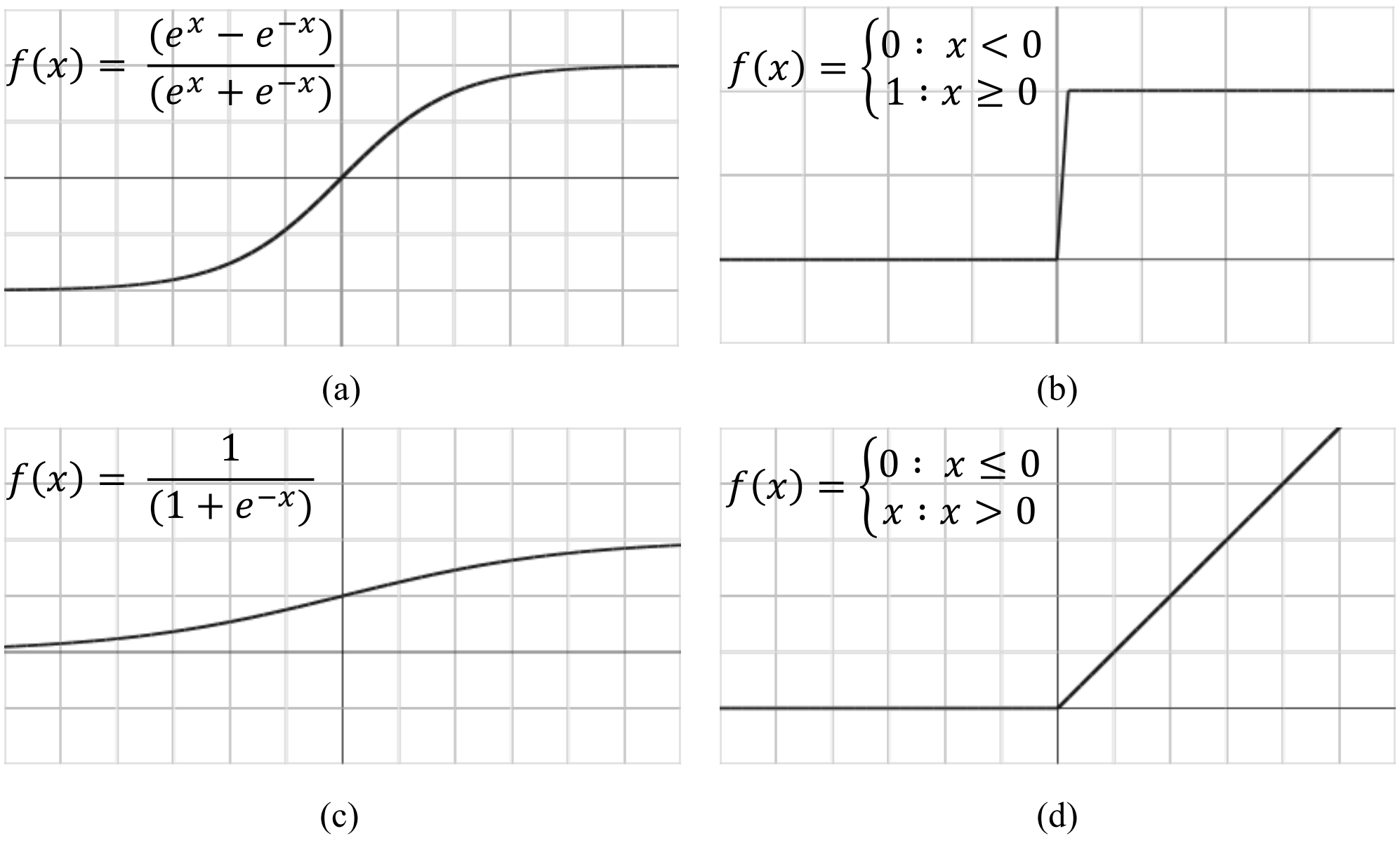

Fig. 3.16. Activation functions (a) hyperbolic tangent (b) binary step (c) Sigmoid (d) rectified linear units

Generally, an ANN can have three types of layers: (1) input layer, (2) intermediate hidden layer(s), and (3) output layer, as shown in Fig. 3.17. The intermediate hidden layers can be inserted between input and output layers to amplify the learning capability of the neural networks. Hence, it can enhance the network prediction accuracy while increasing the computational complexity of the system. ANN can be trained to a specific task with a loss (error) function similar to human continuously learns from their mistakes. The error can be defined as the deviation between the actual value $y$ (ground truth) and the predicted value $\hat{y}$, as shown below:

$$L(w) = y - \hat{y} \tag{3.27}$$

The calculated error is a function of tunable network parameters such as weights and bias, that needs to be minimised to increase the prediction accuracy of a network. During the network training, loss minimisation is achieved through backpropagation, as indicated in Fig. 3.17. The loss in each training iteration is propagated back to the previous layer in order to adjust the weights and bias that can minimise the error. There are different types of loss functions available as listed in Table 3.1, and the choice of loss function should comply with the target problem type such as classification or regression. Typically, network weights are updated based on the gradient of a loss function, as visualised in Fig. 3.18. The ANN training objective is to locate the global minima of the loss function, and this is generally achieved with optimisation methods. Stochastic Gradient Descent (SGD) is a widely used optimisation algorithm to train ANN effectively [14]. It can be formulated as follows:

$$w_{k+1} = w_k - \eta \times \Delta L(w) \tag{3.28}$$

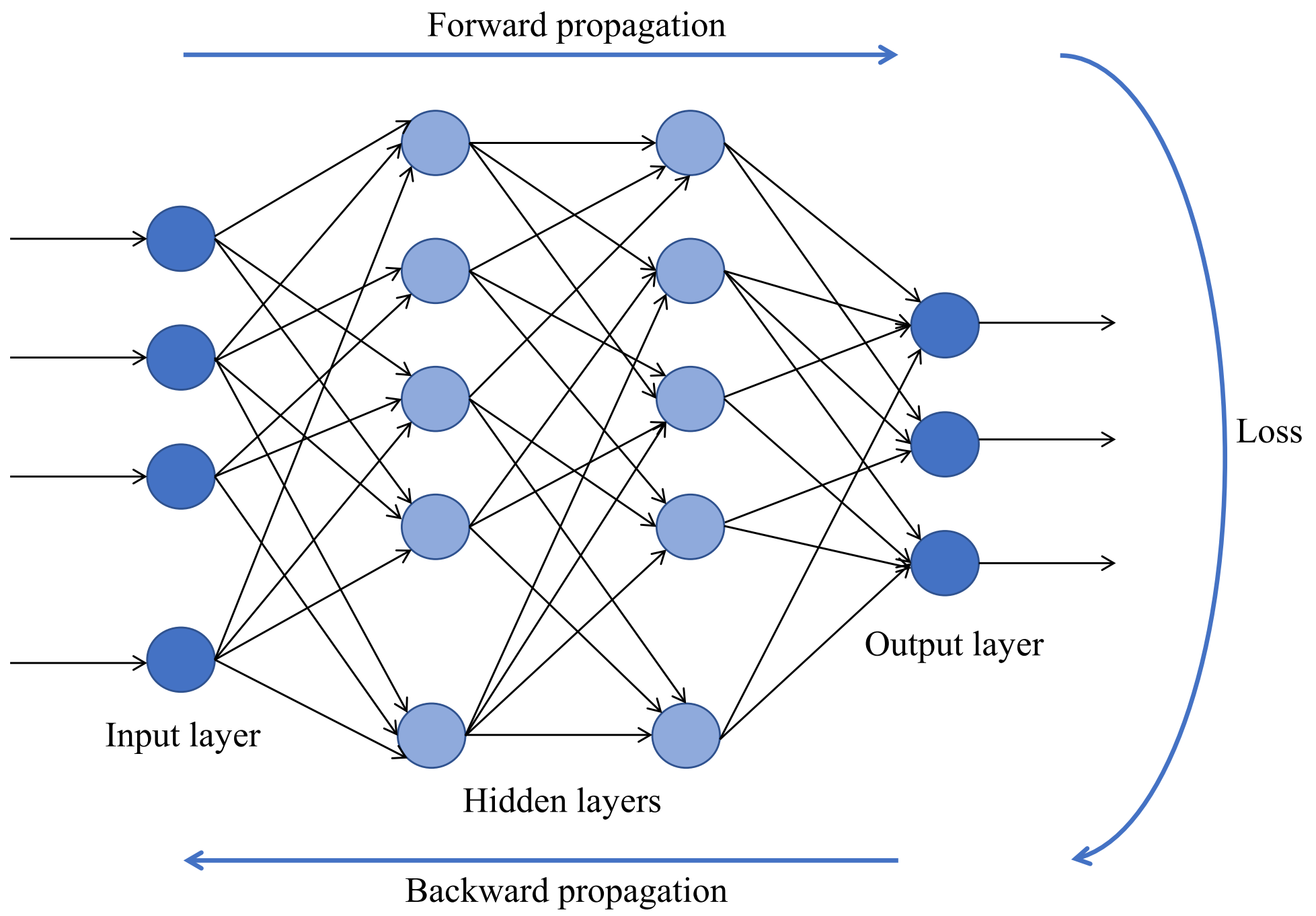

Fig. 3.17. Typical structure of ANN

Table 3.1: Widely used loss functions in ANN

| **Classification** | **Regression** |
|---|---|
| Binary cross entropy<br>$-\sum_{i}^{m}(y_i \log(\hat{y}_i) + (1 - y_i)\log(1 - \hat{y}_i))$<br>where $\hat{y}$ – predicted value, $y$ – actual value, $m$ – number of classes | Mean squared error<br>$\frac{\sum_{i}^{n}(y_i - \hat{y}_i)^2}{n}$<br>where $n$ – number of predicted values |
| Negative log likelihood / cross entropy<br>$-\sum_{i}^{m} y_i \log(\hat{y}_i)$ | Mean absolute error<br>$\frac{\sum_{i}^{n}\lvert y_i - \hat{y}_i \rvert}{n}$ |
| Kullback-Leibler divergence<br>$-\left(\sum_{i}^{m} y_i \log(\hat{y}_i) - \sum_{i}^{m} y_i \log(y_i)\right)$ | Log-Cosh loss<br>$-\sum_{i}^{n}\log(\cosh(\hat{y}_i - y_i))$ |

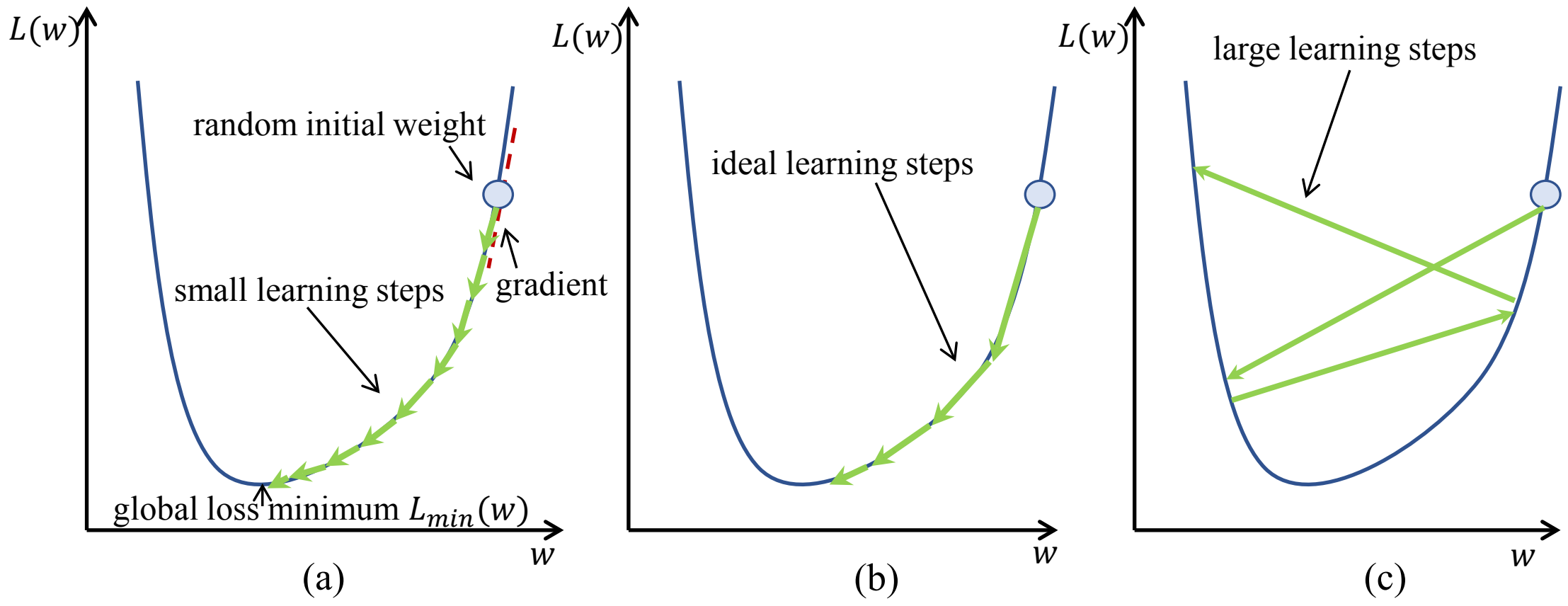

Fig. 3.18. SGD optimization (a) small learning rate (b) ideal learning rate (c) large learning rate

where $\eta$ is the learning rate, and $k$ is the training iteration counter. Learning rate is one of hyperparameter that needs to be appropriately tuned for efficient ANN training. A lower learning rate consumes more number of steps to reach global minima which increases the training time due to slow convergence. On the other hand, a large learning step causes drastic jumps which can slow down the convergence or even lead to divergence.

The main drawback of using SGD is that it requires to define the hyperparameters in advance, but it heavily relies on the problem-specific data. Hence, adaptive gradient descent algorithms such as RMSprop and Adam are introduced, which considerably reduces the effort of manual learning rate modification. RMSprop stands for Root Mean Square propagation, and it divides the learning rate for a specific weight by a moving average of recent gradients for that weight. Adam stands for Adaptive Moment estimation, and it calculates variable learning rates from the gradient mean and variance of each ANN weight. Adam optimiser comparatively works well in practice, and it provides substantial performance gains in terms of training speed. But, its computational complexity is higher than the other optimisers. Thus, the target application should balance the trade-off between the required computational resources and optimum results while choosing an optimisation function.

Since ANN is capable of learning correlations between input features and desired output, it is widely used in several condition monitoring applications such as fault identification, fault-type classification, load identification, energy management, demand forecasting, etc. Fig. 3.19 highlights the main elements of the ANN usage in condition monitoring applications.

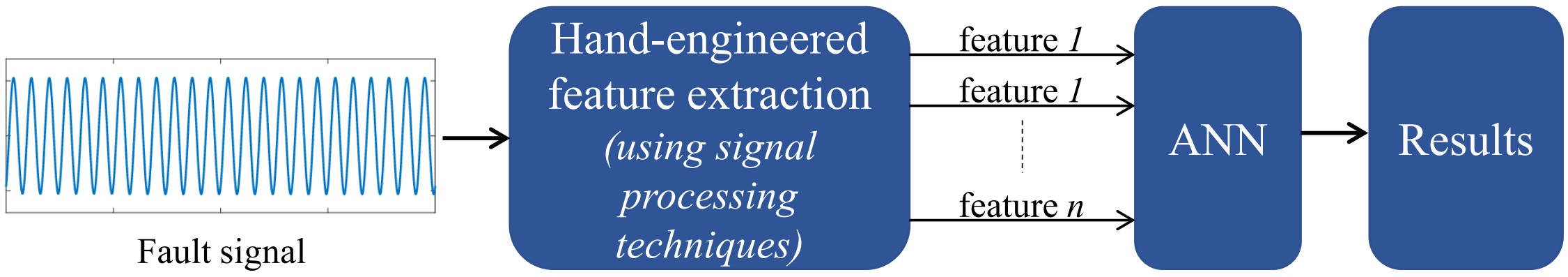

Fig. 3.19. Main elements of ANN application in practice

### 3.3.2. Deep Learning

Deep learning is comparatively a new technology that has been introduced to be the backbone of true AI. Most of the AI tasks can be resolved by determining the right set of features to extract for that task, then feeding the extracted features to a machine learning algorithm. However, for the majority of applications, it is challenging to decide the optimal feature set that needs to be extracted [15]. Deep learning provides the solution to this problem by discovering not only the feature to output mapping but also the feature representation as outlined in Fig. 3.20.

In the early days, deep learning networks were generally believed to be very challenging to train due to lack of available data sets and computational power. The deep learning research began with a breakthrough in the current era of "Big Data" where the automated digital systems are producing massive amounts of data, and sufficient computational resources are readily available. The exponential digitisation of industries drives this growth and requirement of deep learning. Fig. 3.20 highlights the digitisation of modern electricity grids and the deep learning application scope. Since all the measuring devices (sensors) are networked together in the smart electricity grids, it becomes easier to collect the data and accumulate them into a dataset which is relevant for deep learning networks training. Availability of large digital data sets enables the learning algorithms to reach human intelligence on complex tasks with deeper network architectures. The deeper networks can be formed by adding two or more hidden layers to ANN [16]. Modern deep learning introduces variational units in network layers that can represent complex correlation functions. Such variations in the layers and its arrangement lead to different deep learning architectures such as fully connected deep neural networks (DNN), convolutional deep neural networks (CNN), recurrent neural networks (RNN), etc.

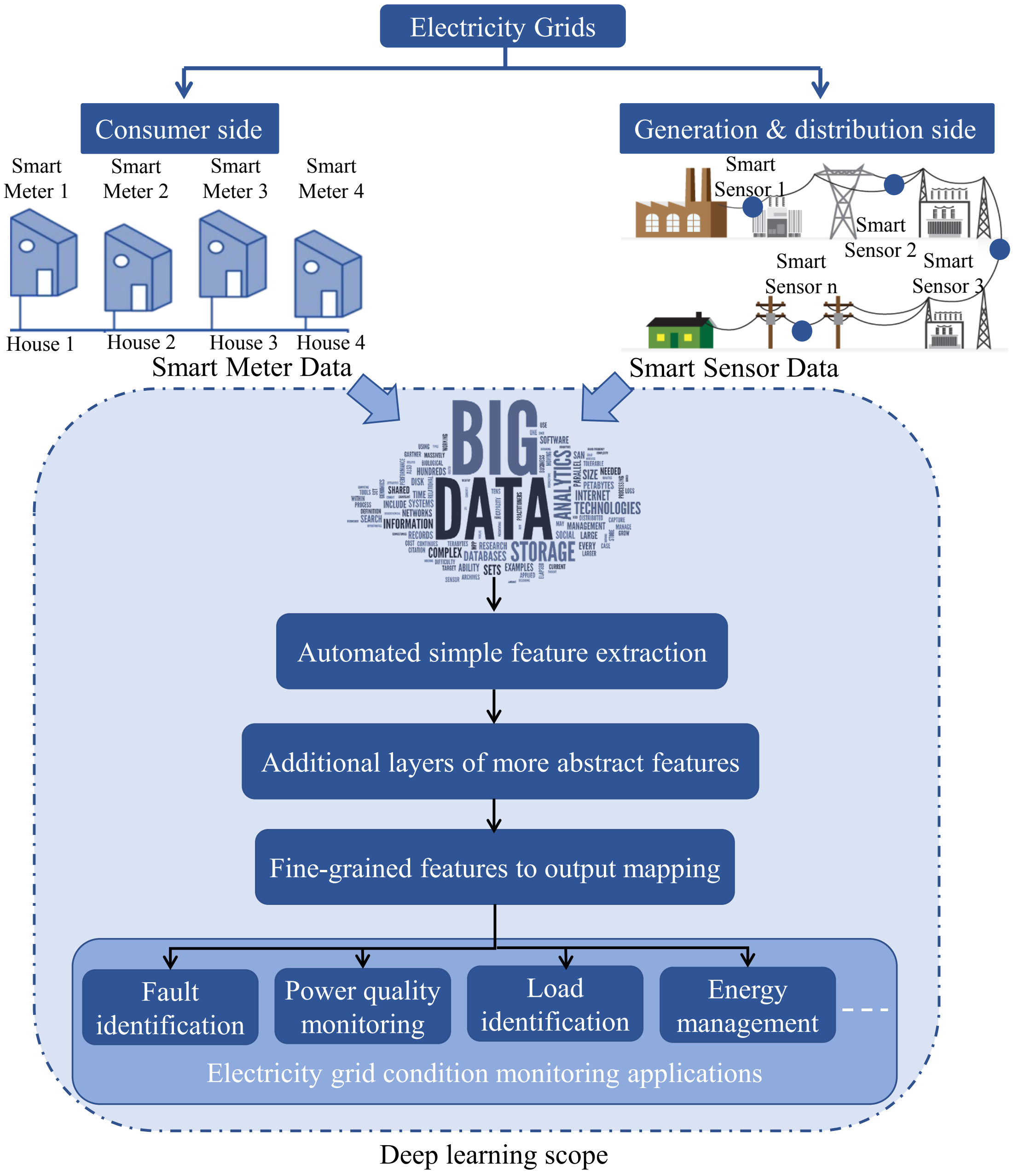

Fig. 3.20. High level schemetic of deep learning in the context of electricity grid condition monitoring

Fully connected DNN is the direct extension of ANN that have more layers between the input and output layers. A fully connected layer is a transfer function from $\mathbb{R}\, m \rightarrow \mathbb{R}\, n$ where $m$ and $n$ are the number of neurons in the same layer and the consecutive layer, respectively. DNN contains a series of fully connected layers, and this architecture is known as universal approximator since it is capable of learning any complicated functions. In fully connected DNN, all neurons in a layer are connected to all neurons in the next layer. Hence, a significant amount of computing power is required to train such a network as it contains more trainable weight parameters $(m \times n)$ compared to other architectures.

CNN is developed from the study of the brain's visual cortex. The neurons in the visual cortex react only to a small local receptive field rather than targeting the entire visual field. Neurons with a smaller region of the visual field only focus on fine-grained details (i.e. overshoots in transient signals) of the input data. In contrast, neurons with larger receptive fields extract more complex patterns (i.e. variation patterns in signals) which are the combinations of the fine-grained details. These observations imply the fact that the output of top-level neurons is derived from the lower-level neighbouring neurons. Furthermore, each neuron is connected with only a few neurons in the previous layer.

Similarly, in CNN, every output unit interacts with specific local input units which also referred to as sparse connectivity. Since the number of input connections is limited to $k\colon k < m$ where $m$ is the total number of neurons in the input layer, the sparsely connected approach only contains $(k \times n)$ trainable weights that are lower than the DNN. Hence, the computational complexity of CNN is lower than DNN.

CNN applies a mathematical operation named convolution to extract the information from the input data. CNN convolution operation is defined as the integral of two functions $(x\,, w)$ which are multiplied after one is reversed and shifted, as shown below:

$$(x * w)[n] = \sum_{m=-\infty}^{\infty} x[m]w[n-m] \tag{3.29}$$

where $x$, $w$ and $(x * w)[n]$ are the input signal, convolution kernel, and feature map, respectively. Each convolutional layer in CNN contains a series of convolution kernels which is also known as filters since they extract informative properties from the input data. In CNN, the feature extraction process is automated through sliding multiple convolution kernels simultaneously over the input, which makes it to identify numerous feature maps, as demonstrated in Fig. 3.21. Convolutional layers derive the most appropriate convolution kernels for its designated task during its training. Furthermore, it also learns to transform the feature maps into more complex patterns that can be directly mapped to a precise output. After a series of convolution operations, fully connected layers are stacked at the end of CNN, which performs the high-level output reasoning and produces the results, as shown in Fig. 3.21. The number of convolution layers and filters are the tunable hyperparameters of CNN that need to be decided based on the complexity of the target application and availability of computation power. Pooling layers can be inserted between two consecutive convolution layers to subsample the results which reduce the computational load and memory requirement. In recent years, CNN has been tremendously successful in practical applications from different domains such as image recognition, video processing, medical image analysis, natural language processing, speech processing, etc.

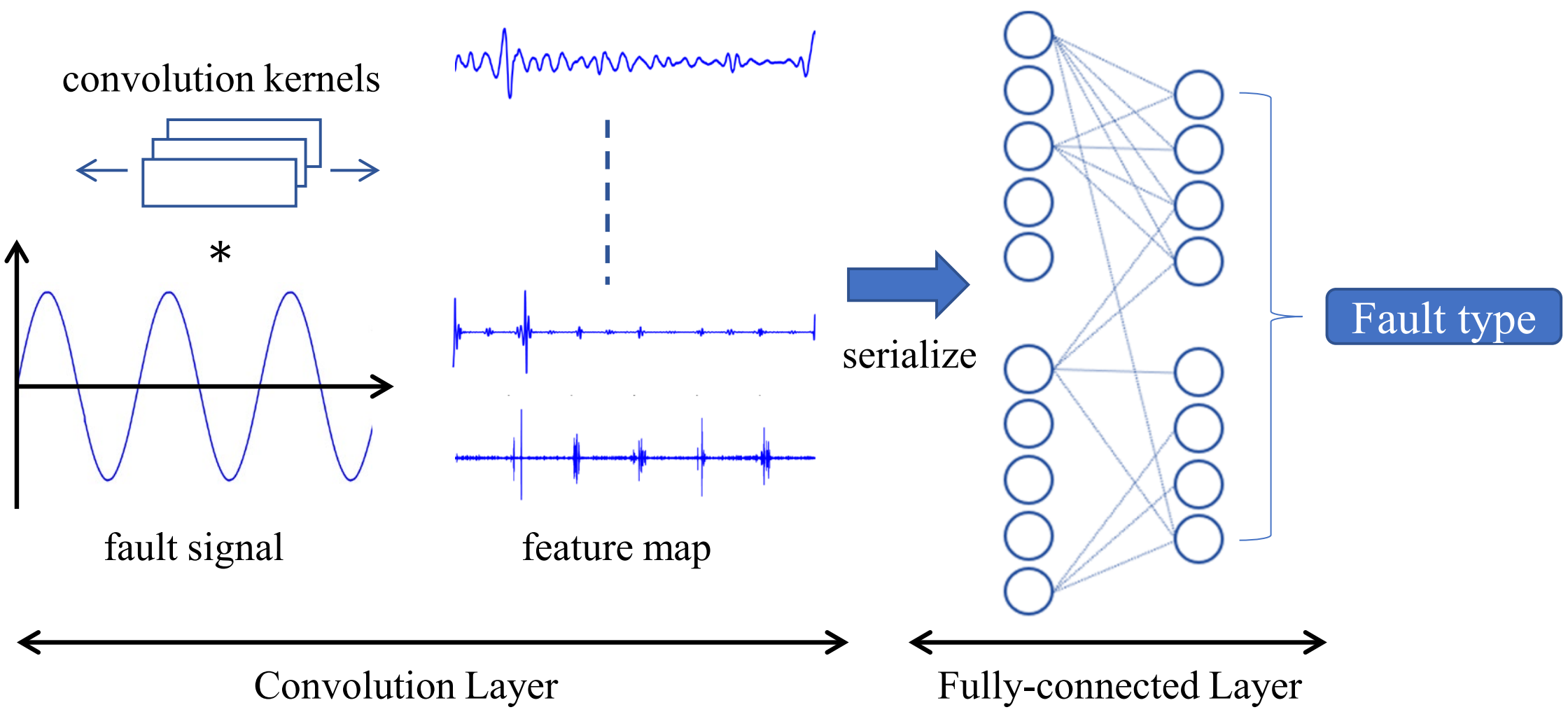

Fig. 3.21. Overview of CNN application for fault type classification

RNN is a class of neural networks that can process the sequential data and predict the future, as shown in Fig. 3.22. In contrast to DNN and CNN, which only works with fixed-sized inputs, RNN can process input sequences of variable lengths. Furthermore, activations flow in feedforward networks is only in one direction, whereas RNN might contain connections pointing in both directions.

RNN holds a form of memory cell where it preserves some internal state across time instances that can be leveraged for the next consecutive input sequences. However, the earlier inputs eventually removed from the memory cell while training an RNN with the long inputs. Hence, the memory cell in RNN gradually forgets the information about the past and fails to detect long-term dependencies in the input sequences. Long Short-Term Memory (LSTM) cell is proposed to overcome this problem by maintaining two separate state vectors for short-term and long-term dependencies. The network can learn to identify the prominent instances and store them into the long-term state vector while leaving the non-essential input to fade away.

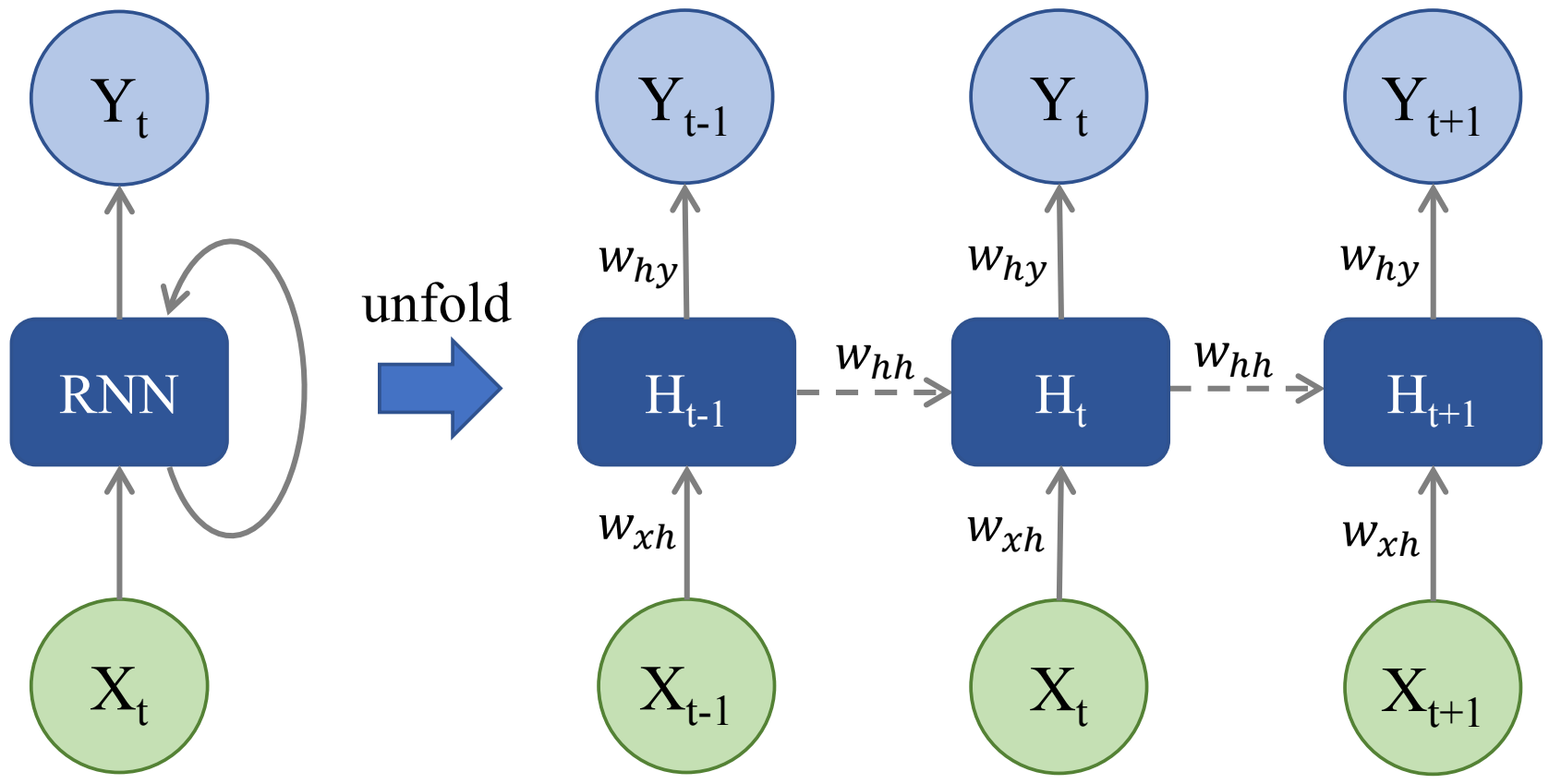

Fig. 3.22. Standard RNN architecture

Even though LSTM can remember long-term sequences, it consumes a significant amount of memory bandwidth and computation power compared to other deep learning architectures such as DNN and CNN. Furthermore, it is not possible to fully utilise the LSTM networks even on top of efficient hardware since they are not scalable. Hence, LSTM based predictions are exceptionally far from becoming practical, especially on resource-constrained embedded hardware [17]. In the context of electricity grid condition monitoring, the developed algorithms need to be executed in real-time on embedded devices such as a pole-mounted monitoring unit. Moreover, recent studies have demonstrated that CNN can outperform both LSTM and attention-based models with reduced memory and computational requirements [18][19]. The memory bandwidth requirements are usually lower for CNN since it uses the same parameter in several convolution operations (parameter sharing) and its weights are relatively small. Therefore, CNN based monitoring solutions are developed in this thesis rather than LSTM/RNN in order to achieve higher accuracy while ensuring the algorithm deployment feasibility on low-power embedded devices.

## 3.4. Real-time implementation requirements and limitations

In the field of electricity grid condition monitoring, each specific application has its application-specific requirements and constraints. Among those, accuracy, latency and input signal granularity requirements are essential for the practical application of a proposed monitoring technique. For instance, if a proposed algorithm fails to meet the application-specific latency requirements and demands very-high input signal granularity requirements, then it is not suitable for real-time implementation in practice. Hence, the essential application-specific constraints need to be considered during algorithm development. Table 3.2 summarises the application-specific requirements.

Table 3.2: Condition monitoring application-specific requirments

| Application | Latency requirement | Input signal granularity |
|---|---|---|
| Power quality monitoring | ~ seconds - minutes | Meduim frequency (up to few kHz) |
| HIF identification | ~ milliseconds | High frequency (kHz to MHz) |
| Consumer load identification | ~ seconds - minutes | High frequency (kHz to MHz) |
| Load specific energy disaggregation | ~ minutes - hours | Low frequency (fraction of Hz) |

The processing of high-frequency (kHz-MHz) data for a time-critical (detection time in milliseconds) application such as HIF identification is a more challenging task. High sampling frequencies create a massive amount of data points in real-time. It is not feasible to transfer the whole raw data to a base station or a processing centre due to communication bandwidth limitations. Hence, the raw data need to be processed near the sensors. Therefore, the proposed algorithms need to be deployed on the pole-mounting monitoring units, that contain resource-constrained embedded hardware. The incoming data stream needs to be processed as early as possible in the embedded device to facilitate real-time applications. Typically, the incoming data is manipulated as frames. Timing constraints in real-time processing is explained in Fig. 3.23.

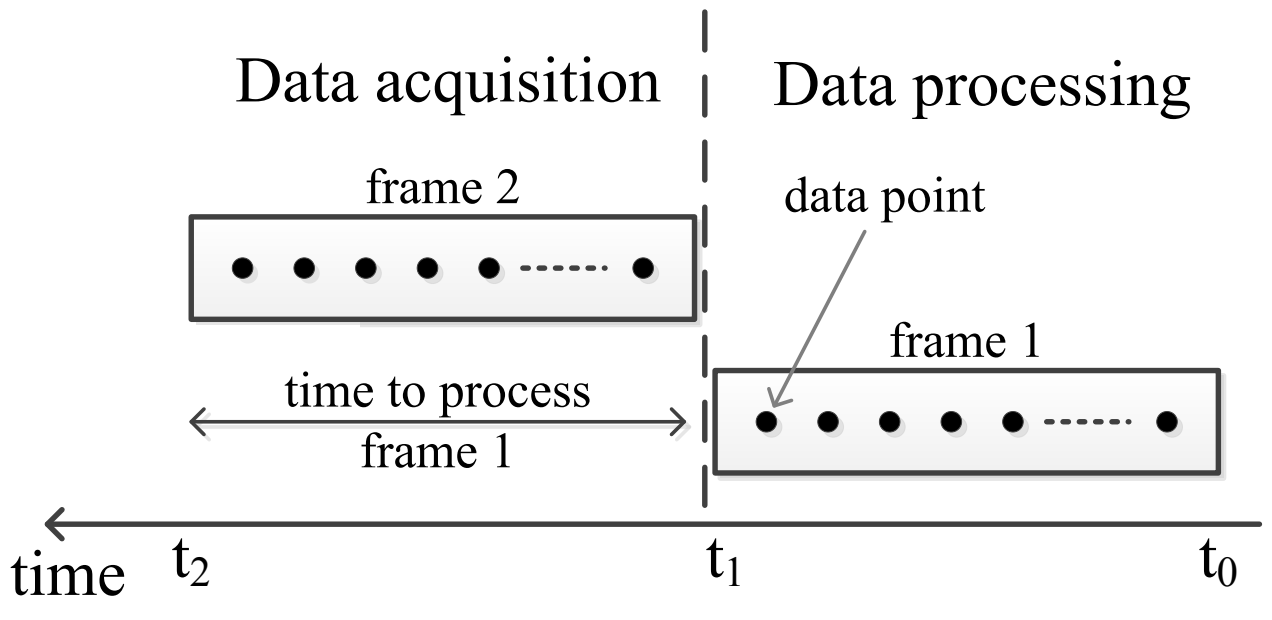

Fig. 3.23. Time constraints on real-time data frame processing

Hence, it is essential to choose a suitable deployment architecture based on the timing and communication bandwidth requirement of an application. In this thesis, a hierarchical data processing architecture is adapted for condition monitoring applications of electricity networks, as shown in Fig. 3.24. High-frequency feature extraction and low-latency real-time data analytics on operational data are performed in edge devices (pole-mounted monitoring devices). This computing paradigm is known as edge computing. This deployment arrangement not only saves the communication bandwidth but also eliminating unnecessary delays. On the other hand, medium latency data analysis with low-frequency data can be done in the intermediate nodes (such as communication gateways or nearby sub-stations) since the time-delay and communication bandwidth are not critical. Transactional data analysis and business intelligence applications can be deployed in the cloud infrastructure since they do not have strict time constraints. Furthermore, those applications demand long-term data and high-performance computing. Fig. 3.25 summarises the hardware distribution in each level of the hierarchy. The devices in the lower layer are comparatively cheaper and more resource-constrained. The middle layer contains medium-scale computing platforms that can be connected to the lower layer devices to receive the pre-processed data for extensive data processing. The top layer represents the cloud infrastructure with high-performance computing capability.

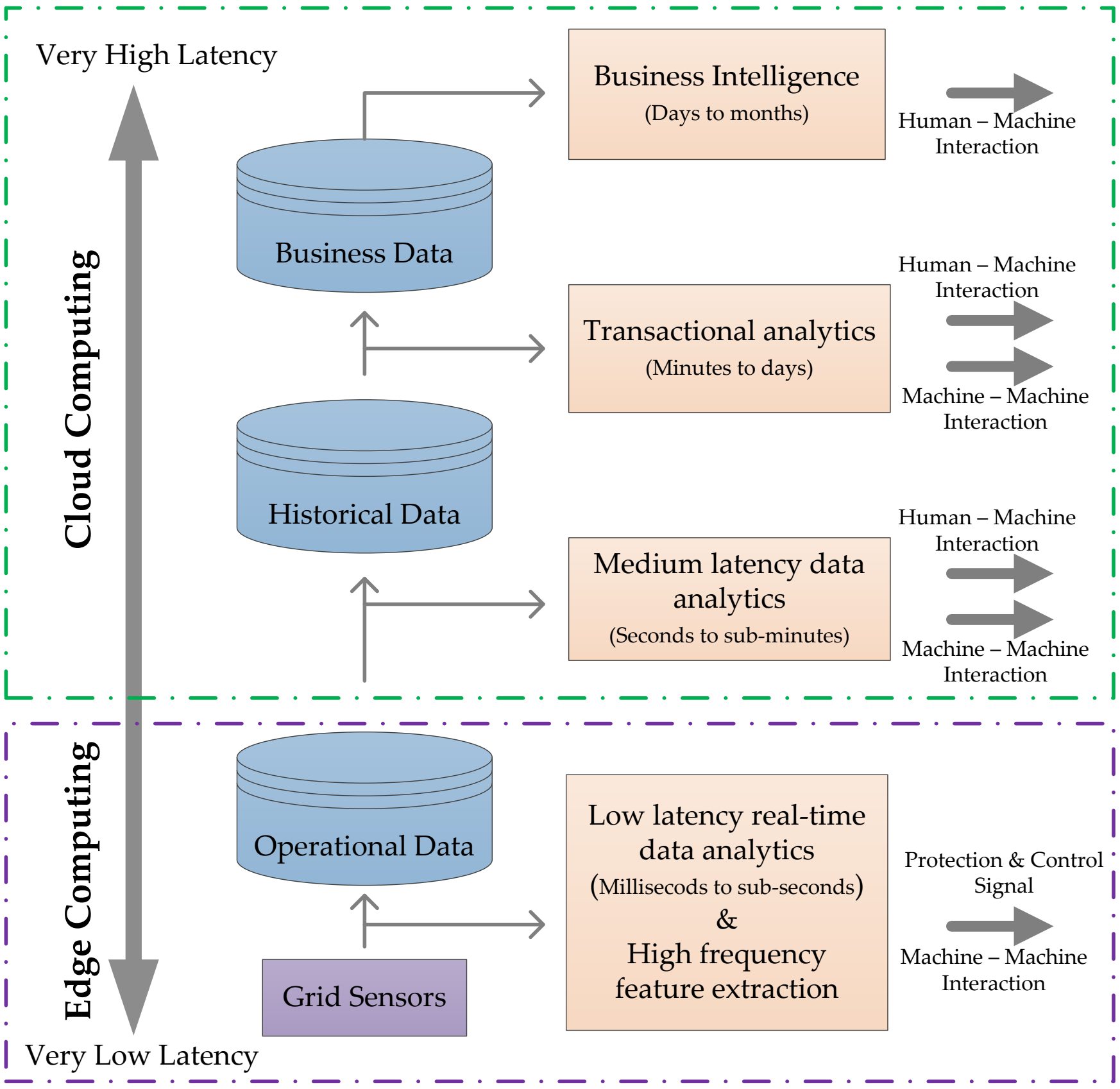

Fig. 3.24. Electricity network data analysis hierarchy

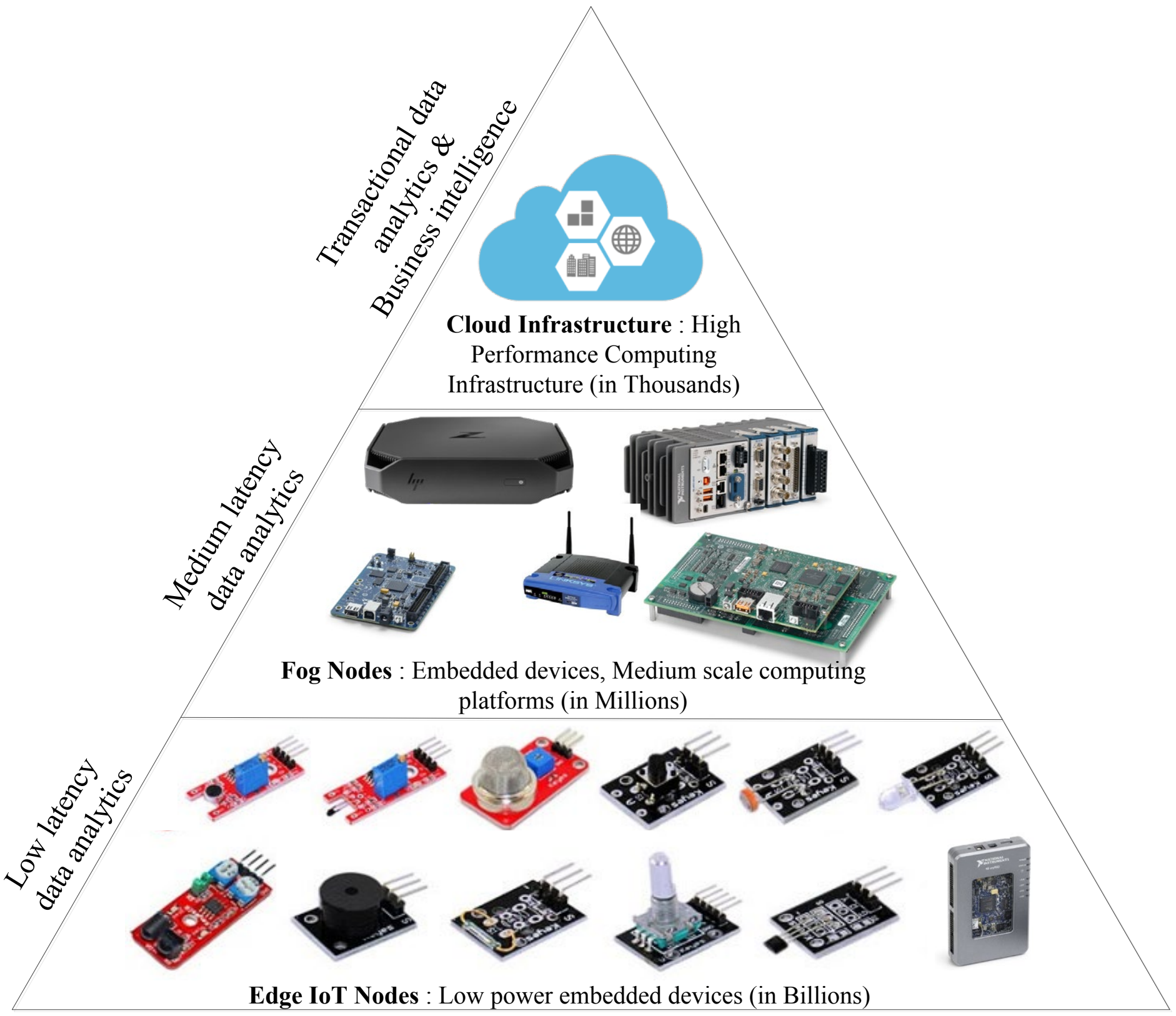

Fig. 3.25. Hardware distribution for electricity grid condition monitoring

The proposed feature extraction and decision making algorithms must obey these application-specific latency constraints during the real-time execution on resource-constrained embedded hardware. However, most of the existing research works mainly focus on the accuracy of the monitoring algorithms and fail to do a feasibility analysis of the proposed techniques against these application-specific constraints. Hence, most of the algorithms in the literature are far away from practical application due to its computational complexities and very high-frequency input signal requirements. This thesis aims to develop AI-based condition monitoring algorithms that can improve state-of-the-art accuracy. Furthermore, the developed algorithms are validated in real-time on low-power pole mounting monitoring units. The proposed solutions are reported in chapter 4, 5 and 6 of this thesis.

## 3.5. CHAPTER SUMMARY

Modern applications of electricity grid condition monitoring rely on digital signal processing and AI-based techniques to extract the intrinsic features from the raw input signals and to make decisions from the extracted features. Signal processing methods used in this thesis are theoretically explained in this chapter, along with the application steps to analyse the electricity grid faulty signals. Besides, AI-based modelling techniques, learning procedures, optimisation steps and decision-making techniques used in this thesis are detailed. Furthermore, real-time implementation requirements of different condition monitoring applications are summarised. Application-specific limitations and the impact of hardware resource constraints in practical applications are highlighted. A hierarchical data processing architecture is adapted for electricity network condition monitoring. The next chapter discusses a newly developed, AI-based condition monitoring framework for overhead power line monitoring. The proposed monitoring platform is validated with the laboratory and field experiments.

# CHAPTER 4

# 4. OVERHEAD POWER LINE MONITORING FRAMEWORK FOR REMOTE ELECTRICITY NETWORKS

## 4.1. Introduction

The SWER feeders in the rural electricity networks tend to have very long spans between power poles, which significantly reduces the infrastructure cost. It uses galvanised steel conductors at high tension. Due to the long spans across the rural areas and high mechanical stresses, it is more likely to get affected by the tree falls, vegetation contacts and vibrations from the wind. It leads to challenging problems such as inadequate voltage regulations and bushfire risks for SWER feeders. This chapter proposes a distributed on-line monitoring framework that incorporates power quality monitoring, real-time HIF identification and transient classification for overhead power lines. An overview of the proposed monitoring framework is shown in Fig. 4.1.

## 4.2. Monitoring unit hardware

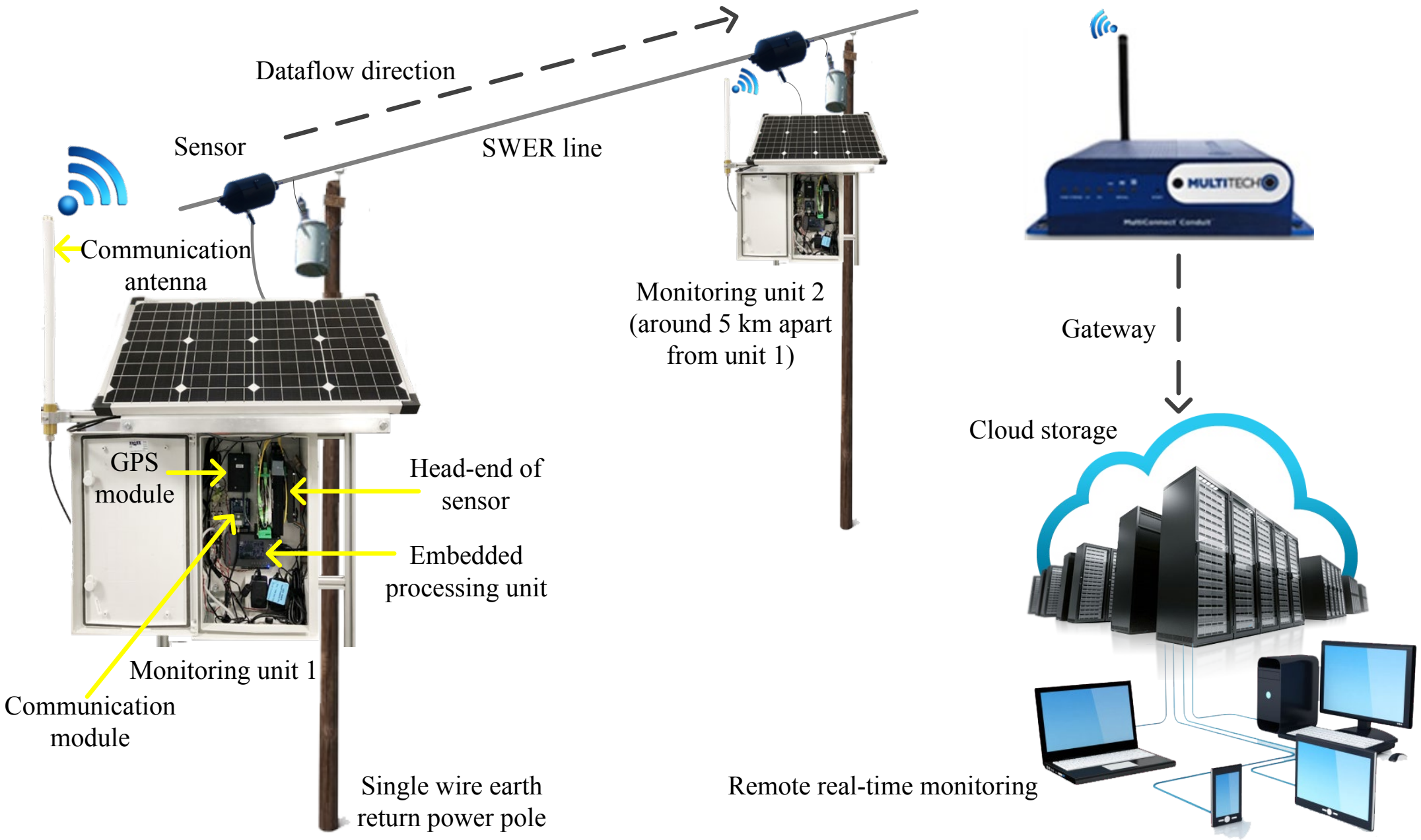

Fig. 4.1. Overview of the proposed overhead power line monitoring framework

The proposed monitoring system contains a passive optical sensor that can capture the voltage and current signals from the SWER line. As shown in Fig. 4.1, head-end of the sensor is attached with the monitoring box, which amplifies the captured signals and feeds them into the DAQ hardware. Low-power embedded processing hardware is included in the monitoring unit to process the operational data for time-critical applications such as HIF detection. The proposed monitoring device relies on its own, standalone communication infrastructure since the communication resources are limited in rural areas. Each monitoring unit accommodates a long-distance communication module that can transmit the processed data and results up to 5 km. It not only sends own data but also relays the signals from adjacent units up to the gateway. Besides, a global positioning system (GPS) receiver is embedded with each device to synchronise the system time and to localise the identified faults and analysis results.

## 4.3. THE INTERNAL ARCHITECTURE OF THE PROPOSED FRAMEWORK

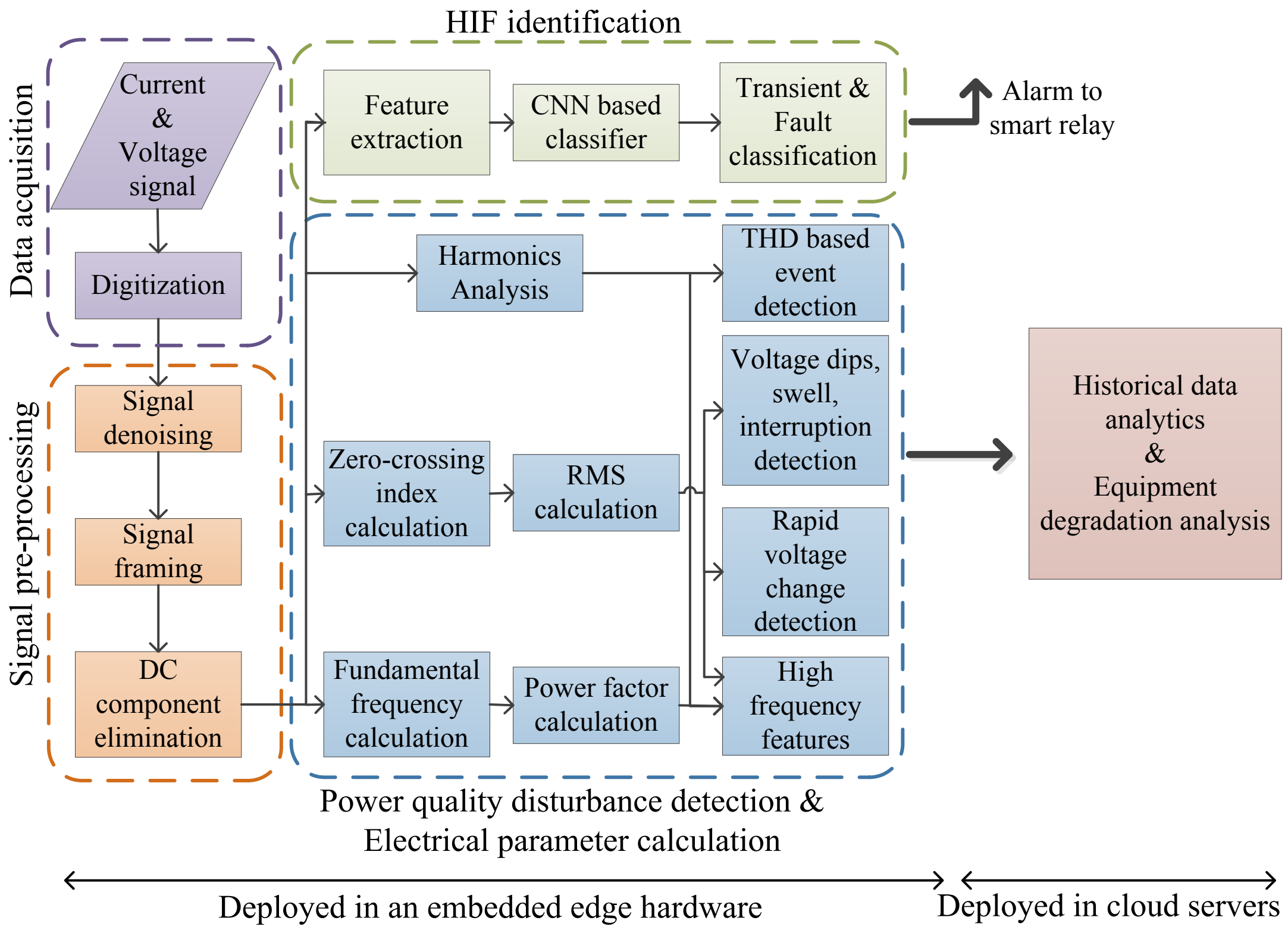

Fig. 4.2. Internal architecture of the proposed overhead power line monitoring framework

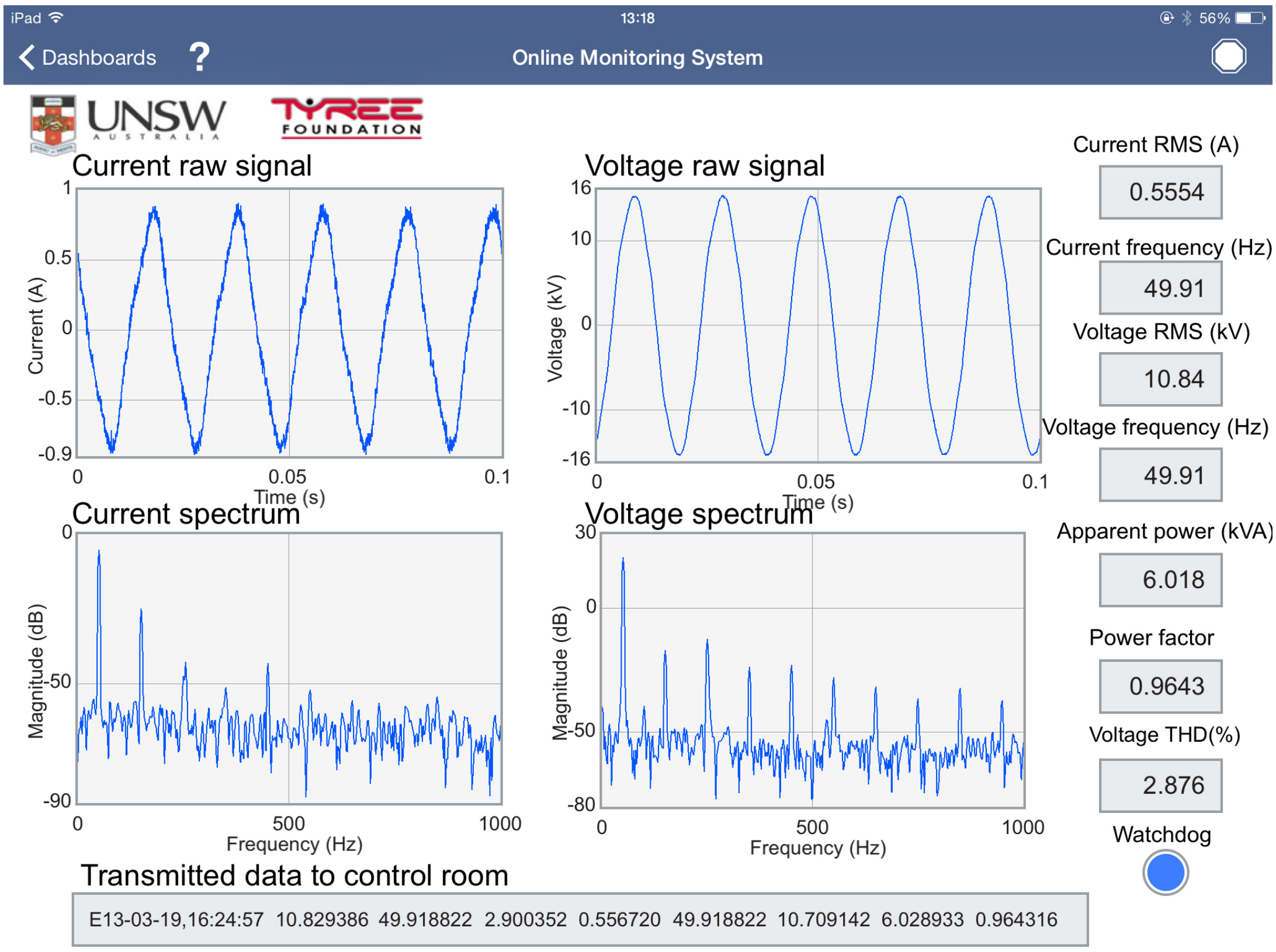

Fig. 4.3. Mobile dashboard of the online monitoring system

The internal architecture is outlined in Fig. 4.2. Input current and voltage signals are digitised at a high sampling rate (20kHz) to capture high-frequency oscillations during the disturbances/faults. Then the digitised signals are pre-processed, that includes the signal denoising with a low-pass filter, framing and direct current (DC) component elimination. After that, basic electrical parameters such as power frequency, RMS values, power factor, and harmonic components are precisely calculated from the high-resolution raw signals. The calculated parameters are visualised in Fig. 4.3. These high-frequency features are periodically transmitted (every 10 seconds) to the cloud server for long-term historical data analysis and equipment degradation analysis. On the other hand, these features are directly used for threshold-based on-site power quality disturbance detection and AI-based HIF detection, as outlined in Fig. 4.2. Before the application-specific experiments, the monitoring unit has been calibrated with the benchmark readings.

## 4.4. MONITORING SYSTEM CALIBRATION

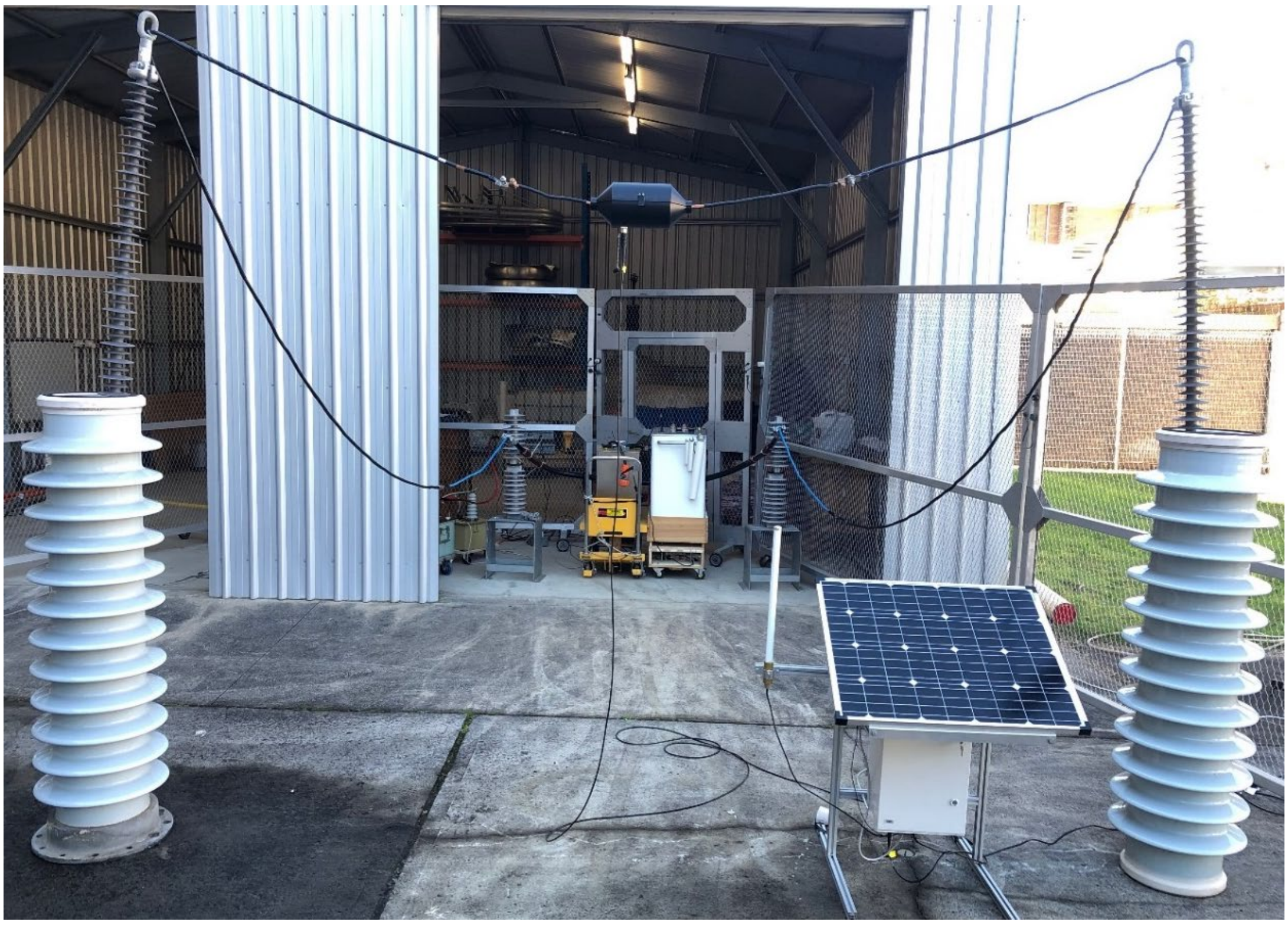

Fig. 4.4. Monitoring system calibration setup

As shown in Fig. 4.4, an experimental setup is arranged at the high voltage laboratory, National Measurement Institute (NMI), which could generate voltage up to 25 kV and current up to 38A via voltage transformer and current transformer respectively. From the generation, conductors and insulators are used to mimic the overhead SWER transmission line, as shown in Fig. 4.4. The optical sensor that can measure the voltage and current signals is attached to the bare conductor. The raw signals are captured through the sensors and transferred to the monitoring unit where they processed and analysed through signal processing techniques. Voltage and current signal are precisely measured through the 8-digit Keysight multimeter and used as the benchmark to calibrate our system. The monitoring unit is calibrated at 15 kV, 10A. Then it is verified at different times (temperature), different voltage (0 kV – 25 kV) and current (0A – 38 A) ranges. The measurement errors are reported in Fig. 4.5.

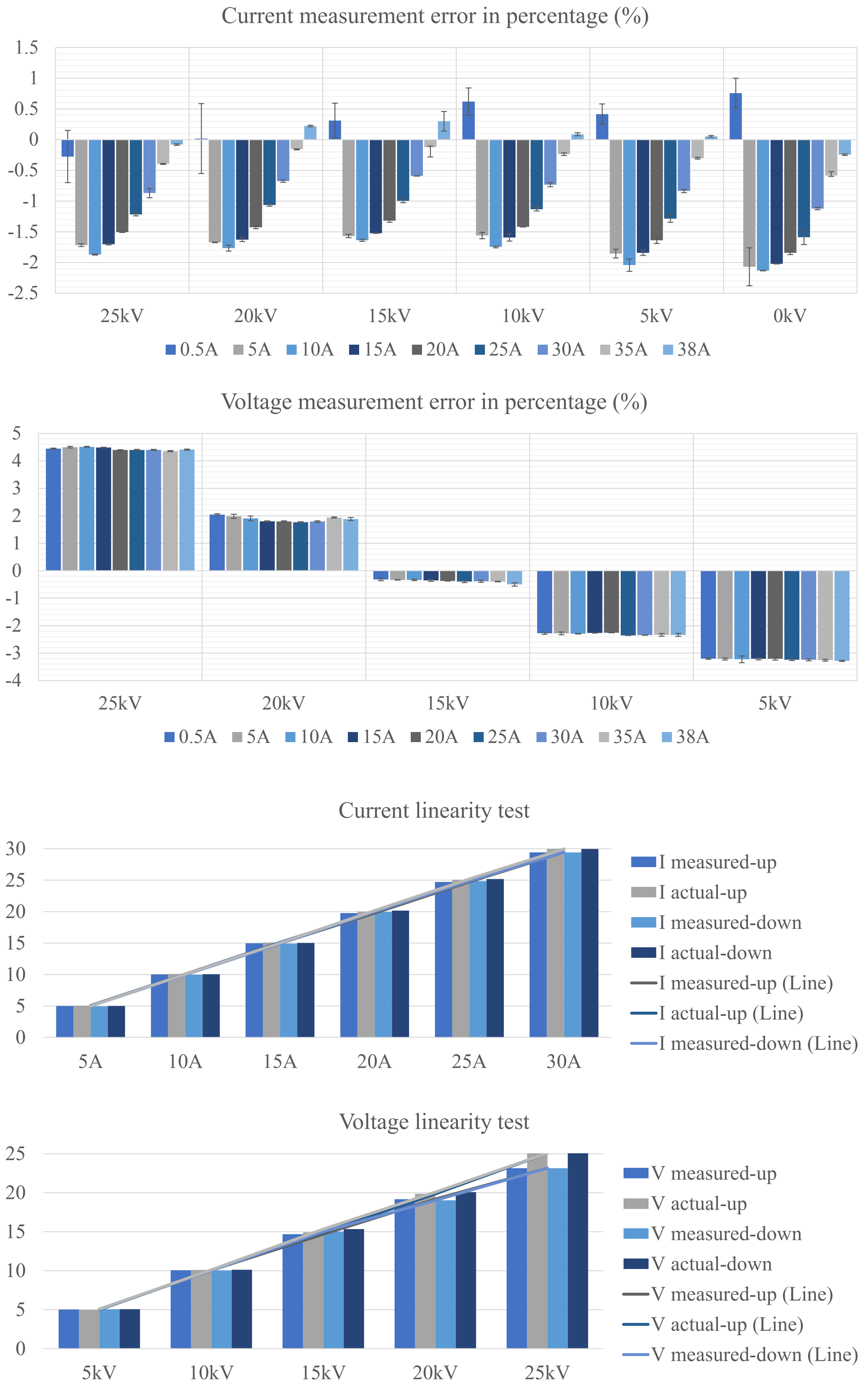

Fig. 4.5. Calibration results at different voltage and current ranges

## 4.5. THRESHOLD-BASED POWER QUALITY DISTURBANCE DETECTION EXPERIMENTS

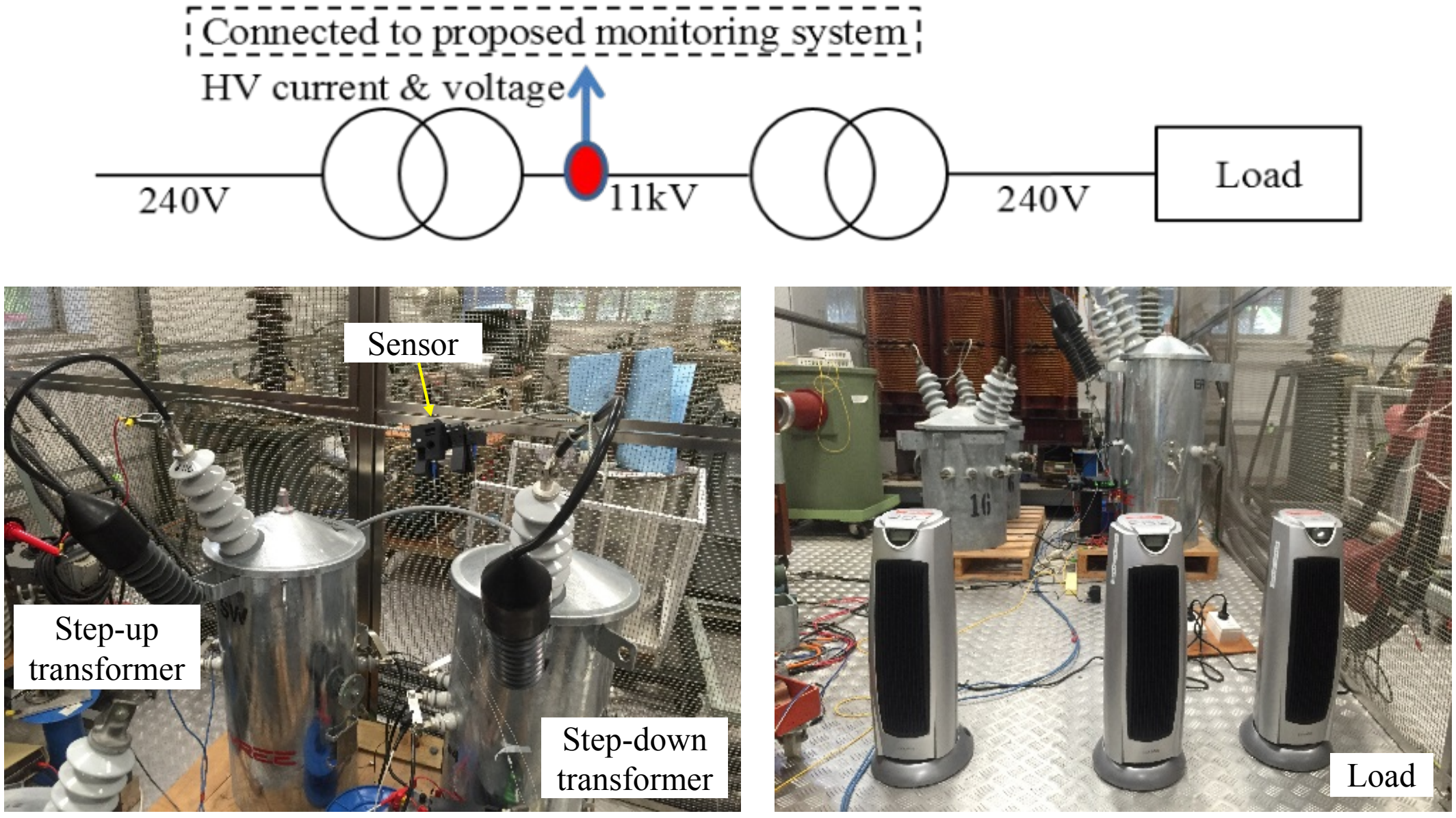

Fig. 4.6. Experimental setup

In order to test the power quality monitoring functionality of the proposed monitoring unit, an experimental SWER line is set up in the laboratory, as shown in Fig. 5. It contains two 12.7 kV SWER distribution transformers with the rating of 25 kVA, SWER conductor and loads (2,000W heaters). The first transformer steps up the AC voltage from 240V to 11kV and feeds to SWER conductor where the sensors are installed to capture the current and voltage signals. The second transformer steps down the AC voltage to 240V and supplies the loads. The sensor readings from the high voltage side are fed into the proposed monitoring unit for power quality monitoring.

There are well-defined international standards that set the threshold values to consider a specific power quality distortion. Thresholds values are chosen from the EN 50160 [1] and AS/NZS 61000.3.6 [2] standards for the proposed monitoring unit. Based on these standards, a voltage swell event begins when the voltage RMS value rises to 110% - 180% of its nominal value and ends when the voltage RMS is equal to or below the lower limit of the swell threshold. Voltage dip event initiates when the voltage RMS value falls to 10% - 90% of its nominal value and ends when the voltage RMS is equal to or above the upper limit of dip threshold. Voltage interruption event triggers when the voltage RMS value falls below 10% of its rated RMS value and ends when the voltage RMS is equal to or greater than the interruption threshold. Furthermore, rapid voltage changes can be experienced during the transition between two steady-state conditions. In such situations, the variations in voltage must not exceed the voltage dip threshold, swell threshold and the rate of change should be less than the minimum rate-of-change threshold value. When these conditions are violated, the rapid voltage change event is triggered. Fig. 4.7 reports some power quality disturbances that are detected during the laboratory experiments.

After the disturbance detection and high-frequency feature extraction, the raw waveforms are discarded due to the resource limitations in the embedded hardware. The detected power quality events and the compressed high-frequency features are transmitted to a cloud server through a communication gateway. Long-range communication from monitoring unit to the gateway is facilitated through the Long-Range Wide Area Network (LoRaWAN) which delivers its optimal performance in the line-of-sight communications. Based on the test results, the packet success ratio is almost 100% when the monitoring unit and the gateway are in the line-of-sight. However, when several buildings block the line-of-sight, up to 11% of data loss is observed. Since this research is focused on overhead power line monitoring in rural electrification schemes, it is very likely to get the line-of-sight among the nodes since they are mounted on top of the power poles and not much obstacles in rural areas. Thus, LoRaWAN based communication is well suited for long-distance data transfer in rural areas.

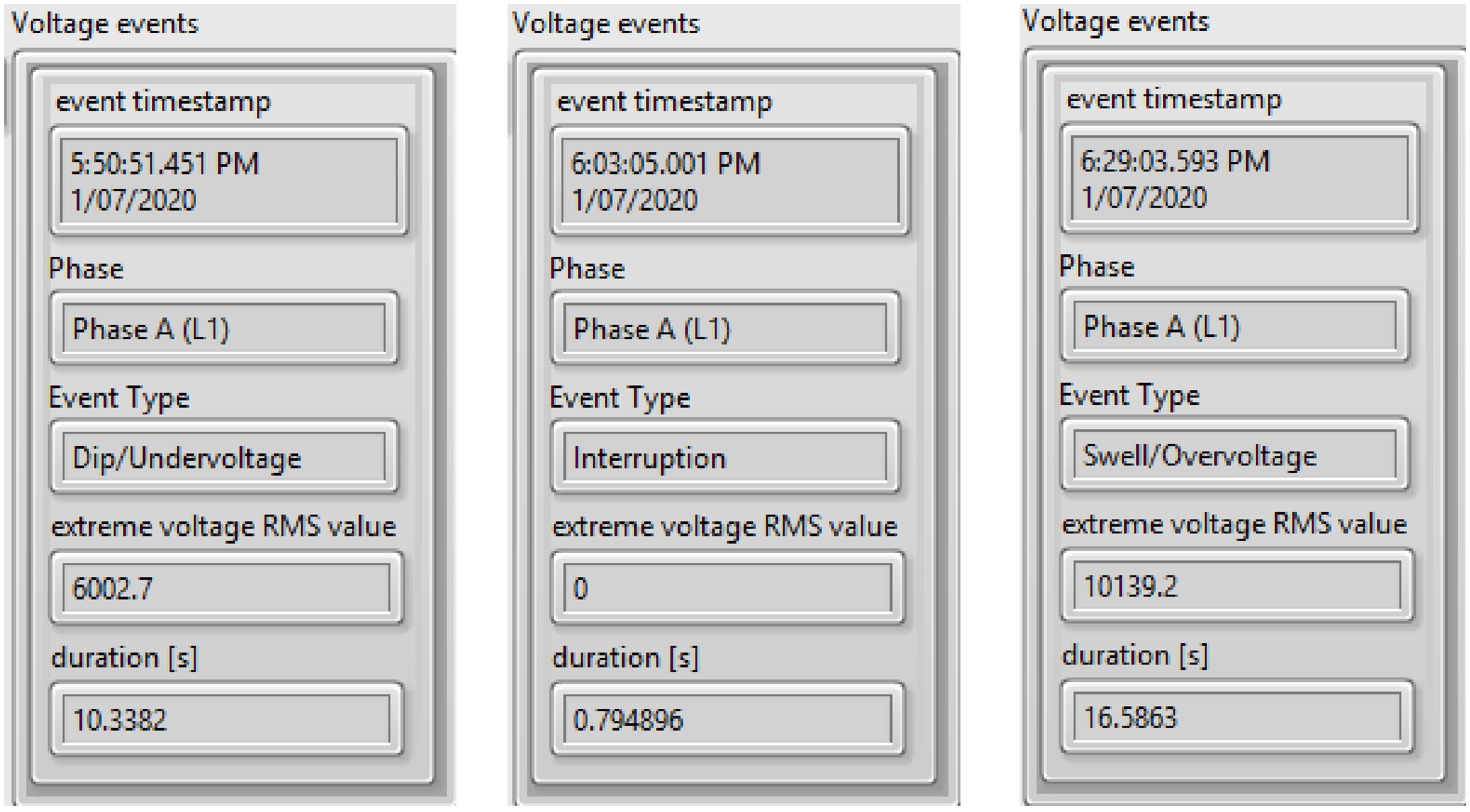

Fig. 4.7. Detected events during the experiment

Such real-time power quality information that is transmitted from multiple units along the overhead transmission lines can be synchronised using system timestamps and analysed in a control room or a cloud server to identify and isolate the power quality disturbances. These threshold-based techniques are a straight forward implementation with insignificant computational complexity and well-suited for power quality disturbance detections. However, these threshold-based detection techniques are not suitable for the faults with time-varying characteristics such as HIF, since it is not easy to calculate a threshold value for different operating conditions, contact surface, and system states. Hence, we have leveraged AI-based techniques to identify the HIF with high accuracy.

## 4.6. PROPOSED AI-BASED HIF IDENTIFICATION

HIF is challenging to be detected due to its low fault current level (the normal load current is in the order of several hundred amps, while the fault current is in the order of several amps to tens of amps). Besides, HIF current exhibits random intermittence and significant variations on its characteristics based on the contact surface and environmental conditions. HIF modelling using its characteristics and behaviours is essentially the first step towards HIF identification, especially in model-based HIF detection process. Faulty condition data need to be collected to construct the fault model. Furthermore, samples during healthy operating conditions and regular transient events also collected to prepare a data set.

### 4.6.1. Data Collection

An experimental setup is created at the laboratory for data collection, as shown in Fig. 4.8. It includes a 1 kVA variable voltage transformer, 5.55 kVA AC power supply, a 16-kVA step-up transformer with the ratio of 240V/11kV, a short length of bare aluminium conductor with 7.5 mm diameter, a 6.25 kΩ current-limiting resistor to protect the transformer and a small measuring resistor (47 Ω). Different high impedance objects such as sand, soil and tree branch are used to collect the HIF condition in various surface humid levels. The HIFs can be generated by a high voltage conductor touching the tree branch or falling to the surface of sand or soil. The applied voltage is varied from 2kV to 11kV, and the HIF current is varying from 0.01A to 0.2A depending on the surface conditions of the high impedance objects. Then, the current signals of three different HIF types are captured for fault analysis. A data acquisition system which comprises a National Instruments (NI) PXIe-1073 is used for the data collection. The output signals are sampled at 20 kHz. An anti-aliasing analog filter is placed before sampling to band-limit the signals at 10 kHz. Examples of different HIF signals are illustrated in Fig. 4.8. Data augmentation process is carried out to expand the dataset since a broad set of data is required to create an AI-based model. It is achieved by superimposing the HIF current signal with a 50 Hz sinusoid signal which has the same phase angle as the HIF signal.

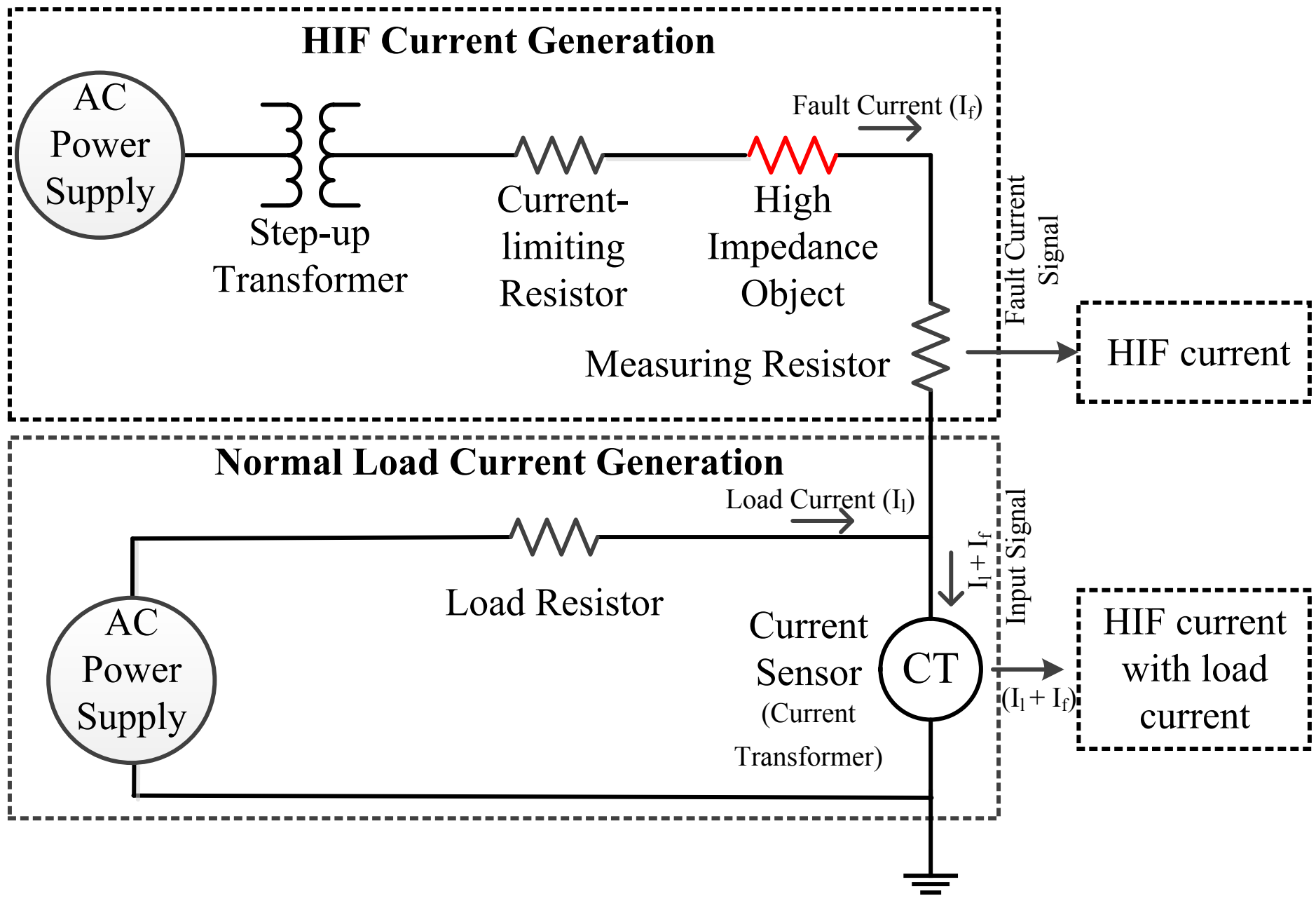

Fig. 4.8. Experiment setup for HIF data collection

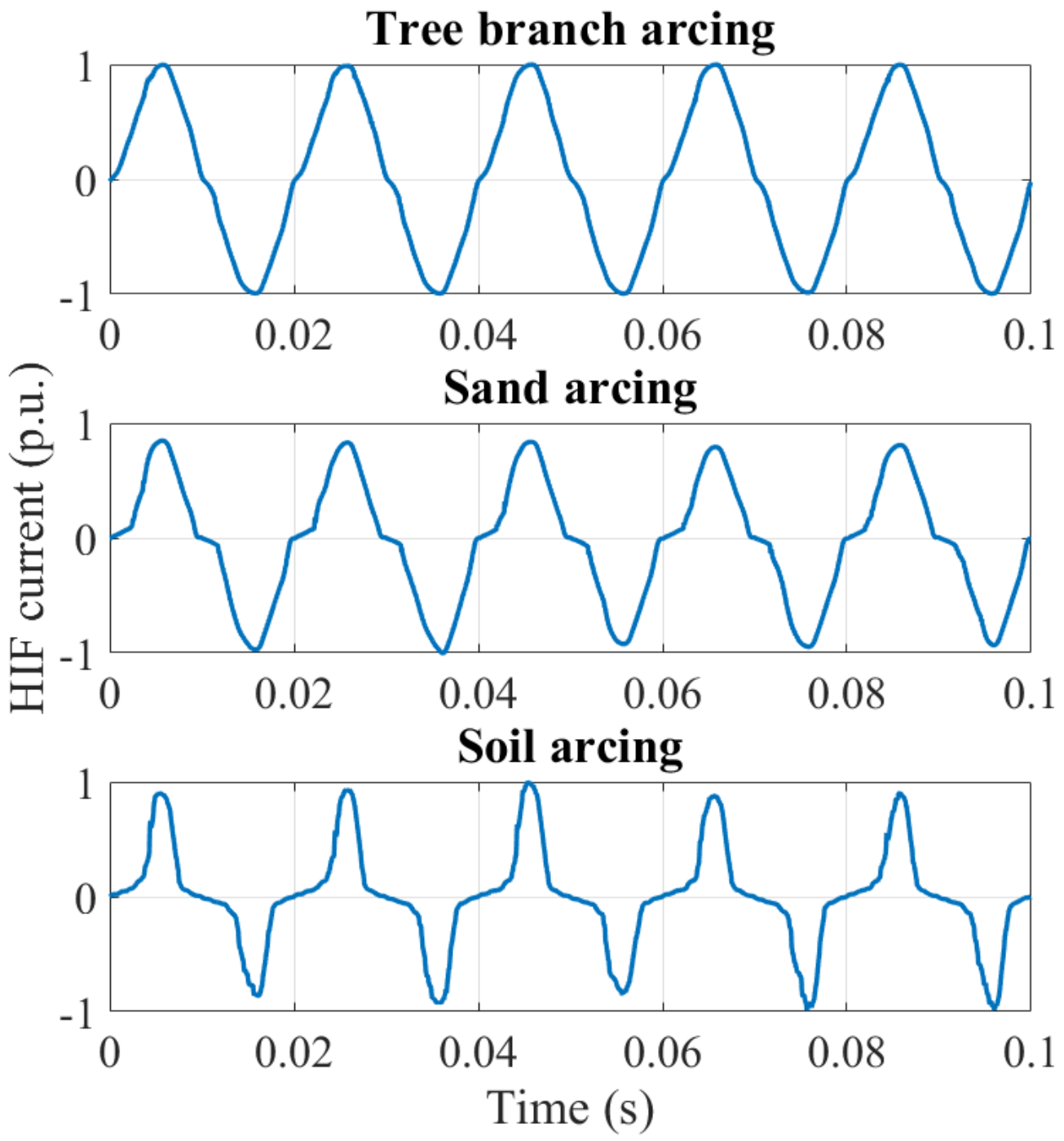

Fig. 4.9. HIF signals at different contact surfaces

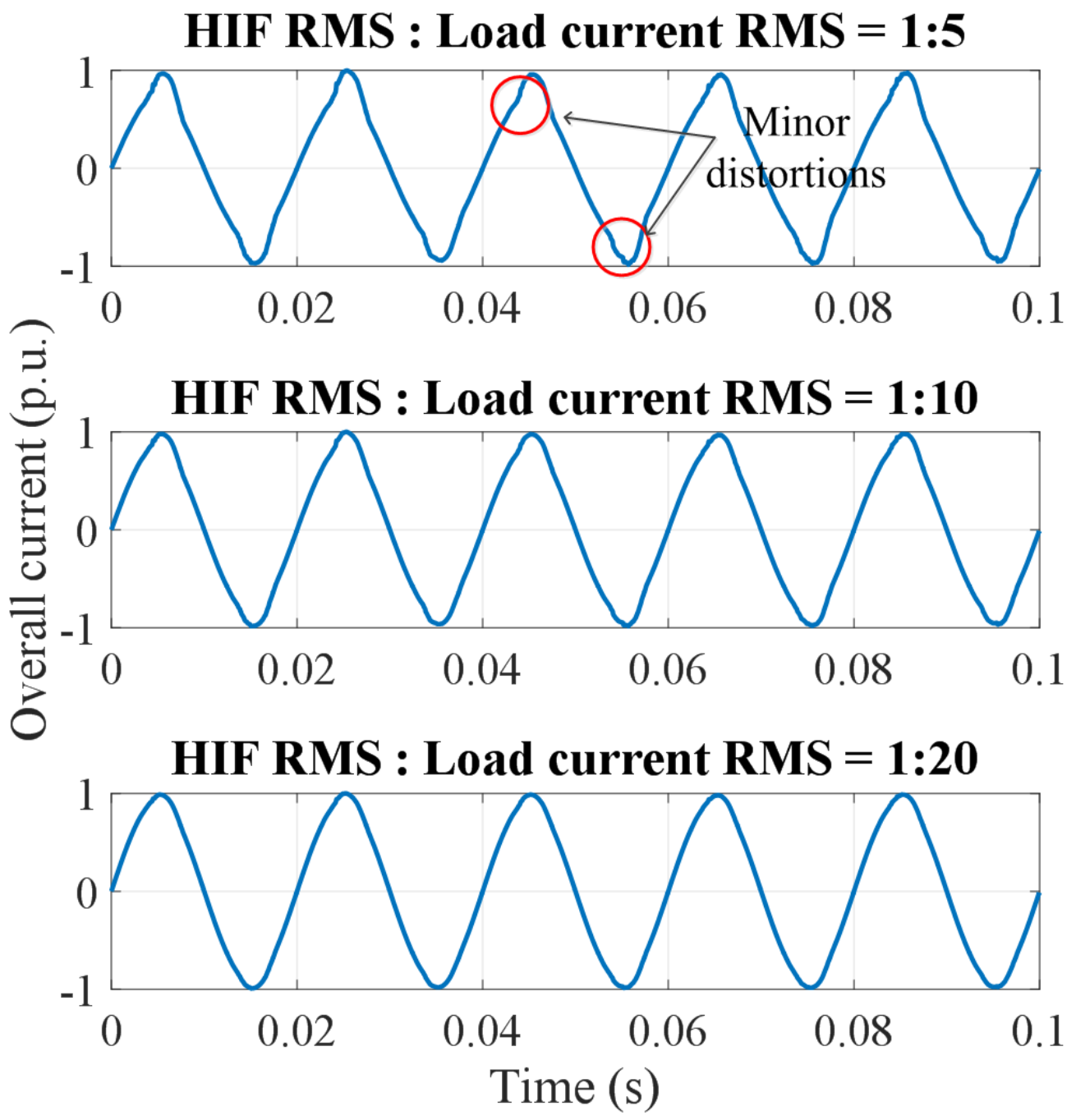

Fig. 4.10. HIF signals superimposed with load signals at different ratios

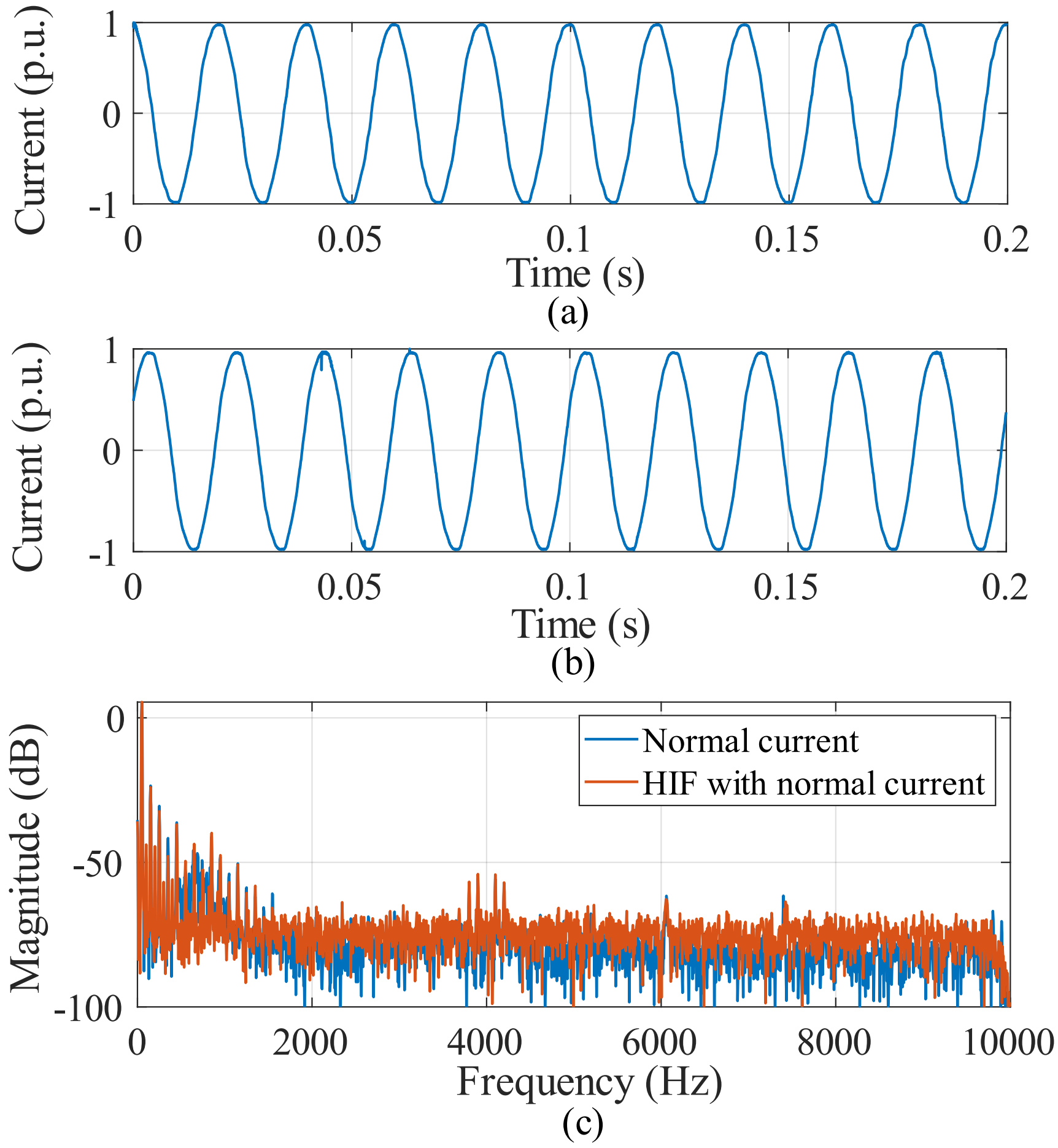

Fig. 4.11. Comparison of waveform and frequency spectrum (a) normal load current (b) HIF with normal load current (c) frequency spectrum of both

The RMS value of the 50 Hz sinusoid signal is set to 5, 10, and 20 times higher than the HIF signal, that are the common ratios in practice. Fig. 4.10 visualises the superimposed signal variations at different rates. There are only small distortions can be observed in the waveform as indicated in red circles for higher HIF current (HIF current: Load current = 1:5). On the other hand, when HIF current is smaller (HIF current: Load current = 1:20), it will ultimately be masked by the large load current, and the distortions are not visible. However, frequency-domain analysis exposes some noticeable variations between HIF conditions and healthy conditions, as shown in Fig. 4.11. It can be seen that the noise intensity increases after HIF occurs in Fig. 4.11 (c).

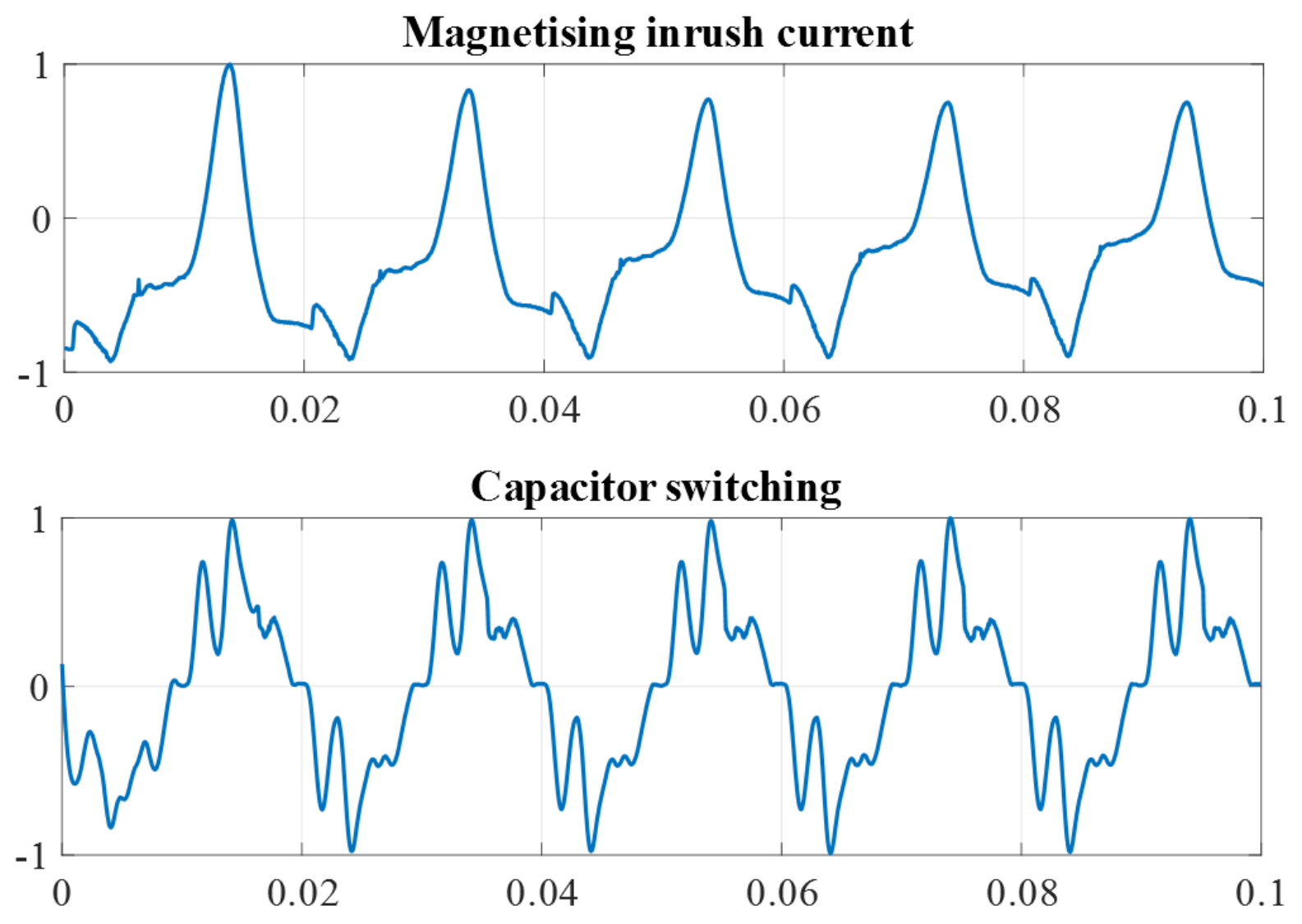

Fig. 4.12. Normal transient signals similar to HIF

Besides the HIF signal, other transient waveforms such as magnetising inrush and capacitor switching current are collected, which show similar properties, as visualised in Fig. 4.12. These transient waveforms also manually superimposed (identical to the ratios shown in Fig. 4.10) with the large load currents, which has the same phase angle. After the data collection, all types of HIF signals are commonly labelled as HIF events, whereas the normal transients and the standard current signals are labelled as healthy events. This dataset is used to generate a model that can discriminate the HIF event from the normal operating conditions.

The second step of the HIF identification process is feature extraction. Since the frequency domain visualisation of HIF and healthy signals exhibits more variations compared to the time-domain, feature extraction is carried out in the frequency domain using digital signal processing techniques such as WPT and FFT, that are detailed in Chapter 3.

### 4.6.2. Wavelet Packet Entropy based feature

The HIF and healthy signals are analysed using WPT to extract the intrinsic properties of those signals. As described in Chapter 3, WPT can decompose both lower and higher frequency band of the signal at each decomposition level, which provides more information compared to DWT. WPT is applied to the input signal, and it results in a set of wavelet coefficients at each decomposition levels and frequency bands. It can be represented as follows:

$$W_m^j = \{w_{m1}^j, w_{m2}^j, \dots \dots . w_{mN}^j\} \tag{4.1}$$

where $j$ , $N$ and $m = 1,2, \dots \dots, (2^{j-1} - 1)$ represent the decomposition level, the number of coefficients and frequency bands, respectively. Frequency components of the HIF waveform show rapid variations due to the intermittence and randomness characteristics of HIF. Hence, the entropy principle is adapted from the information theory, which can quantify the degree of disorder and measure the uncertainty. The entropy value will reach its maximum when all events have the same uncertainty. Entropy is calculated for each frequency band to quantify the variations, that can be used as feature instances. The entropy calculations can be mathematically represented as follows:

$$En_m^j = -\sum_{n=1}^{N} p(w_{mn}^j) \log p(w_{mn}^j) \tag{4.2}$$

$$p(w_{mn}^j) = \frac{|w_{mn}^j|^2}{\sum_{n=1}^{N} |w_{mn}^j|^2} \tag{4.3}$$

where $p(w_{mn}^j)$ is the probability of $w_{mn}^j$, and $\sum_{n=1}^{N} p(w_{mn}^j) = 1$. Then the calculated entropy is normalised as below:

$$En_{norm,m}^j = \frac{En_m^j}{\sum_{m=1}^{2^{j-1}} En_m^j} \tag{4.4}$$

As discussed in chapter 3, there are two critical parameters, such as decomposition levels and mother wavelet, which need to be decided appropriately in the application of WPT. It directly influences the results in terms of accuracy and computational complexity. From the experiments with different decomposition levels and mother wavelets, it has been identified that three levels of decomposition with Daubechies 9 (db9) mother wavelet demonstrates better discrimination between HIF and normal conditions. Each cycle of an input current signal is divided into four consecutive frames, and wavelet packet entropy is calculated for three decomposition levels. The decomposed entropy features from three successive cycles are consolidated to formulate a 2-D feature map, that can be used to discriminate the HIF from the typical power system disturbances. Several classifiers such as ANN, Support vector machines (SVM) and CNN have been used to evaluate the feature, and their HIF identification accuracy is 96.1%, 92.64% and 98.53%, respectively. Even though the wavelet entropy based feature map demonstrates high accuracy in HIF discrimination, the computational complexity of the three-level decomposition is comparatively higher than the FFT based feature extraction techniques. The time complexity of the WPT is $O(L \times N \log N)$ where $N$ and $L$ are input signal length and decomposition level, respectively. During the implementation process, the WPT based feature extraction algorithm failed to execute in real-time on low power embedded hardware due to its time complexity. Hence, an FFT based feature extraction technique with lower time complexity $O(N \log N)$ is formulated, which enables the real-time execution on embedded devices while achieving high HIF identification accuracy.

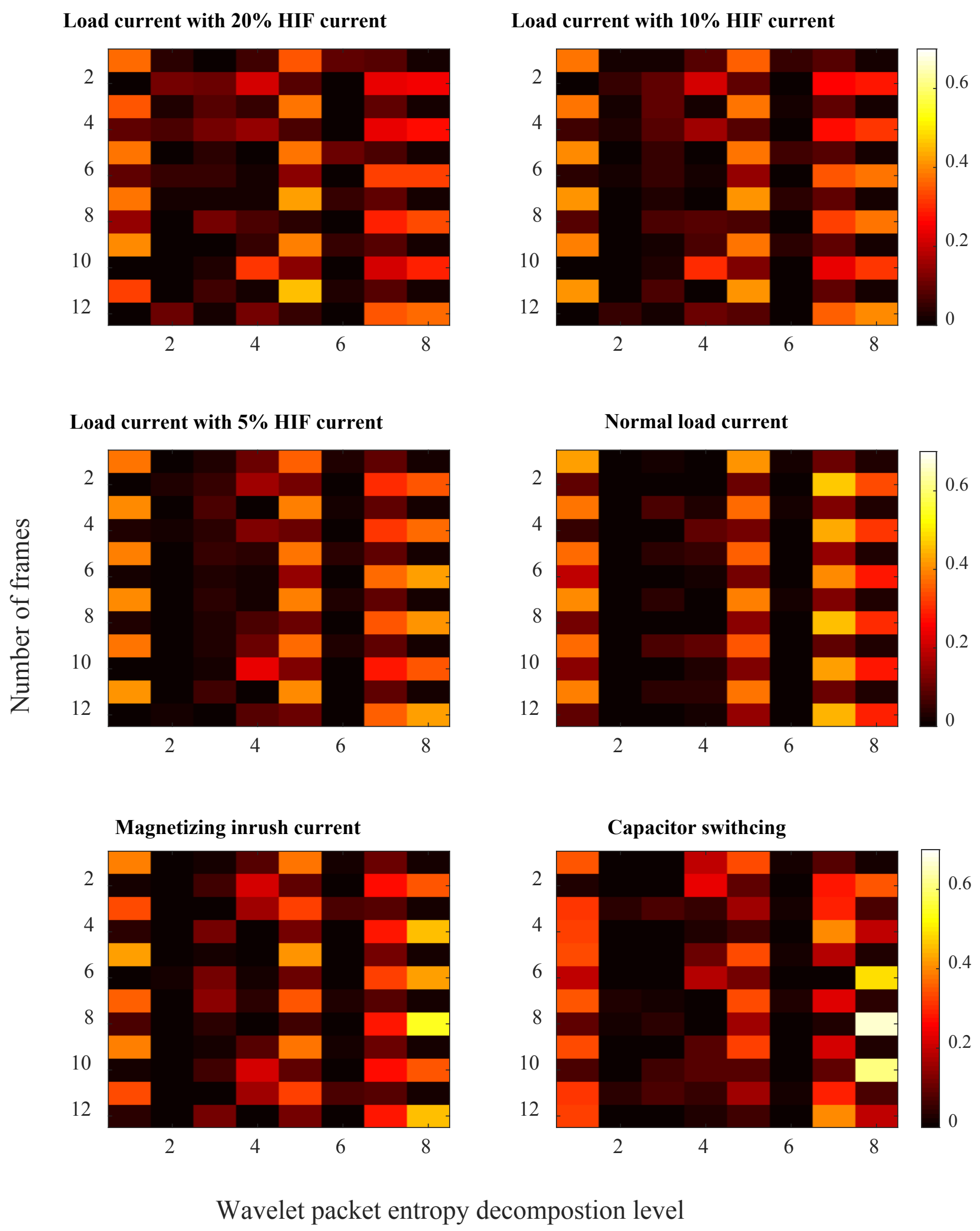

Fig. 4.13. Visualisation of proposed feature maps for different events

### 4.6.3. FFT BASED FEATURE

A short-time FFT based 2-D feature is proposed, which achieves better performance compared to the wavelet entropy based feature. Furthermore, it is optimised to be extracted in real-time with reduced computational requirements on embedded hardware. As the first step, the input signal is framed using a Hanning window to minimise the discontinuities of truncated waveforms when it is framed as a finite-length data. While smoothing out the discontinuities, it attenuates the sample points at the edges of the framed signal to reduce the spectral leakage. This can mask the features near both ends in a frame. To avoid that, the Hanning window is applied to the time-domain data samples with 50% overlapping. The mathematical representation of the Hanning window can be expressed as follows:

$$Hann(j) = \frac{1}{2} - \frac{1}{2} cos\left(\frac{2\pi j}{N}\right) \tag{4.5}$$

where $N$ is the number of sample points in an input frame. This framed signal is fed into the FFT algorithm to compute the frequency components. Higher frequency resolution can be achieved when the number of points $(N)$ in a frame is increased. Therefore, the frequency spectrum gives more accurate results for higher values of $N$.

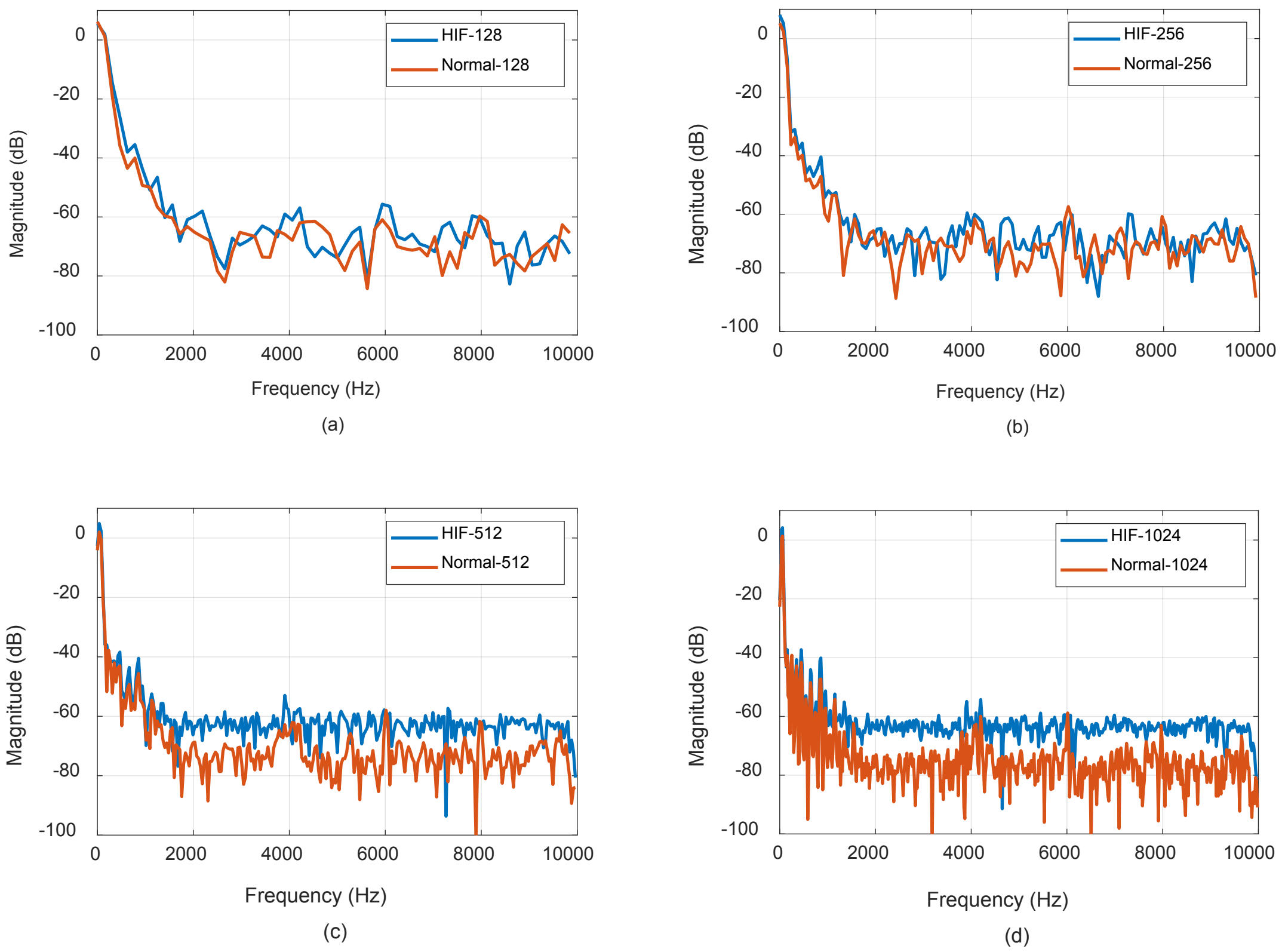

Fig. 4.14. Frequency spectrum comparison between HIF and normal current in case of (a) 128 FFT points; (b) 256 FFT points; (c) 512 FFT points; (d) 1024 points

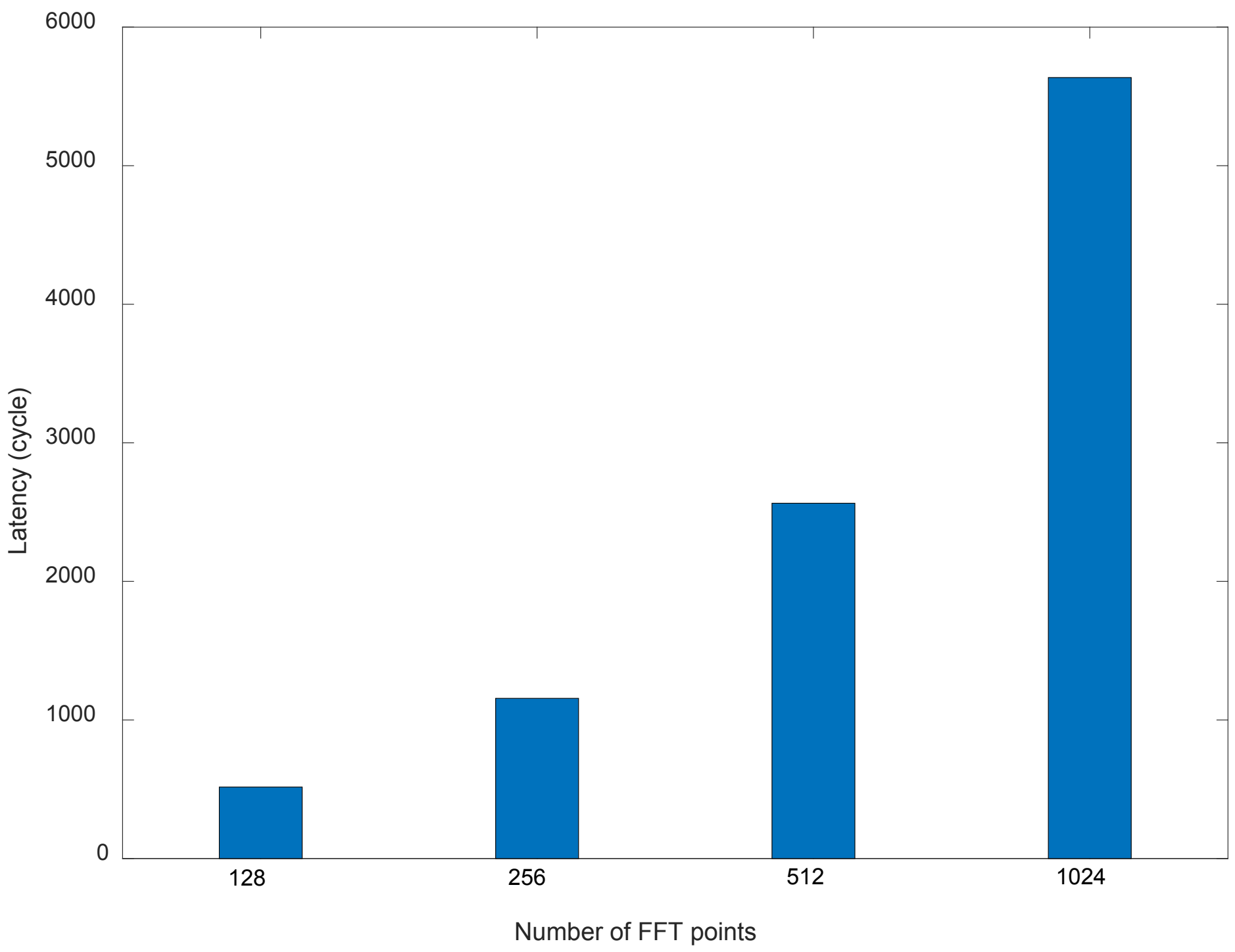

Fig. 4.15. Latency requirements of varied number of FFT points in a frame

An experiment has been conducted to find an optimal frame size for HIF discrimination. The comparison results are visualised in Fig. 4.14. The frequency spectrum of HIF and normal current looks very similar for $N$ = 128 and 256, because of the severe spectral leakage. In contrast, the HIF event can be discriminated from the normal operating condition when $N$ = 512, and there is almost no improvement when further increasing $N$ to 1024, as shown in Fig. 4.14. (c) and (d). Moreover, increasing $N$ negatively affects the computational complexity as well as the latency. Fig. 4.15. highlights the latency requirement for the different number of FFT points in a frame. Since HIF detection is a time-sensitive application, using $N$ = 512 in the proposed algorithm can achieve an appropriate balance across accuracy, latency, and computational complexity.

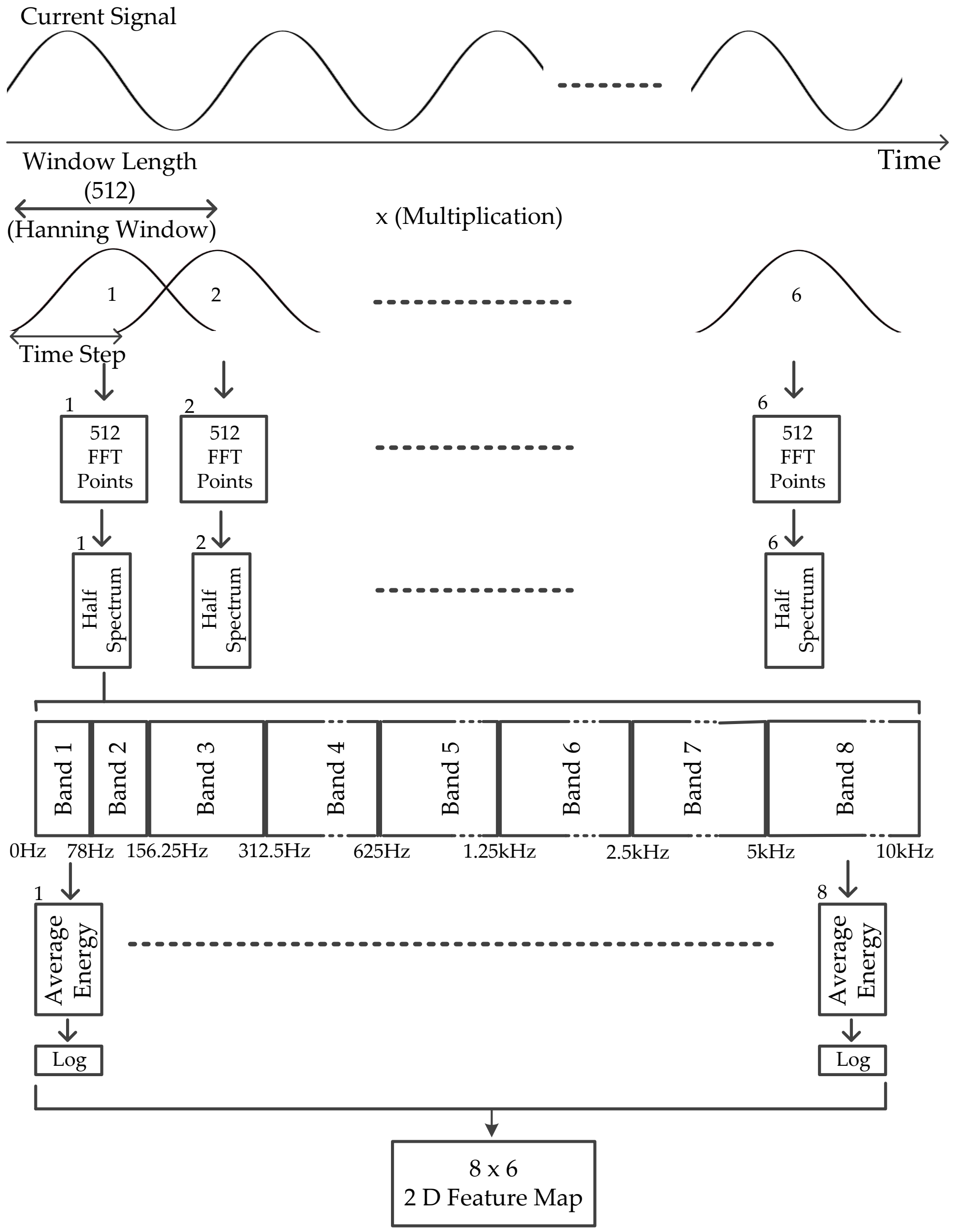

Fig. 4.15. Proposed feature extraction process

Fig. 4.15 outlines the proposed feature extraction methodology. A frame with 512 sample points corresponds to 1.28 cycles (25.6 ms) of 50 Hz current signal. The nature of the fault signal is intermittent and have shown distinctive characteristics in each cycle. So, it is worthwhile to analyse the frequency components cycle-by-cycle to extract the common patterns during the fault. FFT algorithm computes 512 symmetric frequency points ranging from 0 – 2π rad for each framed signal. The calculated frequency points are symmetric such that the first 256 FFT points correspond to the frequency range from 0

Table 4.1: Harmonics information of feature vector

| **Frequency Band** | **Harmonics Information** |
|---|---|
| Band 1 (0Hz – 78Hz) | Fundamental component |
| Band 2 (78Hz – 156Hz) | $2^{nd}$ & $3^{rd}$ Harmonics |
| Band 3 (156Hz – 312Hz) | $4^{th}$ ,$5^{th}$ & $6^{th}$ Harmonics |
| Band 4 (312Hz – 625Hz) | $7^{th}$ - $12^{th}$ Harmonics |
| Band 5 (625Hz – 1.25kHz) | $13^{th}$ - $25^{th}$ Harmonics |
| Band 6 (1.25kHz – 2.5kHz) | $26^{th}$ - $50^{th}$ Harmonics |
| Band 7 (2.5kHz – 5kHz) | $51^{st}$ - $100^{th}$ Harmonics |
| Band 8 (5kHz – 10kHz) | $101^{st}$ - $200^{th}$ Harmonics |

Hz – 10 kHz. Then the whole frequency range is divided into specific ranges of frequencies named as frequency bands. This process is known as sub-band decomposition. An octave scale is used in the proposed feature to decompose the frequencies such that the upper band frequency is twice the lower band frequency.

Table 4.1 summarises the frequency bands decomposition along with the harmonics information. Octave scale decomposition is chosen to yield more selectivity (narrow bandwidth) to the lower order harmonics since they have exhibited more fluctuations. On the other hand, higher-order harmonics demonstrated consistent divergence of HIFs from the healthy system. Larger bandwidths (less selective bands) are suitable to capture consistent patterns in higher-order harmonics. After the decomposition of frequencies with appropriate bandwidths, the average energy is calculated for each frequency bands. After that, a feature vector (8 x 1) is formulated by applying a logarithmic transformation

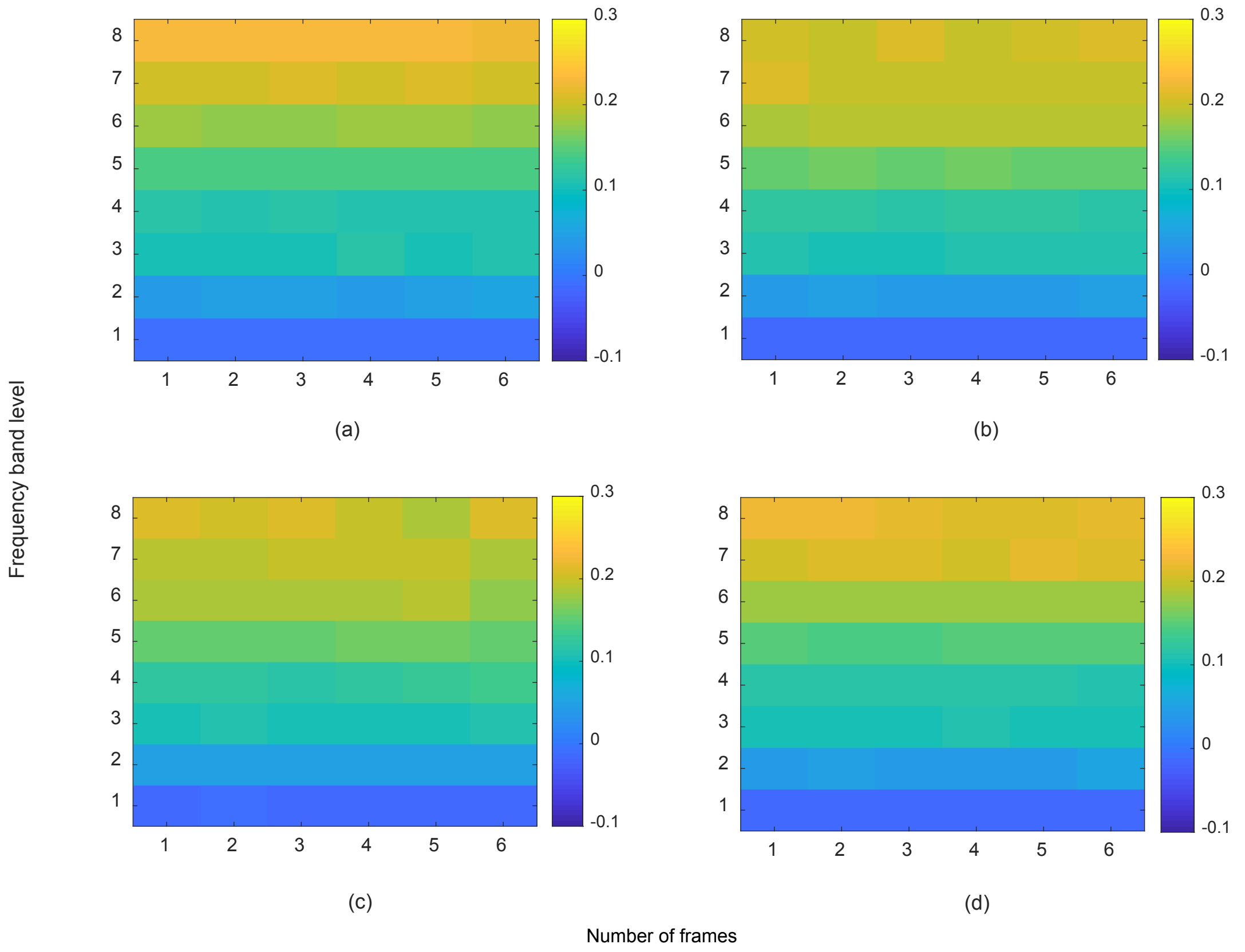

Fig. 4.16. Short time FFT based 2-D Feature Map: (a) Normal Current; (b) HIF generated by tree branch; (c) HIF due to sand contact (d) HIF due to soil contact

to the frequency band average energies. The logarithmic transformation is applied to enhance the visibility of patterns in energy levels.

A key point to note during the feature extraction process is that the extracted feature should be capable enough to identify the faults as well as immune to false-positive results. Since the fault current nature includes intermittency and random variations as its properties, the feature needs to be more reliable to prevent false positives. Six adjacent frames are consolidated into a 2-D feature map to improve the reliability of the proposed feature. Based on that, 4.5 consecutive current waveform cycles are taken into consideration in the 2-D feature map to capture the properties of HIF such as asymmetry, intermittency, and buildup. As a result, eight frequency band average energies of six successive frames are stacked in a 2-D feature map to identify the HIF in real-time. Fig. 4.16. displays the extracted feature maps for different signals. While the variations in higher frequency bands are clearly observable, there is not much difference in the lower band levels of HIF and normal feature maps. These feature maps can be further enhanced by the proposed deep learning based classifier to identify faulty conditions in real-time with higher accuracy.

#### 4.6.4. Deep learning based HIF Classifier

As described in Chapter 3, CNN based classifiers continuously achieve state-of-the-art performances on various tasks in different domains such as image classification, object detection, semantic segmentation and speech recognition. Inspired by the success of CNNs, a light-weight CNN architecture is proposed to classify HIFs in real-time at resource-constrained embedded devices. Fig. 9 outlines the internal structure of the proposed CNN architecture. The extracted 2-D (8x6) feature map is fed into the input layer of the CNN. Then it is forwarded to a convolutional layer which contains four filters of size 2x2 to extract higher-level features from the input. Several combinations of convolutional filters are tested based on the domain-specific intuitions and its computational complexities. Fig. 4.18. reports the comparison results in terms of latency and accuracy. The convolutional layer is structured to perform the computations as convolutions with the sliding filters that can only cover a small neighbourhood of input activations. The convolution operation is accomplished by moving the filters along the input vertically and horizontally and calculating the dot product of the weights *(W)* and the input *(I)* and then adding a bias *(B)* term. The filter movement along the input is determined by the stride *(S)*. The mathematical representation of a 2-D convolutional layer can be defined as follows:

$$O[f][x][y] = \sum_{k=0}^{C-1}\sum_{i=0}^{X-1}\sum_{j=0}^{Y-1} I[k][S*x+i][S*y+j] \times W[f][k][i][j] + B(f) \quad (4.6)$$

where $O$, $f$, $C$, $X$ and $Y$ denote the output matrix, number of filter groups in a convolutional layer, number of channels in the input, filter height and filter width respectively. A filter group in a convolution layer contains a stack of 2-D filters to process different channels in the input. But, the proposed feature map contains only one channel (1 x8 x 6).

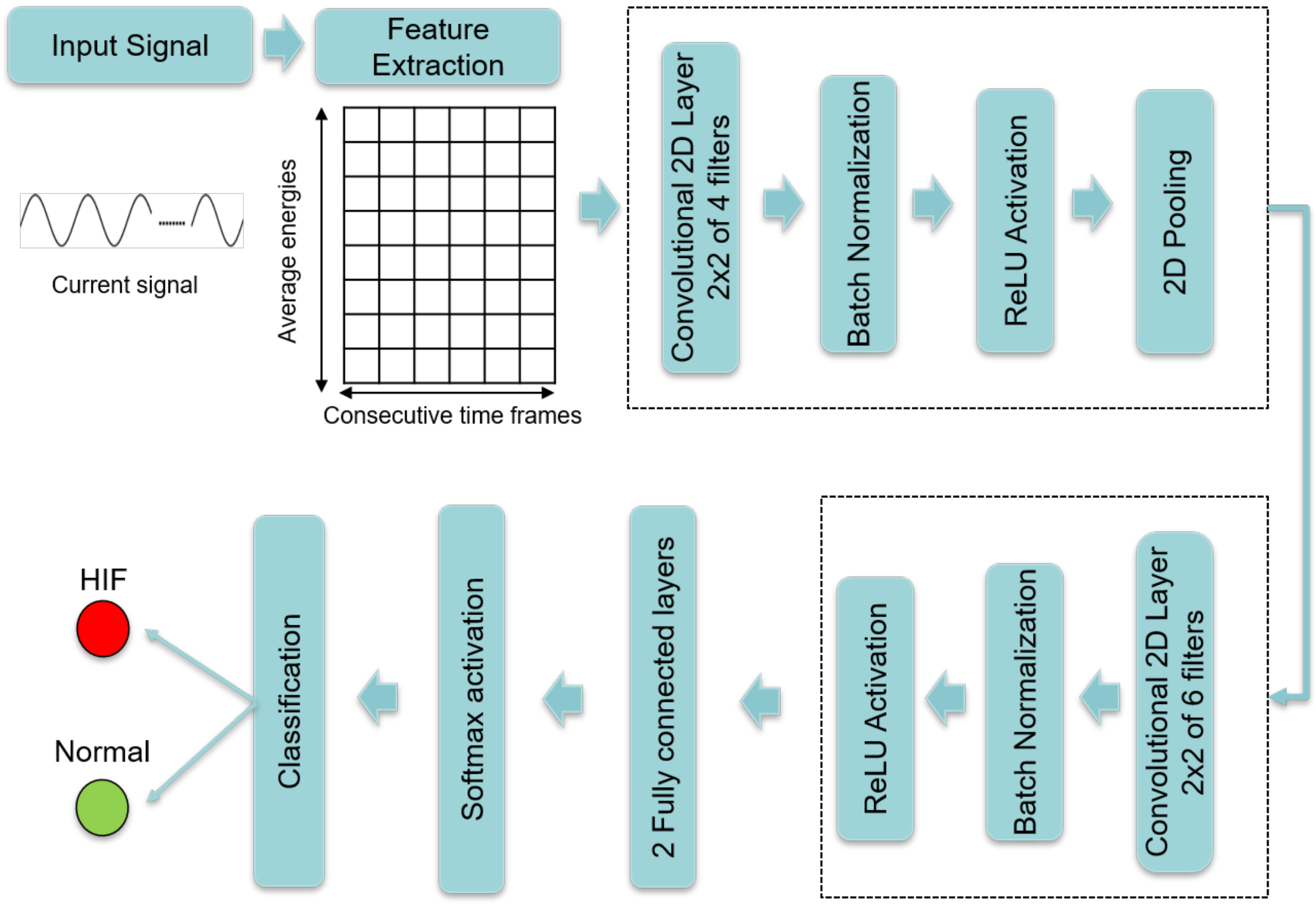

Fig. 4.17. Proposed CNN Architecture

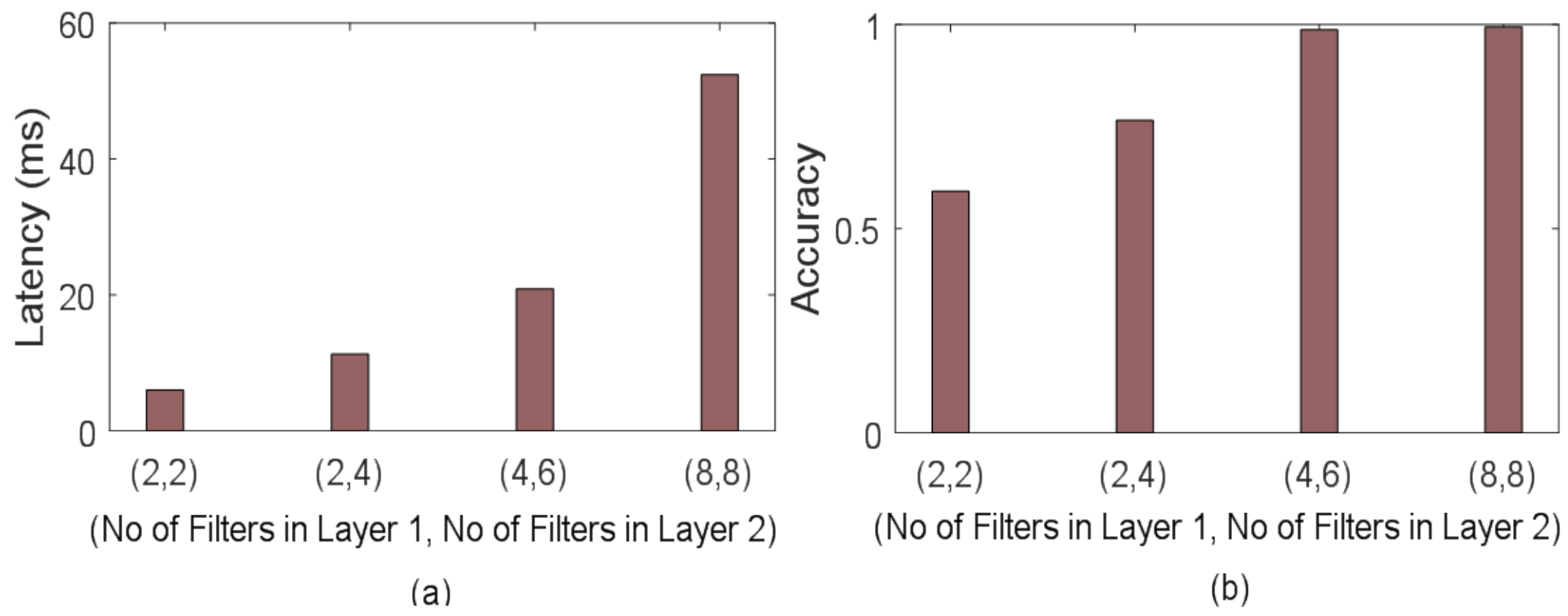

Fig. 4.18. Latency and accuracy results for varied no of convolutional filters: (a) Latency requirement (b) Detection accuracy

The output from the convolutional layer is forwarded to a batch normalisation layer. It normalises each input channel as mini-batches. The computation of a batch normalisation can be defined as:

$$y_i = \gamma\left(\frac{x_i - \mu_B}{\sqrt{\sigma_B^2 + \varepsilon}}\right) + \beta \quad (4.7)$$

where $x_i$, $\mu_B$, $\sigma_B^2$, $\gamma$, and $\beta$ are input, mini-batch mean, mini-batch variance, scale factor and offset. $\varepsilon$ is introduced to improve numerical stability when the minibatch variance is relatively small (closer to 0). Initially, normalisation layer transforms the input to zero mean and unit variance mapping. Then it shifts and scales that mapping with the learnable parameters $(\gamma, \beta)$ to make it optimal for the successive layers in CNN. The output from batch normalisation is forwarded through a threshold operation using Rectified Linear Unit (ReLU) activation. It replaces zero for the values less than zero. The reason for placing normalisation layers and nonlinear activations between convolutional layers is to improve network stability and the learning speed. Followed by the nonlinear activation, the feature map is down-sampled by a 2D max-pooling layer. Down-sampling is performed by dividing the feature map into 2 x 2 pooling regions and extract the maximum of each region. The max-pooling operation transforms the feature map to be robust and invariant to minor fluctuations and distortions. Furthermore, it reduces the dimensionality of the feature map, which can contribute towards the reduction of computational costs in the consecutive layers.

Then, the feature map is forwarded to the 2$^{nd}$ convolutional layer, which contains six filter groups. Each group is a stack of four filters to process the four channels produced by the 1$^{st}$ convolutional layer. The convolution output in each filter groups is summed across all four channels. As shown in Fig. 4.18, this combination can provide a good balance between latency and accuracy. After convolution, the output feature map is normalised by a batch normalisation layer and sent through a ReLU activation. Series of convolutions and nonlinear activations produce more higher-level feature maps. Then those feature maps are forwarded to fully connected layers for classification. Fully connected layer multiplies the flatten higher-level feature map by a weight matrix and then adds a bias vector. The softmax activation function is applied to the fully connected layer output to calculate the probability of HIF. It can be mathematically represented as follows:

$$y_r(x) = \frac{e^{x_r}}{\sum_{j=1}^{k} e^{x_j}} \tag{4.8}$$

where $x$ is a vector of inputs from the last fully connected layer, and $k$ represents the number of output classes. Classification results can be produced based on the probability of the output classes. The deviation between the actual scores and the scores predicted by CNN is defined as a loss. In the proposed network, cross-entropy function is used to calculate the loss, and it can be formulated as follows:

$$E(\theta) = -\sum_{i=1}^{n}\sum_{j=1}^{k} t_{ij} \ln y_j(x_i, \theta) \tag{4.9}$$

where $\theta$ is the parameter vector, $t_{ij}$ denotes the $i^{th}$ sample belongs to the $j^{th}$ class, and $y_j(x_i, \theta)$ is the output for the $i^{th}$ sample. Once the loss function is formulated, CNN is trained to find an optimal set of weights that can minimise the above loss function. Typically, training requires a large dataset with class labels. After learning an optimal set of weights, the proposed CNN can be deployed as a HIF classifier in practical applications.

### 4.6.5. Embedded Hardware for Algorithm Implementation

A portable reconfigurable embedded device (National Instruments myRIO-1900) is used as the embedded hardware for the proof of concept. The integrated device contains three main units: 1. Data acquisition (DAQ) hardware, 2. Field Programmable Gate Arrays (FPGA) and 3. Microprocessor with the real-time operating system. The DAQ includes analog input channels with analog-to-digital converters that can digitise the analog current signals with 12 bits resolution. The signal sampling rate is set to 20 kHz due to the requirement of high-frequency components analysis for fault detection. The microprocessor could not handle DAQ with higher sampling rates. In this case, the acquired signal needs to be processed by dedicated hardware. FPGAs are well suited to process the data from highspeed DAQ systems. The embedded device has a Xilinx Z-7010 type FPGA which accommodates 80 digital signal processor slices (DSPs), 60 blocks of random access memory (RAM), slice registers and look-up tables that can be used for high-speed signal processing. In addition to the FPGA hardware, the embedded device contains a 2-core 667 MHz microprocessor with the Linux-based real-time operating system (RTOS).

#### 4.6.6. OPTIMISATIONS FOR REAL-TIME EDGE PROCESSING

This section focuses on the efficient processing of real-time HIF classification at embedded edge devices. The dominant challenge in real-time data processing at resource-constrained edge devices is to handle the input data stream with minimised latency and higher throughput to deliver real-time results. The delay in the processing of incoming data stream can cause data overwriting due to the limited amount of memory in the edge node. Overwriting a data stream before it is processed will result in data loss that can drastically affect the detection accuracy. Higher throughput and lower latency need to be ensured during data analytics to avoid these circumstances.

Throughput can be defined in two directions, such as incoming throughput and outgoing throughput, to evaluate the system performance. The input load on the system is represented by the incoming throughput [3]. It is commonly expressed as samples per second(S/s) [4]. Since the current signal is sampled at 20kHz, the detection system is expected to analyse 20,000 sample points in a second. On the other hand, outgoing throughput is the measure of the rate at which the system can produce results.

Latency can be described as the amount of time taken to complete an operation. Typically, latency is reported in units of time such as microseconds, milliseconds and seconds. In the context of HIF detection, the faults need to be detected as fast as possible to avoid bushfire hazards. The protection response time is expected to be less than 200 milliseconds to contribute towards the fire risk reduction. In addition to that, the HIF detection system demands the deterministic execution of tasks since it requires to respond to the events within a given time limit. By considering the requirements above, optimisation techniques such as FPGA based parallelism, pipelined execution of tasks, and timed loop execution using RTOS have been adapted to the real-time HIF detection.

#### 4.6.6.1. FPGA BASED TRUE PARALLELISM

FPGA based deployments are well suited for the applications with lower latency requirements. FPGAs are made of programmable hardware logic blocks along with reconfigurable interconnects, which can be programmed to define their functionality. As mentioned in the previous sections, the HIF detection task can be partitioned into data acquisition, feature extraction, and fault identification. Data acquisition task needs to be continuously executed in the interval of 50 microseconds to sample the signal at 20 kHz. Since data acquisition and framing with 50% overlapping requires guaranteed highspeed execution, it is deployed in dedicated FPGA slices. Data acquisition and framing tasks continuously push the data into an internal FIFO buffer which has a limited number of slots, as shown in Fig. 4.19. These data need to be processed at least the same rate of sampling to avoid the buffer overflow in the continuous operation. So, the feature calculation logic is implemented on a different section of FPGA hardware. This implementation approach allows both data acquisition and feature calculation to run exceptionally parallel. Furthermore, it reduces the latency and eliminates jitter. The FPGA resource requirements for this implementation is reported in Table 4.2. After extracting the feature, the raw signal is discarded due to insufficient storage in the edge device. Only the extracted feature is forwarded to the real-time program.

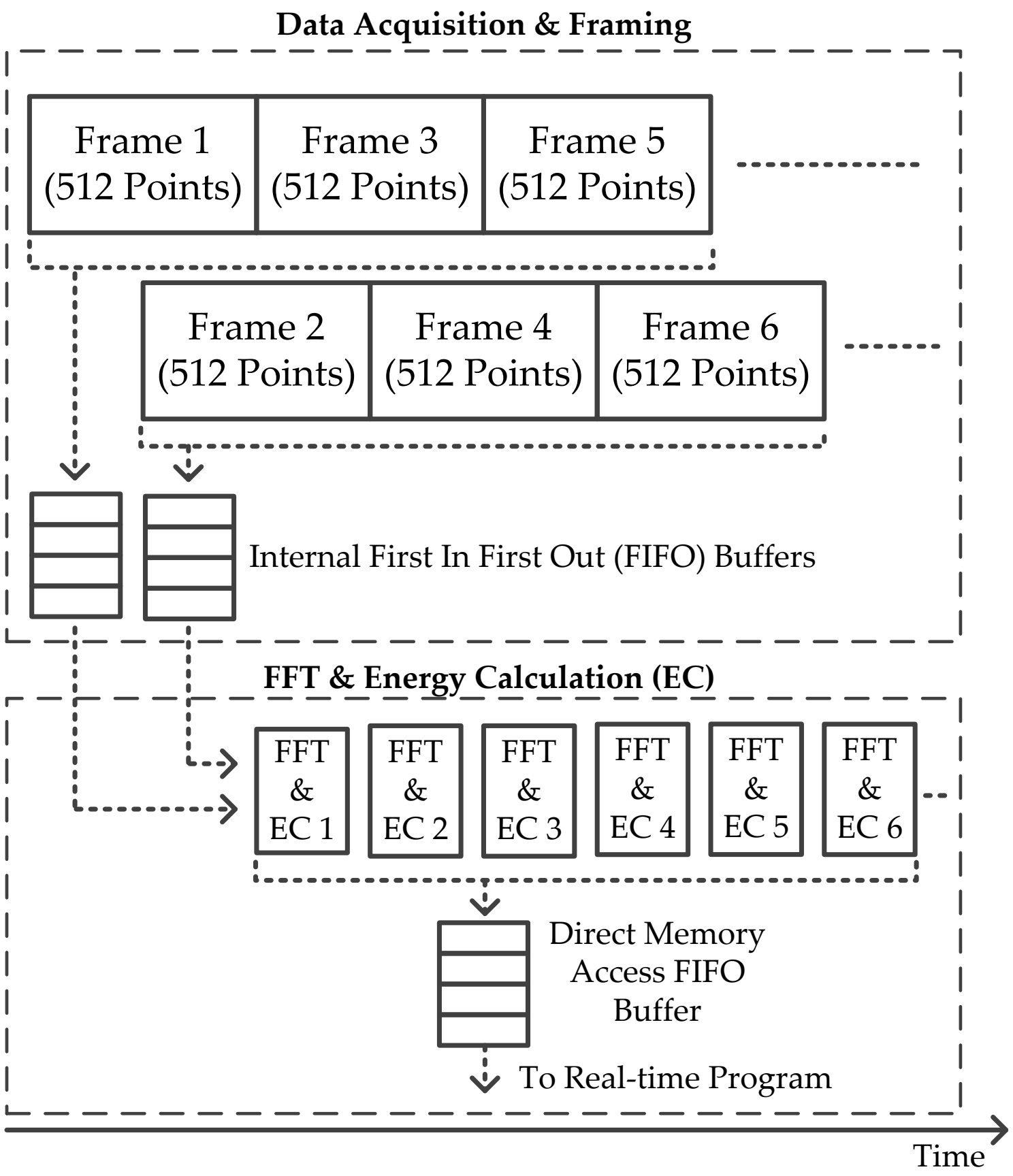

Fig. 4.19. FPGA based true parallel execution

Table 4.2: FPGA resource utilisation

| Resource name | Count (usage%) |
|---|---|
| Slice registers | 14413 (40.9%) |
| Slice Look-up tables | 14678 (83.4%) |
| Block RAMs | 20 (33.3%) |
| DSP 48s | 69 (86.2%) |

#### 4.6.6.2. PIPELINING

Pipelining improves the execution performance by decomposing long latency tasks into several sub-tasks and allowing them to run in parallel. Fig. 4.20. presents the comparison results between sequential execution and pipelined parallel execution. In the proposed HIF detection scheme, 4.48 cycles (89.6 ms) of the current waveform is required to generate a feature map for HIF classification. This signal acquisition task is identified as the most timeconsuming stage in the identification process. Since 4.48 cycles are broken into six frames, it allows to leverage the pipelining concept. Feature calculation task execution can be initiated soon after the first frame of signal acquisition (25.6 ms), rather than waiting for 89. 6 ms. Similarly, once the feature calculation task is executed for the first frame, a feature vector extraction task can be triggered, as shown in Fig. 4.20. (b). Typically, the pipelining technique is used to improve the throughput. Furthermore, most of the optimisations that will enhance throughput often negatively affect latency due to the time-sliced operation of multiple sub-tasks in a clock cycle and the additional data transfers introduced by subtask decompositions. But, real-time HIF detection requires lower latency as well as higher throughput. To accomplish both requirements, the number of pipelined stages are minimised via combining low computationally intensive sub-tasks such as FFT and energy calculation into one pipeline stage. In addition, most of the pipeline stages are implemented in dedicated hardware to avoid time-sliced execution and the data transfer between those stages is facilitated via high-speed direct memory access (DMA) first-in-first-out (FIFO) buffers. Since the HIF classification is based on six consecutive feature vectors, the pipelined parallel execution contributes to lowering the overall detection latency as well as improves the throughput.

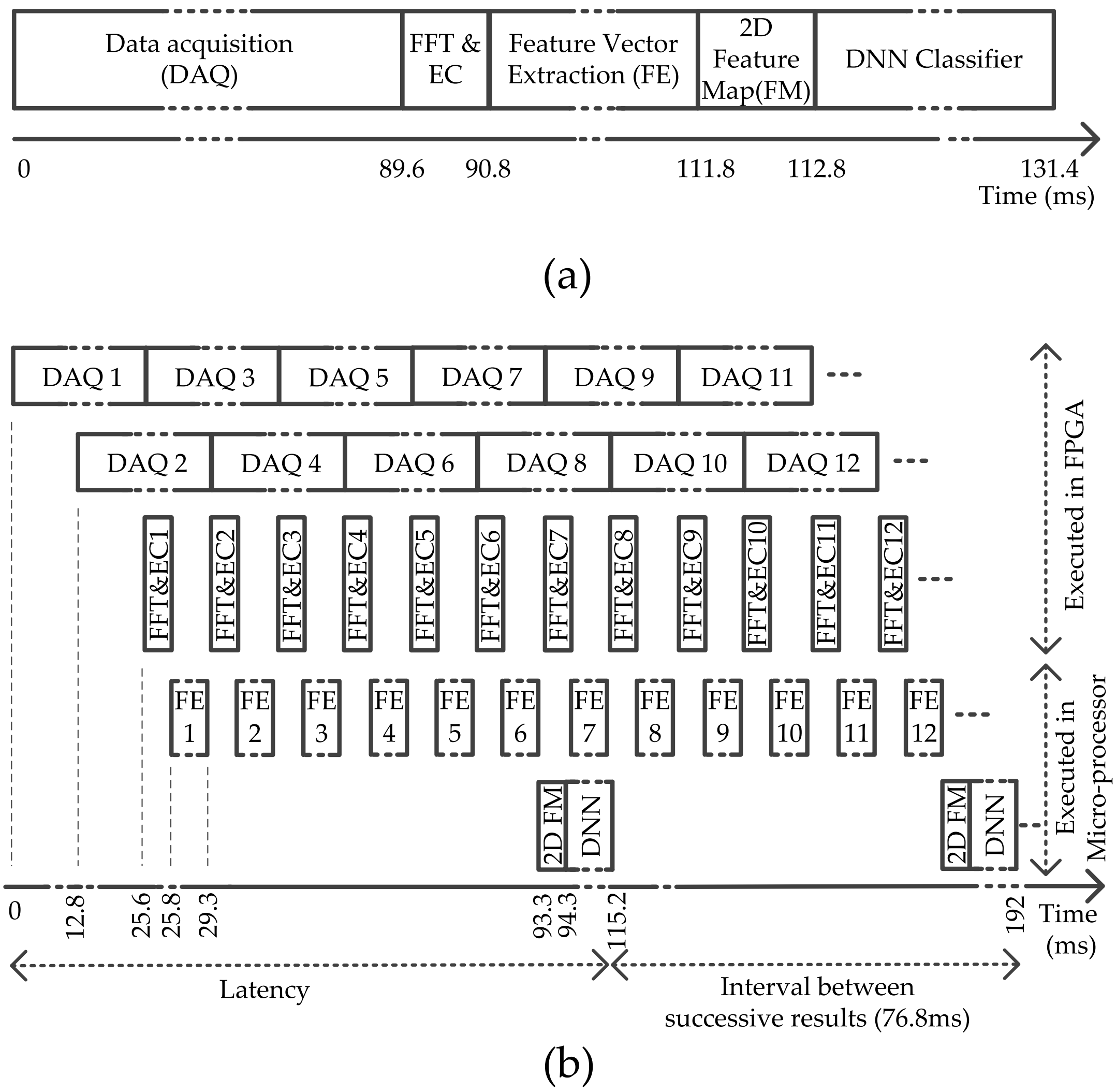

Fig. 4.20. Time comparison between sequential execution and pipelined parallel execution: (a) sequential execution; (b) pipelined parallel execution

#### 4.6.6.3. RTOS BASED TIMED LOOP EXECUTION

The real-time operating system is specially designed to execute tasks with precise timing and high reliability. RTOS is used in the proposed approach to ensure the time synchronisation of HIF detection tasks that are implemented in a microprocessor. There is a possible variation in the latency of each task iteration when multiple tasks are executed in the same processor at the same time. This latency variation is known as jitter, which is primarily introduced due to the contention caused by sharing resources such as processor and memory. The effect of jitter can be clearly observed in general-purpose operating systems such as Windows since they are designed to ensure fairness across all the running programs. So, they are not suitable to achieve guaranteed execution time and periodicity of a specific task. Moreover, the programmer has no control over the task scheduling and prioritisation. In contrast, RTOS allows manipulating the execution loop timing, task scheduling, and prioritisation based on the application requirement. Jitter can be minimal with the optimal choice of loop execution parameters in RTOS. Fig.13 compares the latency variation between general-purpose OS (Windows PC, HP Elitedesk 800 G1) and RTOS (NI Linux Real-Time, myRIO-1900). Loop duration, period, and priority level can be configured in the timed loop structure provided by LabVIEW real-time module for RTOS. While choosing the execution parameters, it is essential to ensure that the scheduled loops do not need to compete for processor cycles. This can be guaranteed via retaining the processor usage well below 100%. A timing budget for each loop is required to estimate the processor usage. Theoretical processor usage can be calculated as follows:

$$Processor\ usage\ (\%) = \sum_{k=1}^{N} \frac{Loop\ duration}{Loop\ period} * 100 \qquad (4.10)$$

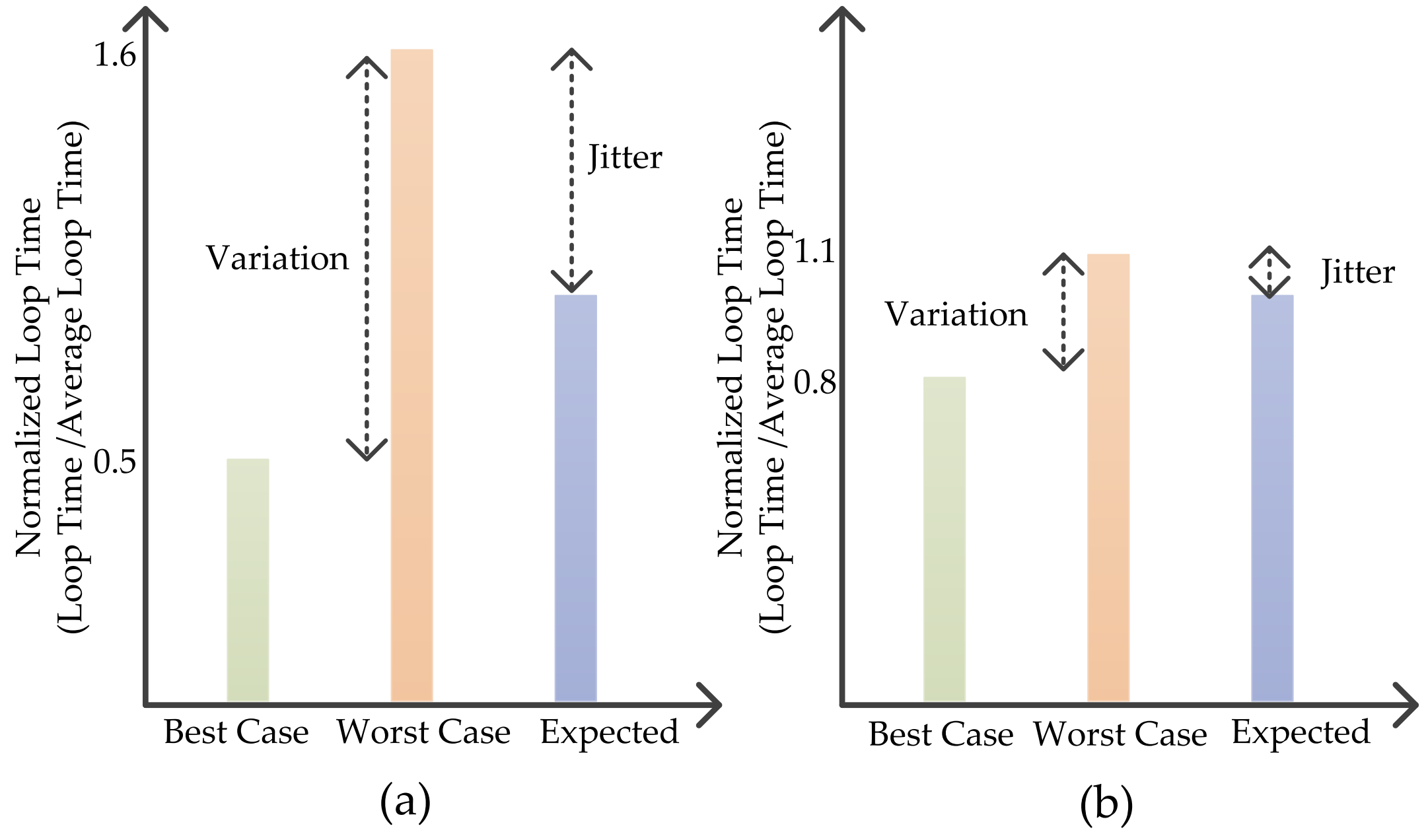

Fig. 4.21. Jitter comparison between general purpose operating system and real-time operating system: (a) Windows operating system; (b) Real-time Linux operating system

Table 4.3. lists the timing requirement of each loop. Based on the reported timings, the estimated processor usage is 55.9%. Since the processor is not overloaded, the expected loop execution timing can be guaranteed in the proposed HIF detection scheme.

Table 4.3: Time budget for loop exectuions

| Loop | Duration (ms) | Period (ms) |
|---|---|---|
| Feature Vector Extraction | 3.5 | 12.8 |
| 2-D feature Map | 1 | 76.8 |
| DNN classifier | 20.9 | 76.8 |
| Estimated processor usage: 55.9% | | |

#### 4.6.7. PROPOSED SYSTEM VALIDATION RESULTS

The proposed short-time FFT feature extraction technique and the deep-learning-based classifier is validated offline with an unseen portion of the dataset. Furthermore, the algorithm is evaluated in real-time to demonstrate the practical potential of the proposed methodology.

##### 4.6.7.1. OFFLINE VALIDATION

A dataset with about 60,000 current samples from normal conditions and HIF conditions in case of different objects is used for the offline system validation. Each current sample in the dataset is corresponding to 4.48 power cycles. The overall detection accuracy of the proposed algorithm can reach 98.67%. Furthermore, some well-known classification techniques such as ANN with one hidden layer and SVM are used with the short-time FFT based feature for comparison, where they can only achieve 90.39% and 91.54% accuracy, respectively.

Besides the accuracy, several criteria, as shown in equation (4.11) - (4.15), are used to evaluate the performance of the algorithms.

$$Accuracy\ (A) = \frac{TP + TN}{TP + TN + FP + FN} \times 100\% \tag{4.11}$$

$$Dependability\ (D) = \frac{TP}{TP + FP} \times 100\% \tag{4.12}$$

$$Security\ (S) = \frac{TN}{TN + FN} \times 100\% \tag{4.13}$$

$$Safety\ (SF) = \frac{TN}{TN + FN} \times 100\% \tag{4.14}$$

$$Sensibility\ (SN) = \frac{TP}{TP + FN} \times 100\% \tag{4.15}$$

Table 4.4: HIF dectection system evaluation results

| **Classifier** | **Accuracy** | **Dependability** | **Security** | **Safety** | **Sensibility** |
|---|---|---|---|---|---|
| SVM | 91.54% | 92.23% | 91.16% | 95.45% | 85.36% |
| ANN | 90.39% | 88.18% | 91.69% | 92.95% | 86.19% |
| Proposed CNN | **98.67%** | **99.70%** | **98.05%** | **99.82%** | **96.84%** |

where TP, TN, FP, FN are true positive (correct HIF detections count), true negative (correct healthy condition detections count), false positive (number of HIF misclassified as healthy conditions), and false negative (number of healthy states misclassified as HIF) in the confusion matrix of the classifier, respectively. Also, *Accuracy*, *Dependability*, *Security*, *Safety* and *Sensibility* are the measures of overall precision, HIF condition detection precision, normal condition detection precision, hazard prevention level, and system sensitivity related to normal conditions, respectively. The evaluation results are summarised in Table 4.4, and the proposed algorithm demonstrates better performance compared to the conventional machine learning techniques in the domain of HIF detection.

#### 4.6.7.2. Real-time system validation

The real-time experimental validation setup is shown in Fig. 4.22. HIF signal is combined with normal load current $(I_l + I_f)$ and fed into the monitoring unit for fault identification. A DSO1004A digital oscilloscope is used for timing validation. Pure HIF signal, combined input signal and the fault identification digital signal are connected to the oscilloscope for precise time measurement of HIF identification. The key measures are reported in Table 4.5. Validation results confirm that the proposed method can detect HIF within six power cycles.

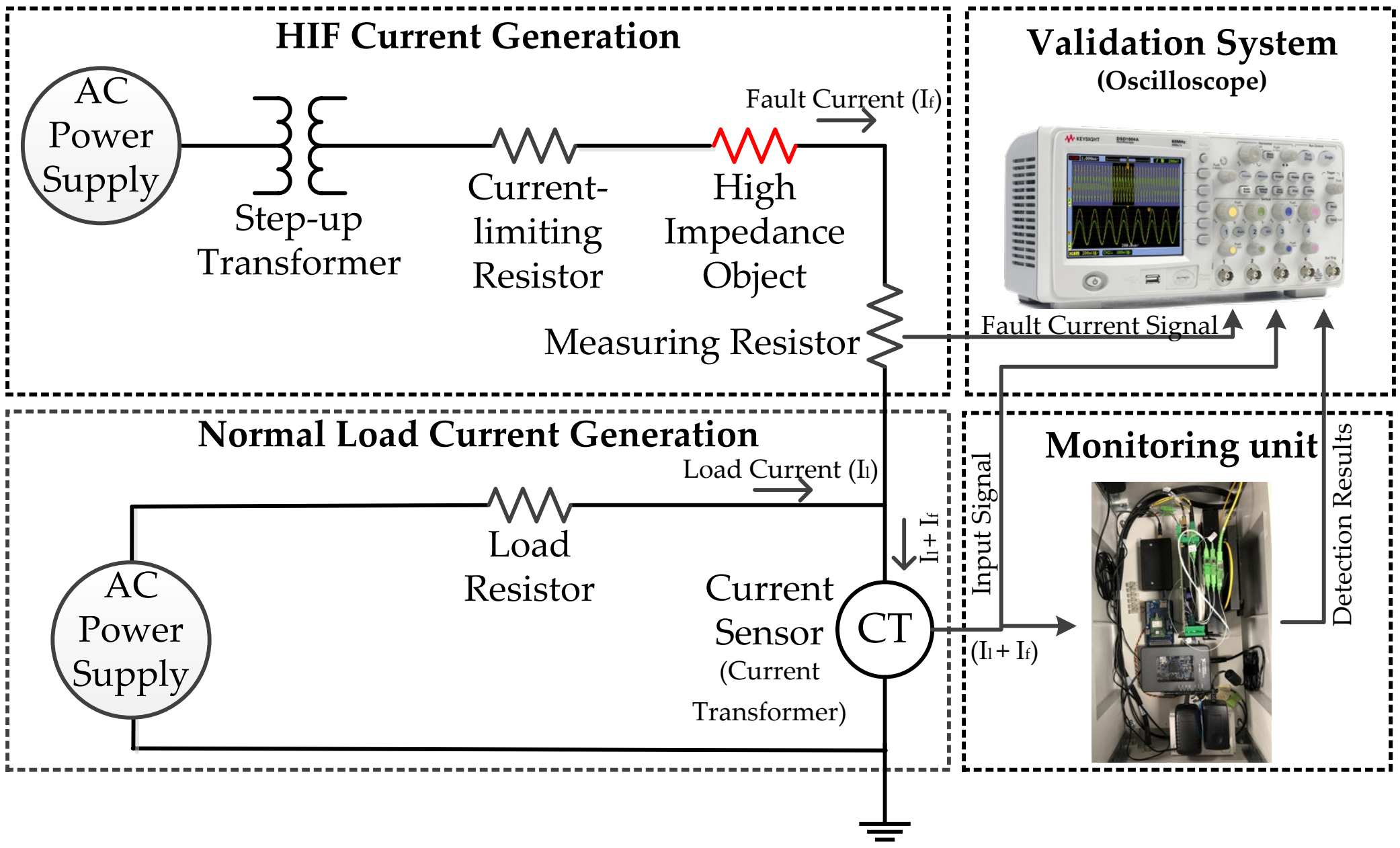

Fig. 4.22. Real-time experimental setup

Table 4.5: Real-time validation results

| **Key measures** | **Value** |
|---|---|
| Worst case latency | 115.2 ms (< 6 cycles) |
| Throughput (incoming) | 20,000 Samples/s |
| Throughput (outgoing) | 13 Detection results/s |
| Maximum memory usage | 156.7/256 MB (61.2%) |
| Worst case power requirement | 14W |
| Average processor usage | 49% |
| Internal FIFO Overflows | Nil (0) |
| DMA FIFO Overflows | Nil (0) |

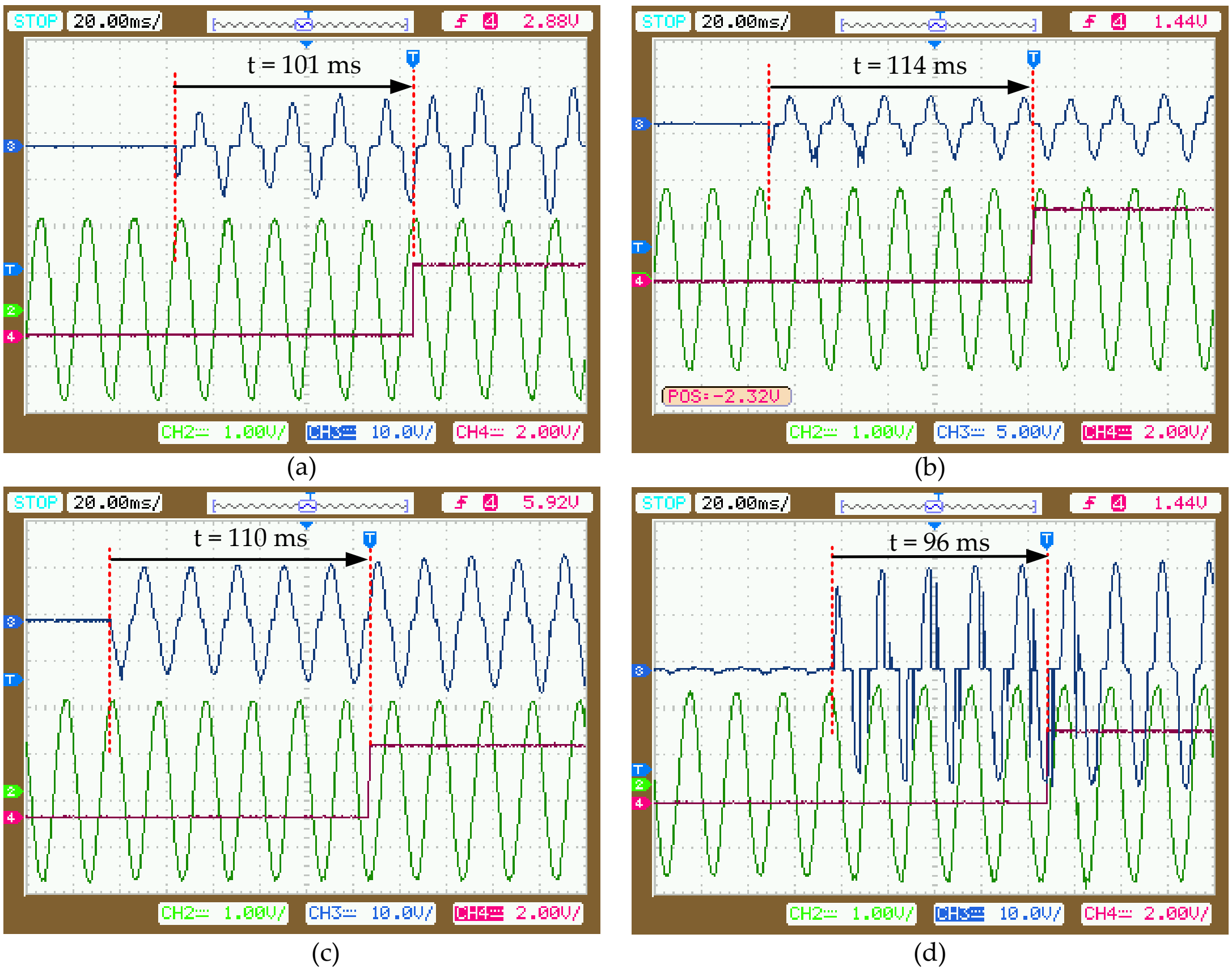

Fig. 4.22. Real-time HIF detection latency validation results (using oscilloscope): (a) HIF due to wet sand contact; (b) HIF due to dry sand contact; (c) HIF due to tree branch contact; (d) HIF due to soil contact. The oscilloscope window shows three different signals; 1) Pure HIF current captured by 47 ohms measuring resistor to indicate the start of the fault (denoted by blue signal); 2) Current signal from experimental setup (denoted by green signal); 3) Fault identification signal (digital) sent out from edge device (denoted by pink signal).

Fig. 4.22. visualises the real-time detection results of different types of HIFs. During the real-time system validation, no FIFO buffer overflows are reported. Therefore, the proposed edge device can process 20,000 sample points in a second without any data losses. As the outcome of data processing, it can consistently produce detection results in every 76.8ms (13 results/s). Furthermore, the reported utilisation of resources such as memory, power and processor ensures the sustainability of the deep learning based fault identification in an embedded monitoring device.

#### 4.6.7.3. EXTENSIVE HIGH POWER HIF EXPERIMENTS AND TRANSIENT CLASSIFICATION

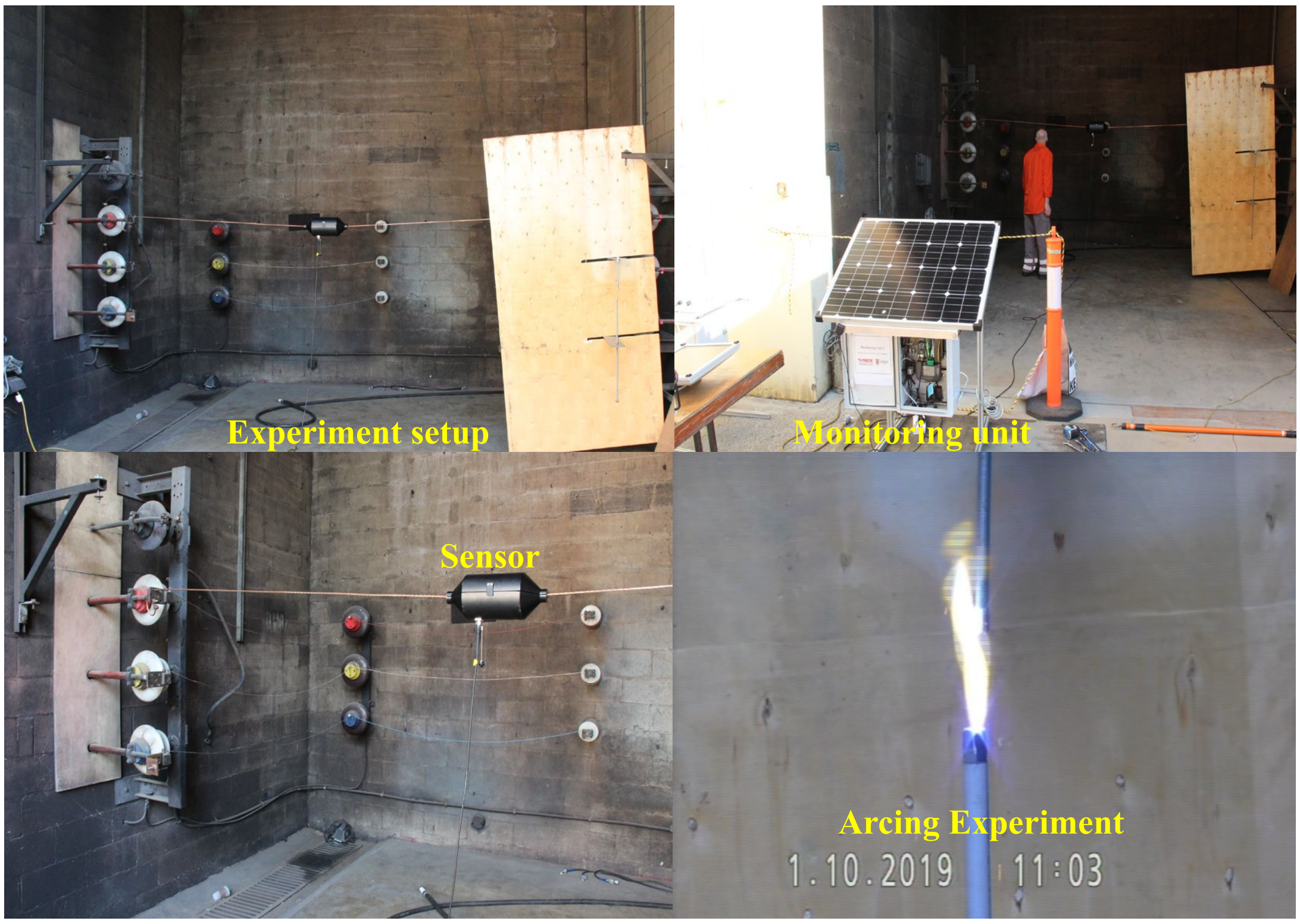

Fig. 4.22. High power arcing experiment setup

An experimental setup, as shown in Fig. 4.22, is arranged in an industrial test site (PLUS ES high power testing station, Sydney), that can generate arcing and switching events such as capacitor switching and resistive load switching. The maximum current is limited to 40 A, and the operating voltage is 20.2 kV and 11.7 kV. The above images show the arrangements to initiate the arcing along the line and arcing generation. During these experiments, the raw waveform is recorded by our monitoring unit. There are two different ratios between the load current, and the arc current are tried during the arcing experiment, such as 5 and 10 (Load current: Arc current = 5:1, 10:1). In total, there are four arc current levels, and two load current levels are used to capture the variations in the arcing characteristics in different current levels and ratios.  Apart from the HIF signals, other transients such as resistive, inductive and capacitive load switchings and 200 kVA SWER transformer inrush current samples are collected to create a transient dataset, that may exhibit similar characteristics as HIF. Furthermore, healthy condition data is also collected at 20.2 kV, 11.7 kV with 40 A, 20 A load current. All three types of data are combined to create a dataset, which is labelled into three classes such as HIF, transient and normal. Proposed CNN classifier is trained with three class labels and validated with a set of unseen data. Fig. 4.22 outlines the extensive system testing process, and the results are summarised in Table. 4.5.

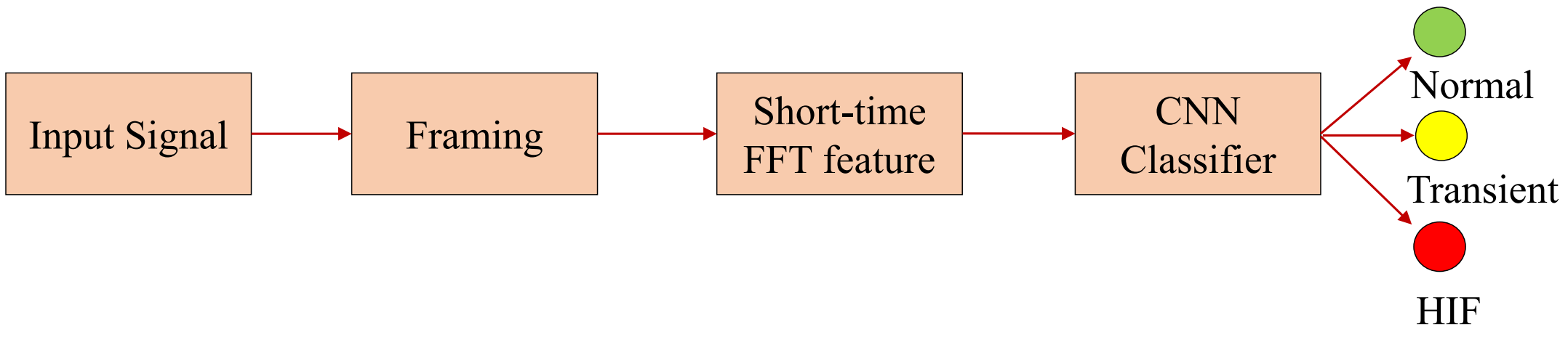

Fig. 4.22. Extensive validation with three class labels

Table 4.6: Extensive validation results

| True labels | Predicted labels | | | |
|---|---|---|---|---|
| | | HIF | Transient | Normal |
| | HIF | 98.48% | 0.88% | 0.64% |
| | Transient | 0.6% | 99.4% | 0% |
| | Normal | 0% | 0% | 100% |

All the system validation results reported above demonstrate the suitability of proposed short-time FFT feature and deep learning based classifier for the HIF discrimination from the normal operating conditions and other typical transients in electricity networks. This HIF identification module is a part of the proposed monitoring framework for overhead transmission lines. In the future, this monitoring framework can be further extended to facilitate the edge computation of other time-sensitive requirements related to power grids.

## 4.7. Chapter Summary

A distributed on-line monitoring framework is proposed for the overhead transmission lines. It facilitates simple threshold-based power quality monitoring, AI-based real-time HIF identification and transient classification. The monitoring unit has been calibrated with an industrial multimeter, and its measurements are verified at different times (temperature), different voltage (0 kV – 25 kV) and current (0A – 38 A) ranges. Internal system architecture, precise electrical parameter calculations from high-resolution signals, threshold-based power quality disturbance detections and communication protocols for event transmission are detailed in this chapter.

It introduces WPT entropy and short-time FFT based features for HIF identification. Computational complexities of these feature extraction techniques are compared for the real-time execution. Furthermore, a light-weight CNN classifier is proposed to identify HIF in real-time at resource-constrained embedded devices. Hardware-level optimisation techniques are adapted for the algorithm deployment on embedded monitoring devices. The end-to-end HIF identification process is validated offline with the data set, which is created from the laboratory experiments. Besides, the HIF identification process is verified in real-time, and the resource consumptions are profiled and reported. Also, the proposed feature and the classifier is validated with an extensive set of data collected from an industrial high power testing station, that contains high power HIF signals, typical switching transients and normal operating conditions. The work detailed in this chapter has led to the following publications [5][6][7].

# CHAPTER 5

# 5. INTELLIGENT EDGE ANALYTICS FOR REAL-TIME CONSUMER LOAD IDENTIFICATION

## 5.1. INTRODUCTION

Global energy demand is predicted to increase in the coming decades. International energy outlook report released by the U.S. Energy Information Administration states that the world energy consumption will grow by nearly 50 per cent between 2018 and 2050 [1]. Hence, the capacity of existing electricity network infrastructures may become inadequate soon. Renewable energy microgrids are identified as a key solution to solve energy poverty, especially for rural electrifications. The microgrid market is estimated to reach USD 47.4 billion by 2025, due to the increasing global deployment of microgrids for remote electrification [2]. In the context of renewable energy microgrids, it is essential to coordinate the consumer load profile along with the generation profile, which can result in better energy efficiency, energy savings and optimal energy storage options. The consumer load shaping and load profiling heavily rely on the performance of real-time load identification.

Real-time load identification adds value to this context since the renewable energy generation profiles changes at different times of the day based on the environmental conditions. For instance, solar power has a generation profile which peaks in the middle of the day and tails off toward darkness [3]. This chapter proposes an edge computing based intelligent load identification framework, that can be deployed in low power embedded hardware such as smart meters for real-time execution. This chapter also investigates the requirement of signal sampling rate and digitisation resolution and their impact on the load identification performance. Furthermore, the proposed load identification techniques are implemented in embedded hardware and experimentally validated in real-time at university buildings.

## 5.2. PROPOSED ARCHITECTURE FOR LOAD IDENTIFICATION ON SMART METERS

Since smart meters provide increased access to customer energy usage data, it can be directly leveraged to the consumer load identification. However, the signal sampling rate and data resolution in existing smart meters are not adequate to achieve precise load identification results. The recent advancements in low-power microprocessors and the availability of cheap, high-speed ADC cards expands the usage of smart meters to numerous potential applications such as demand-side management, consumer load identification and abnormality detection.

As discussed in chapter 2, precise load identification requires high-resolution data (thousands of samples per second) from smart meters since load switching events produce significant variations in higher frequency regions. However, signal acquisition with high sampling rates produces a massive amount of data in smart meters. For instance, the data acquisition of a single-phase consumer unit can create around 500 MB of data in an hour when the signals are sampled at 10 kHz. Transmitting all the data to a centralised server or cloud for load analysis requires an incredible amount of bandwidth as well as resource-intensive processing. Furthermore, this is not a scalable solution since the smart meter market is forecasted to grow further in the coming years across the world [4]. Thus, more optimised architecture is required to facilitate these modern applications.

An edge computing paradigm is proposed to overcome limitations in enabling precise load identification on smart meters. As discussed in chapter 3, it is a distributed computing approach which brings the computational capability to the location where it is relevant. In the context of load identification, the computational process can be brought up to the point of sensing. In other words, it can be integrated with the smart meter using embedded microprocessors. Smart meters are continually keeping track of the energy usage of a consumer unit. At the same time, it is aware of the current energy demand in the microgrid via the bi-directional communication link with the rest of the microgrid components. Therefore, smart meters have been chosen as an ideal place to deploy the load identification algorithm, rather than send all the high-resolution data to the utility providers. This local processing approach avoids a round-trip of data transmission between the centralised server and consumer site as well as the time delay in consumer response. Furthermore, it eliminates the complexities such as location sensitivity and privacy concerns in the smart meter data based load identification.

Fig. 5.1 outlines the proposed architecture for real-time load identification, and it recommends the sub-tasks allocations for embedded edge computing and cloud computing. High-resolution signal acquisition, transient event detection, feature extraction via high-frequency component analysis and load classification are deployed in embedded devices such as smart meters. After the load identification process, raw signals are down-sampled to a manageable level that is suitable for transmission and storage. Since the high-resolution raw signals are processed and down-sampled in the embedded hardware attached to the smart meter, the bandwidth requirement for communication is significantly reduced. It can be managed with the low-power wide-area network (LPWAN) based wireless communication technologies. The following sections give detailed descriptions of each load identification sub-task.

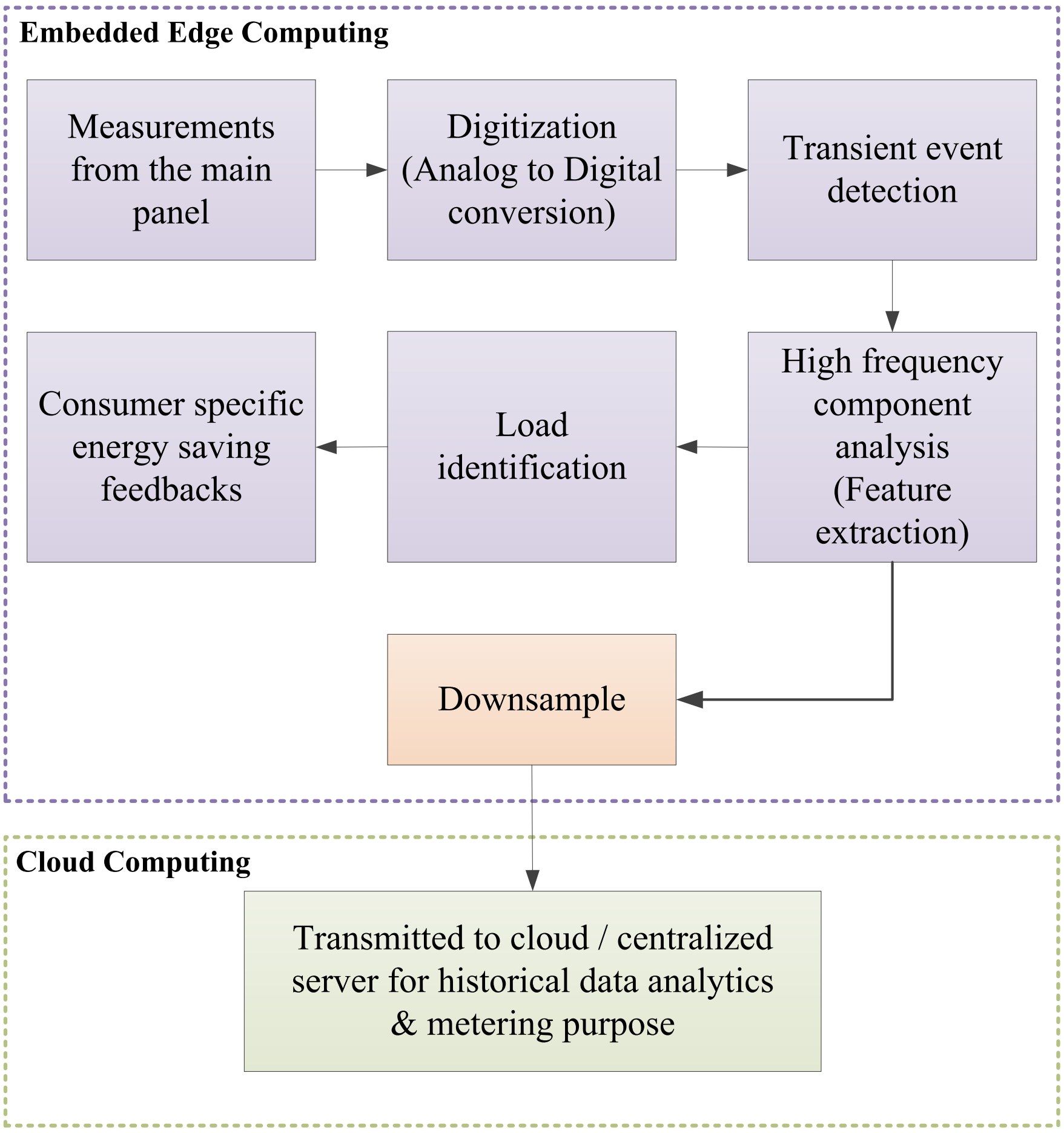

Fig. 5.1. Proposed architecture for load identification on smart meters

## 5.3. DATA ACQUISITION FROM THE MAIN METER PANEL

As discussed in chapter 2, NILM based load identification methodologies require a single point of measurement from the main panel level of a consumer site. It is identified as a low-cost alternative to attaching sensors on individual appliances. The load identification techniques proposed in this chapter is based on NILM concepts. As the first step, DAQ hardware is used to digitalise the incoming analog voltage and current signal into digital representation so that any digital systems can interpret them. Since analog signals continuously vary over time, an ADC is used to take periodic samples of the signal at a predefined rate named as the sampling rate. Another key specification in the selection of a digitiser is the resolution which is the ability to identify discrete voltages within the operating input range of the device. Impact of these parameters of a DAQ hardware has been studied in the context of load identification.

### 5.3.1. EFFECT OF SAMPLING FREQUENCY ON LOAD IDENTIFICATION

The choice of sampling rate is important since it influences the timing precision, analysis accuracy and the data size of raw signals. The Nyquist sampling theorem determines the minimum sampling rates needed in an application to capture all the information from a finite bandwidth. If the upper limit of the relevant frequency components in a system is $f_n$, then the required sampling frequency ($f_s$) should exceed twice the upper limit frequency ($f_s > 2 \times f_n$). Two sampling frequencies, such as 10 kHz and 20 kHz, have chosen in this study. Therefore, the frequency contents up to 5 kHz and 10 kHz can be analysed based on Nyquist theorem. Consumer appliances' turn-on transients are captured using the aforementioned sampling rates, and its spectral components are visualised using spectrogram, as shown in Fig. 2. The captured events are highlighted in the image. It can be concluded from the visualisation that sufficient information regarding the turn-on event can be extracted from the frequency components lower than 5 kHz. Therefore, the sampling rate can be limited as 10 kHz since up to 20 kHz sampling rate is redundant, and it generates an almost double amount of data. Accordingly, the sampling frequency needs to be chosen based on the application requirement and the resource constraints for storing and processing of sampled data.

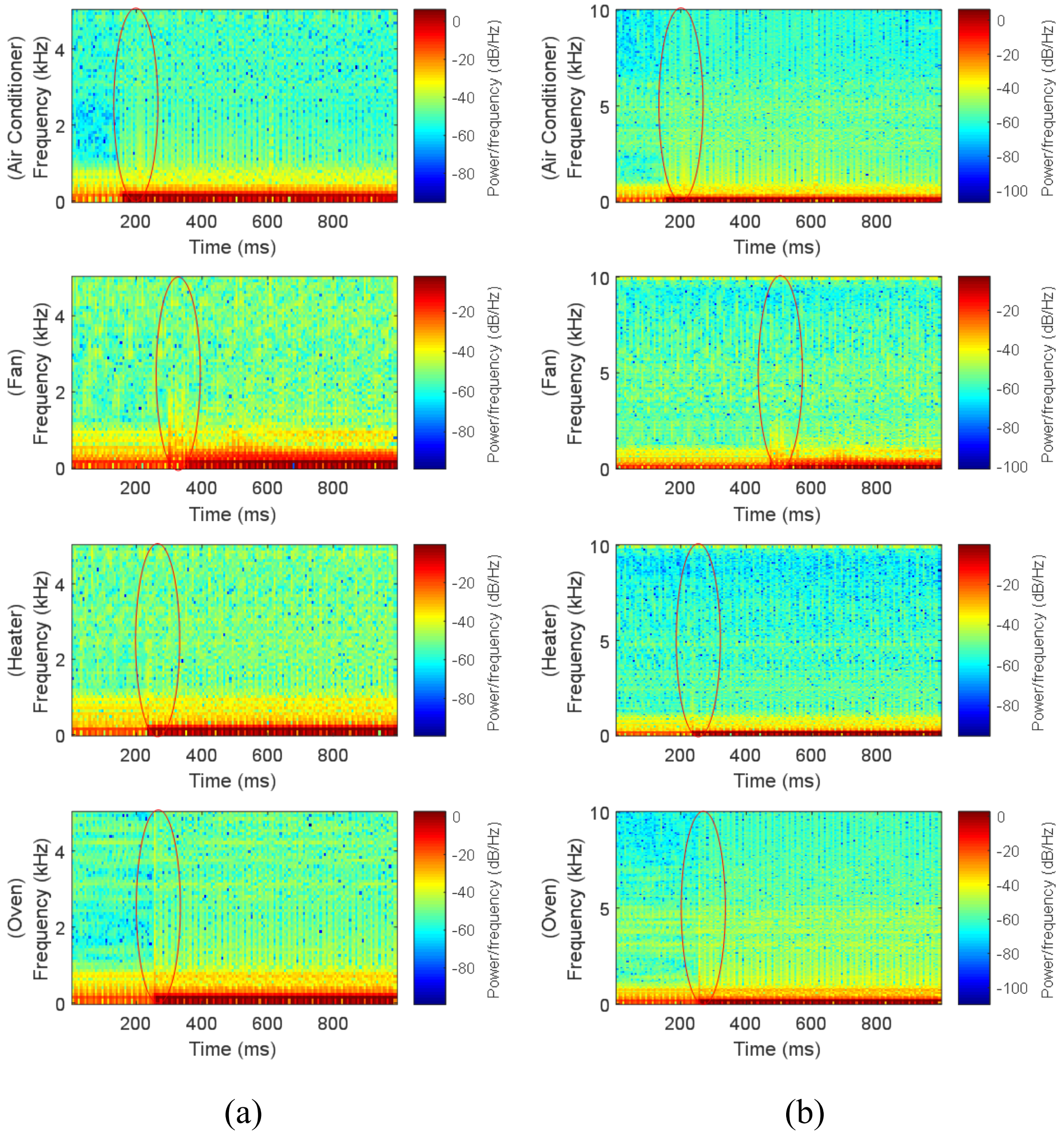

Fig. 5.2. Comparison of sampling frequency (a) 10 kHz (b) 20 kHz

### 5.3.2. Effect of digitisation resolution on load identification

Digitisation resolution is another important specification of a DAQ system where it refers to the voltage of an ADC code. A code can be defined as the digital representation of an analog voltage. The number of available codes is decided by the ADC resolution, which is commonly expressed in bits. Furthermore, the voltage representation of each code, also known as code width, depends on the ADC resolution as well as the operating input range. These can be calculated as follows:

$$Number\ of\ codes = 2^{number\ of\ bits} \tag{5.1}$$

$$Code\ width = \frac{Max\ V - Min\ V}{Number\ of\ codes} \tag{5.2}$$

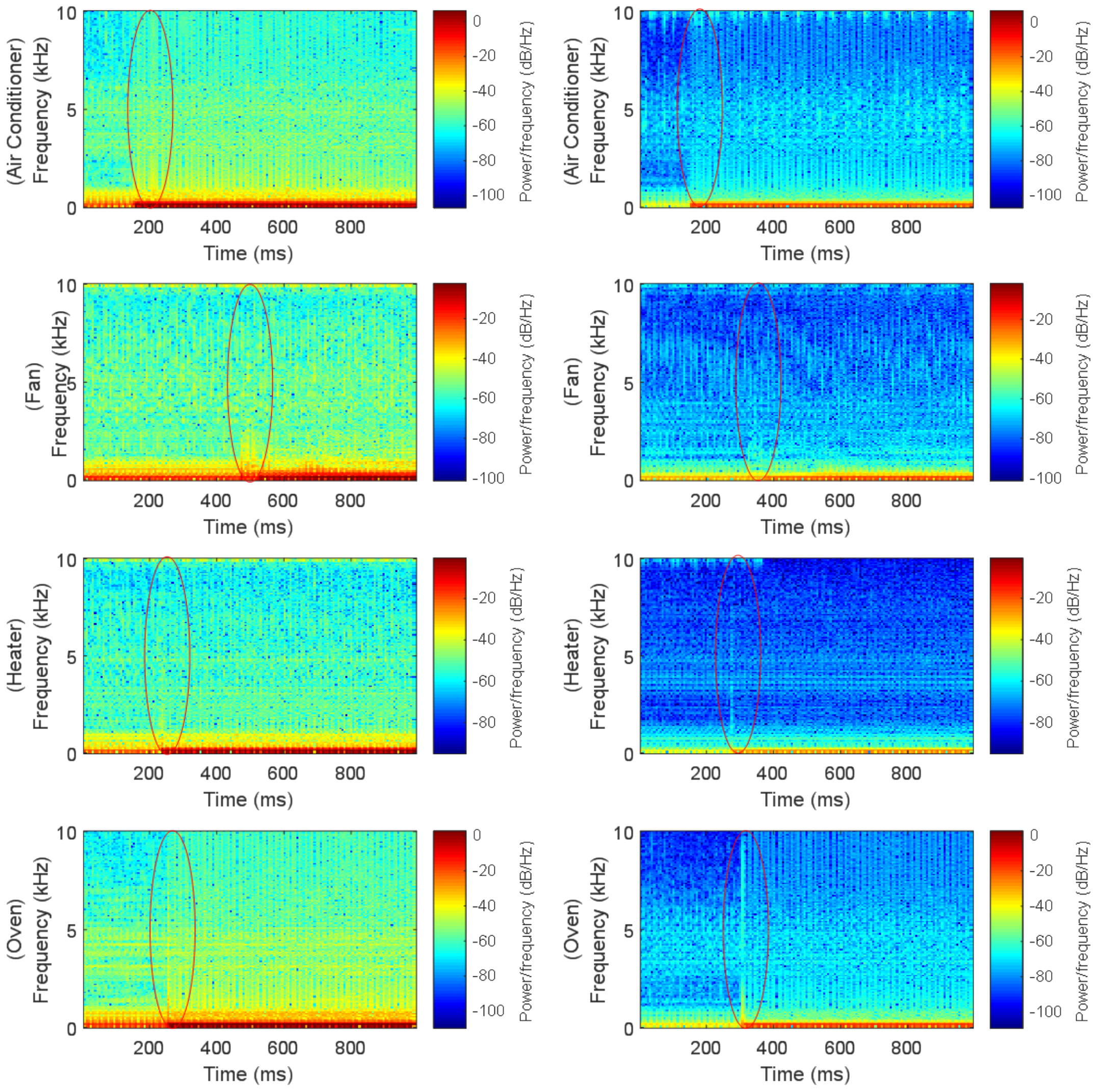

(a) (b)

Fig. 5.3. Comparison of digitization resolution (a) 16 bits (b) 12 bits

For instance, a DAQ device with 16-bit ADC and a +/- 10 V operating input range has a code width of 300 μV. This device failed to capture variations of less than 300 μV. In this experiment, NI sbRIO-9637 and NI myRIO 1990 DAQ devices are used for the comparison which contains 16 bits and 12 bits ADC resolution respectively. The comparison results are visualised using spectrogram in Fig. 5.3. From the comparison results, it is evident that ADC with 16 bits resolution captures apparent variations in the frequency components during an appliance turn-on event, where ADC with 12 bits resolution lagging behind. Furthermore, it is worthwhile to note that whenever the DAQ devices have selectable input ranges, the smallest input range that can satisfy the application requirement need to be chosen. It reduces the code width and improves the sensitivity of the acquired signal based on Eq. (5.2). The visualisation results demonstrate that the correct choice of sampling frequency and the ADC resolution plays an influential role in the load identification.

## 5.4. Transient event detection

After the signal acquisition, the device switching events need to be isolated for detailed analysis. Two different methods are used to capture the appliance switching transients, such as (1) time-domain based event detection, and (2) frequency-domain based event detection.

The time-domain based analysis is leveraged to compute an empirical estimate of the RMS value of the input signal. The transient events are identified from the abnormal changes in the input signal RMS value. Potential transient event points can be located in an input signal frame, as follows:

$$\sum_{j=m}^{n} \Delta\left(x_i;\ \hat{E}\left(\left[x_{k_r} \ldots\ldots x_{k_{r+1}-1}\right]\right)\right) = (n-m+1)log\left(\frac{1}{n-m+1}\sum_{r=n}^{m} x_r^2\right) \quad (5.3)$$

$$E(K) = \sum_{r=0}^{K-1} \sum_{i=k_r}^{k_{r+1}-1} \Delta\left(x_i;\ \hat{E}\left(\left[x_{k_r} \ldots\ldots x_{k_{r+1}-1}\right]\right)\right) + \beta K \quad (5.4)$$

where $\hat{E}, K, \beta, k_0, k_K, m,$ and $n$ represents the empirical estimate of a section in the signal frame, event point, minimum threshold, first sample of the frame, last sample of the window, the first sample of the considered section in a frame and the previous sample of the same part in a frame, respectively.

Based on our experiments, this time-domain based event detection approach demonstrates better performance in the transient identification of energy-hungry appliances such as air conditioners, electric ovens and heaters. However, it struggles to isolate the transients of low-power appliances such as an incandescent lamp, fluorescent lamp, laptop and table fan. Since the power-hungry devices create significant oscillations in the waveform during their start-up, those transients can be identified by the time-domain based technique with reduced computational complexity. On the other hand, the start-up oscillations of the low-power devices can be easily masked in the time-domain signals since the magnitude change is insignificant compared to the background load variations. Hence a frequency-domain based transient isolation process is proposed for low-power appliances.

In the frequency-domain based event detection, wavelet analysis is used to extract the high-frequency contents of the input current signal. Wavelet analysis is chosen for this approach since it is necessary to have both the time and frequency information simultaneously for the precise transient detection. The frequency components from 2.5 kHz to 5 kHz are extracted from the input signal since it is sampled at 10 kHz, as discussed on section 5.3.2. Based on the experimental results visualised in Fig 5.4, the extracted high-frequency contents have the potential to discriminate the turn-on transients from the normal variations in the current signal in the main panel level. The wavelet decomposition is limited with one level to reduce the computational complexity of the proposed algorithm, which plays a critical role in the real-time implementation, as discussed in chapter 4. This high-frequency component extraction can be seen as a band-pass filtering operation.

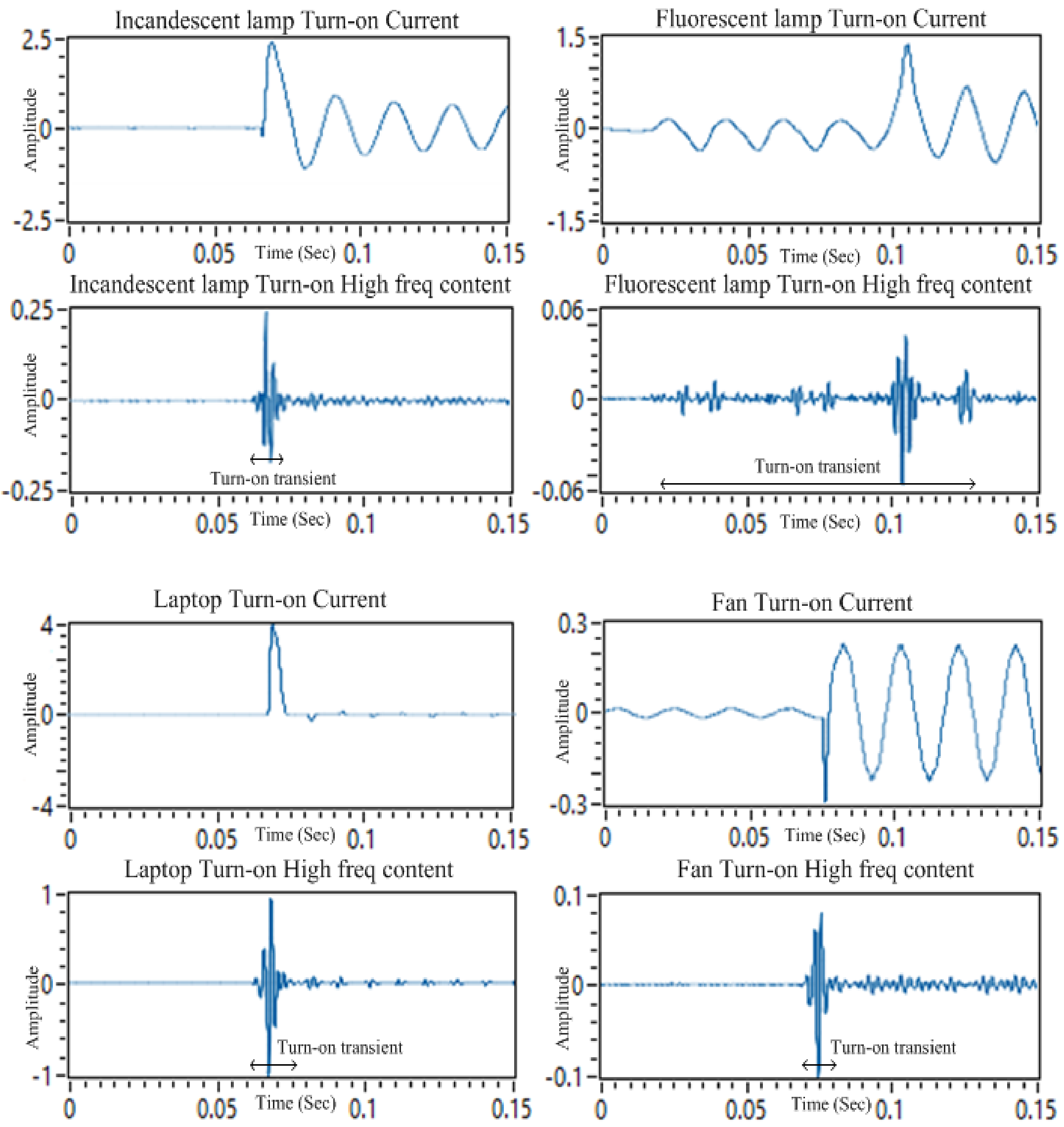

Fig. 5.4. Turn-on transient detection of low-power appliances

The time instant of a transient switching event can be determined from the methodologies as mentioned above. Once it is identified, 512 adjacent time samples from before and after the transient point are extracted. Then the extracted transient signal is further analysed to capture the event-specific features.

## 5.5. Feature extraction

Feature extraction is an essential part of load identification. It refers to the process of extracting useful information from the current and voltage waveforms that can differentiate the loads. The high-frequency component analysis, which is used to identify the transient event detection, can also be applied to separate the different states of a signal, as shown in Fig. 5.5.

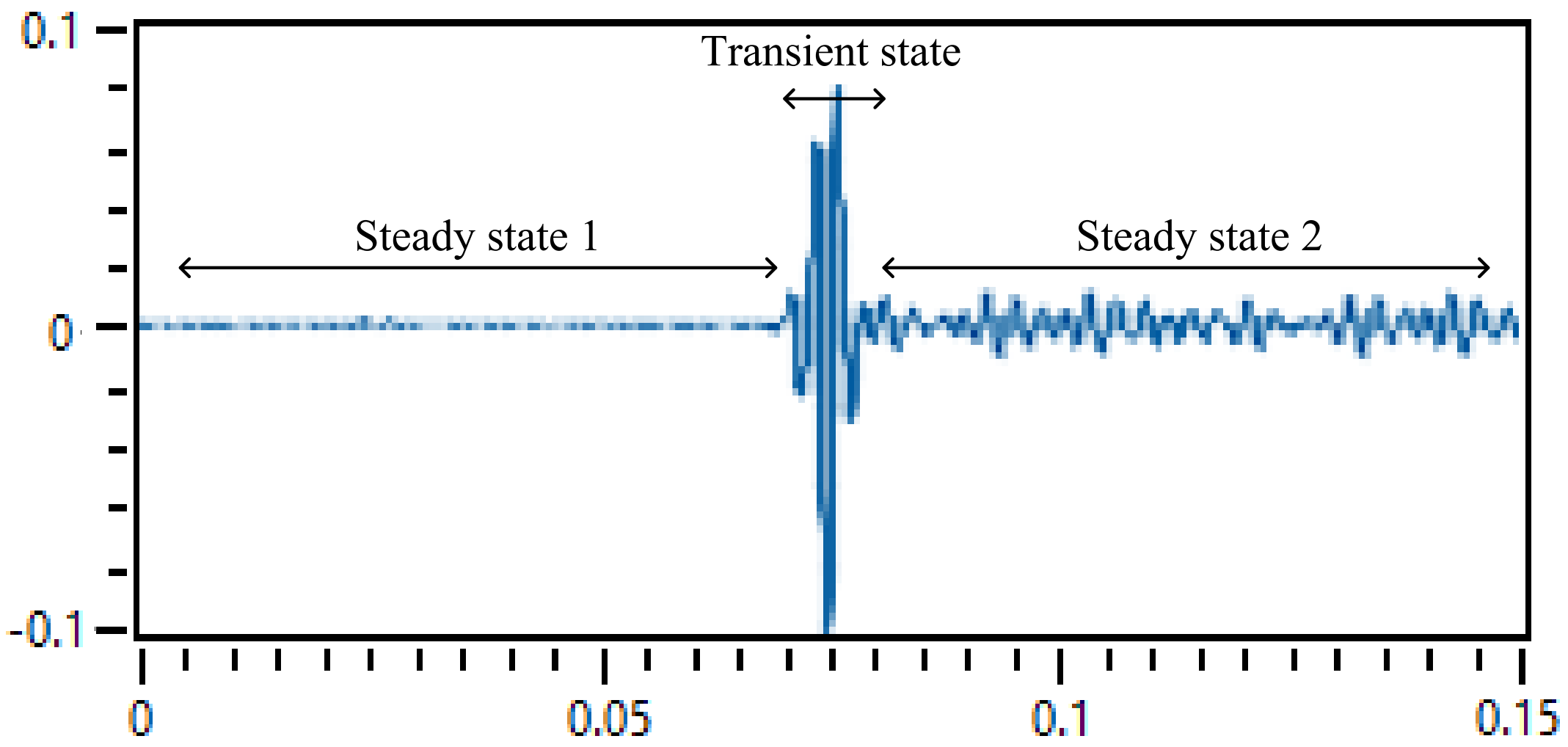

Fig. 5.5. State separation of a signal using high-frequency component analysis

After labelling the states, active and reactive power of each steady-state is calculated using (5.5) and (5.6). The next step is to compute the change in active power $(\Delta P)$ and reactive power $(\Delta Q)$ using (5.7) and (5.8), as represented below:

$$Active\ power\ (P) = \sum_{k=0}^{N} V_k I_k \cos(\varphi_k) \tag{5.5}$$

$$Reactive\ power\ (Q) = \sum_{k=0}^{N} V_k I_k \sin(\varphi_k) \tag{5.6}$$

$$\Delta P = P_{steady\ state2} - P_{steady\ state1} \tag{5.7}$$

$$\Delta Q = Q_{steady\ state2} - Q_{steady\ state1} \tag{5.8}$$

The transient state time-domain signal between two adjacent steady-state signals is decomposed into different frequency bands using FFT. In this case, FFT is used since the time information is not relevant, and its computational complexity is lower than the multi-level wavelet analysis. Based on the spectral component visualisation in Fig. 5.3, the low-frequency range demonstrates more significant variations in amplitude compared to the high-frequency range. Thus, the whole frequency range is decomposed based on the octave scale, as shown in Table. 5.1, such that higher sensitivity is given to the lower frequency range to capture those variations. Each band is said to be an octave scale when the upper band frequency is twice the lower band frequency. Then, the logarithmic transformation is applied to the frequency components to enrich the visibility of patterns in energy levels. After that, average energies for each frequency band are calculated and used as the features to identify the consumer loads. The energy distribution in the decomposed frequency bands is visualised in Fig. 5.6 for different household appliances. Along with the decomposed energies of the transient state, active and reactive power change between two consecutive steady states are consolidated to create a nine-dimensional feature vector for load identification, as outlined in Fig. 5.7.

Table 5.1: Frequency band decomposition based on octave scale

| **Band No** | **Frequency range (Hz)** |
|---|---|
| Band 1 | 2500 Hz – 5000 Hz |
| Band 2 | 1250 Hz – 2500 Hz |
| Band 3 | 625 Hz – 1250 Hz |
| Band 4 | 312.5 Hz – 625 Hz |
| Band 5 | 156.25 Hz – 312.5 Hz |
| Band 6 | 78.125 Hz – 156.25 Hz |
| Band 7 | 39.06 Hz – 78.125 Hz |

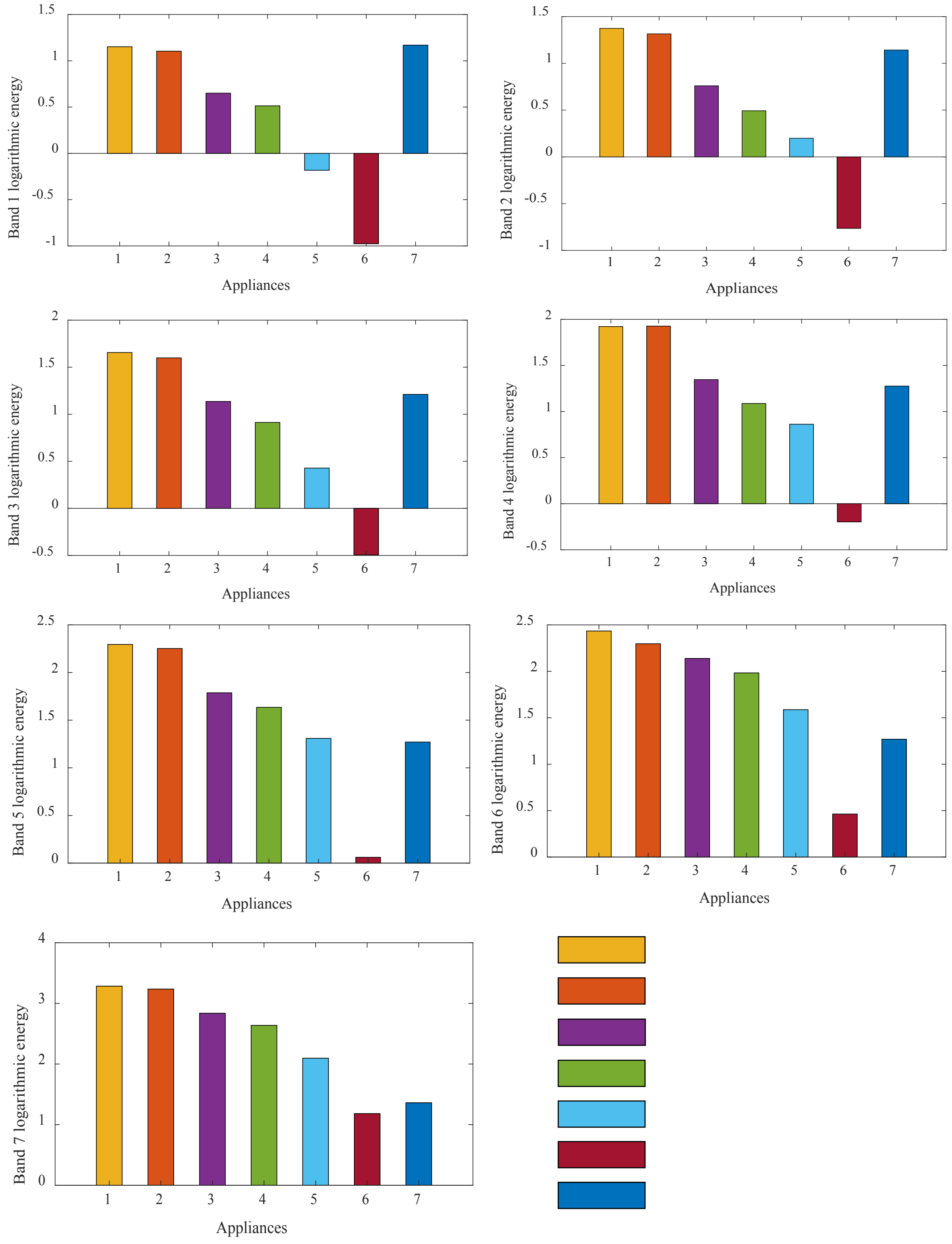

Fig. 5.6. Logarithmic energy comparison of different appliances in decomposed frequency bands

| | | | | | | | | |
|---|---|---|---|---|---|---|---|---|
| Band 1 energy (E1) | Band 2 energy (E2) | Band 3 energy (E3) | Band 4 energy (E4) | Band 5 energy (E5) | Band 6 energy (E6) | Band 7 energy (E7) | $\Delta P$ | $\Delta Q$ |

Fig. 5.7. Nine-dimentional feature vector for load identification

## 5.6. Load Identification

The load identification process can be considered as a pattern classification problem, in which the input feature vector contains the appliance specific patterns, and the classifier is required to distinguish those patterns and identify the loads. A light-weight, three-layer feedforward ANN is proposed to classify the loads from the input feature vector. As discussed in chapter 3, a supervised feedforward ANN is generally divided into three layers: input, hidden and output. Each layer comprises several neurons which are connected by the weighted links that are selected to meet the desired associations between inputs and outputs. Since the ANN input is a nine-dimensional feature vector, the input layer contains nine neurons. On the other hand, neurons number in the output layer is decided by the number of appliances that need to be classified. Hidden layer neurons are determined through the trial-and-error experiments, based on the load identification accuracy and computational complexity of the network. Fig. 5.7 outlines the internal structure of the proposed load identification network. Input features are propagated through the interconnections with different weights to find the non-linear input-output mapping. Each neuron in the output layer computes a numerical value based on the input from the hidden layer and the interconnection weights. The score in the final layer can be represented as $(z_1, z_2 \dots \dots .., z_7)$. In the output layer, the softmax activation function is used to derive the one-hot representation, as outlined below:

$$\arg max\,(z_1, z_2 \dots \dots .., z_7) = (0, \dots 0, 1, 0 \dots, 0) \tag{5.9}$$

where the output result $y_i = 1$ if and only if $z_i$ is the unique maximum value of $(z_1, z_2 \dots \dots .., z_7)$. It provides the load identification results such that the input feature corresponds to the $i^{\text{th}}$ appliance.

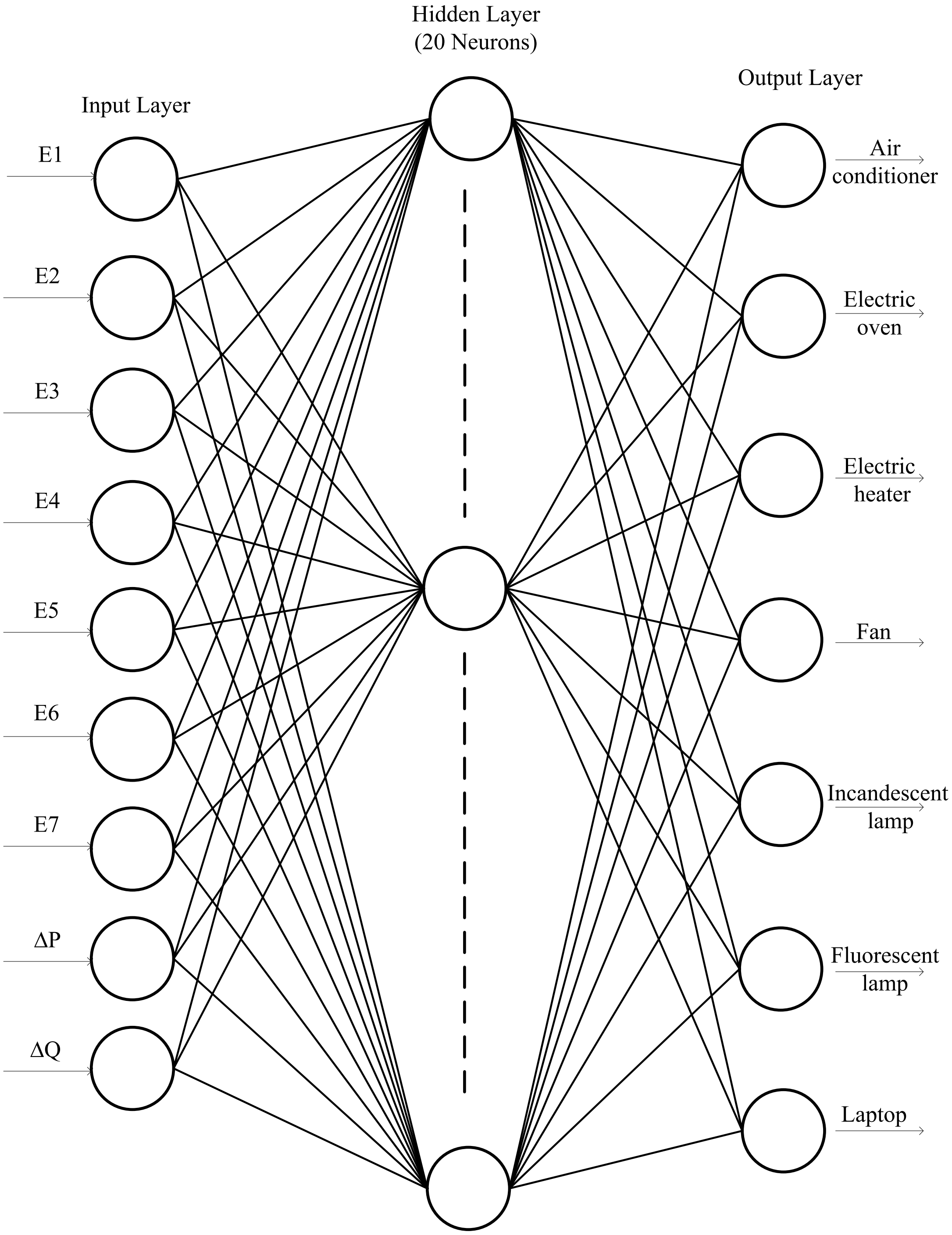

Fig. 5.7. ANN structure

## 5.7. Experimental Validation

The proposed load identification methodology is experimentally validated with domestic appliances. As the first step, turn-on transients of the different household appliances are collected to study device-specific turn-on transient properties. Discriminative appliance-specific features are used to train the supervised feedforward ANN for load identification.

### 5.7.1. Appliance data collection

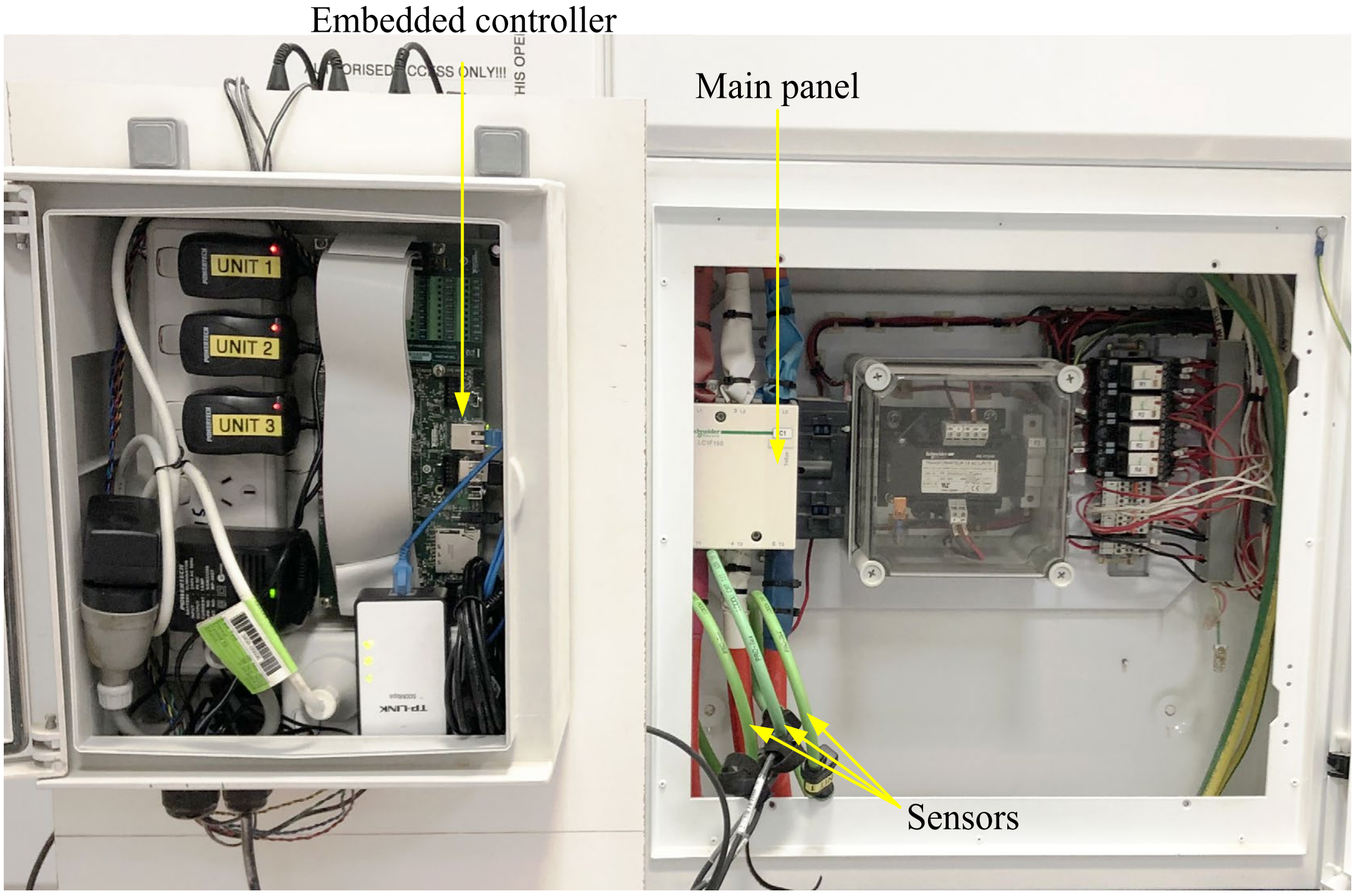

Fig. 5.8. Proposed load identification system

The data collection arrangement and the proposed load identification system are shown in Fig. 5.8. Three flexible AC current probes are used as sensors to measure the current signal from the main switchboard panel. The voltage waveforms are directly captured from the power socket. The input channels with 16 bits ADC resolution are used in the embedded controller to capture signals from the sensors at 10 kHz sampling frequency, according to the analysis results from section 5.3. The turn-on transients from power-hungry appliances and the low-power appliances are collected. Air conditioner, electric oven and electric heater are chosen from the power-hungry load types. Furthermore, fan, incandescent lamp, fluorescent lamp and laptop are considered from the low-power devices.

There are around 200 turn-on samples from each appliance collected for the experimental validation. The data collection is carried out at diverse times in different days, which includes the varying background noise from various devices and loading conditions. Altogether, 1400 turn-on samples are collected from seven different appliances and labelled with seven class labels, which creates a labelled dataset for supervised learning and offline validation.

### 5.7.2. Offline Validation

The experimental dataset is divided into three parts: (1) training dataset (70%) (2) validation data set (5%) and (3) testing dataset (25%). The training dataset is used to create the discriminative model using the proposed ANN. The discriminative approach is used for this load identification process since discriminative models often perform better on classification tasks compared to generative models, as discussed in chapter 3. Moreover, these models make the learning task easier and less computationally intensive since they indent to identify decision boundaries rather than learning the entire data distribution. The categorical cross-entropy loss function and adam optimisation function are used to adjust the network weights during the training based on the deviation between the network predictions and the target values. The validation dataset is used for an unbiased evaluation during the modelling process, which prevents the model from overfitting to the training dataset. Once the load identification model is completely trained using the train and validation sets, it can be evaluated with the unseen test dataset. The test results are reported in Fig.5.9. Based on the test results, the proposed load identification methodology achieves an average accuracy of 98% with the data set collected from different background loading and noisy environment. Besides, it is essential to analyse the real-time implementation feasibility of the proposed approach.

**Predicted labels**

| **True labels** | AC | Oven | Heater | Fan | Incan Lamp | Flouro Lamp | Laptop |
|---|---|---|---|---|---|---|---|
| AC | 98% | 2% | 0% | 0% | 0% | 0% | 0% |
| Oven | 4% | 96% | 0% | 0% | 0% | 0% | 0% |
| Heater | 0% | 0% | 100% | 0% | 0% | 0% | 0% |
| Fan | 0% | 0% | 0% | 98% | 0% | 0% | 2% |
| Incan Lamp | 0% | 0% | 0% | 2% | 98% | 0% | 0% |
| Flouro Lamp | 0% | 0% | 0% | 0% | 0% | 100% | 0% |
| Laptop | 0% | 0% | 2% | 2% | 0% | 0% | 96% |

Fig. 5.9. Load identification results

### 5.7.3. REAL-TIME IMPLEMENTATION AND VALIDATION

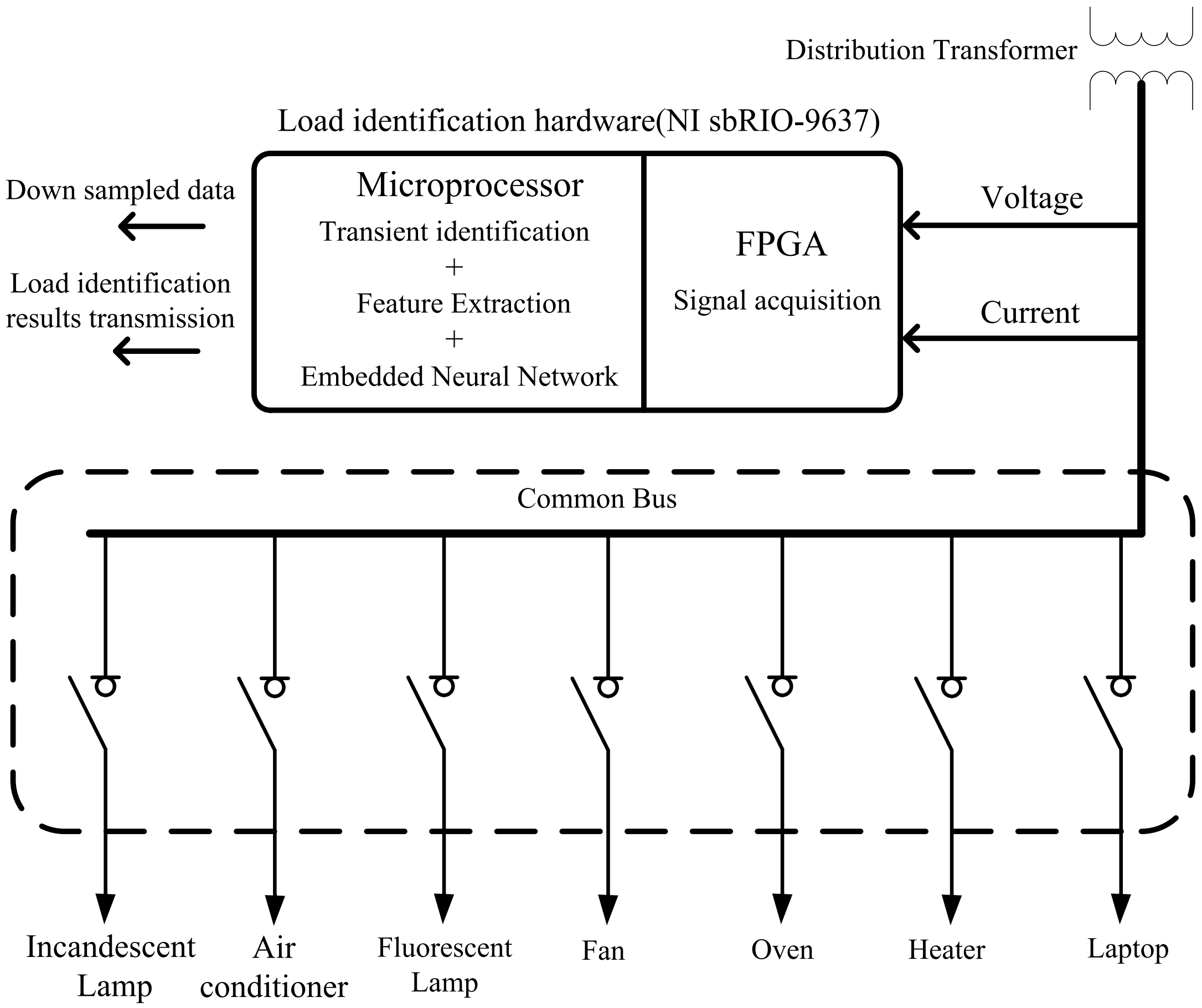

Fig. 5.10. Proposed algorithm deployment in embedded hareware

An embedded load identification system is developed using a NI single-board controller (NI sbRIO-9637) that contains a NI Linux based real-time microprocessor, reconfigurable FPGA, and I/O on a single printed circuit board, as shown in Fig 5.10. The proposed transient identification algorithm, feature extraction logic and the ANN structure are implemented in the microprocessor for the real-time execution. The trained ANN model parameters are stored in the non-volatile memory of the embedded hardware for the real-time inference.

During the real-time system execution, the current and voltage signals are sampled at 10 kHz and forwarded to the microprocessor through memory buffers. In the microprocessor, the transient detection module analyses the signal frame to find the abrupt changes. If any transients are detected, those signal fragments are forwarded to the feature extraction module. The proposed feature vector is extracted at real-time from the raw waveforms and transferred to the ANN classifier for the load classification. This real-time load identification results can be used to generate the consumer-specific feedback based on the live energy demand in a microgrid. On the other hand, the signals are downsampled to 1/6 Hz and transmitted to the utility providers along with the identified load information, for long-term energy usage pattern analysis and metering purpose.

After each signal frame processing, the raw signals are discarded to free up the memory since the embedded processing units have stringent constraints in memory usage. The continuous sampling of voltage and current signals at 10 kHz for a day generates around twelve gigabytes of data. The proposed embedded edge computing based processing approach manipulates this data stream and facilitates the high-frequency feature extraction and real-time load identification. After the data processing in the embedded edge unit, the output data size is significantly reduced to a few megabytes per day. Hence, this approach substantially reduces the communication bandwidth required for data transmission. There are no timing violations or memory overflows reported during the long-runs of the load identification unit. These validation results demonstrate the better performance and the real-time implementation feasibility of the proposed approach.

## 5.8. Chapter Summary

In this chapter, an embedded edge computing based architecture is proposed for real-time consumer load identification based on NILM concepts. It requires only a single point of voltage and current measurement from the main panel level of a residential site to identify consumer appliance usage. The impact of the input signal sampling rate and digitisation resolution are studied in the context of load identification, and the results are reported. Transient identification algorithms are presented, that can extract and isolate the transient segments in the input signal. A nine-dimensional feature vector is derived from the isolated transient signals for precise consumer load identification, and its computational steps are detailed. An ANN-based discriminative model is described, which can classify the loads based on the extracted feature vector.

The ANN-based model is trained with the turn-on transient features from seven different appliances such as air conditioner, electric oven, electric heater, fan, incandescent lamp, fluorescent lamp and laptop to assess the performance of the proposed approach. After the supervised training, the model is evaluated with the unseen turn-on transients from the dataset, and it infers the load identification results with an average accuracy of 98%. Finally, the proposed approach is implemented in embedded hardware and executed in real-time to ensure the viability of the proposed system. The work detailed in this chapter has led to the following publications [5][6].

Next chapter proposes a DNN-based energy disaggregation technique that can leverage the downsampled, low-resolution data from the smart meters to separate the device-specific energy consumptions.

# CHAPTER 6

# 6. DEEP NEURAL NETWORK BASED CONSUMER LOAD-SPECIFIC ENERGY DISAGGREGATION

## 6.1. INTRODUCTION

Load-specific energy disaggregation dramatically contributes to the electricity consumers to identify the significant energy consumers in their home. Energy enthusiastic consumers can pour over this energy disaggregation data to follow the appropriate demand-side management strategies, which leads to substantial energy savings and reduced energy bills. Research studies have shown that energy disaggregation results can be directly leveraged to generate personalised energy-saving recommendations for targeted consumers [1]. In the context of renewable energy microgrids, users can adjust their load usage patterns based on the real-time energy generation statistics, which not only reduces the energy demand in peak periods but also leads to optimal electricity tariff management. Hence, the near-real-time actionable recommendations have the potential to enhance the remote microgrid operations. More consumers can be attracted towards the optimal load management plan through energy breakdown estimates when the accuracy of the energy disaggregation algorithms is improved.

This chapter proposes a near-real-time energy disaggregation framework using DNN. It leverages the downsampled, aggregated energy usage data from the smart meters to estimate the load specific energy consumption patterns. The household appliances with significant energy consumption are chosen for this research since they have substantial potential in energy savings and demand-side management. The energy consumption pattern of each device is independently modelled using their load-specific intrinsic properties. Hence, a generative modelling approach is applied to solve the energy disaggregation problem. A novel DNN based generative architecture is developed for the energy disaggregation by combining CNN and Variational Auto-encoders (VAE). The proposed architecture is evaluated using a real-world dataset UK-DALE along with two standard error measures. The evaluation results have shown that the proposed system outperforms the state-of-the-art performance and shows relatively acceptable performance across different appliances.

## 6.2. PROPOSED ENERGY DISAGGREGATION APPROACH

The primary aim of energy disaggregation is to estimate the individual energy consumption of the power-hungry appliances in a consumer site from the aggregated low-frequency smart meter data. As the first step, the energy disaggregation problem is mathematically formulated, as described below.

### 6.2.1. PROBLEM FORMULATION

The aggregated reading, which is transmitted from the smart meter at time $t = \{1, 2 \ldots.., T\}$ in a consumer site can be represented as $X = \{X_1, X_2 \ldots \ldots., X_T\}$. The task of energy disaggregation algorithm is to approximate the energy contribution $y_t^i$ of an appliance $i \in \{1, 2, \ldots \ldots., N\}$ at time $t$, where $N$ is the number of devices considered for the energy disaggregation task. Hence, the aggregated energy reading at any point in time $t$ can be written as follows:

$$X_t = \sum_{i=1}^{N} y_t^i + \sigma(t) \tag{6.1}$$

where $\sigma(t)$ represents the noise and energy usage from the remaining appliances that are not considered for the energy disaggregation task, especially deficient power devices. The final goal of this energy disaggregation task is to develop an algorithm which produced optimised results while yielding a better generalisation across different consumer site appliances.

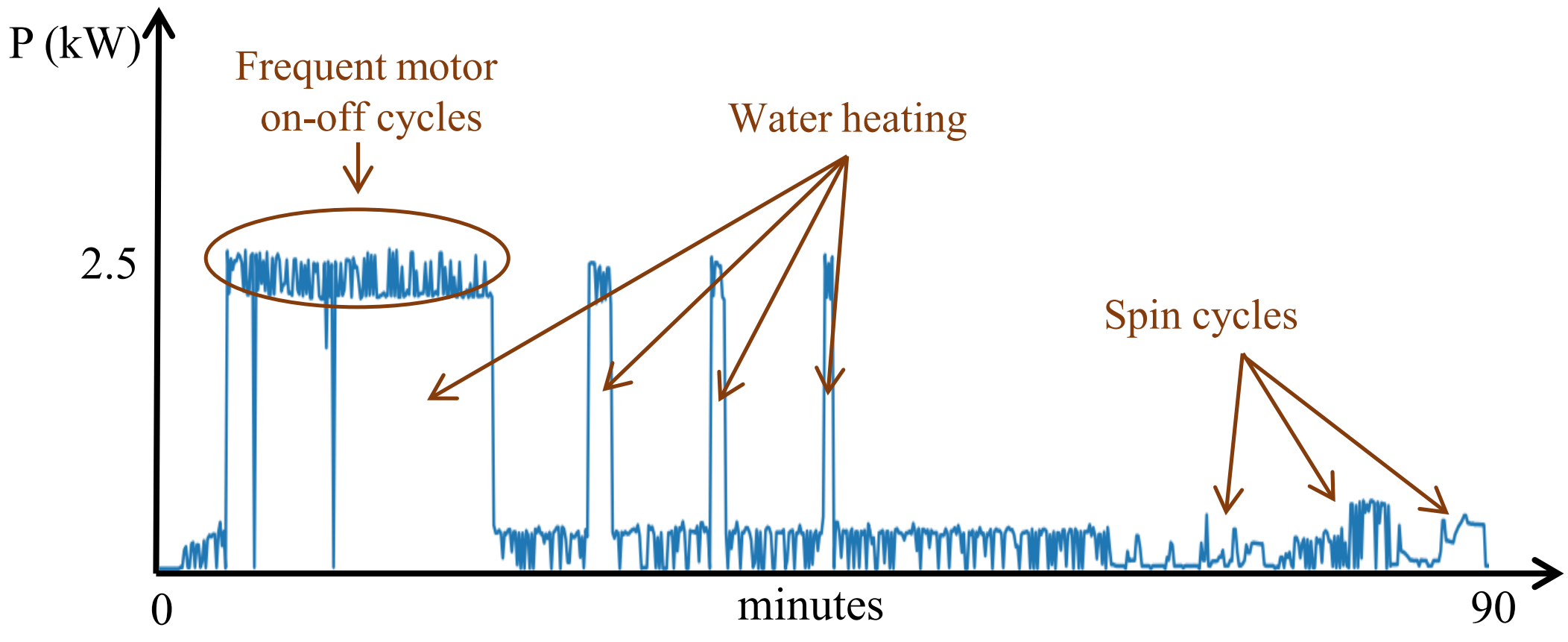

(a)

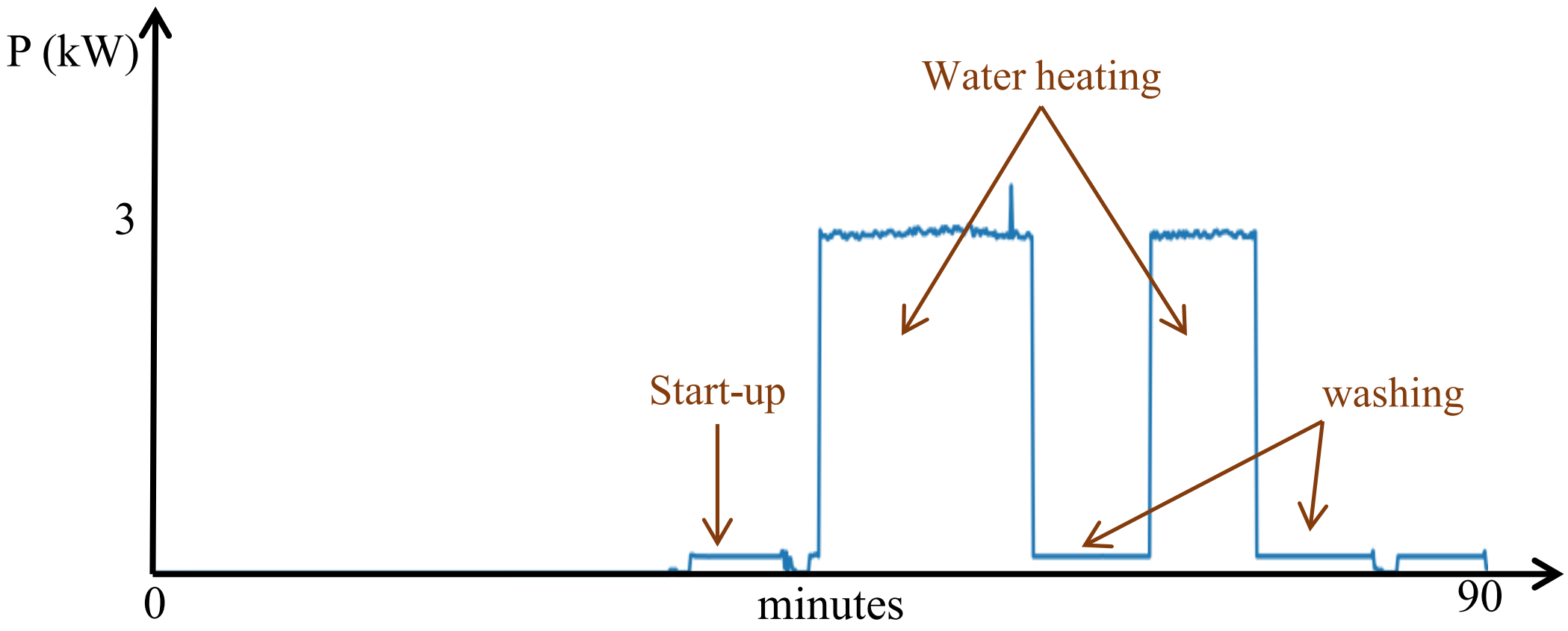

(b)

Fig. 6.1. Energy consumption of multi-state appliances (a) washing machine (b) dish washer

Every appliance contains unique characteristics in their energy consumption patterns. Typically, these energy usage patterns are determined by the number of states associated with the appliance and the switching frequency between those states. Fig. 6.1 visualises the state-specific variations in energy consumption of multi-state appliances. For instance, the water heating, frequent on-off cycles of the motor during washing/rinsing and multiple spinning cycles are the common observations for all washing machines. Hence, these state-specific energy variations can be leveraged to generate more generalised energy disaggregation model, that can estimate the energy consumption from different types of washing machines.

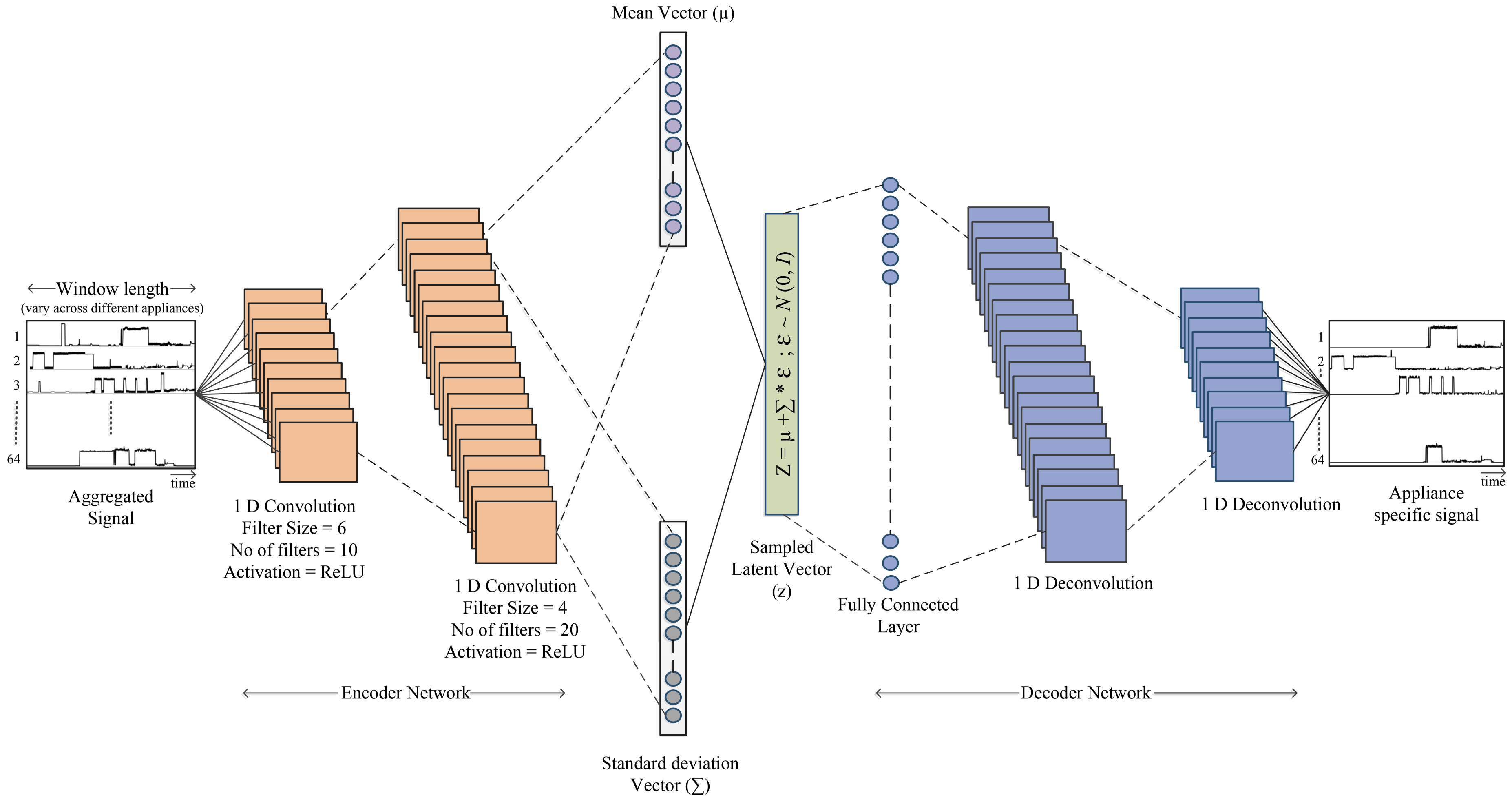

Fig. 6.2. Proposed end-to-end system architecture for energy disaggregation task

These appliance state information not only contributes towards the intra-class similarity (identify different washing machines) but also plays a vital role in inter-class discrimination (differentiate washing machines from other appliances). However, deriving a quantised set of state-specific features for a generalised appliance is a tedious task due to the numerous variations across different appliance brands and inconsistent duty cycles. In such contexts, hand-engineered feature extractors often fail to find an optimal feature set for appliance-specific energy estimation. Moreover, the manual feature extraction techniques are vulnerable to the background noise, which may mislead the disaggregation results. Hence, an end-to-end energy disaggregation framework is proposed, which automates the feature extraction from the raw signals and produces better energy disaggregation results. The internal structure of the proposed framework is detailed in the following sections.

### 6.2.2. SYSTEM ARCHITECTURE

A Convolutional Variational Autoencoder (CVAE) is proposed for energy disaggregation task, which is a combination of VAE and CNN. Fig. 6.2. outlines the internal structure of the proposed architecture. VAE contains a stochastic encoder and generative decoder. The stochastic encoder is a function $Q(z|X)$ which can take aggregated signal ($X$) and give a standard Gaussian distribution over D-dimensional stochastic variable $z$ that is likely to produce appliance specific signal:

$$Q(z|X) = N\left(z|\mu_1(X;\theta), \sum\nolimits_1(X;\theta)\right) \quad (6.2)$$

where $\mu_1, \sum_1$ are arbitrary deterministic functions of $X$ with parameters $\theta$ that can be learned from training data. In the proposed architecture, $\mu$ and $\sum$ are implemented using neural networks and $\sum$ is constrained to be a diagonal matrix in order to reduce the computational complexity. The entire network should be able to forward-pass as well as backpropagate to implement the encoder and decoder as a neural network. Even though the forward-pass of this network works fine, it cannot back-propagate the error through the layer of samples $z$ from $Q(z|X)$, which is a non-continuous operation that has no gradient.

Reparameterisation trick is applied to enable the back-propagation by moving the random sampling to an input layer. The encoding network $Q(z|X)$ roughly follows a normal distribution. Hence, it can be approximated with another normal distribution. Based on that, $z$ is re-parametrised with normally distributed $\varepsilon$ as follows:

$$z = \mu_1(X) + {\sum}_1(X)^{1/2} * \varepsilon \quad (6.3)$$

$$\varepsilon \sim N(0,1) \quad (6.4)$$

The decoder function $P(Y|z)$ which can take sample values of $z$, which represents the latent source of variability related to appliance specific signal $(Y)$, and compute $P(Y)$ just from those as follows:

$$P(Y|z) = N\left(Y|\mu_2(z;\varphi), {\sum}_2(z,\varphi)\right) \quad (6.5)$$

where $\mu_2, \sum_2$ are non-linear deterministic functions of $z$ and $\varphi$ denotes the model parameters.

The proposed encoder network contains two 1-D convolutional layers: the first layer applies ten filters of size 6 x 1; the second layer uses twenty filters of size 4 x 1, as shown in Fig. 6.2. The first convolutional layer filters are responsible for identifying features such as appliance turn-on and turnoff edges. On the other hand, higher layer convolutional filters focus even more abstract elements like the active duration, idle time and energy consumption pattern of an appliance. Rectified Linear Unit (ReLU) is used as the activation function for both convolutional layers in order to threshold the elements at zero.

Followed by the convolutional layers, two standard fully connected layers with no activation function are used to generate the mean ($\mu$) and standard deviation ($\sum$).

On the other hand, the decoder network contains a fully connected layer with ReLU activation function and two 1-D deconvolutional layers, also known as transposed convolutional layers. The purpose of having transposed convolutional layers is to progressively construct appliance specific signal by increasing the spatial size of the input while reducing the number of feature channels. The output of the last layer is a device-specific energy signal that is extracted from the aggregated energy signal with noise.

### 6.2.3. Loss function

An objective function needs to be formulated to train the proposed CVAE with the training data. It enables the network to learn the non-linear relationship between input and target signals. During the training process, the network aims to minimise the error based on the objective function. As such, it is often referred to as the loss function. A combined loss function is adapted to train the network for the energy disaggregation task, as shown in Fig. 6.3. It can be represented as follows:

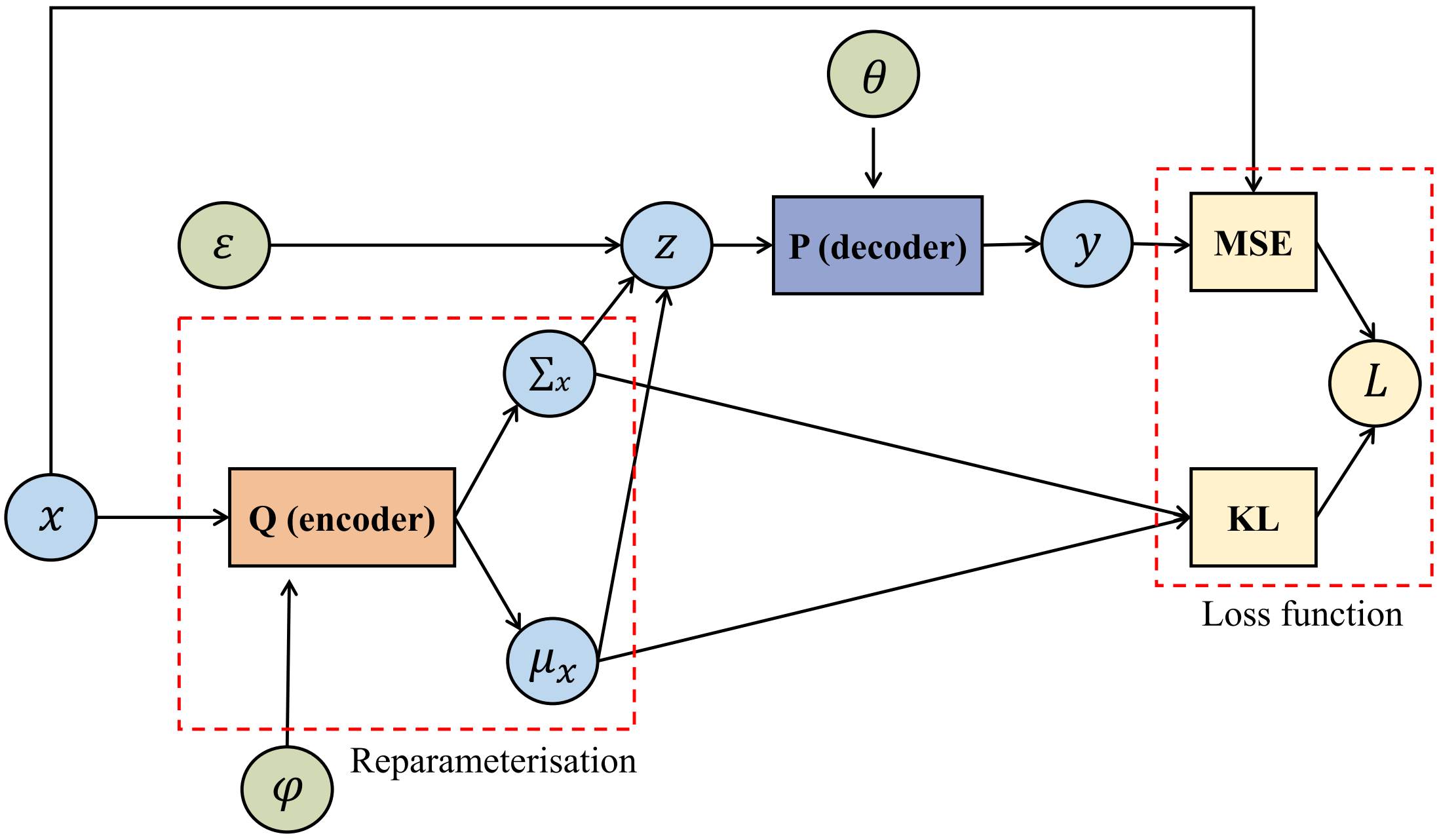

Fig. 6.3. Loss function of the proposed network

$$Loss\ function = estimation\ loss + \lambda * variational\ loss \quad (6.6)$$

$$estimation\ loss = \frac{1}{T}\sum_{t=1}^{T}\left(Y_t - \hat{Y}_t\right)^2 \quad (6.7)$$

$$variational\ loss = KL\left[N\left(\mu_1(X), \sum\nolimits_1(X)\right) || N(0, I)\right] \quad (6.8)$$

where $Y_t$ - the actual power of an appliance at time $t$, $\hat{Y}_t$ - estimated power of that appliance at time $t$, $\lambda$ - hyperparameter which controls the contribution of $KL$ term and $KL$ − Kullback-Leibler divergence which is used to measure the distance between statistical populations in terms of their measure of information. In the proposed loss function, the first term encourages proper estimation of appliance specific signal. Theoretically, $P(z)$ needs to be inferred from $P(z|X)$, which is the probability distribution that projects the input signal into latent space. However, $P(z|X)$ distribution is an unknown. Hence, a similar distribution $Q(z|X)$ is leveraged to estimate $P(z|X)$. Thus, the encoder network needs to learn the distribution $Q(z|X)$ such that it should be as close as possible to the actual distribution $P(z|X)$. Therefore, the variational loss term is added with the loss function, which captures the divergence between the encoder's distribution $Q(z|X)$ and standard Gaussian distributed prior $P(z)$. In other words, it is the measure of how close encoder's latent variables can match the unit Gaussian distribution.

In addition to that, the $KL$ term acts as a regulariser since it adds a constraint on the encoding network. If it is not included in the loss function, the encoder might cheat and represent different regions in Euclidean space for two signals that are produced by the same appliance. It leads to incorrect mappings such that two different brands of the washing machines can be mapped to the different regions of the latent space. As a consequence, the generalisation property of the model is affected. Hence, this learning behaviour is penalised by the KL term.

On the other hand, it is essential to note that optimising the $KL$ term may converge to a solution in which only a small subset of latent units is active. This issue is known as over-pruning, which results in a suboptimal generative model rather than an optimal solution. Moreover, over-pruning makes the model to underfit with respect to training data. Hence, several essential features are failed to be embedded in the generated model. In the proposed approach, over-pruning is controlled by weighting the KL term. A hyperparameter $\lambda$ is used in the aforementioned loss function in order to reduce the contribution of the $KL$ term. The optimal value of $\lambda$ is determined by the experiments.

## 6.3. Evaluation of Proposed Approach and Results

A real-world data set (UK-DALE) is used to evaluate the proposed approach. It contains long-term (655 days) energy consumption data from five different homes in the United Kingdom. As the first step, the consumer loads that significantly contributes to the total household energy consumption are chosen for the evaluation.

### 6.3.1. Choice of Consumer Loads

Based on the energy consumption statistics, five different consumer loads such as (1) kettle, (2) microwave, (3) fridge, (4) dishwasher, and (5) washing machine are considered for the experimental evaluation. Furthermore, each chosen device has different nature in its power consumption. For instance, the kettle's power consumption is more consistent (not many fluctuations) since it is a binary state (on/off) appliance. On the other hand, the multi-state appliance such as fridge, washing machine and dishwasher generate more complex power consumption signatures. Hence, the proposed model can be validated for all types of power signatures.

### 6.3.2. Data Preparation

An open-source toolkit NILMTK [2] is used to prepare the data for the proposed network training, validation and testing. In the UK-DALE dataset, a device activation is defined as the energy consumption of a single device over one complete cycle of that load. As the initial step, the activations of the chosen appliances are extracted from the dataset. Since the activation frame length of each device is different, variable window length is used to capture the complete activation cycle. For instance, 10 - 15 minutes of frame length is enough to capture an activation of a kettle, whereas few hours of frame length is required for dishwasher and washing machine. On the other hand, previous studies have explored that the large frame size significantly affects the disaggregation performance, especially for the devices with short-activation time [3]. Thus, the input signal is framed based on the worst-case activation duration of a load.

An extensive set of training data need to be prepared to train the proposed network. Since the supervised learning approach is followed during the energy consumption pattern modelling, a collection of input and output data (ground truth) is required. In this energy disaggregation context, the input data is the aggregated energy demand from a smart meter, and the ground truth is device-specific energy consumption. The UK-DALE data set contains real aggregate energy usages from main panel level of the selected houses and the sub-metered device-specific power readings sampled at 1/6 Hz. However, these data samples are not adequate to train the proposed network. As a general practice in deep learning, realistic domain-specific transformations are applied to the existing data in order to maximise the dataset, and this process is known as data augmentation.

Data augmentation is relatively easy in the context of energy disaggregation since a massive amount of aggregate input data can be created by randomly superimposing different device activations. The real activations of different devices can be shifted in time-domain and concatenated together to create an extensive input data set along with the ground-truth. While preparing the data set for a model, the target device signature is included only for 50% of the input. The remaining 50% of input signals are randomly chosen from the aggregate window, which does not contain any activations of the target device. Its ground-truths are the vectors of zeros that denotes the absence of the target device. After the dataset preparation, the proposed approach needs to be implemented and trained for the energy disaggregation task.

### 6.3.3. Network Implementation and Training

The proposed system is implemented in Python using Tensorflow, which is an end-to-end open-source machine learning platform. Since a generative modelling approach is proposed, each device-specific energy consumption is modelled with separate networks. Since the UK-DALE data set contains the energy data from five different houses, the signals from house 1, 3, 4, and 5 are used for the model training. An unseen set of data from various brands of the same appliance type is required to test the generalisation property of the model. Thus, the data from house two is reserved for model testing. The input data is sent as a mini-batch of 64 input sequence to train each network since mini-batch learning supports the rapid convergence of the proposed model. During the training, real data and augmented data are used in a 1:1 ratio to improve the learnability and generalisation of the network. Furthermore, each input sequence is normalised to zero mean and unit variance, that can bring all the input sequence into the same range. This enforces the network to learn patterns rather than the scales.

For a DNN-based network, hyperparameter tuning plays an essential role in the estimation results. The network hyperparameters such as convolutional filter size, filter number and $\lambda$ are decided based on the grid search. The grid search builds a model for each combination of the specified hyperparameters and evaluates each model to find the optimal values. Based on the grid search results, the convolutional filter numbers are decided as 10 and 20 in layer one and layer two, respectively. The 1-D filter sizes are determined as 4 and 6 in the first layer and second layer, respectively. Finally, the weight of the KL term $\lambda$ is set to 0.1.

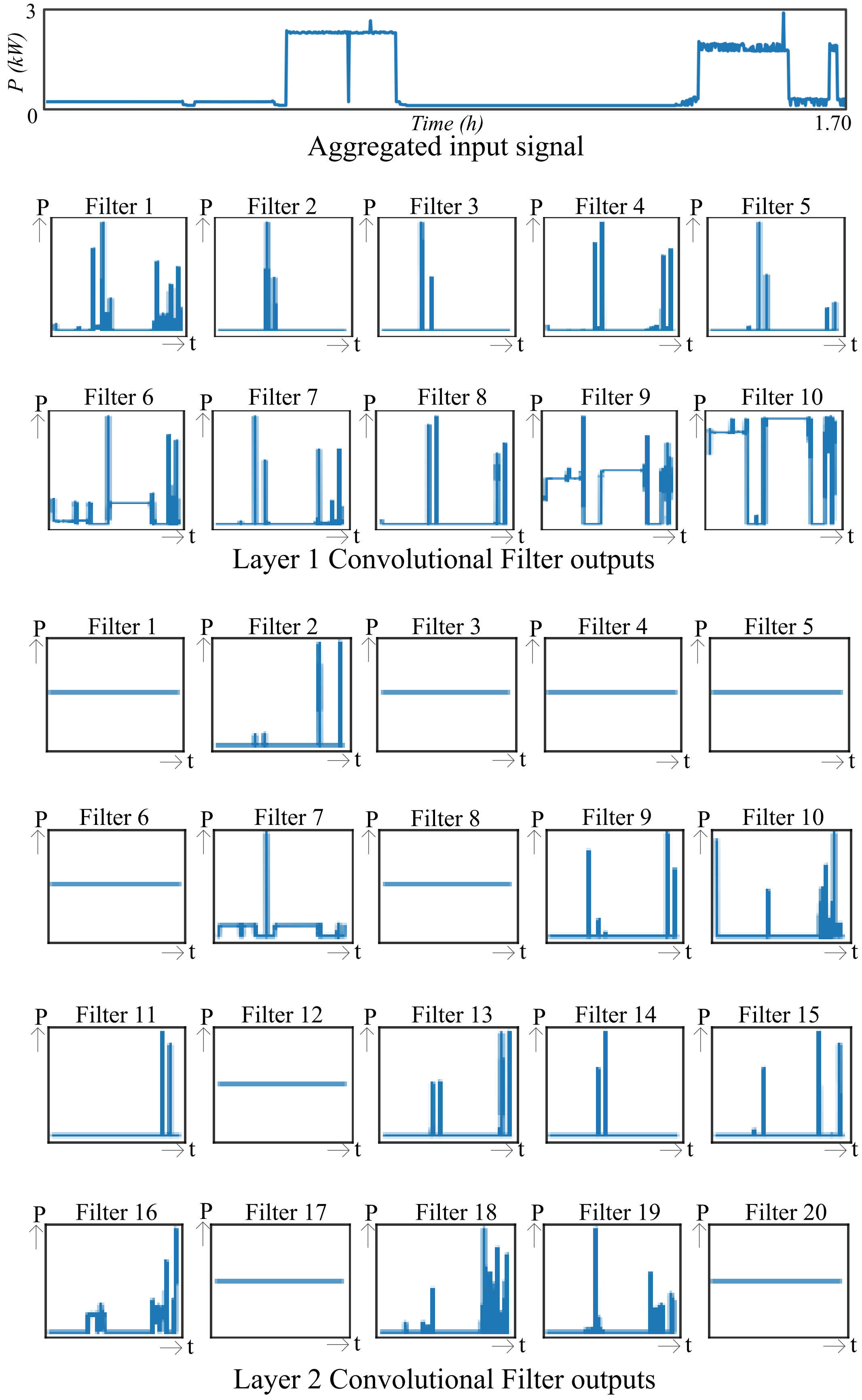

Fig. 6.4. Automated feature extraction using convolutional filters

During the training process, the outputs from the intermediate convolutional filters are visualised to demonstrate the automated feature learning of the proposed network. An aggregate input signal and its intermediate convolutional filter outputs are visualised in Fig. 6.4. The filters in the first convolutional layer capture the transient components such as turn-on and state-changes. On the other hand, the second layer filter visualisation demonstrates more abstract features like state-duration and idle time.

### 6.3.4. Error Measures and Comparison Results

In the energy separation domain, there are two standard error measures such as (1) mean absolute error (MAE) and (2) signal aggregate error (SAE) to evaluate disaggregation performance. The MAE of a specific appliance $i$ can be defined as follows:

$$MAE_i = \frac{1}{T}\sum_{t=1}^{T} \left| y_t^{(i)} - \widehat{y_t^{(i)}} \right| \tag{6.9}$$

where $y_t^{(i)}$ – the actual power of appliance $i$ at time $t$ and $\widehat{y_t^{(i)}}$ – estimated power of appliance $i$ at time $t$. The SAE of a specific appliance $i$ can be mathematically represented as follows:

$$SAE_i = \frac{\left| E_i - \hat{E}_i \right|}{E_i} \tag{6.10}$$

where $E_i$ – total actual energy consumption of appliance $i$ and $\hat{E}_i$ – total predicted energy consumption of appliance $i$.

The proposed approach is compared with other state-of-the-art techniques such as DNN-AFHMM [4], DAE[3], sequence-to-sequence (Seq2seq) [5] learning and sequence-to-point (Seq2point) [5] learning, which achieve better results in the energy disaggregation literature. The detailed descriptions of these state-of-the-art techniques are summarised in chapter 2. Table. 6.1 and Table. 6.2 report the comparison of experimental results based on the error measures mentioned above. Based on the comparison results, the proposed system improves the SAE by 44% and MAE by 19%.

Table. 6.1: Appliance-level mean absolute error (Watt) comparison

| **Methods** | **Kettle** | **Microwave** | **Fridge** | **Dish Washer** | **Washing Machine** | **Overall** |
|---|---|---|---|---|---|---|
| DNN-AFHMM | 47.4 | 21.2 | 42.4 | 199.8 | 103.2 | 82.8 ± 64.5 |
| DAE | 13.0 | 14.6 | 38.5 | 238.0 | 163.5 | 93.5 ± 91.1 |
| Seq2seq | 9.2 | 13.6 | 24.5 | 32.5 | **10.2** | 18.0 ± 9.1 |
| Seq2seq | 7.4 | 8.7 | 20.9 | 27.7 | 12.7 | 15.5 ± 7.7 |
| Proposed approach | **7.0** | **7.5** | **18.1** | **19.6** | 10.8 | **12.6 ± 5.3** |

Table. 6.2: Appliance-level signal aggregate error comparison

| **Methods** | **Kettle** | **Microwave** | **Fridge** | **Dish Washer** | **Washing Machine** | **Overall** |
|---|---|---|---|---|---|---|
| DNN-AFHMM | 1.06 | 1.04 | 0.98 | 4.5 | 8.28 | 3.17 ± 2.88 |
| DAE | 0.085 | 1.348 | 0.502 | 4.237 | 13.831 | 4.001 ± 5.124 |
| Seq2seq | 0.309 | 0.205 | 0.373 | 0.779 | 0.453 | 0.423 ± 0.194 |
| Seq2seq | 0.069 | 0.486 | **0.121** | 0.645 | 0.284 | 0.321 ± 0.217 |
| Proposed approach | **0.063** | **0.181** | 0.132 | **0.317** | **0.206** | **0.18 ± 0.084** |

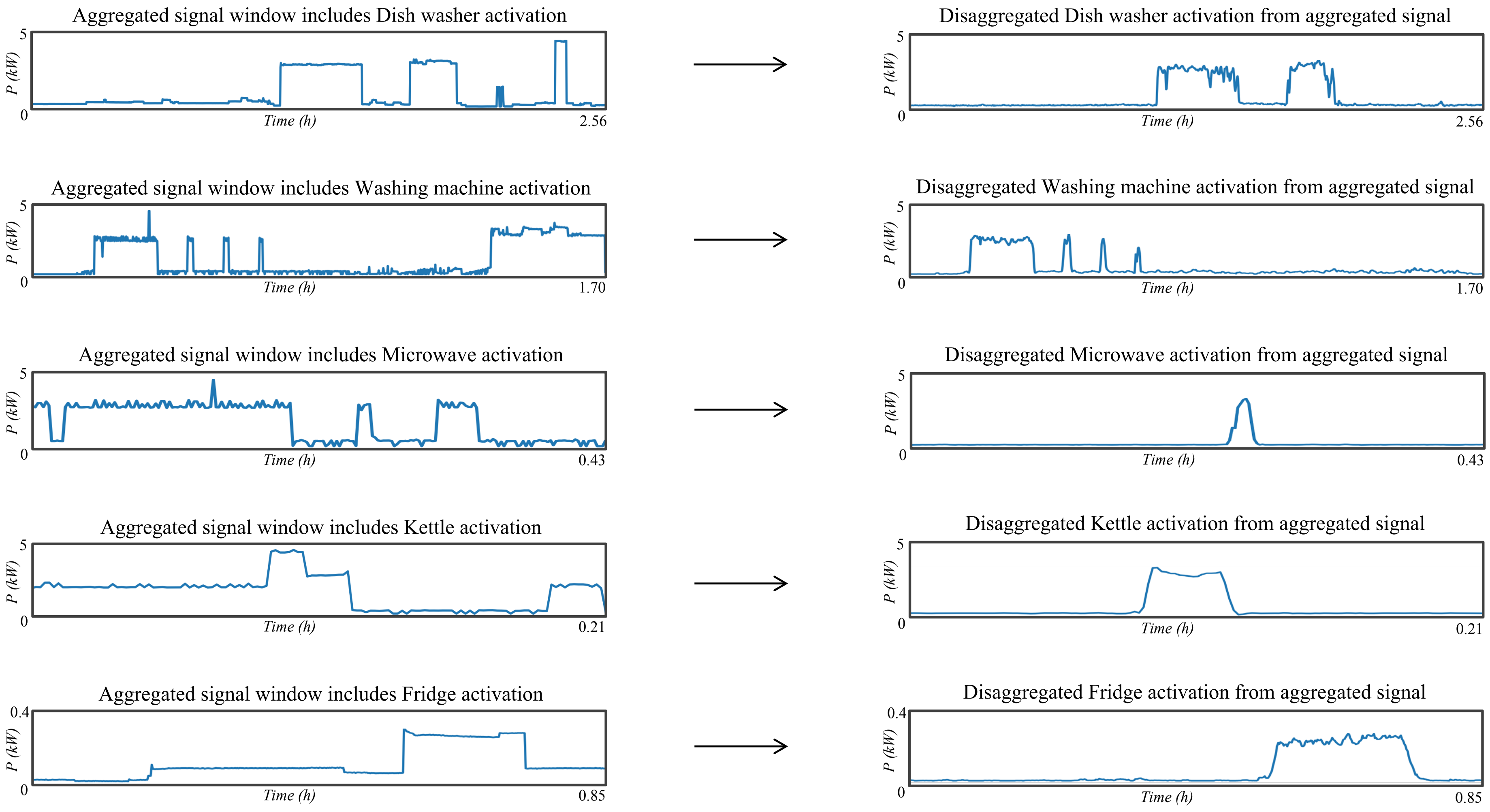

Fig. 6.5. Energy disaggregation results for different appliances

Fig. 6.5 visualises the disaggregated results produced by the proposed approach, along with the corresponding aggregate input signal. The input aggregated signals used in this evaluation are sampled at a shallow rate (1/6 Hz). Thus, the downsampled data stream from the smart meter data processing module proposed in chapter 5 can be used for the energy disaggregation along with the identified load labels. The low-frequency data transmission only consumes a few MBs of data for each house. Hence, the proposed algorithm can be deployed in the cloud infrastructures, which gives greater scalability and reusability. Furthermore, it enables the near-real-time load-specific energy disaggregation. The evaluation results and the output visualisations demonstrate the energy disaggregation capability of the proposed framework.

## 6.4. Chapter Summary

This chapter proposes a DNN based energy disaggregation framework to estimate the appliance specific energy consumption in near-real-time. A CVAE architecture is developed by combining CNN with VAE, which includes the stochastic encoder and generative decoder. A KL divergence based combined loss function is formulated that can account the energy estimation loss as well as the model variation loss. It enforces the network to learn more generalised characteristics which are typical for different appliance brands. The generative modelling approach is applied to model the device-specific energy usage with separate networks. It increases the scalability of the energy disaggregation task when the new devices are introduced.

The proposed framework is evaluated with real-world dataset (UK-DALE). Appliances with significant energy usage such as kettle, microwave, fridge, dishwasher and washing machine are considered for the model evaluation. Based on the evaluation results, the proposed method improves the SAE by 44% and MAE by 19% in comparison with the state-of-the-art techniques. The work detailed in this chapter has led to the following publication [6].

The next chapter concludes this thesis with the conclusions and future research directions.

# CHAPTER 7

# 7. CONCLUSIONS AND FUTURE WORKS

## 7.1. Conclusions

The encompassing goals of this thesis are to develop AI-based condition monitoring algorithms for remote electricity networks and optimise those algorithms for real-time execution on resource-constrained embedded hardware such as pole-mounted monitoring units and smart meters. This work has addressed three novel contributions in the context of electricity network condition monitoring: (1) A distributed online monitoring platform that can monitor the power quality, detect HIF and classify transients in real-time, (2) A consumer load identification methodology to identify the load type from its turn-on transients and (3) A deep neural network based energy disaggregation framework to separate the load specific energy usage from an aggregated smart meter data. The potential importance behind this research is to reduce the potential risks associated with the existing SWER networks and alleviate the increasing energy demand from remote consumers with the optimised energy management of renewable energy microgrids.

The research aims of this thesis were attained in a series of steps, starting with an investigative analysis of the challenges in remote electricity networks and its mitigation techniques, followed by an in-depth review of existing monitoring solutions for remote electricity networks in chapter 2.

Chapter 2 outlines the operating conditions and internal structure of the SWER network along with its advantages in the context of rural electrification. Furthermore, it describes the primary shortcomings of existing SWER lines such as (1) power quality deficiencies, (2) bushfire risks due to downed conductors or live conductor contact with overgrown tree branches, and (3) insufficient current-carrying capacity to address the increasing energy demand from rural areas. It also highlights the possible mitigation techniques such as (1) online power quality monitoring - to detect the power quality issues on time and enable proactive maintenance of power system equipment, (2) real-time HIF identification – to isolate the faulty region before the high-impedance contact ignite the fire, (3) consumer load identification, and (4) consumer load specific energy disaggregation – to address the increasing energy demand by optimal demand-side management with the renewable energy microgrids in rural areas.

The implementation process of the mitigation steps is broadly divided into three stages, such as (1) data acquisition, (2) feature extraction and (3) decision making. Digital signal processing and AI-based techniques are identified as the key domains, which have a massive potential in the enhancement of condition monitoring steps mentioned above.

Chapter 3 provides a detailed description of the digital signal processing techniques and AI-based approaches, which are used in this thesis. It also includes theoretical concepts of signal processing algorithms and AI-based modelling techniques. Furthermore, the application steps of those techniques to the condition monitoring use cases such as faulty signal analysis are outlined. Real-time implementation requirements of the condition monitoring applications are studied, and the limitations with existing monitoring solutions are identified.

It has been realised that most of the previous research works mainly focus on the accuracy of the algorithms and fail to do a feasibility analysis of the proposed techniques against these application-specific constraints. Consequently, most of the existing algorithms are far away from practical applications. Hence, this thesis aims to address application-specific limitations while improving the accuracy of the algorithms.

Chapter 4 introduces a distributed online monitoring platform for overhead powerline monitoring, which facilitates the real-time power quality monitoring, HIF identification and transient classification. The threshold-based detection technique is identified as a computationally efficient solution for power quality disturbance detection. However, the threshold-based approaches are not suitable for the faults with complex characteristics such as HIF. Thus, AI-based method is developed for real-time HIF identification. Signal processing techniques such as WPT and FFT are used to extract features for HIF identification. The real-time implementation of 3-level decomposition of WPT is failed to meet timing requirements on embedded hardware due to its computational complexity of 3-level decomposition of WPT. Hence, the short-time FFT based feature is used, that can provide better discrimination and supports the real-time execution.

A light-weight CNN is employed as a classifier to discriminate the HIF and classify the transients. Optimisation techniques such as parallelism, pipelining, and timed-loop executions are adapted to implement the proposed structure and meet the real-time requirement of time-sensitive monitoring applications such as HIF identification. The proposed framework is experimentally validated, and the results demonstrate the HIF identification accuracy of 98.67% with the worst-case HIF detection latency of 115.2ms. Furthermore, extensive validation is carried out with the data collected from an industrial high-power testing station, and the proposed framework achieves 98.48%, 99.4% and 100% accuracy on identifying HIF, typical transient and normal condition, respectively.

Chapter 5 details intelligent edge analytics architecture for consumer load identification. The impact of input signal sampling frequency and digitisation resolution on load identification are investigated in the context of load identification. Based on the analysis results, 10 kHz sampling frequency and 16-bits of digitisation resolution are chosen for the proposed methodology. Turn-on transients of high-power appliances are extracted using a time-domain based empirical estimate, which is a computationally efficient process compared to the frequency domain methods. However, it fails to retrieve the switching transients of low-power devices. Hence, first-level wavelet decomposition is applied to isolate the transient states of low-power devices, that can separate the steady-states from the transient-states. After the transient isolation, a nine-dimensional feature vector is derived for load identification. A light-weight ANN is formulated to classify the loads from the extracted features. The developed architecture is implemented in an embedded controller and evaluated in real-time. Furthermore, the algorithm performance is verified with seven different consumer loads, including high-power and low-power devices. The experimental results demonstrate an average accuracy of 98% in the load identification task.

Furthermore, the proposed edge analytics architecture not only facilitates real-time on-site load identification on smart meters but also enables the near-real-time energy disaggregation, long term energy demand analysis and demand forecasting applications on the cloud infrastructure via transmitting downsampled data to the cloud. This approach significantly reduces the required communication bandwidth.

Chapter 6 describes the DNN-based consumer load specific energy disaggregation framework. A generative modelling approach is used to model the energy consumption pattern of a consumer load. The power signals are analysed, and the relationships between the energy consumption patterns and the appliance state changes are studied. An automated feature extraction approach is implemented to eliminate the deficiencies in the hand-engineered features. The CVAE network that contains a stochastic encoder and a generative decoder is employed to model the energy consumption. Stochastic encoder maps the aggregated power consumption to a latent space via discarding the irrelevant information. The generative decoder reconstructs the appliance specific energy consumption from the latent space representation. A loss function, which is a combination of estimation loss and variational loss, is used to train the model. This loss function not only enforces the encoder to learn the generalised properties but also indicates the signal reconstruction error from the decoder. The proposed framework is evaluated with the energy consumption data of five different consumer loads from a real-world data set (UK-DALE) and compared with the state-of-the-art techniques. Based on the evaluation results, the developed framework improves the standard measures SAE and MAE by 44% and 19%, respectively.

In summary, this thesis envisions that advancements in condition monitoring applications such as online power quality monitoring, real-time HIF identification and transient classification, consumer load identification and load-specific energy separation can significantly enhance the operation of existing SWER networks and the energy management of rural microgrids. Thus, the development of online condition monitoring algorithms for electricity networks is an active area of ongoing research. In this context, the novel contributions to the condition monitoring applications are documented in this thesis that can enhance the application accuracy and real-time execution on resource-constrained hardware. Furthermore, the documented approaches are experimentally validated and compared with the state-of-the-art techniques. The validation results demonstrate that the proposed methods are up-and-coming to real-world practical applications.

## 7.2. FUTURE WORKS

The research outlined in this thesis involved the development of AI-based sensor data analytics framework for the online condition monitoring of remote electricity networks. Embedded monitoring hardware is built as a proof of concept which consists of a high-resolution DAQ unit, FPGA for high-speed and hugely parallel signal processing, embedded microprocessor, GPS receiver for time synchronisation and a long-distance communication module. It facilitates the real-time execution of time-sensitive applications such as HIF detection with minimised delay. Furthermore, it supports the hierarchical data analytics structure via transmitting the high-frequency, information-rich features to the substation or cloud infrastructure with reduced transmission costs. The long-term, non-time-sensitive applications can be deployed in the cloud with greater scalability. In this context, the following could be some directions for future research.

The transmitted high-resolution features from the embedded monitoring unit can be leveraged in the higher-level applications such as historical energy demand analysis, pattern mining, seasonal load forecasting and equipment degradation analysis. Since these applications are not much time-sensitive, more complex AI-models can be developed to improve accuracy.

Similar to HIF detection, there are several time-sensitive applications in the power system analysis and protection domain. These applications can utilise this embedded monitoring platform. Furthermore, each of these applications has different time constraints on algorithm execution timings. Hence, more light-weight signal processing and AI techniques can be developed to support such use cases.

The load identification approach discussed in chapter 5 requires 10 kHz of input sampling frequency and 16-bits of digitisation resolution. Development of load identification methodologies with low-resolution signals will decrease the burden of smart meters. Also, there is a possibility of accuracy drop with the increased number of target loads since a discriminative approach is used to classify the load types. It is beneficial to develop more scalable load identification methodologies that can be deployed inside the smart meters.

The load specific energy disaggregation approach detailed in chapter 6 leverages a DNN structure to improve accuracy via automated feature extraction and modelling with KL divergence based combined loss function. There are plenty of DNN architectures and variational loss functions that demonstrates better performance on different domains such as image classification, speech recognition and language identification. It is worthwhile to investigate those architectures and techniques to enhance the state-of-the-art energy disaggregation.